\documentstyle[prb,aps,preprint,eqsecnum]{revtex}
\begin{document}
\tighten
\draft

\title
{SUPERCONDUCTIVITY AND LOCALIZATION}

\author{Michael V.\ Sadovskii}

\address
{Institute for Electrophysics, Russian Academy of Sciences, Ural Branch,\\
Ekaterinburg, 620049, Russia}
\maketitle

\begin{abstract}
We present a review of theoretical and experimental works
on the problem of mutual interplay of Anderson localization and
superconductivity in strongly disordered
systems.\ Superconductivity exists close to the metal---insulator transition in
some disordered systems such as
amorphous metals,\ superconducting compounds disordered by fast neutron
irradiation etc.\ High-temperature
superconductors are especially interesting from this
point of view.\ Only bulk systems are considered in this review.\ The
superconductor-insulator transition in purely two-dimensional disordered
systems is not discussed.

We start with brief discussion of modern aspects of localization theory
including
the basic concept of scaling,\ self---consistent theory and interaction
effects.\
After that we analyze disorder effects on Cooper pairing and superconducting
transition
temperature as well as Ginzburg---Landau equations for superconductors which
are close
to the Anderson transition.\ A necessary generalization of usual theory of
``dirty'' superconductors is formulated which allows to analyze anomalies of
the main superconducting properties close to disorder-induced metal---insulator
transition.\ Under very rigid conditions superconductivity may persist even in
the localized phase (Anderson insulator).

Strong disordering leads to considerable reduction of superconducting
transition
temperature $T_{c}$ and to important anomalies in the behavior of the
upper critical field $H_{c2}$.\ Fluctuation effects are also discussed.\
In the vicinity of Anderson transition inhomogeneous superconductivity appears
due to statistical fluctuations of the local density of states.

We briefly discuss a number of experiments demonstrating superconductivity
close to the Anderson transition both in traditional and high---$T_{c}$
superconductors.\ In traditional systems superconductivity is in most cases
destroyed before metal---insulator transition.\ In case of high---$T_{c}$
superconductors a number of anomalies show that superconductivity is apparently
conserved in the localized phase before it is suppressed by strong enough
disorder.
\end{abstract}
\pacs{PACS numbers 74, 72.15.Rn}

\newpage
\tableofcontents

\newpage
\narrowtext
\section{INTRODUCTION}
The concept of electron localization \cite{And58} is basic for the
understanding of electron properties of
disordered systems \cite{Mott74,MottDav}.\ In recent
years a number of review papers had appeared,\ extensively discussing this
problem \cite{Sad81,NagFuk,LeeRam,Sad86}.\ According to this concept
introduction of
sufficiently strong disorder into a metallic system leads to spatial
localization of electronic states near the Fermi level and thus to a transition
to dielectric state (Anderson transition).\ After this transition dc
conductivity (at zero temperature,\ $T=0$) vanishes,\ despite the finite value
of
electronic density of states at the Fermi level (at least in one-electron
approximation).

At the same time it is well-known that even the smallest attraction of
electrons close to the Fermi level leads to
formation of Cooper pairs and the system becomes superconducting at
sufficiently low temperatures \cite{BCS,Genn}.\ It is known that the
introduction of disorder which does not break the time---reversal invariance
(normal,\ nonmagnetic impurities etc.) does not seriously influence
superconding transition temperature $T_{c}$ and superconductivity in general
(Anderson theorem) \cite{AG58,AG59,Gor59,And59}.

Thus a problem appears of the mutual interplay of these two possible electronic
transitions in a disordered system which
leads to quite different (even opposite) ground states
(insulator or superconductor).\ This problem is very important both from
theoretical and experimental points of
view.\ Actually superconducting properties of many compounds
depend strongly on structural disorder.\ In this respect we can mention
amorphous systems (metallic glasses) and
superconductors disordered by different forms of irradiation
by high-energy particles (fast neutrons,\ electrons,\
heavy-ions etc.).\ It appears that in many of these systems
superconductivity is realized when the system in normal
state is quite close to the metal---insulator transition
induced by disorder.\ In this case many anomalies of
superconducting properties appear which cannot be satisfactorily explained
within the standard theory of
``dirty'' superconductors
\cite{Genn,AG58,AG59,Gor59,And59}.\ These include rather
strong dependence of $T_{c}$ on disorder in apparent
contradiction with Anderson theorem,\ as well as some
unusual behavior of the upper critical field $H_{c2}$.

The discovery of high-temperature superconductivity in metallic oxides
\cite{BM86,BM88} has lead to the entirely
new opportunities in the studies of strong disorder effects in
superconductors.\ Very soon it had been
established that high---$T_{c}$ superconductors are
quite sensitive to structural disordering which leads to
rather
fast destruction of superconductivity and metal---insulator transition.\
However,\ the high values of
initial $T_{c}$,\ as well as a small size of Cooper
pairs and quasi-two dimensional nature of electronic
states in these systems are very appropriate for the
studies of the mutual interplay of localization and
superconductivity \cite{Sad89}.\ It may be stated with
some confidence that in these systems superconductivity
can be observed even in the region of localization (Anderson insulator).

This review is mainly concerned with theoretical aspects
of localization and superconductivity close to Anderson
transition.\ However,\ we shall pay some attention to
experiments demonstrating the importance of localization
phenomena for the correct analysis of superconductivity
in strongly disordered systems.\ Special emphasis will
be on the experiments with high---$T_{c}$ superconductors.\ We shall limit
ourselves with discussing
only three-dimensional and quasi-two-dimensional (in case
of HTSC) systems,\ practically excluding any discussion of purely
two-dimensional systems,\ which are quite special both
in respect to localization and superconductivity.\ In
this case we refer a reader to recent reviews \cite{Ramak,Belev,LiuGold}
which are specifically concerned with two-dimensional case.

We must stress that the material presented in this review
is concerned mainly with the personal interests of its
author and we apologize to those people whose important
contributions in this field would not be discussed in
detail or even would be missed because of the lack of space.

The usual theory of ``dirty superconductors'' \cite{Genn,AG58,AG59,Gor59,And59}
is a cornerstone of our understanding of superconducting properties of
disordered metals.\ It is based on the following main statements:
\begin{enumerate}
\item As impurity concentration (disorder) grows a
transition takes place from the ``pure'' limit,\ when
the electron mean-free path $l$ is much larger than the
superconducting coherence length $\xi_{0}$: $l\gg\xi_{0}=
\hbar v_{F}/\pi\Delta_{0}$ to a ``dirty'' superconductor
with $\xi_{0}\gg l\gg \hbar/p_{F}$ (Here $v_{F},p_{F}$---are Fermi velocity and
momentum,\ $\Delta_{0}$---is the zero---temperature energy gap).\
Transition temperature $T_{c}$ change only slightly,\ mainly due to
small changes of Debye frequency $\omega_{D}$ and of
pairing constant $\lambda_{p}$,\ which are due to relatively small changes
in the electronic density of states under disordering.\
Transition from the free electron motion to diffusive
one does not change $T_{c}$ at all (Anderson's theorem).\
These statements ignore any disorder dependence of microscopic pairing
interaction,\ which is assumed to
be some constant as in the simplest BCS model.
\item Superconducting coherence length $\xi$ (at $T=0$)
determining the spatial scale of superconducting order-parameter (the size of a
Cooper pair) diminishes with
$l$ so that $\xi\approx \sqrt{\xi_{0}l}$ in the limit
of $\hbar/p_{F}\ll l \ll \xi_{0}$.
\item As $\xi$ diminishes the critical region near
$T_{c}$ where thermodynamic fluctuations are important
widens and is of the order of $\tau_{G}T_{c}$,\ where
$\tau_{G}\sim [T_{c}N(E_{F})\xi^{3}]^{-2}$ is the so called Ginzburg's
parameter
($N(E_{F})$ is electronic density of states at the Fermi level $E_{F}$).\
For ``pure'' superconductors $\tau_{G}\sim (T_{c}/E_{F})^{4}\ll 1$
and as $l$ drops $\tau_{G}$ grows as $\xi$ drops.\
However,\ in the limit of $l\gg \hbar/p_{F}$ the value
of $\tau_{G}$ still remains very small.
\end{enumerate}
Theory of ``dirty'' superconductors is the basis of our understanding of
superconducting properties of many disordered alloys.\ However,\ the main
results  of this theory must be modified for the
mean-free path values  $l$ of the order of inverse Fermi momentum $\hbar/p_{F}$
(i.e. \ of the
order of interatomic distance).\  In three---dimensional systems the growth of
disorder leads to
destruction of diffusive motion of electrons and transition from extended to
localized states at
critical disorder determined by $l_{c}\approx \hbar/p_{F}$,\ i.e.\ to
transition to Anderson insulator.\
This metal---insulator transition is reflected in a continuous drop to zero of
the static metallic
conductivity (at $T=0$) as $l\rightarrow l_{c}$.\ For $l\gg l_{c}$ conductivity
is determined by
the usual Drude formula $\sigma_{0}\sim l$,\ while for $l\rightarrow l_{c}$ it
drops as
$\sigma\sim (l-l_{c})^{\nu}$,\ where $\nu$ is some critical exponent.\
Transition from diffusion
to localization is realized at the conductivity scale of the order of the
so-called ``minimal
metallic conductivity'' $\sigma_{c}\approx (e^{2}p_{F}/\pi^{3}\hbar^{2})\approx
(2-5) 10^2
Ohm^{-1}cm^{-1}$ .\  The usual theory of ``dirty'' superconductors does not
consider localization
effects and is valid for conductivities in the interval
$(E_{F}/T_{c})\sigma_{c}\gg \sigma\gg
\sigma_{c}$.

At present the following results are well established for superconductors close
to localization
transition (i.e.\  $\sigma\leq \sigma_{c}$):
\begin{enumerate}
\item  Assuming independence of the density of states at the Fermi level
$N(E_{F})$  and of the
pairing constant $\lambda_{p}$  from the value of the mean-free path $l$
(disorder) we can show
that  $T_{c}$ drops as disorder grows due to respective growth of Coulomb
pseudopotential
$\mu^{*}$.\ This effect is due to the growth of retardation
effects of Coulomb interaction within the Cooper pair as
diffusion coefficient drops close to Anderson transition
\cite{AMR}.\   $T_{c}$ degradation starts even for $\sigma\gg \sigma_{c}$ and
becomes
fast for $\sigma < \sigma_{c}$  \cite{Bul1,Bul2}.\ The growth of spin
fluctuations and changes
in the density of states due to interaction effects may also lead to the drop
of $T_{c}$,\ though
these mechanisms were not analyzed in detail up to now.
\item Close to Anderson transition the usual expression for superconducting
coherence length
for a ``dirty'' limit  $\xi=\sqrt{\xi_{0}l}$ should be replaced by $\xi\approx
(\xi_{0}l^{2})^{1/3}$  and it remains finite even below Anderson transition
(i.e. in insulating
phase)\cite{Bul1,Bul2,KK1,KK2,LeeMa},\  signalling the possibility of
superconductivity in
Anderson insulator.\ Obviously these results are valid only in case of
finite $T_{c}$ close to Anderson transition,\ which is possible only if very
rigid conditions are
satisfied.
\item  The growth of disorder as system moves to Anderson transition leads to
the growth
of different kinds of fluctuations of superconducting order-parameter both of
thermodynamic
nature and due to fluctuations of electronic characteristics of the system.
\end{enumerate}
In our review we shall present  an extensive discussion of these and some of
the other problems
concerning the interplay of superconductivity and localization.\  However,\
first of all we shall
briefly describe the main principles of modern theory of electron localization
and physics of
metal---insulator transition in disordered systems,\ which will be necessary
for clear understanding
of the main problem under discussion.\ After that we shall give rather detailed
presentation
of theoretical problem of superconductivity close to the Anderson transition.\
Finally,\ we shall
describe the present experimental situation.\ We shall briefly describe some of
the experiments with traditional superconductors,\ but our main emphasis will
be on high---$T_{c}$ oxides.\ We shall concentrate on the experiments with
high---temperature---superconductors disordered by fast neutron irradiation
which we consider one of the best methods to introduce disorder in a controlled
fashion without any chemical (composition) changes.\ In this sense our review
of experiments is also far from being complete,\ but we hope that it is full
enough to claim that high---$T_{c}$ systems are especially good for testing
some
of the main theoretical ideas,\ expressed throughout this review.\ Also we
believe that better understanding of their properties under disordering may be
important for the development of the general theory of high---temperature
superconductivity.\ The preliminary version of this review has been published
in Ref.\cite{Sadov93}.

\newpage
\section{ANDERSON LOCALIZATION AND METAL-INSULATOR  TRANSITION
IN DISORDERED SYSTEMS}


\subsection{Basic Concepts of Localization}

In recent years a number of review papers had appeared dealing with basic
aspects of Anderson
localization \cite{Sad81,NagFuk,LeeRam,Sad86,Th74,Efr78,VW90}.\ Here we shall
remind the
main points of this theory and introduce the accepted terminology.

In 1958 Anderson \cite{And58} has shown for the first time that the wave
function of a quantum
particle in a  random potential can qualitatively change its nature if
randomness becomes large
enough.\  Usually,\ when disorder is small,\ the particle (e.g.\ electron) is
scattered randomly and
the wave function changes at the scale of the order of mean free path $l$.\
However,\ the wave
function remains extended plane---wave---like (Bloch wave---like) through the
system.\ In case of
strong enough disorder,\ the wave function becomes localized,\ so that its
amplitude (envelope) drops
exponentially with distance from the center of localization ${\bf r}_{0}$:
\begin{equation}
|\psi({\bf r})|\sim exp(|{\bf r-r}_{0}|/R_{loc})
\end{equation}
where $R_{loc}$ is localization length.\ This situation is shown qualitatively
in Fig. \ref{fig1}.\
The physical meaning of Anderson localization is relatively simple:  coherent
tunneling of
electrons is possible only between energy levels with the same energy (e.g.\
between equivalent
sites in crystalline lattice).\ However,\ in case of strong randomness the
states with the same
energy are too far apart in space for tunneling to be effective.

At small disorder dc conductivity of a metal at $T=0$ is determined by Drude
expression:
\begin{equation}
\sigma_{0}=\frac{ne^2}{m}\tau=\frac{ne^2}{p_{F}}l  \label{Drude}
\end{equation}
where $\tau$ --- is the mean free time,\  $n$ --- is electron density and $e$
--- its charge.\ Usual
kinetic theory can be applied if
\begin{equation}
\frac{p_{F}l}{\hbar}\gg 1 \mbox{ or } \frac{E_{F}\tau}{\hbar}\gg 1
\label{weak}
\end{equation}
which is a condition of weak scattering (disorder).\   From Eq. (\ref{Drude})
and Eq. (\ref{weak}),\  taking into
account $n=p_{F}^3/(3\pi ^2\hbar^3)$,\  we can estimate the lower limit of
conductivity for which
Drude approximation is still valid:
\begin{equation}
\sigma_{0}=\frac{e^2p_{F}}{3\pi^2\hbar^2}(\frac{p_{F}l}{\hbar})\gg
\frac{e^2p_{F}}{3\pi^2\hbar^2}
\end{equation}
The conductivity value:
\begin{equation}
\sigma_{c}\approx \frac{e^2p_{F}}{3\pi^2\hbar^2}    \label{mmc}
\end{equation}
is usually called the ``minimal metallic conductivity'' \cite{Mott74,MottDav}.\
As disorder grows the mean free
path diminishes and becomes of the order of lattice
spacing $a$,\ so that we reach $p_{F}l/\hbar\sim 1$,\ and the
usual kinetic theory based upon Boltzmann equation becomes inapplicable.\ This
was first noted by Ioffe and
Regel \cite{IR},\ who observed that at such disorder the
qualitative form of wave function must change,\ transforming from extended to
localized accompanied by
metal---insulator transition.\ From Eq.(\ref{mmc}) it is
clear that this transition takes place at the conductivity scale of the order
of $\sigma_{c}\sim
(2-5) 10^2 Ohm^{-1}cm^{-1}$ for typical $\hbar/p_{F}\sim a\sim (2-3)
10^{-8} cm$.

Qualitative form of energy spectrum near the band---edge
of a disordered system is shown in Fig. \ref{fig2}.\ When
the Fermi level lies in the high---energy region electronic states close to it
are slightly distorted
plane waves.\ As Fermi energy moves towards the band---edge (or with the growth
of disorder) the critical energy
$E_{c}$ (mobility edge) separating extended and localized states crosses the
Fermi level.\ If $E_{F}$ belongs to the region of localized states the system
becomes insulating,\ conductivity is possible only for $T>0$ or
by exciting the carriers by alternating electric field.\
The appearance of these hopping mechanisms of conductivity signals Anderson
transition\cite{Mott74,MottDav}.

One of the main problems is the qualitative behavior of
conductivity when the Fermi level $E_{F}$ crosses the
mobility edge $E_{c}$ (at $T=0$).\ While Mott assumed
the discontinuous drop of conductivity from $\sigma_{c}$
to zero \cite{Mott74,MottDav} modern approach\cite{Sad81,NagFuk,LeeRam,VW90}
based
mainly on the scaling theory to localization \cite{AALR} demonstrates {\em
continuous} transition.\
Experiments at low temperatures clearly confirm this type
of behavior \cite{LeeRam},\ and $\sigma_{c}$ acts as a
characteristic conductivity scale close to transition.\
Static conductivity of a metal at $T=0$ close to Anderson
transition within this approach is written as:
\begin{equation}
\sigma=A\frac{e^2}{\hbar\xi_{loc}}\approx
\sigma_{c}\left|\frac{E_{F}-E_{c}}{E_{c}}\right|^{(d-2)\nu}
\label{scal}
\end{equation}
where $A$ --- is a numerical constant,\ $d$ --- is space
dimension,\ and $\sigma_{c}\approx Ae^2/(\hbar a^{d-2})$.\ Here we introduced
the correlation length of scaling theory diverging at the transition:
\begin{equation}
\xi_{loc}\approx
\frac{\hbar}{p_{F}}\left|\frac{E_{F}-E_{c}}{E_{c}}\right|^{-\nu}
\label{xiloc}
\end{equation}
Critical exponent $\nu$ determines this divergence.\ In
one---electron approximation and in the absence of magnetic scattering $\nu\sim
1$ \cite{LeeRam,Sad86,VW90,WV82}.\ In the region of
localized states (i.e.\ for $E_{F}<E_{c}$) $\xi_{loc}$
coincides with localization length of electrons $R_{loc}$.\ In metallic region
$\xi_{loc}$ determines the
effective size of a sample at which ``Ohmic'' behavior
appears,\ i.e.\ conductivity becomes independent of a
sample size \cite{LeeRam,Im80}.\ ``Minimal metallic
conductivity'' $\sigma_{c}$ determines,\ as we noted,\
the conductivity scale close to a transition.

In the vicinity of Anderson transition conductivity
acquires an important frequency dependence\cite{Weg76,ShAbr}.\
For $E_{F}=E_{c}$ i.e.\ at the transition we have:
\begin{equation}
\sigma(\omega)\approx \sigma_{c}(i\omega\tau)^{\frac{d-2}{d}}    \label{frc}
\end{equation}
which is valid also close to the transition (from either
side) for frequencies $\omega\gg \omega_{c}\sim [N(E_{F})
\xi_{loc}^{d}]^{-1}$.\ For $d=3$ this is sometimes refered to as Gotze's
\cite{Got81} law $\omega^{1/3}$,\
although this particular derivation was later acknowledged to be wrong
\cite{BelGold}.

The spatial dimension $d=2$ is the so called ``lower
critical dimensionality''\cite{Sad81,NagFuk,LeeRam,Sad86}
.\ For $d=2$ all electronic states are localized for
infinitesimal disorder\cite{AALR},\ and there is no
Anderson transition.\

Quasi---two---dimensional systems are especially interesting,\ mainly because
most of high---$T_{c}$ oxides demonstrate strongly anisotropic electronic
properties.\ Here we shall make the
simplest  estimates for such systems on the line of Ioffe---Regel approach.\
Consider a system
made of highly---conducting ``planes'' where the current carriers are
``nearly---free'',\  while the
interplane tunneling is possible only due to some small transfer integral $w\ll
E_{F}$ ($E_{F}$ ---
is the Fermi energy of two---dimensional gas within the plane).\ Conductivity
within the plane is
determined for small disorder as:
\begin{equation}
\sigma_{\|}=e^2D_{\|}N(E_{F})      \label{inpl}
\end{equation}
where $D_{\|}=v_{F}^2\tau/2$,\ $N(E_{F})=m/(\pi a_{\bot} \hbar^2)$,\ $a_{\bot}$
--- is interplane spacing,\ which is noticeably larger than interatomic
distance within the plane.\
Interplane conductivity is given by:
\begin{equation}
\sigma_{\bot}=e^2D_{\bot}N(E_{F})     \label{interpl}
\end{equation}
where $D_{\bot}=(wa_{\bot})^2\tau/\hbar^2$.\ The appropriate mean free paths
are $l_{\|}=v_{F}\tau$,\ $l_{\bot}=wa_{\bot}\tau/\hbar$.\ Ioffe-Regel criterion
for a quasi---two---dimensional system can be written as:
\begin{equation}
l_{\bot}=wa_{\bot}\tau/\hbar\sim a_{\bot}        \label{IR2d}
\end{equation}
which is equivalent to $w\tau/\hbar\sim 1$ --- the condition of breaking of
coherent tunneling between the planes.\ Elementary estimate shows that this
corresponds to:
\begin{equation}
\sqrt{\sigma_{\|}\sigma_{\bot}}\sim \frac{e^2}{\sqrt{2}\pi\hbar a}
\sim\sigma_{c}  \label{mmc2d}
\end{equation}
where $a$ --- is interatomic distance {\em within the planes}.\ In isotropic
case this reduces to Eq.(\ref{mmc}).\
For strongly anisotropic system when $\sigma_{\|}\gg \sigma_{\bot}$ it is clear
that Eq.(\ref{mmc2d}) can be
satisfied even for $\sigma_{\|}\gg \sigma_{c}$,\ because of
small values of $\sigma_{\bot}$.\ Formally,\ for $\sigma_{\bot}\rightarrow 0$,\
critical value of $\sigma_{\|}$ diverges,\ that reflects on this elementary
level the tendency towards complete localization for purely
two---dimensional case.

The important property of energy spectrum in the region of
localized states is its local discretness.\ As we noted above,\ the physical
meaning of localization itself leads
to a picture of close energy levels being far apart in
space,\ despite the continuous nature of average density of
states.\ Due to exponential decay of the localized wave
functions it leads to the absence of tunneling \cite{And58}.\ The energy
spacing between levels of electrons localized within the sphere of the radius
of the
order of $R_{loc}(E)$ can be estimated \cite{Mott74,MottDav} as:
\begin{equation}
\delta_{E_{F}}\approx [N(E_{F})R_{loc}^d]^{-1}
\label{spac}
\end{equation}
As the metallic system moves toward Anderson transition,\ i.e.\ as the mean
free path drops to interatomic distance and conductivity becomes less than
$\sim 10^3 Ohm^{-1}cm^{-1}$ there appear the well known anomalies like the
negative
temperature coefficient of resistivity \cite{IR,Moo}.\
These anomalies are apparently closely connected with localization phenomena
\cite{LeeRam}.

Up to now we discussed Anderson transition,\ neglecting
electron interactions.\ Its importance for the problem of
metal---insulator transition in disordered systems was
known for a long time \cite{Mott74}.\ In recent years
there was a serious progress in the general approach to
a theory of ``dirty'' metals,\ based on the analysis of
interference of impurity scattering and Coulomb interactions
\cite{AltAr79,AltAr82,AltAr85}.\ Later we
shall review its implications for the general picture
of Anderson transition.\ Apparently the continuous nature
of metal---insulator transition is not changed though
interaction lead to a number of specific effects,\ e.g.\
in the behavior of the density of states at the Fermi
level,\ as well as to the growth of magnetic (spin)
fluctuations.\  Here we shall briefly describe
the concept of ``soft'' Coulomb gap appearing below the transition in the
region of localized states \cite{EfrShkl75,Efr76,ShklEfr79,EfrShkl85}.\ Coulomb
interaction between localized electrons can be estimated as $e^2/\epsilon
R_{loc}$,\
and it is obviously important if this energy is comparable with the local level
spacing $[N(E_{F})R_{loc}^3]^{-1}$ (for three---dimensional system).\ As a
result a Coulomb pseudogap appears at the
Fermi level with the width:
\begin{equation}
\Delta_{c}\approx (e^3/\epsilon^{3/2})[N(E_{F})]^{1/2}
\label{cgap}
\end{equation}
where $\epsilon$ is the dielectric constant.\ We shall
see later that close to the Anderson transition $\epsilon\approx 4\pi e^2
N(E_{F})R_{loc}^2$ and accordingly:
\begin{equation}
\Delta_{c}\approx [N(E_{F})R_{loc}^3]^{-1}\approx \delta_{E_{F}}
\label{gapest}
\end{equation}
so that in this case Coulomb effects are comparable with the effects of
discretness of energy spectrum in localized phase. At the moment there is no
complete
theory connecting the localization region with metallic
phase within the general approaches of interaction theory.



\subsection{Elementary Scaling Theory of Localization}
The behavior of electronic system close to  the Anderson
transition can be described by a scaling theory similar
to that used in the theory of critical phenomena \cite{Kad,WK,PP}.\ The main
physical idea of this approach is based upon a series of scale transformations
from smaller to larger ``cells'' in coordinate space
with appropriate description of a system by transformed
parameters of initial Hamiltonian.\ These transformations
are usually called renormalization group.\ In the theory
of critical phenomena this approach is usually motivated
by the growth of correlation length of order---parameter
fluctuations near the critical point \cite{Kad}.\ This
is analogous to the growth of localization length on the
approach of mobility edge from Anderson insulator.

The accepted scaling approach to localization problem
was proposed by Abrahams,\ Anderson,\ Licciardello and
Ramakrishnan \cite{AALR}.\  In this theory localization is described in terms
of  {\em conductance} $g$ as a function of a sample size $L$.\ For a small
disorder  $(p_{F}l/\hbar\gg 1)$ the system is in a metallic state and
conductivity $\sigma$ is determined by Eq. (\ref{Drude})
and is independent of a sample size if this size is much larger than the mean
free path,\ $L\gg l$.\
Conductance is determined in this case just by Ohm law and for a
$d$---dimensional hypercube
we have:
\begin{equation}
 g(L)=\sigma L^{d-2}
\label{gmet}
\end{equation}
If electronic states near the Fermi level are localized,\ conductivity of an
infinite system at $T=0$ is
zero and matrix elements for transitions between different electronic states
drop exponentially on
distances of the order of $R_{loc}$.\  Then it can be expected that for $L\gg
R_{loc}$,\ the effective conductance becomes exponentially small:
\begin{equation}
g(L)\sim exp(-L/R_{loc})
\label{gloc}
\end{equation}
Elementary scaling theory of localization assumes that in general case the
conductance of a
hypercube of a size $L$ satisfies the simplest differential equation of a
renormalization group:
\begin{equation}
\frac{d ln g(L)}{d ln L}=\beta_{d}(g(L))
\label{GML}
\end{equation}
Most important assumption here is the dependence fo $\beta_{d}(g)$  only on one
variable $g$
(one---parameter scaling).\ Then the qualitative behavior of $\beta_{g}$ can be
analyzed in a
simplest possible way interpolating between limiting forms given by Eq.
(\ref{gmet}) and Eq. (\ref{gloc}).\ For metallic phase (large $g$) from Eq.
(\ref{gmet}) and Eq.(\ref{GML}) we get:
\begin{equation}
\lim_{g \to \infty} \beta_{d}(g) \rightarrow  d-2
\label{betamet}
\end{equation}
For insulator $(g\rightarrow 0)$ from Eq. (\ref{GML}) and Eq. (\ref {gloc}) it
follows that:
\begin{equation}
\lim_{g \to 0} \beta_{d}(g)\rightarrow ln \frac{g}{g_{c}}
\label{betaloc}
\end{equation}
Assuming the existence of two perturbation expansions over the ``charge'' $g$
in the limits of
weak and strong ``couplings'' we can write correction to
Eq. (\ref{betaloc}) and Eq. (\ref{betamet}) in the following form:
\begin{eqnarray}
\beta_{d}(g\rightarrow 0)=ln \frac{g}{g_{c}}(1+bg+ \cdots )
\label{betaloc1}    \\
\beta_{d}(g\rightarrow \infty)=d-2 - \frac{\alpha}{g} + \cdots  \qquad \alpha>0
  \label{betamet1}
\end{eqnarray}
Following these {\em assumptions} and
supposing now monotonous and continuous form of $\beta_{d}(g)$ it is easy to
plot it
qualitatively for all $g$,\ as shown in Fig. \ref{fig3}.\ All the previous
equations are written for
dimensionless conductance,\ which is measured in natural units of
$e^2/\hbar\approx 2.5 10^{-4} Ohm^{-1}cm^{-1}$.\ We see that $\beta_{d}(g)$
definitely has no zeros for $d<2$.\  If expansion Eq. (\ref{betamet1})  is
valid there is no zero also for $d=2$.\ For $d>2$  $\beta_{d}$
--- function must have a zero: $\beta_{d}(g_{c})=0$.\ It is clear that
$g_{c}\sim 1$ and no form of perturbation theory is valid near that zero.\ The
existence of a zero of $\beta_{d}(g)$ corresponds
to existence of an unstable fixed point of Eq. (\ref{GML}).\ The state of a
system is supposedly determined by disorder at microscopic distances of the
order of interatomic spacing $a$,\ i.e.\ by $g_{0}=g(L=a)$.\ Using $g_{0}$ as
an initial value and
integrating Eq. (\ref{GML}) it is easy to find that for
$g_{0}>g_{c}$ conductivity $\sigma_{L}=g(L)L^{2-d}$ tends
for $L\rightarrow\infty$ to a constant (metallic) value.\
For $g<g_{c}$ in the limit of $L\rightarrow\infty$ we get
insulating behavior.\ Using for $g\sim g_{c}$ an approximation (shown with
circles in Fig. \ref{fig3}):
\begin{equation}
\beta_{d}(g)\approx \frac{1}{\nu}ln\frac{g}{g_{c}}\approx
\frac{1}{\nu}\frac{g-g_{c}}{g_{c}}   \label{betappr}
\end{equation}
we obtain from Eq. (\ref{GML}) for $g_{0}>g_{c}$ the
following behavior of conductivity for $L\rightarrow\infty$:
\begin{equation}
\sigma\approx
A\frac{e^2}{\hbar}\frac{g_{c}}{a^{d-2}}\left(ln\frac{g_{0}}{g_{c}}\right)^{(d-2)\nu}\approx
A\frac{e^2}{\hbar}\frac{g_{c}}{a^{d-2}}\left(\frac{g_{0}-g_{c}}{g_{c}}\right)^{(d-2)\nu}
\label{wcond}
\end{equation}
where $A=const$ and we have explicitly introduced the
conductivity scale of the order of $\sigma_{c}$.\ (Cf.\
Eq. (\ref{mmc})).\ We see that the existence of a fixed
point leads to the existence of mobility edge,\ and
behavior of $\beta_{d}(g)$ close to its zero determines
the critical behavior at the Anderson transition.\ Under
these assumptions conductivity continuously goes to zero
for $g_{0}\rightarrow g_{c}$,\ and the value of $\sigma_{c}\approx e^2/(\hbar
a^{d-2})$ is characteristic
scale of conductivity at the metal---insulator transition.\ To get a
discontinuous drop of conductivity
at the mobility edge $\beta_{d}(g)$ must be nonmonotonous
as shown by dashed line for $d=2$ in Fig. (\ref{fig3}).\ This behavior seems
more or less unphysical.

Integrating Eq. (\ref{GML}) with $\beta_{d}(g)$ from
Eq. (\ref{betappr}) with initial $g_{0}<g_{c}$ gives:
\begin{equation}
g(L)\approx g_{c}
exp\left\{-A\left|ln\frac{g_{0}}{g_{c}}\right|^{\nu}\frac{L}{a}\right\}
\label{glocal}
\end{equation}
{}From here it is clear (Cf. Eq. (\ref{xiloc})) that:
\begin{equation}
R_{loc}\sim a\left|\frac{g_{0}-g_{c}}{g_{c}}\right|^{-\nu}      \label{rloc}
\end{equation}
and $\nu$ is the critical exponent of localization length.\ For $d=2$ we  have
$\beta_{d}(g)<0$ in the whole
interval of $g$,\ so that $\sigma_{L\rightarrow\infty}\rightarrow 0$ for any
initial
value of $g$,\ so that there is no mobility edge and all
states are localized.

For $d>2$ limiting ourselves by those terms of perturbation expansion in
$g^{-1}$ shown in Eq. (\ref{betamet1}) we can solve $\beta_{d}(g_{c})=0$
to find:
\begin{equation}
g_{c}=\frac{\alpha}{d-2}    \label{gcrit}
\end{equation}
We can see that for $d\rightarrow 2$ the mobility edge goes to infinity which
corresponds to complete localization in two---dimensional case.\ Now we have:
\begin{equation}
\beta_{d}(g\sim g_{c})\approx (d-2)\left(\frac{g_{0}-g_{c}}{g_{c}}\right)
\label{betad-2}
\end{equation}
and for the critical exponent of localization length we
get (Cf.\ Eq.\ (\ref{betappr})):
\begin{equation}
\nu=\frac{1}{d-2}      \label{crexp}
\end{equation}
which may be considered as the first term of $\varepsilon$---expansion near
$d=2$ (where $\varepsilon=d-2$),\ i.e. near ``lower critical dimension'' for
localization \cite{AALR,Weg79a,ShAbr81a}.\ Note that
the expansion Eq. (\ref{betamet1}) can be reproduced in
the framework of standard perturbation theory over
impurity scattering \cite{GLK79,AR80}.\ For $d=3$ this
gives $\alpha=\pi^{-3}$ (Cf.Ref.\cite{LeeRam}).

Let us define now correlation length of localization
transition as:
\begin{equation}
\xi_{loc}\sim a\left|\frac{g_{0}-g_{c}}{g_{c}}\right|^{-\nu}
\label{corleng}
\end{equation}
For $g_{0}<g_{c}$ this length coincides with localization
length $R_{loc}$.\ It is easy to see that Eq. (\ref{wcond}) can be written as:
\cite{Weg76}
\begin{equation}
\sigma\approx Ag_{c}\frac{e^2}{\hbar\xi_{loc}^{d-2}}
\label{weg}
\end{equation}
It follows that for $g>g_{c}$ correlation length $\xi_{loc}$ determines
behavior of conductivity close
to the mobility edge,\ when this length becomes much
larger than interatomic distance and mean free path.

Let us consider three---dimensional case in more details.\ Integrating Eq.
(\ref{GML}) with $\beta_{3}(g)=1-g_{c}/g$ where $g_{c}=\alpha$ gives
$g(L)=(\hbar/e^2)\sigma_{L}L=(\hbar/e^2)\sigma+g_{c}$
so that for a finite sample close to the mobility edge
($\xi_{loc}\gg l$) we obtain:
\begin{equation}
\sigma_{L}=\sigma+\frac{e^2g_{c}}{\hbar L}
\label{condfin}
\end{equation}
where in correspondence with Eq. (\ref{weg})
\begin{equation}
\sigma\approx Ag_{c}\frac{e^2}{\hbar\xi_{loc}}
\label{3dcond}
\end{equation}
It follows that for $L\gg\xi_{loc}\gg l$ conductivity
$\sigma_{L}\rightarrow \sigma$ while for $l\ll L\ll \xi_{loc}$ conductivity
$\sigma_{L}$ and the appropriate
diffusion coefficient,\ determined by Einstein relation
$\sigma=e^{2}DN(E_{F})$ are equal to:
\begin{eqnarray}
\sigma_{L}\approx \frac{e^2g_{c}}{\hbar L}  \label{condL}\\
D_{L}\approx \frac{g_{c}}{N(E_{F})}\frac{1}{\hbar L}
\label{difL}
\end{eqnarray}
where $N(E_{F})$ is the electron density of states at
the Fermi level.\ Thus in this latest case conductivity
is not Ohmic,\ diffusion of electrons is ``non---classical''
\cite{AMR,LeeRam}.\ From this discussion it
is clear that the characteristic length $\xi_{loc}$ in
metallic region determines the scale on which conductivity becomes independent
of sample size.\ Close
to the mobility edge when $\xi_{loc}\rightarrow\infty$
only the samples with growing sizes $L\gg\xi_{loc}$
can be considered as macroscopic.\ These considerations allow to understand the
physical meaning of diverging
length $\xi_{loc}$ of scaling theory in metallic region
\cite{Im80}.\ Close to the mobility $\xi_{loc}$ is considered as the only
relevant length in the problem
(with an exception of a sample size $L$) and the scaling
hypothesis is equivalent to the assumption of:
\begin{equation}
g(L)=f\left(\frac{L}{\xi_{loc}}\right)   \label{sclg}
\end{equation}
where $f(x)$---is some universal (for a given dimensionality $d$) function.\ In
metallic region for
$L\gg\xi_{loc}\gg l$ it is obvious that $f(x)\sim x^{d-2}$ which reproduces Eq.
(\ref{weg}).

For finite frequencies $\omega$ of an external electric
field a new length appears in the system \cite{ShAbr}:
\begin{equation}
\L_{\omega}=\left[\frac{D(\omega)}{\omega}\right]^{1/2}
\label{Lomeg}
\end{equation}
where $D(\omega)$---is the frequency dependent diffusion
coefficient.\ $L_{\omega}$ is a length of electron diffusion during one cycle
of an external field.\ Close
to the mobility edge $\xi_{loc}$ is large and for $L_{\omega}<\xi_{loc},\ L$
and $L_{\omega}$ become the
relevant length scale.\ In general,\ for finite $\omega$
localization transition is smeared,\ a sharp transition is realized only for
$L^{-1}=L_{\omega}^{-1}=0$.\ Thus
for the finite frequency case the scaling hypothesis of
Eq. (\ref{sclg}) can be generalized as: \cite{ShAbr}
\begin{equation}
g(L,\omega)=f\left(\frac{L}{\xi_{loc}},\frac{L_{\omega}}{\xi_{loc}}\right)
 \label{frscal}
\end{equation}
where $g$ denotes a real part of conductance.\ In metallic phase for
$L\gg\xi_{loc}$ we have $g\sim L^{d-2}$ so that:
\begin{eqnarray}
\sigma(\omega)=\frac{e^2}{\hbar} L^{2-d}
f\left(\frac{L}{\xi_{loc}},\frac{L_{\omega}}{\xi_{loc}}\right) \rightarrow
\frac{e^2}{\hbar}\xi_{loc}^{2-d} f\left(\infty,\frac{L_{\omega}}{\xi_{loc}}
\right)  \nonumber\\
\equiv \frac{e^2}{\hbar\xi_{loc}^{d-2}}
F\left(\frac{\xi_{loc}}{L_{\omega}}\right)
\end{eqnarray}
For small frequencies,\ when $L_{\omega}\gg\xi_{loc}$,\  
we can write down the universal function $F(x)$ as
$F(x)\approx Ag_{c}+Bx^{d-2}$ which  reproduces Eq. (\ref{weg}) and the small
frequency
corrections found earlier in \cite{GLK79}.\ For $L_{\omega}\ll \xi_{loc}$ i.e.\
for high frequencies or
close to mobility edge the relevant length is $L_{\omega}$ and frequency
dependent part of conductivity
is dominating.\ In particular at the mobility edge itself
the length $\xi_{loc}$ drops out and must cancel in
Eq. (\ref{frscal}) which leads to:
\begin{equation}
\sigma(\omega,E_{F}=E_{c})\sim L^{2-d}_{\omega}\sim
\left[\frac{\omega}{D(\omega)}\right]^{\frac{d-2}{2}}
\label{omegcond}
\end{equation}
On the other hand,\ according to Einstein relation we
must have $\sigma(\omega)\sim D(\omega)$.\ Accordingly,\
from $[\omega / D(\omega)]^{(d-2)/2} \sim D(\omega)$ we get at
the mobility edge:
\begin{equation}
\sigma(\omega,E_{F}=E_{c})\sim D(\omega)\sim \omega^{\frac{d-2}{d}}
\label{diffreqd}
\end{equation}
For $d=3$ this leads \cite{Weg76,Got81} to $\sigma(\omega)\sim D(\omega)\sim
\omega^{1/3}$.\ The crossover between
different types of frequency dependence occurs for $L_{\omega}\sim \xi_{loc}$
which determines characteristic frequency: \cite{ShAbr}
\begin{equation}
\omega_{c}\sim \frac{1}{\hbar\xi_{loc}^{d}N(E_{F})}
\label{omegc}
\end{equation}
The $\omega^{(d-2)/d}$---behavior is realized for $\omega\gg \omega_{c}$,\
while for $\omega\ll \omega_{c}$
we get small corrections of the order of $\sim\omega^{(d-2)/2}$ to Eq.
(\ref{weg}).

Finally we must stress that for finite temperatures
there appear {\em inelastic} scattering processes which
destroy the phase correlations of wave functions at distances greater than a
characteristic length of the order of $L_{\varphi}=\sqrt{D\tau_{\varphi}}$ ,\
where $D$ is the diffusion coefficient due to {\em elastic}
scattering processes considered above and $\tau_{\varphi}$ is the ``dephasing''
time due to inelastic processes \cite{AltAr82}.\ For $T>0$ this length
$L_{\varphi}$ effectively replaces the sample size
$L$ in all expressions of scaling theory when $L\gg L_{\varphi}$,\ because on
distances larger than
$L_{\varphi}$ all information on the nature of wave
functions (e.g.\ whether they are localized or extended)
is smeared out.\ Taking into account the usual low---temperature dependence
like $\tau_{\varphi}\sim T^{-p}$
(where $p$ is some integer,\ depending on the mechanism
of inelastic scattering) this can lead to a non---trivial
temperature dependence of conductivity,\ in particular to
a possibility of a negative temperature coefficient of
resistivity of ``dirty'' metals \cite{Im80} which are
close to localization transition.\ It is important to stress that similar
expressions determine the temperature
dependence of conductivity also for the localized phase
until $L_{\varphi}<R_{loc}$.\ Only for $L_{\varphi}>R_{loc}$ the localized
nature of wave functions starts to signal itself in temperature dependence of
conductivity and the transition to exponentially activated hopping behavior
takes place,\
which becomes complete for $T<[N(E_{F})R_{loc}^d]^{-1}$.



\subsection{Self---Consistent Theory of Localization}

\subsubsection{Isotropic Systems}
It is obvious that qualitative scaling picture of Anderson transition described
in the previous section
requires microscopic justification.\ At the same time we need a practical
method of explicit calculations for any physical characteristic of electronic
system close to the mobility edge.\
Here we shall briefly describe the main principles of so called
self---consistent theory of localization
which while leaving aside some important points,\ leads to an effective
scheme for analysis of the
relevant physical characteristics important for us.\ This
approach,\ first formulated
by Gotze \cite{Got79,Got81} was later further developed by Vollhardt and Wolfle
and other
authors \cite{VW80,VW82,WV82,MS82,KotSad83,Sad86,VW90}.

Complete information concerning Anderson transition and transport in a
disordered system is
contained in the two---particle Green's function:
\begin{equation}
\Phi^{RA}_{{\bf pp'}}(E\omega{\bf q})=-\frac{1}{2\pi i}<G^{R}({\bf
p_{+}p'_{-}}E+\omega)G^{A}({\bf p'_{-}p_{-}}E)>
\label{phipp}
\end{equation}
where  ${\bf p_{+-}=p^{+}_{-}}(1/2){\bf q}$,\ in most cases below $E$ just
coincides with the Fermi energy $E_{F}$.\ Angular brackets denote averaging
over disorder.\ Graphically this Green's function is shown in Fig. \ref{fig4}.\
 It is well known that this Green's function is determined by the
Bethe---Salpeter
equation also shown graphically in Fig . \ref{fig4} \cite{Edw58,AGD,VW80}:
\begin{equation}
\Phi^{RA}_{{\bf pp'}}(E{\bf q}\omega)=G^{R}(E+\omega{\bf p_{+}})G^{A}(E{\bf
p_{-}}) 
\left\{-\frac{1}{2\pi i}\delta({\bf p-p'})+\sum_{{\bf p''}}U^{E}_{{\bf
pp''}}({\bf q}\omega)\Phi^{RA}_{{\bf p''p'}}(E{\bf q}\omega)\right\}
\label{BS}
\end{equation}
where $G^{R,A}(E{\bf p})$ --- is the averaged retarded (advanced)
one---electron Green's
function,\ while irreducible vertex part $U^{E}_{{\bf pp'}}({\bf q}\omega)$ is
determined by
the sum of all diagrams which can not be cut over two electron lines (Cf.\ Fig.
\ref{fig4}).

In fact,\ two---particle Green's function Eq. (\ref{phipp}) contains even some
abundant information and for the complete description of Anderson transition it
is sufficient to know the
two---particle Green's function summed over ${\bf pp'}$ \cite{VW80}:
\begin{equation}
\Phi_{E}^{RA}({\bf q}\omega)=-\frac{1}{2\pi i} \sum_{{\bf pp'}}<G^{R}({\bf
p_{+}p'_{+}}E+\omega)G^{A}({\bf p'_{-}p_{-}}E)>         \label{phi}
\end{equation}
Using Bethe---Salpeter equation Eq. (\ref{BS}) and exact Ward identities we can
obtain a closed equation for $\Phi^{RA}_{E}({\bf q}\omega)$
\cite{VW80,WV82,Sad86},\ and for  small $\omega$ and ${\bf q}$ the solution of
this equation has a typical diffusion---pole form:
\begin{equation}
\Phi^{RA}_{E}({\bf q}\omega)=-N(E)\frac{1}{\omega+iD_{E}({\bf q}\omega)q^2}
\label{phidiff}
\end{equation}
where $N(E)$---is electron density of states at energy $E$ and the {\em
generalized} diffusion
coefficient $D_{E}({\bf q}\omega)$ is expressed through the so called
relaxation kernel
$M_{E}({\bf q}\omega)$ :
\begin{equation}
D_{E}({\bf q}\omega)=i\frac{2E}{dm}\frac{1}{M_{E}({\bf
q}\omega)}=\frac{v_{F}^2}{d}\frac{i}{M_{E}({\bf q}\omega)}
\label{diffM}
\end{equation}
where $v_{F}$ is Fermi velocity of an electron.\ The retarded density---density
response function at small $\omega$ and ${\bf q}$ is given by:
\begin{equation}
\chi^{R}({\bf q}\omega)=\omega\Phi_{E}^{RA}({\bf q}\omega)+N(E)+O(\omega,q^{2})
\label{chiret}
\end{equation}
or from Eq. (\ref{phidiff}):
\begin{equation}
\chi^{R}({\bf q}\omega)=N(E)\frac{iD_{E}({\bf
q}\omega)q^{2}}{\omega+iD_{E}({\bf q}\omega)q^{2}}
\label{denden}
\end{equation}

For relaxation kernel $M_{E}({\bf q}\omega)$ (or for generalized diffusion
coefficient)
a self---consistency equation can be derived,\ which is actually the main
equation of the theory \cite{VW80,VW90,WV82}.\ The central point in this
derivation is some approximation for
the irreducible vertex part $U^{E}_{{\bf pp'}}({\bf q}\omega)$ in
Bethe---Salpeter equation.\
The approximation of Vollhardt and Wolfle is based upon the use for
$U^{E}_{{\bf pp'}}({\bf q}\omega)$ of the sum of ``maximally-crossed'' graphs
shown in Fig. \ref{fig5}.\  This series is
easily summed and we get the co called ``Cooperon'' \cite{GLK79,VW80}:
\begin{equation}
U^{EC}_{{\bf pp'}}({\bf q}\omega)=\frac{2\gamma\rho V^{2}}{D_{0}({\bf
p+p'})^{2}+i\omega}         \label{cooperon}
\end{equation}
where
\begin{equation}
D_{0}=\frac{E}{md\gamma}=\frac{1}{d}v^{2}_{F}\tau         \label{diffbare}
\end{equation}
is the classical (bare) diffusion coefficient determining Drude conductivity
Eq. (\ref{Drude}).\ For point scatterers randomly distributed with spatial
density
$\rho$ ($V$ is scattering amplitude) we have:
\begin{equation}
\gamma=\frac{1}{2\tau}=\pi\rho V^{2} N(E_{F})   \label{gamma}
\end{equation}
These "maximally crossed" diagrams lead to the following quantum correction
to diffusion coefficient:
\begin{equation}
\frac{\delta D(\omega)}{D_{0}}=-\frac{1}{ \pi N(E) }
\sum_{|{\bf k}|<k_{0}} \frac{1}{-i\omega + D_{0}k^2}
\label{dDLNl}
\end{equation}
Appropriate correction to relaxation kernel can be expressed via the
correction to diffusion coefficient as:
\begin{equation}
\delta M_{E}(\omega)=-i\frac{2E_{F}}{dm}\frac{\delta D(\omega)}{D(\omega)^{2}}=
-\frac{M_{E}(\omega)}{D(\omega)}\delta D(\omega)
\label{dM}
\end{equation}
Considering the usual Drude metal as the zeroth approximation we get:
\begin{equation}
\delta M_{E}(\omega)=-\frac{M_{0}}{D_{0}}\delta D(\omega)
\label{dM0}
\end{equation}
The central point of the self-consistent theory of localization\cite{Got79}
reduces to the replacement of Drude diffusion coefficient $D_{0}$ in the
diffusion pole of Eq.(\ref{dDLNl}) by the generalized one $D(\omega)$. Using
this relation in Eq.(\ref{dM0}) we obtain the main equation of
self-consistent theory of localization determining the
relaxation kernel $M(0\omega)$ (for ${\bf q}=0$)\cite{VW80,WV82}:
\begin{equation}
M_{E}(\omega)=2i\gamma\left\{1+\frac{1}{\pi N(E)}\sum_{|{\bf
k}|<k_{0}}\frac{i}{\omega+\frac{2E}{dm}
\frac{k^2}{M_{E}(\omega)}}\right\}      \label{M}
\end{equation}
or the equivalent equation for the generalized diffusion coefficient itself:
\begin{equation}
\frac{D_{0}}{D_{E}(\omega)}=1+\frac{1}{\pi N(E)}\sum_{|{\bf
k}|<k_{0}}\frac{1}{-i\omega+D_{E}(\omega)k^2}       \label{D}
\end{equation}
Cut---off in momentum space in Eqs. (\ref{dDLNl}),\ (\ref{M}),\ (\ref{D}) is 
determined by the limit of applicability of
diffusion---pole approximation of Eq. (\ref{phidiff}) or
Eq. (\ref{cooperon}) \cite{Sad86}:
\begin{equation}
k_{0}\approx Min\{p_{F},l^{-1}\}    \label{cut}
\end{equation}
Close to the mobility edge $p_{F}\sim l^{-1}$.\
Note, that from here on we are generally using natural units with Planck
constant $\hbar=1$,\ however in some of the final expressions we shall write
$\hbar$ explicitly.

Conductivity can be expressed as:\cite{VW80,WV82}
\begin{equation}
\sigma(\omega)=\frac{ne^2}{m}\frac{i}{\omega+M_{E}(\omega)}
\rightarrow e^2D_{E}(\omega)N(E)  \mbox{ for }\omega\rightarrow 0
\label{condM}
\end{equation}
where we have used $n/N(E)=2E/d$.\ It is clear that for
metallic phase $M_{E}(\omega\rightarrow 0)=i/\tau_{E}$,\
where $\tau_{E}$ is generalized mean free time.\ Far from
Anderson transition (for weak disorder) $\tau_{E}\approx
\tau$ from Eq. (\ref{gamma}) and Eq. (\ref{condM}) reduces to standard Drude
expression.

If  the frequency behavior of relaxation kernel leads to the existence of a
limit  $lim_{\omega\to 0}\omega M_{E}({\bf q}\omega)$ a singular contribution
appears in Eq. (\ref{phidiff}) for $\omega\rightarrow 0$ :\cite{Got81,Sad86}
\begin{equation}
\Phi_{E}^{RA}({\bf q}\omega)\approx -\frac{N(E)}{\omega}
\frac{1}{1-\frac{2E}{md}\frac{q^2}{\omega M_{E}({\bf q}\omega)}}\approx
-\frac{N(E)}{\omega}\frac{1}{1+R_{loc}^{2}q^{2}}
\label{phising}
\end{equation}
where we have defined:
\begin{equation}
R_{loc}^2(E)=-\frac{2E}{md}lim_{\omega\to 0}\frac{1}{\omega M_{E}(\omega)}
\label{Rlocal}
\end{equation}
According to the general criterion of localization \cite{BG79,Sad86} (Cf.\
Appendix A) this behavior corresponds to the region of localized states.\ Using
Eq. (A 16)  we immediately
obtain from Eq. (\ref{phising}) the singular contribution to
Gorkov---Berezinskii
spectral density (Cf.\ Eqs. (A 8),\ (A 9)):
\begin{equation}
\ll\rho_{E}\rho_{E+\omega}\gg^{F}_{{\bf q}}=\frac{1}{\pi
N(E)}Im\Phi_{E}^{RA}({\bf q}\omega)=
A_{E}({\bf q})\delta(\omega)                 \label{BGsdsing}
\end{equation}
where
\begin{equation}
A_{E}({\bf q})=\frac{1}{1+R_{loc}^{2}(E)q^{2}}\rightarrow 1-R_{loc}^{2}(E)q^{2}
\mbox{ for }q\rightarrow 0
\label{Aq}
\end{equation}
{}From here and from Eq. (A 11) we can see that $R_{loc}(E)$ as defined in Eq.\
(\ref{Rlocal})
is actually the localization length.\ It is useful to define a characteristic
frequency\cite{VW80}:
\begin{equation}
\omega_{0}^2(E)=-lim_{\omega\to 0}\omega M_{E}(\omega)>0
\label{omega}
\end{equation}
so that
\begin{equation}
R_{loc}(E)=\sqrt{\frac{2E}{md}}\frac{1}{\omega_{0}(E)}
\label{rlocal}
\end{equation}
Thus,\ the localization transition is signalled by the
divergence of relaxation kernel for $\omega\rightarrow 0$
\cite{VW80},\ so that two characteristic types of it behavior for ${\bf q}=0$
and $\omega\rightarrow 0$ appear:
\begin{equation}
M_{E}(0\omega)\approx
\left\{ \begin{array}{l}
\frac{i}{\tau_{E}} \mbox{ for } E\geq E_{c} \\
\frac{i}{\tau_{E}}-\frac{\omega_{0}^{2}(E)}{\omega} \mbox{ for } E\leq E_{c}
\end{array}
\right.
\label{Mtau}
\end{equation}
The frequency $\omega_{0}(E)$ is in some crude sense analogous
to the order parameter in the usual theory of phase transitions.\ It appears in
the localized phase signalling about Anderson transition.

{}From Eq. (A 16) neglecting nonsingular for
$\omega\rightarrow 0$ and ${\bf q}=0$ contribution from
$Im\Phi_{E}^{RR}({\bf q} \omega)$ we can get explicit
expression for Berezinskii---Gorkov spectral density
which is valid for small $\omega$ and ${\bf q}$ \cite{KazSad84,Sad86}:
\begin{equation}
\ll\rho_{E}\rho_{E+\omega}\gg_{{\bf q}}^{F}=
\left\{ \begin{array}{l}
\frac{1}{\pi}\frac{D_{E}q^{2}}{\omega^{2}+(D_{E}q^{2})^{2}}  \mbox{ (Metal) }
\\
A_{E}({\bf q})\delta(\omega)+\frac{1}{\pi}\frac{D_{E}q^{2}}
{\omega^{2}+[\omega_{0}^{2}(E)\tau_{E}+D_{E}q^{2}]^{2}}
\mbox{ (Insulator) }
\end{array}
\right.
\label{rorodiff}
\end{equation}
where we have introduced renormalized diffusion coefficient,\ determined by
relaxation time $\tau_{E}$:
\begin{equation}
D_{E}=\frac{2E}{dm}\tau_{E}=\frac{1}{d}v_{F}^{2}\tau_{E}
\label{rendiff}
\end{equation}
Substituting Eq. (\ref{Mtau}) into self---consistency equation Eq. (\ref{M}) we
can obtain equations for $\tau_{E}$ and $\omega_{0}(E)$ \cite{VW82,MS82,Sad86}
and
thus determine all the relevant characteristics of the
system.\ For $d>2$ Eq. (\ref{M}) and Eq. (\ref{D}) do really describe
metal---insulator transition \cite{VW82,MS82,Sad86,VW90}.\ For $d=2$ all
electronic
states are localized \cite{VW80}.

Below we present some of the results of this analysis
which will be important for the following.\ For $2<d<4$ a correlation length
similar to that of Eq. (\ref{xiloc}) and Eq. (\ref{corleng}) appears:
\begin{equation}
\xi_{loc}(E)\sim \frac{1}{p_{F}}\left|\frac{E-E_{c}}{E_{c}}\right|^{-\nu}
\mbox{ for } E\sim E_{c}
\label{corlengsc}
\end{equation}
where $\nu=1/(d-2)$ .\ The position of the mobility edge is determined by a
condition:
\begin{equation}
\left.\frac{E}{\gamma}\right|_{E=E_{c}}=\frac{d}{\pi(d-2)}      \label{mobed}
\end{equation}
which follows if we assume the cut---off $k_{0}=p_{F}$ in Eq. (\ref{M}) and Eq.
(\ref{D}).\ Static
conductivity in metallic phase ($E>E_{c}$) is given by (Cf. Eq. (\ref{weg}):
\begin{equation}
\sigma=\frac{\sigma_{0}}{[p_{F}\xi_{loc}(E)]^{d-2}}          \label{sigWeg}
\end{equation}
where $\sigma_{0}=(ne^2/m)\tau$ is usual Drude conductivity.\ In particular,\
for $d=3$ :
\begin{equation}
\left.\frac{E}{\gamma}\right|_{E=E_{c}}=\left.p_{F}l\right|_{E=E_{c}}=\frac{3}{\pi}        \label{mobedd}
\end{equation}
in complete accordance with Ioffe---Regel criterion ,\ and
\begin{equation}
\sigma=\frac{\sigma_{0}}{p_{F}\xi_{loc}(E)}            \label{sig3d}
\end{equation}
Critical exponent $\nu=1$.\ Mean free path which follows from Eq.
(\ref{mobedd}) corresponds to Drude conductivity :
\begin{equation}
\sigma_{c}=\left.\frac{ne^2}{m}\tau\right|_{E=E_{c}}=\frac{e^{2}p_{F}}{3\pi^{2}\hbar^{2}}
\left.\left(\frac{p_{F}l}{\hbar}\right)\right|_{E=E_{c}}=\frac{e^{2}p_{F}}{\pi^{3}\hbar^{2}}
\label{mmcond}
\end{equation}
which is equivalent to elementary estimate of Eq. (\ref{mmc}).

Eq. (\ref{sig3d}) can also be rewritten as \cite{Bul2} :
\begin{equation}
\sigma=\sigma_{0}\left\{1-\frac{\sigma_{c}}{\sigma_{0}}\right\}=\sigma_{0}-\sigma_{c}
\label{sigmaBS}
\end{equation}
where Drude conductivity $\sigma_{0}$ is now the measure of disorder.\ It is
obvious that for
small disorder (large mean free path) $\sigma_{0}\gg \sigma_{c}$ and Eq.
(\ref{sigmaBS})
reduces to $\sigma\approx \sigma_{0}$.\ As disorder grows (mean free path
drops) conductivity
$\sigma\rightarrow 0$ for $\sigma_{0}\rightarrow \sigma_{c}$.

In dielectric phase ($E<E_{c}$)  we have $\xi_{loc}(E)=R_{loc}(E)$ and finite
$\omega^{2}_{0}(E)$ from Eq. (\ref{omega}) which tends to zero as $E\rightarrow
E_{c}$ from below.\  This frequency determines dielectric function of
insulating phase \cite{Sad86} :
\begin{equation}
\epsilon(\omega\rightarrow 0)=1+\frac{\omega^{2}_{p}}{\omega_{0}^{2}(E)}=
1+\kappa^{2}_{D}R_{loc}^{2}(E)\sim \left|\frac{E-E_{c}}{E_{c}}\right|^{-2\nu}
\label{epsil}
\end{equation}
where $\omega_{p}^{2}=4\pi ne^{2}/m$ is the square of plasma frequency,\
$\kappa^{2}_{D}=4\pi e^{2}N(E)$ is the square of inverse screening length of a
metal.

Thus the main results of self---consistent theory of localization coincide with
the main predictions
of elementary scaling theory of localization.\ Vollhardt and Wolfle had shown
\cite{VW82,WV82} that  equations of this theory and especially the main
differential equation of renormalization
group Eq. (\ref{GML}) for conductance may be explicitly derived from
self---consistency equations Eq. (\ref{M}) and Eq. (\ref{D}) reformulated for a
finite system by introduction of low---momentum cut---off at $k\sim 1/L$,\
where L is the system size.

The results considered up to now are valid for $\omega\rightarrow 0$.\
Self---consistent theory of
localization allows to study the frequency dependence of conductivity
(generalized diffusion coefficient) \cite{WV82}.\ At finite frequency the main
Eq. (\ref{D}) for the generalized
diffusion coefficient for $d=3$ can be rewritten as \cite{BelGold,WV82}:
\begin{equation}
\frac{D_{E}(\omega)}{D_{0}}=1-\left(\frac{E_{c}}{E}\right)^{1/2}+\frac{\pi}{2}\left(\frac{E_{c}}{E}\right)^{1/2}
\left\{-\frac{i\omega}{2\gamma}\frac{D_{0}}{D_{E}(\omega)}\right\}^{1/2}
\label{D3fr}
\end{equation}
which can be solved explicitly.\ With sufficient for our aims accuracy this
solution may be written
as:
\begin{equation}
D_{E}(\omega)\approx
\left\{ \begin{array}{l}
D_{E} \qquad \omega\ll\omega_{c} \quad E\geq E_{c} \mbox{ (Metal) } \\
D_{0}\left(-\frac{i\omega}{2\gamma}\right)^{1/3} \qquad \omega\gg\omega_{c}
\mbox{  (Metal and Insulator) }  \\
D_{E}\frac{-i\omega}{-i\omega+\frac{3D_{E}}{v_{F}^{2}}\omega_{0}^{2}(E)}
\qquad \omega\ll\omega_{c}\quad E<E_{c} \mbox{ (Insulator) }
\end{array} \right.
\label{difffreq}
\end{equation}
where (Cf. Eq. (\ref{omegc})):
\begin{equation}
\omega_{c}\sim 2\gamma[p_{F}\xi_{loc}]^{-d}\sim \frac{1}{N(E)\xi_{loc}^{d}}
\label{omegac}
\end{equation}
Here the renormalized diffusion coefficient:
\begin{equation}
D_{E}=\frac{D_{0}}{p_{F}\xi_{loc}(E)}
\label{D3dgen}
\end{equation}
At the mobility edge itself $\xi_{loc}(E=E_{c})=\infty$,\ so that
$\omega_{c}=0$ and we get the $\omega^{1/3}$---behavior (Cf. Eq.
(\ref{diffreqd})):
\begin{equation}
D_{E}(\omega)=D_{0}\left(-\frac{i\omega}{2\gamma}\right)^{1/3}
\label{diffreq3d}
\end{equation}
Note that $\omega_{c}$ is in fact determined by $D_{E}(\omega_{c})\sim
D_{E}\sim D_{0}(\omega_{c}/2\gamma)^{1/3}$.\ The meaning of the limit
$\omega\rightarrow 0$ used above (Cf.\ e.g.\ Eq. (\ref{Mtau})) is just that
$\omega\ll \omega_{c}$.\
In particular,\ the expression Eq. (\ref{rorodiff}) for
Gorkov---Berezinskii spectral density is valid only for
$\omega\ll \omega_{c}$.\ For $\omega_{c}\leq \omega\leq
2\gamma$,\ using Eq. (\ref{diffreq3d}) in Eq. (\ref{phidiff}) we get from Eq.
(A 16):
\begin{equation}
\ll\rho_{E}\rho_{E+\omega}\gg^{F}_{{\bf q}}=
\frac{\sqrt{3}}{2\pi}\frac{\alpha^{2/3}\omega^{1/3}q^{2}}
{\omega^{2}+\alpha^{2/3}\omega^{4/3}q^{2}+\alpha^{4/3}
\omega^{2/3}q^{4}}
\label{rorome}
\end{equation}
where $\alpha=D_{0}v_{F}/2\gamma=D_{0}l\sim [N(E)]^{-1}$,\ where the last
estimate is for $l\sim p_{F}^{-1}$.\ Eq. (\ref{rorome}) is valid also at the
mobility edge itself
where $\omega_{c}=0$.\ Obviously the correct estimate can
be obtained from Eq. (\ref{rorodiff}) by a simple replacement $D_{E}\rightarrow
D_{0}(\omega/\gamma)^{1/3}$.\ It should be noted that the self-consistent
theory approach to the frequency dependence of conductivity is clearly
approximate.\ For example
it  is unable to reproduce the correct $Re \sigma(\omega)\sim
\omega^{2}ln^{4}\omega$ dependence for $\omega\rightarrow 0$ in the insulating
state\cite{MottDav}.\ This is apparently related to its inability to take the
correct account of locally discrete nature of energy levels in Anderson
insulators (Cf.\ below).\ However this is unimportant for our purposes while
the general nature of frequency dependence at the mobility edge is apparently
correctly
reproduced.

In the following analysis we will also need a correlator
of {\em local} densities of states defined in Eq. (A 3).\ This correlator can
be expressed via two---particle
Green's function as in Eq. (A 15).\ Neglecting nonsingular for small $\omega$
and $q$ contribution from
the second term of Eq. (A 15) and far from the Anderson
transition (weak disorder) we can estimate the most
important contribution to that correlator from the diagram shown in Fig.
\ref{fig6}.\cite{BulSad86} The
same contribution comes from the diagram which differs
from that in Fig. \ref{fig6} by direction of electron lines in one of the
loops.\ Direct calculation gives:
\begin{eqnarray}
\ll\rho_{E}\rho_{E+\omega}\gg^{H}_{{\bf q}}\sim
\frac{N(E)}{\gamma^{2}}(\rho V^{2})^{2} Re \int d^{d}
{\bf Q}\frac{1}{-i\omega+D_{0}{\bf Q}^{2}}\frac{1}{-i\omega+D_{0}({\bf
Q+q})^{2}}  \nonumber \\
\sim
\frac{1}{N(E)}Re\frac{1}{D_{0}^{d/2}}\frac{1}{(-i\omega+D_{0}q^{2})^{2-d/2}}
\label{locdencor}
\end{eqnarray}
For the first time similar result for this correlator was
found for some special model by Oppermann and Wegner \cite{OppWeg79}.\ For
$d=3$ from Eq. (\ref{locdencor}) we
find:
\begin{equation}
\ll\rho_{E}\rho_{E+\omega}\gg^{H}_{{\bf q}}\sim
\frac{1}{N(E)D_{0}^{3/2}}\left\{\frac{D_{0}q^{2}}{\omega^{2}+(D_{0}q^{2})^{2}}+[\omega^{2}+(D_{0}q^{2})^{2}]^{-1/2}\right\}^{1/2}
\label{locdencor3d}
\end{equation}
It is obvious that for the estimates close to the mobility edge we can in the
spirit of self---consistent
theory of localization replace $D_{0}$ in Eq. (\ref{locdencor}) and Eq.
(\ref{locdencor3d}) by the
generalized diffusion coefficient $D(\omega)$.\ In
particular,\ for system at the mobility edge ($\omega_{c}=0$) $D_{0}\rightarrow
D_{0}(\omega/\gamma)^{1/3}$ in Eq. (\ref{locdencor3d}).

Surely,\ the self---consistent theory of localization is
not free of some difficulties.\ Apparently the main is
an uncontrollable nature of self---consistency procedure
itself.\ In more details these are discussed in Refs.\cite{Sad86,VW90}.\ Here
we shall concentrate only
on some problems relevant for the future discussion.\
{}From the definition of generalized diffusion coefficient
in Eq. (\ref{diffM}) it is clear that it may be a function of both $\omega$ and
${\bf q}$,\ i.e. it can
also possess spatial dispersion.\ Self---consistent
theory of localization deals only with the limit of
$D_{E}({\bf q}\rightarrow 0 \omega)$.\ At present it is
not clear whether we can in any way introduce spatial
dispersion into equations of self---consistent theory.\
Using scaling considerations the $q$---dependence of
$D_{E}({\bf q}\omega\rightarrow 0)$ can be estimated as
follows.\cite{LeeRam,Lee82} We have seen above that for
the system of finite size of $L\ll \xi_{loc}$ elementary
scaling theory of localization predicts the $L$---dependent diffusion
coefficient $D_{E}\approx (g_{c}/N(E))/L^{d-2}$ (Cf.\ Eq.\ (\ref{difL}) for
$d=3$).\ From simple dimensional considerations we can
try the replacement $L\rightarrow q^{-1}$ and get:
\begin{equation}
D_{E}(\omega\rightarrow 0 {\bf q})\approx
\left\{ \begin{array}{l}
D_{E} \mbox { for } q\xi_{loc}\ll 1 \\
\alpha q^{d-2} \mbox { for } q\xi_{loc}\gg 1
\end{array} \right.
\label{Dq}
\end{equation}
where $\alpha\sim g_{c}/N(E)\sim D_{0}l$ and $E\sim E_{c}$,\ $l^{-1}\sim
p_{F}$.\
Obviously an attempt to
incorporate such $q$---dependence into equations of
self---consistent theory of localization (like Eq. (\ref{M}) and Eq. (\ref{D}))
will radically change its structure.\ At the same time the $L$---dependence
like
$D_{E}\sim \alpha/L^{d-2}$ (for $L\ll \xi_{loc}$) can be
directly derived from Eq. (\ref{D}) as equations of
elementary scaling theory are derived from it \cite{VW82,WV82,VW90}.\ Thus the
foundations for the simple
replacement $L\rightarrow q^{-1}$ like in Eq. (\ref{Dq})
are not completely clear.\ More detailed analysis of wave number dependence of
diffusion
coefficient leading to Eq. (\ref{Dq}) was given by Abrahams and Lee
\cite{AbLee86} within the scaling approach.\ However,\ the complete solution of
this problem is apparently still absent. In a recent paper\cite{Suslov} it was
shown that Eq.(\ref{Dq}) actually contradicts the general localization criterion
of Berezinskii and Gorkov, from which it follows directly that at the 
localization transition the static diffusion coefficient $D(\omega=0,{\bf q})$
vanishes for all $q$ simultaneously. The detailed analysis performed in
Ref.\cite{Suslov} demonstrates the absense of any significant spatial dispersion
of diffusion coefficient on the scale of $q\sim \xi^{-1}$, while its presence
on the scale of $q\sim p_{F}$ is irrelevant for the critical behavior of the
system close to the Anderson transition. In fact in Ref.\cite{Suslov} it is 
claimed that the {\em exact} critical behavior at the mobility edge coincides
with that predicted by the self-consistent theory of localization. 

Finally we should like to stress that self---consistent
theory of localization can not be applied ``deep'' inside
localization region.\ Its derivation is based on a kind
of extrapolation of ``metallic'' expressions and it does
not take into account local discreteness of energy spectrum in the region of
localized states as discussed
in previous section.\ This is reflected in the form of
one---particle Green's function used in self---consistent
theory \cite{VW80,WV82,VW90,Sad86}.\ It does not describe
the effects of local level repulsion,\ though it does not
contradict it.\cite{Ohk82} Thus self---consistent theory
of localization can be applied within localized region
only until local energy spacing given by Eq. (\ref{spac})
is much smaller than other relevant energies of the problem under
consideration.\ In fact this always leads
to a condition of sufficiently large localization length
$R_{loc}$,\ i.e. the system must be in some sense close
to the mobility edge.

  \subsubsection{Quasi-Two-Dimensional Systems}
Self---consistent theory of localization for quasi---two---dimensional systems
was first analyzed by Prigodin and
Firsov \cite{PrFir84}.\ The electronic spectrum
of a quasi---two---dimensional system can be modelled by nearly---free
electrons
within highly conducting planes and tight---binding approximation for
interplane
electron transfer:
\begin{equation}
E({\bf p})-E_{F}=v_{F}(|{\bf p}_{\|}|-p_{F})-
w\varphi(p_{\bot})
\label{2dspectr}
\end{equation}
Here $w$ is the interplane transfer integral and
$\varphi(p_{\bot})=cos p_{\bot}a_{\bot}$,\ where $-\pi/a{\bot}\leq p_{\bot}
\leq \pi/a_{\bot}$.\ Then the
equations of self---consistent theory of localization
for anisotropic generalized diffusion coefficient take the following form
\cite{PrFir84}:
\begin{equation}
D_{j}(\omega)=D^{0}_{j}-\frac{1}{\pi N(E_{F})}\int \frac{d^{3}{\bf
q}}{(2\pi)^{3}} \frac{D_{j}(\omega)}
{-i\omega+D_{\|}(\omega)q_{\|}^{2}+D_{\bot}(\omega)(1-\varphi(q_{\bot}))}
\label{Dq2d}
\end{equation}
where $j={\|},{\bot}$,\ and $D^{0}_{\|}=v_{F}^{2}\tau/2,\
D^{0}_{\bot}=(wa_{\bot})^{2}\tau$ are inplane and interplane bare Drude
diffusion coefficients,\ $\tau$ is
the mean free time due to elastic scattering (disorder).\  This approach is in
complete correspondence with the analysis of  Wolfle  and Bhatt \cite{WB84} who
has shown that
the effects of anisotropy can be completely absorbed into anisotropic diffusion
coefficient.\
It can be seen that the initial anisotropy of diffusion
coefficient does not change as disorder grows up to the
Anderson transition and in fact we have only to find one
unknown ratio:
\begin{equation}
d(\omega)=\frac{D_{j}(\omega)}{D_{j}^{0}}=
\frac{\sigma_{j}(\omega)}{\sigma_{j}^{0}}
\label{alpha}
\end{equation}
which is determined by algebraic equation following from
Eq. (\ref{Dq2d}):
\begin{equation}
d(\omega)=1-\frac{1}{2\pi E_{F}\tau}ln \frac{2}
{[-i\omega\tau/d(\omega)]+(w\tau)^{2}+
[(-i\omega\tau/d(\omega))(-i\omega/d(\omega)+2w^{2}\tau^{2})]^{1/2}}
\label{alphalg}
\end{equation}
Due to a quasi---two---dimensional nature of the system
there is no complete localization for any degree of
disorder which is typical for purely two---dimensional
system.\ However the tendency for a system to become
localized at lower disorder than in isotropic case is
clearly seen.\  All states at the Fermi level become localized only for
$w<w_{c}$,\ where
\begin{equation}
w_{c}=\sqrt{2}\tau^{-1} exp(-\pi E_{F}\tau)
\label{wc}
\end{equation}
Thus the condition of localization is actually more
stringent than given by the simplest Ioffe---Regel type
estimate as in Eq. (\ref{IR2d}).\
For fixed $w$ the mobility edge appears at:
\begin{equation}
E_{F}=E_{c}=\frac{1}{\pi\tau}ln\left(\frac{\sqrt{2}}{w\tau}\right)
\label{mobed2d}
\end{equation}
Thus in case of strong anisotropy when $w\tau\ll 1$
localization can in principle take place even in case of $E_{F}\gg\tau^{-1}$,\
i.e. at relatively weak disorder.\
These estimates are in qualitative accordance with  Eq.(\ref{IR2d}),\ which is
valid in case of relatively
strong disorder $E_{F}\tau\sim 1$.

In the metallic phase close to the Anderson transition:
\begin{equation}
\sigma_{j}=\sigma^{0}_{j}\frac{E_{F}-E_{c}}{E_{c}}
\label{sigq2d}
\end{equation}
For $w\rightarrow 0$ we have $E_{c}\rightarrow \infty$
which reflects complete localization in two dimensions.\
We can also define inplane Drude conductivity at $E_{F}=E_{c}$ as a kind of a
``minimal metallic conductivity'' in this case as a characteristic conductivity
scale at the transition:
\begin{equation}
\sigma_{\|}^{c}=e^{2}N(E_{F})D^{0}_{\|}(E_{F}=E_{c})=
\frac{1}{\pi^{2}}\frac{e^{2}}{\hbar a_{\bot}}
ln\left(\frac{\sqrt{2}\hbar}{w\tau}\right)\approx
\frac{1}{\pi^{2}}\frac{e^{2}}{\hbar a_{\bot}}ln\left(\frac{E_{F}}{w}\right)
\label{mmcq2d}
\end{equation}
where we have used $N(E_{F})=m/(\pi a_{\bot}\hbar^{2})$,\
$m$ is inplane effective mass,\ and the last equality is
valid for $E_{F}\tau/\hbar\sim 1$,\ i.e. for a case of
sufficiently strong disorder.\ For the time being
we again use $\hbar$ explicitly.\ From these estimates it
is clear that inplane ``minimal conductivity'' is
logarithmically enhanced in comparison with usual estimates (Cf.\ Eq.
(\ref{mmc})).\ This logarithmic
enhancement grows as the interplane overlap of electronic
wave functions diminishes.\ Accordingly in case of small
overlap $(w\tau/\hbar\ll 1)$ this conductivity scale may
be significantly larger than $(3-5) 10^2 Ohm^{-1}cm^{-1}$
which is characteristic for isotropic systems.\ Thus in
quasi---two---dimensional case Anderson transition may take place at relatively
high values of inplane conductivity.\ For a typical estimate in a
high---$T_{c}$
system we can take something like $E_{F}/w > 10$ so that
the value of $\sigma^{c}_{\|}$ may exceed $10^{3} Ohm^{-1}cm^{-1}$.\ Obviously
these estimates are in qualitative
accordance with elementary estimates based upon Ioffe---Regel criterion of Eq.
(\ref{IR2d}) and Eq. (\ref{mmc2d}).\ Similar conclusions can be deduced from
the analysis presented in Ref.\cite{Econ89} where it was
shown by a different method that in case of anisotropic Anderson model the
growth of anisitropy leads to a significant drop of a critical 
disorder necessary to
localize all states in a conduction band.

Now let us quote some results for the frequency dependence of generalized
diffusion coefficient in
quasi---two---dimensional case which follow from the
solution of Eq. (\ref{alphalg})\cite{PrFir84}.\ We shall
limit ourselves only to the results valid close to the
mobility edge in metallic phase:
\begin{equation}
d(\omega)\approx
\left\{\begin{array}{l}
\frac{E_{F}-E_{c}}{E_{c}} \qquad \omega\ll \omega_{c} \\
(2\pi E_{F}w\tau^{2})^{-2/3}(-i\omega\tau)^{1/3} \qquad \omega_{c}\ll\omega\ll
w^{2}\tau \\
1-\frac{1}{2\pi E_{F}\tau} ln\left(\frac{1}{-i\omega\tau}\right) \qquad
w^{2}\tau\ll\omega\ll \tau^{-1}
\end{array} \right.
\label{alphafreq}
\end{equation}
where
\begin{equation}
\omega_{c}\approx [2\pi
E_{F}w\tau^{2}]^{2}\frac{1}{\tau}\left|\frac{E_{F}-E_{c}}{E_{c}}\right|^{3}
\label{omegstar}
\end{equation}
{}From these expressions we can see the crossover from
$\omega^{1/3}$---behavior typical for isotropic three---dimensional systems to
logarithmic dependence on frequency which is characteristic for
two---dimensional
systems.

\subsubsection{Self-Consistent Theory of Localization in Magnetic Field}
Early version of self-consistent theory of localization as proposed by
Vollhardt
and Wolfle was essentially based upon time-reversal invariance
\cite{VW80,WV82}.\ This property is
obviously absent in the presence of an external magnetic field.\ In this case
in addition to Eq. (\ref{phi}) we have to consider two---particle Green's
function in
particle---particle (Cooper) channel:
\begin{equation}
\Psi_{E}^{RA}({\bf q},\omega)=-\frac{1}{2\pi i}\sum_{{\bf p_{+}p'_{-}}}
<G^{R}({\bf p_{+},p'_{+}}, E+\omega)G^{A}({\bf -p'_{-},-p_{-}}, E>
\label{psi}
\end{equation}
which for small $\omega$ and ${\bf q}$ again has
diffusion---pole form like that of Eq. (\ref{phidiff}),\
but with {\em different} diffusion coefficient.\  Appropriate generalization of
self-consistent theory of localization was proposed by Yoshioka,\ Ono and
Fukuyama \cite{YOF81}.\ This theory is based on the following system of coupled
equations for relaxation kernels $M_{j}({\bf q},\omega)$,\ corresponding to
diffusion
coefficients in particle---hole and particle---particle
channels:
\begin{equation}
M_{1}=2i\gamma\left\{1-\frac{1}{\pi  N(E)}\sum_{n=0}^{N_{0}}\frac{2}{\pi
L_{H}}\int_{0}^{\sqrt{q_{0}^{2}-4m\omega_{H}(n+1/2)}}\frac{dq_{z}}{2\pi}
\frac{1}{\omega-\frac{D_{0}}{\tau
M_{2}}[q_{z}^{2}+4m\omega_{H}(n+1/2)]}\right\}
\label{M1}
\end{equation}
\begin{equation}
M_{2}=2i\gamma\left\{1-\frac{1}{\pi N(E)}\sum_{|{\bf
q}|<q_{0}}\frac{1}{\omega-D_{0}q^{2}/(\tau M_{1})}\right\}
\label{M2}
\end{equation}
Here $\omega_{H}=eH/mc$ is cyclotron frequency,\
$L_{H}=(c/eH)^{1/2}$ is magnetic length and $N_{0}=q_{0}^{2}/4m\omega_{H}$.\
These equations form the basis of self-consistent theory of localization in the
absence of time-reversal invariance and
were extensively studied in Refs.\cite{YOF81,OYF81,Kot88,Kot89,KS91a}.\
Alternative formulations of self---consistent theory in magnetic
field were given in Refs. \cite{Ting82,Ting85a,Ting85b,Bel84,Alba91}.\ All
these
approaches lead to qualitatively similar results.\ Here
we shall concentrate on formulations given in Ref. \cite{KS91a}.

Let us introduce the dimensionless parameter $\lambda=
\gamma/\pi E$ as a measure of disorder and generalized
diffusion coefficients in diffusion and Cooper channels
$D_{1}$ and $D_{2}$ defined as in Eq. (\ref{diffM}) with
$M$ replaced by $M_{1}$ and $M_{2}$ respectively.\ We
shall use dimensionless $d_{j}=D_{j}/D_{0}$ ($j=1,2$) in
the following.

We are mainly interested in diffusion coefficient in the Cooper channel, which
as we shall see defines the upper critical field of a superconductor.\ Both
this coefficient as well as the usual one are determined by the following
equations which follow from Eq. (\ref{M1})  and Eq. (\ref{M2}) after the use of
Poisson summation over Landau levels in the first equation which allows one to
separate the usual diffusion coefficient independent of magnetic field and the
field---dependent part:
\begin{equation}
\left\{ \begin{array}{ll}
d_{1}=(1+\frac{3\lambda  -\delta_{2} -\Delta_{2}}{d_{2}})^{-1} & \\
d_{2}=(1+\frac{3\lambda  -\delta_{1}}{d_{1}})^{-1}
\end{array}
\right.
\label{d1d2}
\end{equation}
where
\begin{equation}
\delta_{j}=(3/2\pi\lambda)^{3/2}(-i\omega/E)^{1/2}d_{j}^{-1/2}
\label{delta}
\end{equation}
and
\begin{equation}
\Delta_{2}=-3\lambda  \sum_{p=1}^{\infty}(-1)^{p}\int_{0}^{1}dx
2\int_{0}^{\sqrt{1-x}}dy\frac{\cos(2\pi
px_{0}^{2}/c^{2})}{y^{2}+x+3/2\pi\lambda(-i\omega/E)/(d_{2}x_{0}^{2})}
\label{Delta}
\end{equation}
where $c=(2\omega_{H}/E)^{1/2}$.In the following we have to solve Eqs.
(\ref{d1d2}) for the case of small $\delta_{j}$ and $\Delta_{2}$.\ Limiting
ourselves to terms linear in $\delta_{1}$,\ $\delta_{2}$ and $\Delta_{2}$ we
obtain:
\begin{equation}
\frac{d_{1}}{d_{2}}=1+\frac{\Delta_{2}}{1+3\lambda }
\label{d1/d2}
\end{equation}
Using Eq. (\ref{d1/d2}) in Eqs. (\ref{d1d2}) we get an
equation for diffusion coefficient in Cooper channel:
\begin{equation}
d_{2}=1-3\lambda +\delta_{2}+\frac{3\lambda }
{1+3\lambda }\Delta_{2}
\label{d2}
\end{equation}
Introducing $\Delta_{1}$ which differs from $\Delta_{2}$
by the replacement of $d_{2}$ by $d_{1}$ we can write
down also the approximate equation for the usual diffusion coefficient:
\begin{equation}
d_{1}=1-3\lambda  +\delta_{1}+\frac{1}{1+3\lambda}\Delta_{1}
\label{d1}
\end{equation}
In the absence of magnetic field ($\Delta_{1}=\Delta_{2}=0$) Eq. (\ref{d2}) and
Eq. (\ref{d1}) are the same and lead to standard results of
self---consistent theory quoted above.\ Eq. (\ref{d2})
can be written as:
\begin{equation}
2mD_{2}=^{+}_{-}(\frac{\omega_{c}}{E})^{1/3}+(-\frac{i\omega}{E})^{1/2}(2mD_{2})^{-1/2}+\frac{3\lambda}{1+3\lambda}\Delta_{2}
\label{D2}
\end{equation}
where $+$ corresponds to metallic,\ and $-$ to insulating
phases,\ while characteristic frequency
\begin{equation}
\omega_{c}=\left(\frac{|1-3\lambda|}{(3/2)\pi\lambda}\right)^{3}E
\label{omegacr}
\end{equation}
can be considered as a measure of disorder and separate
regions with different frequency dependencies of diffusion coefficient.

Neglecting in Eq. (\ref{Delta}) terms oscillating with
magnetic field (these oscillations are connected with
sharp cut---off in momentum space used above and disappear for smooth
cut---off) we get:
\begin{equation}
\Delta_{2}=-(2\omega_{H}/E)^{1/2}\sum_{p=1}^{\infty}\frac{(-1)^{p}}{p^{1/2}}f(2\pi\rho\kappa)
\label{Delta2}
\end{equation}
where
\begin{equation}
f(y)=\sqrt{2/\pi}\int_{0}^{\infty} \frac{cos(t)dt}{\sqrt{t+y}}; \qquad
\kappa=\frac{-i\omega/E}{2\omega_{H}/E}\frac{1}{2mD_{2}}
\end{equation}
This gives:
\begin{equation}
\Delta_{2}=
\left\{ \begin{array}{l}
W(2\omega_{H}/E)^{1/2} \qquad |\kappa|\ll 1 \\
\frac{1}{48}\left((-i\omega/E)\frac{1}{2mD_{2}}\right)^{-3/2}(2\omega_{H}/E)^{2} \qquad |\kappa|\gg 1
\end{array} \right.
\label{DD2}
\end{equation}
where $W=-\sum_{p=1}^{\infty}(-1)^{p}/p^{1/2}\approx 0.603$.

Solutions of Eq. (\ref{D2}) for different limiting cases
can be found in Ref. \cite{KS91a}.\ Comparison of Eq. (\ref{d1}) and Eq.
(\ref{d2}) shows that the usual diffusion coefficient $D_{1}$ is given by the
same
expressions as $D_{2}$ with the replacement of the
coefficient $3\lambda/(1+3\lambda)$ before the field---dependent correction by
$1/(1+3\lambda)$.\ Here we only
quote the results for $D_{2}$ in case of $\omega_{c}/E\ll
(\omega_{H}/E)^{3/2}$, valid close to the transition in the absense of
magnetic field:
\begin{equation}
D_{2}=\frac{1}{2m}\left\{^{+}_{-}(\omega_{c}/E)^{1/3}+
\left[\frac{3\lambda}{1+3\lambda}\right]W(2\omega_{H}/E)^{1/2}\right\}\approx
\frac{1}{4m}W(2\omega_{H}/E)^{1/2}
\qquad \omega\ll \omega_{c}^{\star}
\label{D22}
\end{equation}
\begin{equation}
D_{2}=\frac{1}{2m}\left\{(-i\omega/E)^{1/3}+\frac{2}{3}\left[\frac{3\lambda}{1+3\lambda}\right]\frac{1}{48}\frac{(2\omega_{H}/E)^{2}}{(-i\omega/E)}\right\} \qquad \omega\gg \omega_{c}^{\star}
\label{D222}
\end{equation}
where $\omega_{c}^{\star}=(W/2)^{3}(2\omega_{H}/E)^{3/2}E$.

Note that for high frequencies larger than $\omega_{c}^{\star}$ the correction
term becomes quadratic in field which differs from usual square root
behavior at low frequencies.

It is easy to see that in the absence of the external magnetic field these
equations reduce to the usual self-consistency equation as derived by Vollhardt
and Wolfle with a single relaxation kernel.\

Let us finally quote some results for the purely two-dimensional case
\cite{KS93}.
Self-consistent equations for the diffusion coefficients
take now the following form:
\begin{eqnarray}
\frac{D_{0}}{D_{2}}=1+\frac{1}{\pi N(E)}\sum_{|{\bf q}| <
q_{0}}\frac{1}{\omega+D_{1}q^{2}} \nonumber\\
\frac{D_{0}}{D_{1}}=1+\frac{1}{\pi N(E)}\sum_{|{\bf k}| <
q_{0}}\frac{1}{\omega+D_{2}k^{2}}
\label{sisdif}
\end{eqnarray}
where $k^{2}=4m\omega_{H}(n+\frac{1}{2})$,\ and we assume that
$\omega$ here is imaginary (Matsubara) frequency,\ which simplifies the
analysis.\ Actually only the dependence on the Matsubara's frequencies are
important for further applications to superconductivity.

Introduce again the dimensionless diffusion coefficients
$d_{1}=\frac{D_{1}}{D_{0}}$, $d_{2}=\frac{D_{2}}{D_{0}}$,
so that Eqs.(\ref{sisdif}) are rewritten as:
\begin{eqnarray}
\frac{1}{d_{2}}&=&1+\frac{\lambda}{d_{1}}ln(1+d_{1}\frac{1}{2 \omega\tau})
\nonumber\\ \frac{1}{d_{1}}&=&1+\frac{\lambda}{d_{2}}\sum_{n=0}^{N_{0}}
\frac{1}{n+\frac{1}{2}+\frac{\omega}{4m\omega_{H}D_{0}} \frac{1}{d_{2}}}
\label{sisdif1}
\end{eqnarray}
where $N_{0}=\frac{1}{8m\omega_{H}D_{0}\tau}$---is the number of Landau
levels below the cutoff.
We assume that the magnetic field is low enough,\ so that $N_{0}\gg1$,\ i.e.
\begin{equation}
H\ll\frac{\Phi_{0}}{D_{0}\tau}
\label{litlH}
\end{equation}

With sufficient for further use accuracy we can write down the following
solution for the diffusion coefficient in Cooper channel:

For weak magnetic field  $\omega_{H}\ll\frac{\lambda e^{-1/\lambda}}{\tau}$
\begin{equation}
d_{2}=
\left\{\begin{array}{ll}
1 & \mbox{ for } \omega\gg\frac{e^{-1/\lambda}}{2\tau} \\
2\omega\tau e^{1/\lambda} & \mbox{ for }\omega\ll\frac{e^{-1/\lambda}}{2\tau}
\end{array}
\right.
\label{d2litl}
\end{equation}
and we can neglect the magnetic field influence upon diffusion.

For larger fields $\omega_{H}\gg\frac{\lambda e^{-1/\lambda}}{\tau}$
\begin{equation}
d_{2}=
\left\{\begin{array}{ll}
1 & \mbox{ for } \omega\gg\frac{e^{-1/\lambda}}{2\tau} \\
\frac{1}{\lambda ln(1/2\omega\tau)} & \mbox{ for
}\frac{e^{-1/\lambda^{2}lnQ}}{2\tau}\ll\omega\ll\frac{e^{-1/\lambda}}{2\tau} \\
2\omega\tau\lambda lnQ e^{1/\lambda^{2}lnQ} & \mbox{ for
}\omega\ll\frac{e^{-1/\lambda^{2}lnQ}}{2\tau}
\end{array}
\right.
\label{d2full}
\end{equation}
where $Q=\frac{\pi\gamma\lambda}{\tau\omega_{H}}$, $\gamma\approx 1,781$.

Here we neglect the magnetic field corrections small in comparison to the
$d_{2}$ value in the absence of magnetic field given by Eq.(\ref{d2litl}).



\subsection{Phase Transition Analogy and Scaling for Correlators}

Scaling description of a system close to Anderson transition can be developed
also on the basis
of some analogies with usual phase transitions \cite{Sad81,Sad86,LeeRam}.\ Most
successful in
this respect is an approach initially proposed by Wegner
\cite{Weg79a,Weg79b,Weg82}.

Let us consider Eq. (\ref{rorodiff}) and Eq. (\ref{locdencor}) which define
basic electronic correlators (spectral densities) in a disordered system.\ For
the metallic region we can write:
\begin{equation}
K_{F}({\bf q}\omega)\equiv N(E)\ll \rho_{E}\rho_{E+\omega}\gg^{F}_{{\bf q}}\sim
Re \frac{N(E)}{-i\omega+D_{E}q^{2}}
\label{Kf}
\end{equation}
\begin{equation}
K_{H}({\bf q}\omega)\equiv N(E)\ll \rho_{E}\rho_{E+\omega}\gg^{H}_{{\bf q}}\sim
Re\frac{D_{E}^{d/2}}{(-i\omega+D_{E}q^{2})^{2-d/2}}
\label{Kh}
\end{equation}
Wegner has noted \cite{Weg79b,OppWeg79} that these expressions are in some
sense similar
to analogous expressions for transversal and longitudinal susceptibilities of a
ferromagnet \cite{PP}:
\begin{equation}
\chi_{\bot}({\bf q})=\frac{M}{H+\rho_{s}q^{2}}
\label{chitr}
\end{equation}
\begin{equation}
\chi_{\|} ({\bf q})\sim \frac{1}{(H+\rho_{s}q^{2})^{2-d/2}}
\label{chilong}
\end{equation}
where $M$ is magnetization,\  $H$ is external magnetic field and $\rho_{s}$ is
the spin---stiffness
coefficient.\ Comparing Eqs. (\ref{Kf}) with Eq. (\ref{chitr}) and Eq.
(\ref{Kh}) with Eq. (\ref{chilong}) we can write down a correspondence between
electron diffusion in a random
system and a ferromagnet as given in Table.\ I.

Now we can use the main ideas of scaling approach in the theory of critical
phenomena \cite{Kad,WK,PP,SKM} and formulate similar  expressions for
electronic system close to the Anderson transition.\ As was noted above
scaling theory is based upon an assumption that a singular behavior of physical
parameters of a system
close to a phase transition appears due to large scale
(long wave---length) fluctuations of order---parameter (e.g.\ magnetization)
close to critical temperature $T_{c}$.\ Scaling hypothesis claims that singular
dependencies on $T-T_{c}$ reflect the divergence of
correlation length of these fluctuations $\xi$ and this
length is the only relevant length---scale in the critical region.\ Scaling
approach is based upon an idea
of scale transformations and dimensional analysis.\ Under
the scale transformation the spatial interval $\Delta x$
changes to $\Delta x'$,\ according to:
\begin{equation}
\Delta x \rightarrow \Delta x'=s^{-1}\Delta x
\label{scl}
\end{equation}
Accordingly for the wave---vector:
\begin{equation}
{\bf q}\rightarrow {\bf q'}=s{\bf q}
\label{sclq}
\end{equation}
Scaling dimension \cite{SKM} of a physical quantity $A$ is equal to $\lambda$
if under scale transformations
defined by Eq. (\ref{scl}) and Eq. (\ref{sclq}) we get:
\begin{equation}
A\rightarrow A'=As^{\lambda}
\label{Alamb}
\end{equation}
Scaling dimensions for the main characteristics of a
ferromagnet are given in terms of standard critical
exponents \cite{SKM} in Table II.

Correlation length of the theory of critical phenomena
behaves like:
\begin{equation}
\xi\sim |T-T_{c}|^{-\nu}
\label{xicorr}
\end{equation}
The knowledge of scaling dimension of a given physical
quantity allows to determine its dependence on $\xi$,\
i.e.\ on $T-T_{c}$.\ For example magnetization $M$ behaves according to Table
II as:
\begin{equation}
M\sim \xi^{-1/2(d-2+\eta)}\sim |T-T_{c}|^{\beta}
\label{magnet}
\end{equation}
where the critical exponent of magnetization equals
\begin{equation}
\beta=\frac{1}{2}\nu(d-2+\eta)
\label{betexp}
\end{equation}
Magnetic susceptibility is given by:
\begin{equation}
\chi({\bf q}, T-T_{c})=\xi^{2-\eta}g(q\xi)
\label{suscep}
\end{equation}
where $g(x)$ is some universal function,\ such that
$g(0)\sim const$,\ $g(x\rightarrow \infty)\sim x^{-(2-\eta)}$.\ From Eq.
(\ref{suscep}) we get standard results:
\begin{equation}
\chi(0, T-T_{c})\approx \xi^{2-\eta}g(0)\sim |T-T_{c}|^{-\gamma}
\label{susc}
\end{equation}
where $\gamma=(2-\eta)\nu$ is the susceptibility exponent.\ Analogously:
\begin{equation}
\chi({\bf q},T=T_{c})\sim q^{-2+\eta}
\label{susq}
\end{equation}
Here $\eta$ is sometimes called Fisher's exponent.

It is easy to see that Eq. (\ref{suscep}) is equivalent
to scaling relation ($H$---dependence is taken from
Table II):
\begin{equation}
\chi(s{\bf q},s^{-1}\xi,s^{1/2(d+2-\eta)}H)=s^{-(2-\eta)}\chi({\bf q},\xi,H)
\label{suscscal}
\end{equation}
It is convenient to make transformation $|T-T_{c}|\rightarrow b|T-T_{c}|$ so
that $\xi\rightarrow b^{-\nu}\xi$ which is equivalent to the choice of
$s=b^{\nu}$.\ Then Eq. (\ref{suscscal}) transforms to:
\begin{equation}
\chi(b^{\nu}{\bf q},b^{-\nu}\xi,b^{\nu(d+2-\eta)/2}H)=
b^{-\gamma}\chi({\bf q},\xi,H)
\label{sussclg}
\end{equation}
Finally note that close to Curie point the spin---stiffness coefficient
$\rho_{s}$ satisfies the so---called Josephson relation \cite{PP}:
\begin{equation}
\rho_{s}\sim |T-T_{c}|^{(d-2)\nu}
\label{Jos}
\end{equation}
and tends to zero as $T\rightarrow T_{c}$ from within the
condensed phase.

Consider now the analogy formulated in Table I.\ Density
of states $N(E)$ is nonsingular at the mobility edge \cite{Th74,Sad86}.\ Then
considering $N(E)$ as an
analog of magnetization $M$ we have to assume $\beta=0$,\
i.e.\ at the localization transition:
\begin{equation}
\eta=2-d
\label{eta}
\end{equation}
and the ``order---parameter'' $N(E)$ is nonsingular  at
the transition $E=E_{c}$.\ Accordingly we have $\gamma=d\nu$.\ Josephson
relation Eq. (\ref{Jos}) now
takes the form:
\begin{equation}
D_{E}\sim |E-E_{c}|^{(d-2)\nu}
\label{WegDiff}
\end{equation}
i.e.\ in fact is equivalent to Wegner's relation for
conductivity given by Eq. (\ref{weg}).\ Correlation length exponent $\nu$
remains unknown.

For electronic correlators of Eq. (\ref{Kf}) and Eq. (\ref{Kh}) from Eq.
(\ref{sussclg}) we obtain scaling
relations \cite{Weg79b,Weg82}:
\begin{equation}
K_{F,H}(b^{\nu}{\bf q},b^{d\nu}\omega,b(E-E_{c}))=
b^{-d\nu}K_{F,H}({\bf q},\omega,E-E_{c})
\label{sclK}
\end{equation}
Taking $\nu=1/(d-2)$ form Eq. (\ref{crexp}) for $d=3$ and
$E=E_{c}$ (i.e.\ at the mobility edge itself) we get from
Eq. (\ref{sclK}):
\begin{equation}
K_{F,H}(b{\bf q},b^{3}\omega)=b^{-3}K_{F,H}({\bf q}\omega)  \label{Kscl}
\end{equation}
which is equivalent to:
\begin{equation}
K_{F,H}({\bf q}\omega)=L_{\omega}^{3}F_{F,H}(qL_{\omega})
\label{Ksclg}
\end{equation}
where $F_{F,H}(x)$ is some universal function and we
introduced characteristic length:
\begin{equation}
L_{\omega}=[\omega N(E)]^{-1/3}
\label{Lomg}
\end{equation}
Note that the same scaling dependence follows e.g.\ for
$K_{H}({\bf q}\omega)$ from Eq. (\ref{locdencor}) or
Eq. (\ref{locdencor3d}) after a simple replacement of
$D_{0}$ by a diffusion coefficient given by:
\begin{equation}
D_{E=E_{c}}({\bf q}\omega)=L^{-1}_{\omega}f(qL_{\omega})
\label{Dscl}
\end{equation}
where $f(x\rightarrow 0)\rightarrow 1$ and $f(x\rightarrow \infty)\rightarrow
x$.\ In particular in
the limit of $qL_{\omega}\rightarrow 0$ we get $F(x)=(1+x^{4})^{-1/4}$ and the
replacement $D_{0}\rightarrow D_{0}(\omega/\gamma)^{1/3}$ mentioned
in connection with Eq. (\ref{locdencor3d}) is valid.\ On
the other hand from Eq. (\ref{eta}) it follows that at
$\omega=0$ we get from Eq. (\ref{susq}):
\begin{equation}
K({\bf q},\omega=0,E=E_{c})\sim q^{-d}
\label{Kmobed}
\end{equation}
which is equivalent to Eq. (\ref{rorodiff}) if we take
$D_{E=E_{c}}(\omega=0,{\bf q})=\alpha q^{d-2}$ (Cf.\ Eq. (\ref{Dq})).

Microscopic justification of this scaling hypothesis can be done with one or
another variant of field---theory
approach based upon nonlinear $\sigma$---model \cite{Weg79a,Weg79b,Weg82}.\
There exist
several alternative schemes of ``mapping'' of the problem of an electron in a
random field onto
field---theoretic formalism of nonlinear $\sigma$---models
\cite{Weg79,ELK80,SchafWeg80,McS81,Schaf82,Ef83}.\ The main physical
justification of this
approach is to represent an effective Hamiltonian of an electronic system in a
form similar to
analogous Hamiltonian of Heisenberg ferromagnet below Curie point:
\begin{equation}
{\cal H}=\frac {1}{2}\left(\frac{\partial {\bf M}}{\partial
x_{\alpha}}\right)^{2}-{\bf HM};
\qquad {\bf M^{2}}=const
\label{sigmafer}
\end{equation}
As a result an effective Hamiltonian for an electron in a random field in terms
of interacting modes
responsible for the critical behavior close to mobility edge appears.\
Following Ref.\cite{ELK80}
we can introduce an ``order---parameter'' as a $2n\times 2n$ matrix $\hat Q$
($n$---integer).\
Every matrix element of $\hat Q$ can be represented as:
\begin{equation}
Q_{ij}=
\left( \begin{array}{cc}
D_{ij} & \Delta_{ij} \\
-\Delta_{ij}^{\star} & D_{ij}^{\star}
\end{array} \right)
\label{Qij}
\end{equation}
where $D_{ij}=D_{ji}^{\star}$ and $\Delta_{ij}=-\Delta_{ji}^{\star}$,\
i.e.\ are elements of Hermitian and antisymmetric matrices respectively.\
Analogously ${\bf M}^{2}=const$ in a ferromagnet $\hat Q$---matrix must satisfy
the condition:
\begin{equation}
\hat Q^{2}=1; \qquad Tr \hat Q=0
\label{TrQ}
\end{equation}
Effective Hamiltonian for diffusion modes takes the following form
\cite{Weg79a,Weg79b}:
\begin{equation}
{\cal H} = D_{0} Tr(-i\nabla \hat Q)^{2}-i\omega Tr\hat \Lambda\hat Q
\label{Heff}
\end{equation}
Here $\hat\Lambda$ is the diagonal matrix with first $n$
elements equal to $1$ and the remaining $n$ are $-1$.\
Correlation function of $D$---elements corresponds to
diffuson,\ while that of $\Delta$---elements to Cooperon.\ Parameter $n$ should
be put equal to zero at
the end of calculations in the spirit of famous ``replica
trick'' in the theory of disordered systems \cite{SKM,Sad81}.

This formalism is useful also for the analysis of different kinds of external
perturbations,\ such as
external magnetic field,\ magnetic impurities,\ spin---orbital scattering
etc.\cite{ELK80}.\ Standard methods of renormalization group using perturbation
theory over
$(p_{F}l)^{-1}\ll 1$ reproduces all the main results of
elementary scaling theory of localization,\ including the qualitative form of
$\beta$---function as in Fig. \ref{fig3}.\ However the formalism of
$\sigma$---model
approach is quite complicated and practically does not
allow to get explicit expressions for physical characteristics of the system,\
especially in localized phase.

Many problems of fundamental nature still remain unresolved.\ Most important
are questions concerning the
role of nonperturbative contributions close to the
mobility edge \cite{Sad81,Sad86,KL85,Ef83,AKL86}.\ Note
especially strong criticism of one---parameter scaling in
Refs.\cite{KL85,AKL86}.\ Among a lot of results obtained
within $\sigma$---model approach we wish to mention an
important paper by Lerner \cite{Ler88} where a distribution function for local
density of states in a
system close to Anderson transition was determined and
shown to be essentially non---Gaussian.

For our future analysis it is important to stress
that in most cases the results of $\sigma$---model approach practically
coincide with predictions of self---consistent theory of localization which
also neglects all
nonperturbative  effects,\ except those determined by  some infinite
resummation of diagrams.\ It must be stressed that self---consistent theory is
based upon some uncontrollable ad hoc assumptions and in this respect it is not
as well justified as $\sigma$---model approach.\
However this simple theory as we have seen above allows practical calculation
of any interesting characteristic of an electronic system close to mobility
edge including the localized phase.



\subsection{Interaction Effects and Anderson Transition}

The main unsolved problem of the theory of metal---insulator transition in
disordered systems is the role
of electron---electron interactions.\ The importance of
interactions for this problem is known for a long time \cite{Mott74}.\ In
recent years the decisive importance
of interactions was revealed in the theory of ``dirty metals''
\cite{AltAr79,AltAr82,AltAr85},\ as well as in
the concept of Coulomb gap at the Fermi level of strongly localized electrons
\cite{EfrShkl75,Efr76,ShklEfr79,EfrShkl85}.\
We have already briefly discussed Coulomb gap.\ It appears for strongly
localized
states.\ In case of ``dirty metals'' diffusive nature of electronic transport
leads to special interference effects between Coulomb interaction and disorder
scattering \cite{AltAr79,AltAr85}.\ Most important is an appearance of some
kind of a precursor to Coulomb gap already in metallic state.\ It is connected
with simple exchange correction to electron self---energy  
(cf. Fig.\ref{fig7}) which leads to the
following cusp---like correction to one---particle density of states in case of
the screened Coulomb interaction in three---dimensional system \cite{AltAr79}:
\begin{equation}
\delta N(E) = \frac{|E-E_{F}|^{1/2}}{2\sqrt{2}\pi^{2}D_{0}^{3/2}}
\label{AAcusp}
\end{equation}
where $D_{0}$ is the usual Drude diffusion coefficient.\ In two---dimensional
case this correction is logarithmic \cite{AltArLee,AltAr85}.\ General belief is
that this cusp
somehow transforms into Coulomb gap as system moves from metal to insulator.\
However,\ up to now there is no complete solution for this problem.

Early attempt to describe electron---electron interactions in Anderson
insulators in a Fermi---liquid like scheme was undertaken in Ref.
\cite{FlAn80}.\ Simple generalization
of the theory of ``dirty metals'' \cite{AltAr79,AltAr82,AltAr85} along the
lines of self---consistent theory of localization was proposed in
Refs.\cite{KazSad83,KazSad84,Sad86}.\ However the most
general approach to this problem was introduced by McMillan \cite{McM81} who
proposed to describe the
metal---insulator transition in a disordered system by
a scaling scheme similar in spirit to elementary scaling
theory of localization of noninteracting electrons discussed above.\ He
formulated a simple system of coupled differential equations of renormalization
group
for two effective ``charges'':\ dimensionless conductance
$g$ and single---particle density of states $N(E)$.\ Later it was realized that
this simple scheme can not be
correct because it assumed for conductivity the relation
like Eq. (\ref{condM}) with density of states while the
correct Einstein relation for interacting system contains
electron compressibility $dn/d\zeta$ ($\zeta$ is chemical
potential)\cite{Kubo57,Lee82,Fin83},\ which is not renormalized close to the
metal---insulator transition as opposed to density of states.\ The most
comprehensive
approach to a scaling description of metal---insulator
transition in disordered systems was formulated by Finkelstein
\cite{Fin83,Fin84a,Fin84b,Fin90}.\ Unfortunately
more or less explicit solutions were only obtained neglecting the scattering
and interaction processes in
Cooper channel which are mainly responsible,\ as we have
seen above,\ for localization itself.\ Some attempts in this direction were
undertaken
only in Ref.\cite{Fin84b}.\ This approach is
still under very active discussion
\cite{DiCast84,KotLee87,Kotl87,KirkBel89,BelKirk89,KirkBel90,BelKirk90,BelKirk91,BelKirk94} and demonstrate fundamental
importance of interactions.\ However the problem is still
unresolved and most of these works consider only the
metallic side of transition with no serious attempts to analyze the insulating
state.

Below we consider only some qualitative results of this
approach,\ following mainly Refs.\cite{KotLee87,Kotl87}.\
Fermi liquid theory survives the introduction of disorder
\cite{Noz66},\ although with some important corrections
\cite{AltAr79,AltAr85},\ and is actually valid up to metal---insulator
transition \cite{Fin83,Fin84a,KotLee87,Kotl87}.

In the absence of translation invariance there is no
momentum conservation and we have to use some unknown exact eigenstate
$\phi_{\nu}({\bf r})$ representation for electrons in random field to
characterize quasiparticles with energies $\varepsilon_{\nu}$ (Cf.\
Ref.\cite{Migd}).\ The free energy as a functional
of quasi---particle distribution function $n_{s}(\varepsilon_{\nu},{\bf r})$
($s$---spin variable) is written as in usual Fermi liquid theory:
\begin{equation}
F\{n_{s}(\varepsilon_{\nu},{\bf r})\}=\sum_{s,\nu}\int d{\bf r}
n_{s}(\varepsilon_{\nu}{\bf
r})(\varepsilon_{\nu}-\zeta)+\frac{1}{2}\sum_{ss'}\int d^{d}{\bf r} \delta
N_{s}({\bf r})\delta N_{s'}({\bf r})f_{ss'} \label{FLen} \end{equation} where
$N_{s}=\sum_{\nu}n_{s}(\varepsilon_{\nu}{\bf r})$ is the total density per
spin and $f_{ss'}=f^{s}+ss'f^{a}$ is the quasi---particle interaction
function.\ The angular dependence of $f$---function in dirty case can be
neglected,\ because $n_{s}(\varepsilon_{\nu}{\bf r})$ is assumed to describe
electrons on distances larger than mean free path there
only $s$---wave scattering is important and Fermi---liquid interaction
becomes point---like.\ In an external spin dependent field $V_{s}$ the
quasi---particle distribution function obeys a kinetic equation:
\begin{equation}
\frac{\partial}{\partial t}n_{s}-D\nabla^{2}n_{s}+
\left(\frac{\partial n_{s}}{\partial \varepsilon}\right)
(-D\nabla^{2})[V_{s}+\sum_{s'}f_{ss'}N_{s'}]=0
\label{FLkin}
\end{equation}
where $D$ is quasi---particle diffusion coefficient.\
Eq. (\ref{FLkin}) is obtained from usual Fermi---liquid
kinetic equation \cite{Migd} by replacing $v_{F}\partial/\partial {\bf r}$ by
$-D\nabla^{2}$ which
reflects a crossover from ballistic to diffusive transport in disordered
system.\ Solving Eq. (\ref{FLkin}) for density---density and spin---spin
response functions one gets:\cite{Fin83,Fin84a,DiCast84}
\begin{equation}
\chi_{\rho}({\bf
q}\omega)=\frac{dn}{d\zeta}\frac{D_{\rho}q^{2}}{D_{\rho}q^{2}-i\omega}
\label{FLpol}
\end{equation}
\begin{equation}
\chi_{s}({\bf q}\omega)=\frac{\chi D_{s}q^{2}}{D_{s}q^{2}-i\omega}
\label{FLchi}
\end{equation}
where $dn/d\zeta=N(E_{F})/(1+F^{s}_{0})$,\
$\chi=N(E_{F})\mu_{B}^{2}/(1+F^{a}_{0})$ ($\mu_{B}$ is Bohr's magneton) and
\begin{equation}
D_{\rho}=D(1+F^{s}_{0})
\label{Dro}
\end{equation}
\begin{equation}
D_{s}=D(1+F^{a}_{0})
\label{Ds}
\end{equation}
Landau parameters $F^{s,a}_{0}$ are defined by
\begin{equation}
N(E_{F})f^{s}=F^{s}_{0} \qquad N(E_{F})f^{a}=F^{a}_{0}
\label{FF}
\end{equation}
Here $N(E_{F})$ is quasi-particle density of states at the Fermi level (for
both spin directions).\ If we neglect Fermi---liquid renormalization
effects Eq. (\ref{FLpol}) reduces to Eq. (\ref{denden}).\ Conductivity is
given now by $\sigma=e^{2}D(dn/d\zeta)$.

As system moves towards metal---insulator transition Hubbard---like interaction
of electrons close to a given
impurity site becomes more and more important.\ It is known for a long time
\cite{Mott74,Sad86} that this interaction leads to the appearance of a band of
single---occupied states just below the Fermi level of
a system on the dielectric side of Anderson transition.\
These states actually simulate paramagnetic centers and
lead to Curie---like contribution (diverging as temperature $T\rightarrow
0$)\cite{Mott74,Sad86}.\ Thus on the metallic side of transition static
magnetic susceptibility $\chi$ is expected to diverge since it is infinite
(at $T=0$) on the insulating side.\ At the same time $dn/d\zeta$ remains
finite.\ Therefore $D_{s}/D_{\rho}=(dn/d\zeta)/\chi$ goes to zero,\ i.e.\
{\em spin diffusion is much slower than charge diffusion} close to
metal---insulator transition.\ This fact was first noted in Ref.
\cite{Fin84b} where it was assumed that it leads to a possibility of local
magnetic appearing in metallic phase before a transition.\ It is interesting
to note that the slowing down of spin diffusion due to ineractions was
actually discovered long before \cite{FulLut} it appeared in the context of
interaction picture of metal---insulator transition.\ This idea was further
elaborated in Refs. \cite{KirkBel90,BelKirk90,BelKirk91},\ where extensive
discussion of this magnetic transition was given.\ There is an intersting
problem why these localized moments are not quenched by the Kondo effect.\ This
can apparently be explained by the local fluctuations of Kondo temperature due
to fluctuations of local density of states induced by disorder\cite{Dobr92} .\
The resulting distribution of Kondo temperatures is shown to be singular enough
to induce diverging magnetic susceptibility as $T\rightarrow 0$.

The idea of paramagnetic moments appearing already in
metallic phase apparently can much simplify the analysis
of metal---insulator transition and allow its description
by equations of elementary scaling theory of localization
\cite{AltAr83a,AltAr83b,AltAr85}.\ In general case electron interactions in
diffusion channel can be classified by total spin of an electron and hole $j$
\cite{AltAr85}.\ It can be shown that all interaction
corrections with $j=0$ do not depend on electron---electron coupling constant
(charge) and are universal
\cite{AltAr85}.\ If paramagnetic scattering is operating
in the system it dumps scattering processes in Cooper (localization) channel
\cite{Lee80} as well as interaction processes in diffusion channel with $j=1$
\cite{AltAr85}.\ In this case only interaction processes
with $j=0$ determine corrections to classical (Drude)
conductivity.\ Due to universal nature of these corrections (independence of
electronic charge) their
structure is actually coincide with that of localization
corrections (Cooperon).\cite{AltAr83a,AltAr83b} This
means that renormalization group has only one effective
``charge'' --- dimensionless conductance $g$.\ In this
case differential equation for the conductance of a finite system is again
given by Eq. (\ref{GML}) with the
same asymptotic forms of $\beta_{d}(g)$.\ This approach
is valid for systems with linear size $L<L_{T}=\sqrt{\hbar D/T}$ .\ This length
$L_{T}$ replaces in the theory of interacting electrons characteristic length
of phase coherence $L_{\varphi}$ of noninteracting theory.\ The appearance of
this new length
is due to the fact that characteristic time of interaction processes
\cite{AltAr85} is $\sim \hbar/T$.\
We must stress that these arguments are probably oversimplified as
Refs.\cite{Fin83,Fin84a,Fin84b,DiCast84} had demonstrated the relevance of
interaction in the sense of appearance
of additional coupling constants (``charges'').\ Also it
is in no way clear that local moments appearing within
this approach are acting just as usual paramagnetic scatterers.\ However,\ the
simple scheme following from
Refs.\cite{AltAr83a,AltAr83b} seems to be too attractive
on physical grounds just to be neglected.

As in noninteracting case for $d=3$ Eq. (\ref{GML}) again
possess unstable fixed point responsible for the existence of mobility edge and
absence of minimal metallic conductivity at the metal---insulator transition.\
However,\ in this case there are no special
reasons to believe that the critical exponent $\nu$ of localization correlation
length $\xi_{loc}$ will
coincide with its value for noninteracting theory.\ At
finite temperatures as in usual scaling picture conductivity  for $d=3$ is
given by:\cite{AltAr83a,AltAr83b,AltAr85}
\begin{equation}
\sigma\approx \frac{e^{2}}{\hbar
\xi_{loc}}f\left(\frac{\xi_{loc}}{L_{T}}\right)
\label{AAcond}
\end{equation}
As system approaches insulating phase $\xi_{loc}\rightarrow\infty$.\ For
$\xi_{loc}\ll L_{T}$
we have $f(\xi_{loc}/L_{T})=A+B(\xi_{loc}/L_{T})$,\ where
$A$ and $B$ are some numerical constants.\ Thus in this
region conductivity corrections are proportional to
$\sqrt{T}$ \cite{AltAr79}.\ In case of $\xi_{loc}\gg L_{T}$,\ i.e. very close
to transition:
\begin{equation}
\sigma\approx C\frac{e^2}{\hbar L_{T}}=C\frac{e^2}{\hbar}
\sqrt{T/D\hbar}
\label{condLt}
\end{equation}
where again $C\sim 1$.\ Using Einstein relation \cite{Kubo57}
$\sigma=e^{2}D(dn/d\zeta)$ we immediately obtain:  \begin{equation}
D=\frac{C^{2/3}}{\hbar}T^{1/3}\left(\frac{dn}{d\zeta}\right)^{-2/3}
\label{AAdiff}
\end{equation}
and
\begin{equation}
\sigma=C^{2/3}\frac{e^2}{\hbar}\left(T\frac{dn}{d\zeta}\right)^{1/3}
\label{AAsigma}
\end{equation}
which is valid for $L_{T}<\xi_{loc}$,\ where
$L_{T}=[C/(Tdn/d\zeta)]^{1/3}$.

In case of a system in alternating electric field with
frequency $\omega\gg T/\hbar$ the relevant length becomes
$L_{\omega}=[D/\omega]^{1/2}$ as in Eq. (\ref{Lomeg}).\
Accordingly for $L_{\omega}\ll\xi_{loc}$ instead of
Eq. (\ref{AAsigma}) we get:
\begin{equation}
\sigma(\omega)\approx
\frac{e^2}{\hbar}\left(\omega\frac{dn}{d\zeta}\right)^{1/3}
\label{AAsigomeg} \end{equation} which is analogous to Eq. (\ref{diffreqd})
and Eq. (\ref{diffreq3d}).\ However we must note that this result can not be
considered very reliable since the dynamical critical exponent in general
case is an independent one \cite{Fin84a,Fin84b}.

The metal---insulator transition can be viewed as a gradual breakdown of the
Fermi---liquid state \cite{Kotl87}.\ As we approach the transition different
Fermi---liquid parameters,\ such as $D$,\ $N(E_{F})$,\
$\chi$ etc.\ change continuously and at a critical point
some of these may either diverge or go to zero.\ This
behavior is related to the divergence of correlation length $\xi_{loc}$
characterized by a critical exponent
$\nu$.\ On the insulating side of the transition this
length can be also  interpreted as the scale inside which
a Fermi liquid description of the system still holds.

At present we are in a need of some kind of new approach to the theory of
interacting electrons in disordered systems which probably may be
formulated along the lines of the self-consistent theory of localization.
The hope is to provide an effective formalism to calculate the basic physical
properties of the system in an interpolating scheme from metallic to
insulating state.
Below we briefly describe an attempt to construct such
self-consistent approach\cite{KSSE95}.

The basic idea is in equal footing (additive) treatment of both localization
and interaction corrections to the current relaxation kernel defining the
generalized diffusion coefficient in Eq.(\ref{diffM}).
As a zeroth approximation we take the
Drude metal and consider the simplest localization and interaction
corrections,\ so that the relaxation kernel takes the following form:

\begin{equation}
M(\omega)=M_{0}+\delta M(\omega)
\label{Md}
\end{equation}
where $\delta M(\omega)=\delta M_{l}(\omega)+\delta M_{c}(\omega)=
-\frac{M_{0}}{D_{0}}(\delta D_{l}(\omega)+\delta D_{c}(\omega))$. Here the
localization correction to diffusion coefficient $D_{l}(\omega)$ is defined by
the usual sum of "maximally crossed" diagramms which yields:
\begin{equation}
\frac{\delta D(\omega)}{D_{0}}=-\frac{1}{ \pi N_{0}(E_{F}) }
\sum_{|{\bf q}|<k_{0}} \frac{1}{-i\omega + D_{0}q^2}
\label{dDLN}
\end{equation}
while the Coulomb correction $D_{c}(\omega)$ is given by
\begin{equation}
\frac{\delta D_{c}(\omega)}{D_{0}}=\frac{\delta \sigma(\omega)}
{2e^{2}N_{0}(E_{F})D_{0}}=
\label{dD1}
\end{equation}
$$=\frac{8i}{\pi d}\mu D_{0}\frac{1}{\pi N_{0}(E_{F})}
\int\limits_{\omega}^{\infty}d\Omega\int \frac{d^{d}q}{(2\pi)^{d}}
\frac{q^{2}}{(-i(\Omega +\omega)+D(\Omega +\omega)q^{2})(-i\Omega+
D(\Omega)q^{2})^{2}}$$
where $\mu=N_{0}(E_{F})v_{0}$ is the dimensionless point-like interaction 
with $N_{0}(E_{F})$ now denoting the single-spin density of states at the
Fermi level for noninteracting case.
Conductivity
correction $\delta\sigma$ due to interactions was defined by the lowest-order
diagramms shown in Fig.\ref{fig7a} which were for the first time analyzed in
Ref.\cite{AKLL1980},\ neglecting localization corrections.
It was shown in Ref.\cite{AKLL1980} that the total contribution of
diagrams (a), (b) and (c) is actually zero and conductivity correction
reduces to that determined by diagrams (d) and (e). Here we neglect also the
so called Hartree corrections to conductivity\cite{AltAr85,AKLL1980}, which
is valid in the limit of $2k_{F}/\kappa_{D} \gg 1$, where $\kappa_{D}$ - is the
inverse screening length. This inequality, strictly speaking, is valid for
systems with low electronic density, which are most interesting for
experimental studies of disorder induced metal-insulator transitions. Also,
if we remember the divergence of screening length at the
metal-insulator transition, we can guess that this approximation becomes
better as we approach the transition.  The point-like interaction model used
above has to be understood only in this sense.

Self-consistency procedure is reduced to the replacement of
$D_{0}$ by the generalized diffusion coefficient in all diffusion denominators.
As a result we obtain the following {\it integral} equation for the
generalized diffusion equation:

\begin{equation}
\frac{D_{0}}{D(\omega)}=  1+ \frac{1}{ \pi N_{0}(E_{F}) }\int \frac{d^{d}q}
{(2\pi)^{d}} \frac{1}{-i\omega + D(\omega)q^2}-
\label{I}
\end{equation}
$$-\frac{8i}{\pi d}\mu D_{0}\frac{1}{\pi N_{0}(E_{F})}
\int\limits_{\omega}^{\infty}d\Omega\int \frac{d^{d}q}{(2\pi)^{d}}
\frac{q^{2}}{(-i(\Omega +\omega)+D(\Omega +\omega)q^{2})(-i\Omega+
D(\Omega)q^{2})^{2}}$$
This equation forms the basis of the proposed self-consistent approach.\ In
the absense of interactions $(\mu=0)$ it obviously reduces to the usual
self-consistent theory of localization.
Let us transform it to dimensionless imaginary Matsubara frequencies which is
the only case we need for further applications to superconducting state:  $
 \frac{-i\omega}{D_{0}k_{0}^{2}}\rightarrow \omega,\ \frac{-i\Omega}
{D_{0}k_{0}^{2}}\rightarrow \Omega $,
and also introduce the dimensionless diffusion coefficient $d(\omega)=
\frac{D(\omega)}{D_{0}}$. In these notations integral equation (\ref{I})
takes the following form:

$$\frac{1}{d(\omega)}=  1+\frac{1}{d(\omega)}d\lambda x_{0}^{d-2}\int
\limits_{0}^{1}\frac{dyy^{d-1}}{y^{2}+\frac{\omega}{d(\omega)}}+$$
\begin{equation}
+\frac{8}{\pi}\mu\lambda x_{0}^{d-2}\int\limits_{\omega}^{\infty}
\frac{d\Omega}{d(\omega +\Omega)d^{2}(\Omega)}
\int\limits_{0}^{1}\frac{dyy^{d+1}}{\left(y^{2}+\frac{\omega+\Omega}
{d(\omega+\Omega)}\right)\left(y^{2}+\frac{\Omega}{d(\Omega)}\right)^{2}}
\label{In}
\end{equation}
where $\lambda=\gamma/\pi E_{F}=1/2\pi E_{F}\tau$ is the usual disorder
parameter. In the following we shall limit ourselves only to the case of
spatial dimension d=3. Diffusion coefficient of the usual self-consistent
theory of localization (\ref{difffreq}) in these notations reduces to:

\begin{equation}
d(\omega)= \left\{
\begin{array}{lll}
\alpha=1-3\lambda x_{0}\approx \frac{E_{F}-E_{c}}{E_{c}} &
\quad \omega \ll \omega_{c}, \quad \alpha >0  & \quad \mbox{Metal}   \\
\left(\frac{\pi}{2}3\lambda x_{0}\right)^{\frac{2}{3}}\omega^{\frac{1}{3}} &
\quad \omega \gg \omega_{c} & \quad \mbox{Metal and Insulator} \\
\frac{\left(\frac{\pi}{2}3\lambda x_{0}\right)^{2}}{\alpha^{2}}\omega
=(\xi_{loc} k_{0})^{2}\omega &
\quad \omega \ll \omega_{c}, \quad \alpha <0 & \quad \mbox{Insulator}
\end{array}\right.
\label{ds}
\end{equation}
where $\omega _{c}=\frac{|\alpha |^{3}}{\left(\frac{\pi}{2}3\lambda x_{0}
\right)^{2}}$ and $\xi_{loc}$ - is the localization length,\ $x_{0}$ - the
dimensionless cutoff.
Let us introduce $K(\omega)=\frac{\omega}{d(\omega)}$ and analyze Eq.(\ref{In})
assuming that $K(\omega)$, $K(\Omega)$ and $K(\omega +\Omega)\ll 1$.
Expanding the right-hand side of Eq.(\ref{In}) over these small parameters
we obtain:

$$\frac{\alpha}{d(\omega)}=1-\frac{\pi}{2}\frac{3\lambda x_{0}}{d(\omega)}
K^{1/2}(\omega)+$$
\begin{equation}
+2\mu\lambda x_{0}\int\limits_{\omega}^{\infty}\frac{d\Omega}{d(\omega
+\Omega)d^{2}(\Omega)}
\frac{K^{1/2}(\Omega)+2K^{1/2}(\omega+\Omega)}{\left(K^{1/2}(\Omega)
+K^{1/2}(\omega +\Omega)\right)^{2}}
\label{Inl}
\end{equation}
Consider the metallic phase and look for diffusion coefficient $d(\omega)$
solution in the following form:

\begin{equation}
d(\omega)= \left\{
\begin{array}{ll}
d & \quad \omega \ll \omega_{c} \\
d{\left(\frac{\omega}{\omega_{c}}\right)}^{\frac{1}{3}} & \quad \omega \gg
\omega_{c}
\end{array}\right.
\label{dm}
\end{equation}
Substituting (\ref{dm}) into Eq.(\ref{Inl}) we find $d$ and $\omega_{c}$
and for the diffusion coefficient we obtain:

\begin{equation}
d(\omega)= \left\{
\begin{array}{ll}
\alpha-\alpha^{\ast} & \quad \omega \ll \omega_{c} \\
\left(\frac{\pi}{2}3\lambda x_{0}\right)^{\frac{2}{3}}
\omega^{\frac{1}{3}} & \quad \omega \gg \omega_{c}
\end{array}\right.
\label{dml}
\end{equation}
where $\omega _{c}=\frac{|\alpha -\alpha^{\ast}|^{3}}{\left(\frac{\pi}{2}3
\lambda x_{0}\right)^{2}}$,
$\alpha ^{\ast}=c\mu $, $c\approx 0.89$

Thus for the metallic phase we come to very simple qualitative
conclusion --- Anderson transition persists and the conductivity exponent
remains $\nu=1$.  The transition itself has shifted to the region of weaker
disorder $\alpha=\alpha^{\ast}=c\mu $---interaction facilitates transition
to insulating state.  The frequency behavior of diffusion coefficient in
metallic phase is qualitatively similar to that in the usual self-consistent
theory of localization (\ref{ds}). In the region of high frequencies $\omega
\gg \omega_{c}$ the behavior of diffusion coefficient remains unchanged after
the introduction of interelectron interactions.

Consider now the insulating phase. In the region of high frequencies
$\omega \gg \omega_{c}$ the diffusion coefficient obviously possess the
frequency dependence like $d(\omega) \sim \omega ^{1/3}$. Assume that for small
frequencies it is also some power of the frequency:

\begin{equation}
d(\omega)= \left\{
\begin{array}{ll}
d{\left(\frac{\omega}{\omega_{c}}\right)}^{\delta} & \quad \omega
\ll \omega_{c} \\
d{\left(\frac{\omega}{\omega_{c}}\right)}^{\frac{1}{3}} & \quad \omega
\gg \omega_{c}
\end{array}\right.
\label{dd}
\end{equation}
where $\delta$ is some exponent to be determined.

Substituting (\ref{dd}) into (\ref{Inl}) and considering the case of
$\alpha <0$ (insulating phase of the usual self-consistent theory of
localization) and $|\alpha |\gg \alpha^{\ast}$,\ we get:

\begin{equation}
d(\omega)= \left\{
\begin{array}{ll}
\frac{\left(\frac{\pi}{2}3\lambda x_{0}\right)^{2}}{\alpha^{2}}\omega
= (\xi_{loc} k_{0})^{2}\omega &
\quad \omega ^{\ast}\ll \omega \ll \omega_{c} \\
\left(\frac{\pi}{2}3\lambda x_{0}\right)^{\frac{2}{3}}\omega^{\frac{1}{3}} &
\quad \omega \gg \omega_{c}
\end{array}\right.
\label{ddl}
\end{equation}
where  $\omega _{c}=\frac{|\alpha |^{3}}{\left(\frac{\pi}{2}3\lambda x_{0}
\right)^{2}}$,\ while $\omega ^{\ast}\approx 0.1\mu \frac{\alpha^{2}}
{\left(\frac{\pi}{2}3\lambda x_{0}\right)^{2}}=0.1\frac{\mu}{(\xi_{loc}
k_{0})^{2}}$ ---is some new characteristic frequency defined by the
interactions. Note that $\omega ^{\ast}\rightarrow 0$ as we approach the
transition point when $\xi_{loc}\rightarrow\infty$.

Thus, sufficiently deep inside the insulating phase when
$\alpha <0$ ¨ $|\alpha |\gg \alpha^{\ast}$ and for the frequencies
$\omega \gg \omega ^{\ast}$, the diffusion coefficient remains the same as in
the self-consistent theory of localization,\ i.e. at small frequencies it is
linear over frequency,\ while for the higher frequencies it is
$\sim \omega^{1/3}$.

The analysis of Eq.(\ref{Inl}) shows that for the frequencies
$\omega \ll \omega ^{\ast}$ it is impossible to find the power-like dependence
for $d(\omega)$, i.e.\ the diffusion coefficient in the insulating phase is
apparently can not be represented in the form of
$d(\omega)=d\frac{\omega ^{\ast}}{\omega_{c}}{\left(\frac{\omega}
{\omega ^{\ast}}\right)}^{\delta}$,
where $\delta$ - is some unknown exponent. Because of this we were unable to
find any analytical treatment of Eq.(\ref{Inl}) in the region of
$\omega \ll \omega ^{\ast}$ within the insulating phase.

Consider now the system behavior not very deep inside the insulating phase
when  $\alpha -\alpha ^{\ast}<0$ while $\alpha >0$, that is when the system
without interaction would be within the metallic phase.
Let us assume that the frequency behavior of the diffusion coefficient for
$\omega\ll \omega _{c}$ possess the power-like form,\ i.e. the diffusion
coefficient is defined by the expression (\ref{dd}).
Substituting (\ref{dd}) into (\ref{Inl}) we get $\delta =\frac{1}{3}$.
As a result for the diffusion coefficient we get:

\begin{equation}
d(\omega)= \left\{
\begin{array}{ll}
\left(4,2\frac{\mu\lambda x_{0}}{\alpha}\right)^{\frac{2}{3}}
\omega^{\frac{1}{3}} &
\quad \omega \ll \omega_{c} \\
\left(\frac{\pi}{2}3\lambda x_{0}\right)^{\frac{2}{3}}\omega^{\frac{1}{3}} &
\quad \omega \gg \omega_{c}
\end{array}\right.
\label{dd0}
\end{equation}
where $\omega _{c}=\frac{|\alpha -\alpha ^{\ast}|^{3}}{\left(\frac{\pi}{2}3
\lambda x_{0}\right)^{2}}$.
Naturally, the exact solution for the diffusion coefficient should show
the continuous change of frequency behavior around  $\omega \sim \omega _{c}$.

Thus, within the insulating phase close to transition point,\ where the
system without interactions should have been metallic,\ the diffusion
coefficient behaves as $\sim \omega ^{1/3}$,\ everywhere,\ though for the low
frequency region the coefficient of $\omega ^{1/3}$ differs from that of
the usual self-consistent theory of localization and explicitly depends upon
the interaction constant.

We have also performed the numerical analysis of the integral equation
(\ref{In}) for the wide region of frequencies, both for metallic
(Fig.\ref{fig7b}) and insulating phases (Fig.\ref{fig7c}).
Solution was achieved by a simple iteration procedure using the results of
the usual self-consistent theory of localization as an initial approximation.
Numerical data are in good correspondence with our analytical estimates.
In the region of high frequencies, both for metallic and insulating phases,
the frequency behavior of diffusion coefficient is very close to that defined
by the usual self-consistent theory of localization.
In the region of small frequencies within the metallic phase diffusion
coefficient $d(\omega)$ diminishes as interaction grows. Dependence of static
generalized diffusion coefficient on disorder for $\mu =0.24$ is shown at the
insert of Fig.\ref{fig7b}, and is practically linear. Metal-insulator
transition in this case is observed at $\alpha =\alpha^{\ast}= á\mu$, where
$c\approx 0,5$, which is also in good correspondence with our qualitative
analysis.
Within the insulating phase for the region of small frequencies
($\omega \ll \omega^{\ast}$) we observe significant deviations from predictions
of the usual self-consistent theory of localization. Diffusion coefficient is
apparently nonanalytic in frequency here and we clearly see the tendency to
formation of some kind of effective gap for the frequencies $\omega \ll
\omega^{\ast}$,\ with this "gap" closing as interctions are turned off.

Our numerical analysis was performed in Matsubara frequency
region, which was used in writing down the Eq.(\ref{In}). Analytical
continuation of our numerical data to the real frequencies was not
attempted,\ but as we stressed above Matsubara frequency behavior is
sufficient for our studies of superconducting state below.

In Ref.\cite{KSSE95} we also were able to study the gradual evolution of the
tunneling density of states from metallic to insulating region,\ demonstrating
the continuous transformation of a cusp singularity of Eq.(\ref{AAcusp}) in a
metal into a kind of interaction induced pseudogap at the Fermi level in an
insulator,\ which is in some respects similar to the Coulomb gap of
Refs.\cite{EfrShkl75,Efr76,ShklEfr79,EfrShkl85}.

For high---$T_{c}$ superconductors problems of interplay
of localization and interactions become especially important because of unusual
nature of normal state of
these systems.\ In the absence of accepted theory of this
normal state we shall limit ourselves only to few remarks
on one specific model.\ The so called ``marginal'' Fermi---liquid theory
\cite{Varma89,Varma90} is a promising semi---phenomenological description of
both normal and
superconducting properties of these systems.\ We shall
see that localization effects are apparently greatly
enhanced in this case \cite{KotVar90,Var90}.

Basically the idea of ``marginal'' Fermi---liquid is
expressed by the following form of one---particle Green's
function \cite{Varma89}:
\begin{equation}
G(E{\bf p})=\frac{Z_{{\bf p}}}{\varepsilon-\xi_{{\bf p}}-i\gamma_{{\bf
p}}}+G_{incoh}
\label{FLGr}
\end{equation}
where $\xi_{{\bf p}}$ is renormalized quasi---particle
energy,\ $\gamma_{{\bf p}}\sim Max[\varepsilon,T]$ is
anomalous (linear) decay---rate for these quasiparticles
which is quite different from quadratic in $\varepsilon$
or $T$ decay---rate of the usual Fermi---liquid theory
\cite{Migd}.\ The concept of ``marginality'' arises due
to peculiar behavior of quasiparticle residue:
\begin{equation}
Z_{{\bf p}}^{-1}\approx ln\frac{\tilde\omega_{c}}{|\xi_{{\bf p}}|}\approx
ln\frac{\tilde\omega_{c}}{|\varepsilon|}
\label{residue}
\end{equation}
where $\tilde\omega_{c}$ is characteristic frequency scale of
some kind of electronic excitations,\ which is the phenomenological parameter
of the theory.\ From Eq. (\ref{residue}) it is clear that quasiparticle
contribution to Green's function Eq. (\ref{FLGr}) vanishes precisely at the
Fermi level,\ while exists
close to it though with logarithmically reduced weight.\
Note that in the case of usual Fermi---liquid $Z_{{\bf p}}\approx
1$\cite{Migd}.

For disordered system we can estimate the impurity
contribution to the scattering rate of quasi---particles
as \cite{KotVar90}:
\begin{equation}
\gamma=2\rho V^{2}Z_{{\bf p}}Im\sum_{{\bf p}}\Lambda^{2}({\bf p+q,p})G({\bf
p+q}\varepsilon)\approx
2\pi\rho V^{2}Z^{2}\Lambda^{2}({\bf q}\rightarrow 0)N(E_{F})\approx
Z\Lambda^{2}\gamma_{0}
\label{gammFL}
\end{equation}
where $\Lambda $ is the appropriate vertex---part renormalized by
Fermi---liquid effects,\ $\rho $ again is
impurity concentration,\ $V$ is impurity potential and
$N(E_{F})=Z^{-1}N_{0}(E_{F})$ is the renormalized density
of states in Fermi---liquid.\ Here $N_{0}(E_{F})$ is density of states for
noninteracting electrons at the Fermi level,\ $\gamma_{0}$ is scattering rate
for noninteracting case.\ To get the last relation in Eq. (\ref{gammFL}) a weak
dependence of vertices and self---energy on momentum was assumed.\ Now we can
use the Ward
identity for $\Lambda({\bf q}\rightarrow 0\omega=0)$
vertex of disordered Fermi---liquid theory \cite{Noz66,KotLee87,Kotl87}:
\begin{equation}
\Lambda({\bf q}\rightarrow 0\omega=0)=(1+F^{s}_{0})^{-1}Z^{-1}
\label{FLWard}
\end{equation}
where $F^{s}_{0}$ is Landau parameter introduced above.\
As a result we can easily get a simple relation between
the mean free paths of interacting and noninteracting
quasiparticles \cite{KotVar90,Var90}:
\begin{equation}
l=(p_{F}/m^{\star})\gamma^{-1}=(p_{F}/m)\gamma^{-1}_{0}/\Lambda^{2}({\bf
q}\rightarrow 0)=l_{0}(1+F^{s}_{0})^{2}Z^{2}
\label{FLfreepath}
\end{equation}
Here $m^{\star}=Z^{-1}m$ is the effective mass of quasiparticle.\ Assuming
$F^{s}_{0}\approx const<1$ and
using Eq. (\ref{residue}) we get at $T=0$:
\begin{equation}
l=l_{0}/\left[ln\frac{\tilde\omega_{c}}{|\varepsilon|}\right]^{2}
\label{freepathMarg}
\end{equation}
Then from usual Ioffe---Regel criterion for localization
$p_{F}l\approx 1$ we obtain that all quasiparticle state
within the region of the order of
\begin{equation}
|\varepsilon_{c}|\approx \tilde\omega_{c} exp(-\sqrt{p_{F}l})
\label{Margloc}
\end{equation}
around the Fermi---level in high---$T_{c}$ oxides are
localized even for the case of weak impurity scattering
$p_{F}l\gg 1$.\ For realistic estimates of $\tilde\omega_{c}\approx 0.1-0.2 eV$
\cite{Varma89} and
$p_{F}l<5$ the width of this localized band may easily be
of the order of hundreds of degrees $K$,\ while for
$p_{F}l\approx 10$ and $\tilde\omega_{c}\approx 1000 K$ we get
$|\varepsilon_{c}|\approx 40 K$.\ Obviously this band
grows with disorder as the mean free path $l_{0}$ drops.\
We can safely neglect this localization for $T\gg |\varepsilon_{c}|$,\ but for
low enough temperatures
localization effects become important and all states are
localized in the ground state.\

Of course,\ the formal divergence of the mean free path denominator in
Eq.(\ref{freepathMarg}) is unphysical.\ Single-impurity scattering cannot
overcome the so called unitarity limit\cite{KotVar90},\ so that we must
always have:
\begin{equation}
l\geq \frac{p_{F}^2}{4\pi\rho}
\label{unit}
\end{equation}
In a typical metal with $p_{F}\sim a^{-1}$ this leads to $l\geq 1/4\pi\rho a^2$
and Ioffe-Regel criterion $l\leq a$ can be easily satisfied for large impurity
concentrations $\rho\sim a^{-3}$.\ Thus the singularity in
Eq.(\ref{freepathMarg})
does not mean that localization can appear for arbitrarily low concentration of
impurities.\ We can safely speak only about the significant enhancement of
localization effects in marginal Fermi liquids.\
These ideas are still at this elementary level and we may quote only
one paper attempting to put them on more sound basis of
scaling theory of metal---insulator transition of interacting electrons
\cite{Ng91a}.


\newpage
\section{SUPERCONDUCTIVITY AND LOCALIZATION: STATISTICAL MEAN---FIELD APPROACH}


\subsection{BCS Model and Anderson Theorem}
We shall start our analysis of superconductivity in strongly disordered systems
within the framework of
simple BCS---model \cite{BCS,Genn} which assumes the
existence of some kind of effective electron---electron
attraction within energy region of the order of $2<\omega>$ around the Fermi
level.\ In usual superconductors $<\omega>\sim \omega_{D}$,\ where $\omega_{D}$
is Debye frequency,\ because pairing is
determined by electron---phonon mechanism,\ however we
shall use some effective $<\omega>$ as an average frequency of some kind of
Bose---like excitations responsible for pairing,\ e.g.\ in high---$T_{c}$
superconductors.\ At the moment we shall not discuss
microscopic nature of this attraction which in general
case is determined by the balance of attraction due to
Boson---exchange and Coulomb repulsion.\ Here we just
assume (as always is done in simple BCS---approach) that
this effective attraction is described by some interaction constant $g$,\ which
is considered just as
a parameter.\ More detailed microscopic approach will be given in later
sections.

Nontrivial results concerning superconductivity in disordered systems were
obtained very soon since the
discovery of BCS---theory \cite{AG58,AG59,Gor59,And59}.\
The concept of ``dirty'' superconductor described the
experimentally very important case of the mean free path
$l$ short in comparison with superconducting coherence
length $\xi_{0}\sim \hbar v_{F}/T_{c}$,\ i.e. the case
when:
\begin{equation}
\xi_{0}\gg l \gg \hbar/p_{F}
\label{dirty}
\end{equation}
Already in this case of not so strongly disordered (in the sense of closeness
to metal---insulator transition)
system Cooper pairing takes place not between electrons
with opposite momenta and spins as in regular case,\ but
between time---reversed exact eigenstates of electrons in
disordered system \cite{And59,Genn}:
\begin{equation}
({\bf p}_{\uparrow},-{\bf p}_{\downarrow})\Longrightarrow
(\phi_{\nu}({\bf r})_{\uparrow},\phi_{\nu}^{\star}({\bf r})
_{\downarrow})
\label{pair}
\end{equation}
In the following we consider only singlet isotropic ($s$-wave) pairing.\
Some aspects of anisotropic pairing are analyzed in Appendix C.\ The
underlying physics is simple: in disordered systems such
as e.g.\ an alloy the electron momentum becomes badly
determined due to the lack of translational invariance.\
However,\ in random potential field we can always define
exact eigenstates $\phi_{\nu}({\bf r})$,\ which are just
solutions of Schroedinger equation in this random field
(for a given configuration of this field).\ We don't need
to know the explicit form of these eigenstates at all,\
the pairing partner of $\phi_{\nu}({\bf r})$ is being
given by time---reversed $\phi_{\nu}^{\star}({\bf r})$.\
This leads to a relative stability of a superconducting
state with respect to disordering in the absence of scattering
mechanisms which break the time---reversal invariance such as
e.g.\  magnetic impurities.

Within standard Green's function approach superconducting system
is described by Gorkov equations
\cite{AGD,Gor58} which in coordinate representation take the form:
\begin{equation}
{\cal G}_{\uparrow}({\bf rr'}\varepsilon_{n})=
G_{\uparrow}({\bf rr'}\varepsilon_{n})-\int d{\bf r''}
G_{\uparrow}({\bf rr''}\varepsilon_{n})\Delta({\bf r''})
{\cal F}({\bf r''r'}\varepsilon_{n})
\label{GorG}
\end{equation}
\begin{equation}
{\cal F}({\bf rr'}\varepsilon_{n})=\int d{\bf r''}
G_{\downarrow} ^{\star}({\bf rr''}\varepsilon_{n})
\Delta^{\star}({\bf r''}){\cal G}_{\uparrow}({\bf r''r'}\varepsilon_{n})
\label{GorF}
\end{equation}
where $G({\bf rr'}\varepsilon_{n})$ is an exact one---electron Matsubara
Green's function of the normal state and superconducting order---parameter
(gap) $\Delta({\bf r})$ is determined by self---consistent gap equation:
\begin{equation}
\Delta({\bf r})=gT\sum_{\varepsilon_{n}}{\cal F}^{\star}({\bf
rr}\varepsilon_{n})
\label{gap}
\end{equation}
where ${\cal F}({\bf rr'}\varepsilon_{n})$ is (antisymmetric over spin
variables) anomalous Gorkov Green's function,\ $\varepsilon_{n}=(2n+1)\pi T$.

If we consider temperatures close to superconducting
transition temperature $T_{c}$,\ when $\Delta({\bf r})$
is small,\ ${\cal F}({\bf rr'}\varepsilon_{n})$ can be
obtained from the linearized equation:
\begin{equation}
{\cal F}({\bf rr'}\varepsilon_{n})=\int d{\bf r''}
G^{\star}_{\downarrow}({\bf rr''}\varepsilon_{n})\Delta^{\star}({\bf
r''})G_{\uparrow}({\bf r''r'}\varepsilon_{n})
\label{linF}
\end{equation}
Then the linearized gap equation determining $T_{c}$
takes the form:
\begin{equation}
\Delta({\bf r})=gT\int d{\bf r'}\sum_{\varepsilon_{n}}
K({\bf rr'}\varepsilon_{n})\Delta({\bf r'})
\label{lingap}
\end{equation}
where the kernel:
\begin{equation}
K({\bf rr'}\varepsilon_{n})=G_{\uparrow}({\bf
rr'}\varepsilon_{n})G^{\star}_{\downarrow}({\bf r'r}\varepsilon_{n})
\label{kern}
\end{equation}
is formed by exact one---electron Green's functions of
a normal metal.\ Now we can use an exact eigenstate representation for an
electron in a random field of a disordered system to write (Cf.\ Eq. (A13)):
\begin{equation}
G_{\uparrow}({\bf rr'}\varepsilon_{n})=\sum_{\nu}
\frac{\phi_{\nu\uparrow}({\bf r})\phi^{\star}_{\nu\uparrow}({\bf
r'})}{i\varepsilon_{n}-\varepsilon_{\nu}}
\label{Gex}
\end{equation}
where $\varepsilon_{\nu}$ are exact energy levels of
an electron in disordered system.\ Then
\begin{equation}
K({\bf rr'}\varepsilon_{n})=Tg\sum_{\mu\nu}\frac{\phi_{\nu\uparrow}({\bf
r})\phi_{\nu\uparrow}^{\star}({\bf r'})\phi_{\mu\downarrow}^{\star}({\bf
r'})\phi_{\mu\downarrow}({\bf
r})}{(i\varepsilon_{n}-\epsilon_{\nu})(-i\varepsilon_{n}+\epsilon_{\mu})}
\label{Kern}
\end{equation}
In the following for brevity we shall drop spin variables
always assuming singlet pairing.\ In case of a system with time---reversal
invariance (i.e.\ in the absence of
an external magnetic field,\ magnetic impurities etc.)
Eq. (\ref{Kern}) can be rewritten as:
\begin{equation}
K({\bf rr'}\varepsilon_{n})=G({\bf rr'}\varepsilon_{n})G({\bf
r'r}-\varepsilon_{n})=
\sum_{\mu\nu}\frac{\phi_{\nu}({\bf r})\phi^{\star}_{\nu}({\bf
r'})\phi_{\mu}({\bf r'})\phi^{\star}_{\mu}({\bf
r})}{(i\varepsilon_{n}-\varepsilon_{\nu})(-i\varepsilon_{n}-\varepsilon_{\mu})}
\label{Kern1}
\end{equation}
Averaging over disorder we get:
\begin{equation}
<\Delta({\bf r})>=gT\int d{\bf r'}\sum_{\varepsilon_{n}}
<K({\bf rr'}\varepsilon_{n})\Delta({\bf r'})>
\label{Gap}
\end{equation}
Practically in all papers on the superconductivity in
disordered systems it is assumed that we can make simplest decoupling in Eq.
(\ref{Gap}) to get the
following linearized equation for the average order---parameter:
\begin{equation}
<\Delta({\bf r})>=gT \int d{\bf r'}\sum_{\varepsilon_{n}}K({\bf
r-r'}\varepsilon_{n})<\Delta({\bf r'})>
\label{GAP}
\end{equation}
where the averaged kernel in case of time---invariance
is given by:
\begin{eqnarray}
K({\bf r-r'}\varepsilon_{n})=K^{\star}({\bf r-r'}\varepsilon_{n})=<K({\bf
rr'}\varepsilon_{n})>= \nonumber \\
=<\sum_{\mu\nu}\frac{\phi_{\nu}({\bf r})\phi^{\star}_{\mu}({\bf
r})\phi_{\mu}({\bf r'})\phi^{\star}_{\nu}({\bf
r'})}{(i\varepsilon_{n}-\varepsilon_{\nu})(-i\varepsilon_{n}-\varepsilon_{\mu})}>=
\nonumber \\
=\int_{-\infty}^{\infty} dE N(E)\int_{-\infty}^{\infty}
d\omega \frac{\ll\rho_{E}({\bf r})\rho_{E+\omega}({\bf
r'})\gg^{F}}{(i\varepsilon_{n}+E)(E+\omega-i\varepsilon_{n})}
\label{Kernel}
\end{eqnarray}
where we have introduced Gorkov---Berezinskii spectral
density \cite{BG79} (Cf.\ Eq. (A2)):
\begin{equation}
\ll\rho_{E}({\bf r})\rho_{E+\omega}({\bf r'})\gg^{F}=
\frac{1}{N(E)}<\sum_{\mu\nu}\phi^{\star}_{\nu}({\bf r})\phi_{\mu}({\bf
r})\phi^{\star}_{\mu}({\bf r'})\phi_{\nu}({\bf
r'})\delta(E-\varepsilon_{\nu})\delta(E+\omega-\varepsilon_{\nu'})>
\label{BGsd}
\end{equation}
Here $N(E)$ is an exact electron density of states per {\em one spin direction}
as it always appears in superconductivity theory (above,\ while discussing
localization we always used density of states for both spin directions).

Usually the decoupling procedure used in Eq. (\ref{Gap})
to reduce it to Eq. (\ref{GAP}) is justified by the
assumption that the averaging of $\Delta({\bf r})$ and
of Green's functions in Eq. (\ref{Gap}) forming the kernel can be performed
independently because of essentially different spatial scales \cite{Gor59}:
$\Delta({\bf r})$ changes at a scale of the order of coherence length (Cooper
pair size) $\xi$,\ while $G({\bf rr'}\varepsilon_{n})$ are oscillating on the
scale of interatomic distance $a\sim \hbar/p_{F}$,\ and we always have $\xi\gg
a$.\ Actually it is clear that this decoupling is valid only if the
order---parameter is
{\em self---averaging} (i.e.\ in fact nonrandom) quantity:\ $\Delta({\bf
r})=<\Delta({\bf r})>, \quad
<\Delta^{2}({\bf r})>=<\Delta({\bf r})>^{2}$.\ Below we
shall see that for a system close to mobility edge the
property of self---averageness of $\Delta({\bf r})$ is
absent and situation is actually highly nontrivial.\ In
this case the so called {\em statistical} fluctuations \cite{BulSad86} leading
to inequality of $<\Delta^{2}({\bf r})>$ and $<\Delta ({\bf r})>^{2}$ become
quite important.\ However,\ we shall start with what we call statistical
mean---field approach
which completely neglects these fluctuations and allows
the simple analysis using Eq. (\ref{GAP}),\ as a necessary
first step to understand superconductivity in strongly disordered systems,\
which will allow to find most of the important deviations from the usual theory
of ``dirty''
superconductors.\ The role of statistical fluctuations
will be analyzed later.

If we look for the solution of Eq. (\ref{GAP}) $\Delta({\bf r})=const$
(homogeneous gap),\ we immediately obtain the following equation for
superconducting transition temperature $T_{c}$:
\begin{eqnarray}
1=gT_{c}\int d{\bf r}\sum_{\varepsilon_{n}}K({\bf r-r'}\varepsilon_{n})=
\nonumber \\
=gT_{c}\int d{\bf r}\sum_{\varepsilon_{n}}\int_{-\infty}^{\infty}dE N(E)
\int_{-\infty}^{\infty}d\omega \frac{\ll\rho_{E}({\bf r})
\rho_{E+\omega}({\bf r'})\gg^{F}}{(E+i\varepsilon_{n})
(E+\omega-i\varepsilon_{n})}
\label{TC}
\end{eqnarray}
Using the general sum---rule given in Eq. (A5) \cite{BG79}:
\begin{equation}
\int d{\bf r} \ll\rho_{E}({\bf r})\rho_{E+\omega}({\bf
r'})\gg^{F}=\delta(\omega)
\label{sumrule}
\end{equation}
we immediately reduce Eq. (\ref{TC}) to a standard BCS
form:
\begin{equation}
1=gT_{c}\int_{-\infty}^{\infty} dE N(E)
\sum_{\varepsilon_{n}}\frac{1}{E^{2}+\varepsilon_{n}^{2}}
=g\int_{0}^{<\omega>} dE N(E) \frac{1}{E}th\frac{E}{2T_{c}}
\label{TcBCS}
\end{equation}
where we introduced the usual cut---off at $\varepsilon_{n}\sim 2<\omega>$.\
Note that $N(E)$ here
is an exact one-particle density of states (per one spin direction) in a normal
state of a disordered system.\
{}From Eq. (\ref{TcBCS}) we get the usual result:
\begin{equation}
T_{c}=\frac{2\gamma}{\pi}<\omega>exp\left(-\frac{1}{\lambda_{p}}\right)
\label{BCSTc}
\end{equation}
where $\lambda_{p}=gN(E_{F})$ is dimensionless pairing constant,\
$ln\gamma=C=0.577...$ is Euler constant.\ This
is the notorious Anderson theorem:\ in the absence of
scattering processes breaking time---reversal invariance
disorder influence $T_{c}$ only through the possible
changes of the density of states $N(E_{F})$ under disordering (which are
usually relatively small).

Due to the sum---rule of Eq. (\ref{sumrule}) all singularities of
Berezinskii---Gorkov spectral density reflecting possible localization
transition do not appear
in equation determining $T_{c}$:\ there is no explicit
contribution from $\delta(\omega)$ term of Eq. (A8) and
Eq. (\ref{TcBCS}) has the same form both in metallic and
localized phases (Cf.\ Ref.\cite{TK82}).

The only limitation here which appears on the physical
grounds is connected with the local discreteness of
electronic spectrum in localized phase discussed above.\
It is clear that Cooper pairing is possible in localized phase only between
electrons with centers of localization
within the distance of the order of $\sim R_{loc}(E)$,\
because only in that case their wave functions overlap \cite{Bul1,Bul2}.\
However,\ these states are splitted in energy
by $\delta_{E}$ defined in Eq. (\ref{spac}).\ Obviously,\
we have to demand that superconducting gap $\Delta$ (at $T=0$,\ $\Delta\sim
T_{c}$) be much larger than this $\delta_{E}$:
\begin{equation}
\Delta\sim T_{c}\gg \delta_{E}\sim \frac{1}{N(E)R^{3}_{loc}(E)}
\label{BScrit}
\end{equation}
i.e.\ on the energy interval of the order of $\Delta\sim T_{c}$ there must be
many discrete levels,\ with centers of localization within distance $\sim
R_{loc}(E)$ from
each other.\ In this case the problem of Cooper pairs formation within $\sim
R_{loc}(E)$ is qualitatively the
same as in metallic state,\ e.g.\ we can replace summation over discrete levels
$\varepsilon_{\nu}$ by
integration.\ Analogous problem was considered previously
in case of Cooper pairing of nucleons in finite nuclei \cite{Migd} and also of
Cooper pairing of electrons in small metallic particles (granular metals)
\cite{IS81,MSD72}.\
For strongly anisotropic high---$T_{c}$ systems we must similarly have
\cite{Sad89}:
\begin{equation}
\Delta\sim T_{c}\gg [N(E)R^{a}_{loc}R^{b}_{loc}R^{c}_{loc}]^{-1}
\label{BScritanis}
\end{equation}
where we have introduced the appropriate values of localization lengths along
the axes of an orthorhombic lattice.

Obviously Eq. (\ref{BScrit}) is equivalent to a condition of large enough
localization
length:
\begin{equation}
R_{loc}(E)\gg [N(E)\Delta]^{-1/3}\sim (\xi_{0}/p_{F}^{2})^{1/3}\sim
(\xi_{0}l^{2})^{1/3}
\label{critBS}
\end{equation}
i.e.\ the system must be close enough to mobility edge
or just slightly localized.\ Here we used the usual estimate of mean free path
close to Anderson transition
$l\sim p_{F}^{-1}$.\ Below we shall see that Eq. (\ref{critBS}) is just a
condition that Cooper pairs must
be much smaller than localization length,\ only in this
case Cooper pairing is possible in localized phase \cite{Bul1,Bul2}.



\subsection{$T_{c}$ Degradation}

In usual BCS model discussed above pairing interaction $g$ is assumed to be a
given constant
in the vicinity of the Fermi level.\ In more realistic approach this
interaction is determined by the
balance of interelectron attraction,\ due e.g.\ to electron---phonon coupling
(as in traditional superconductors) or some other Boson---exchange mechanism
(as is apparently the case in
high---$T_{c}$ superconductors),\ and Coulomb repulsion.\ It is clear that in
strongly disordered
system all these interactions can,\ in principle,\ be strongly renormalized in
comparison with
``pure'' case.\ The aim of this section is to discuss these effects on the
approach
to metal---insulator transition induced by disorder.

Usually the Coulomb repulsion within Cooper pair is strongly reduced in
comparison with electron---phonon attraction due to a retarded nature of
electron---phonon coupling \cite{Genn}.\  Characteristic time of
electron---phonon interaction is of the order of $\omega_{D}^{-1}$,\ while
for Coulomb interaction in ``pure'' metal it is determined by $\sim
\hbar/E_{F}$---the time during
which electrons ``pass'' each other in the pair.\ Due to metallic screening
both interactions are
more or less point---like.\ However,\ in a disordered metal ballistic transport
changes to diffusion
and as disorder grows electron motion becomes slower effectively leading to the
growth of Coulomb repulsion within Cooper pair and the appropriate drop of
$T_{c}$ as was first claimed by Anderson,\ Muttalib and Ramakrishnan
\cite{AMR}.\ Actually electron---phonon interaction can
also change under disordering but a common belief is that these changes are
less significant
than in case of Coulomb interaction \cite{KS76,FKK80}.\ This problem is still
under active discussion and some alternative points of view were expressed
\cite{Bel87a,Bel87b,Bel89}.\
However,\ the general agreement is that some kind of diffusion renormalization
of effective
interaction of electrons within Cooper pair provides effective mechanism of
$T_{c}$ degradation under disordering.\ Below we shall mainly use the approach
of Ref. \cite{Bul2},\ with the main aim to study a possibility of
superconductivity surviving up to Anderson transition.

Later in this section we shall also consider the possible mechanisms of $T_{c}$
degradation under disordering due to magnetic fluctuations (or local moments)
which appear close to metal---insulator transition.\ Possible relation of these
mechanisms to enhanced Coulomb effects will also be discussed.

The general problem of $T_{c}$ degradation under disordering becomes much more
complicated
in case of high---temperature superconductors because of unknown nature of
pairing in these systems.\ However,\ we believe that  the mechanism based upon
the growth of Coulomb
repulsion within Cooper pair is also operational here,\ while of course it is
difficult to say anything
about disorder effects upon attractive interactions leading to Cooper pair
formation in these systems.

If we assume some kind of spin-independent Boson---exchange
(phonons,\ excitons etc.) model of
pairing interaction,\ the $T_{c}$ can be obtained
from the generalized Eliashberg equations and thus be given by the
famous Allen---Dynes expression \cite{AD}:
\begin{equation}
T_{c}=\frac{f_{1}f_{2}}{1.20}\omega_{log}
exp\left\{-\frac{1.04(1+\lambda)}{\lambda-\mu^{\star}(1+0.62\lambda)}\right\}
\label{ADTc}
\end{equation}
where
\begin{eqnarray}
f_{1}=[1+(\lambda/\lambda_{1})^{3/2}]^{1/3}; \qquad
f_{2}=1+\frac{[<\omega^{2}>^{1/2}/\omega_{log}-1]\lambda^{2}}{\lambda^{2}+\lambda_{2}^{2}} \nonumber \\
\lambda_{1}=2.46(1+3.8\mu^{\star}); \qquad
\lambda_{2}=1.82(1+6.3\mu^{\star})\frac{<\omega^{2}>^{1/2}}{\omega_{log}}
\label{ffll}
\end{eqnarray}
Here $\omega_{log}$ is the mean logarithmic frequency and $<\omega>^{2}$ is the
mean square frequency of Bosons responsible for pairing (the averaging is over
the spectrum of these Bosons),\ $\mu^{\star}$ is the Coulomb
pseudopotential,\ $\lambda$ is the dimensionless pairing
constant due to Boson---exchange.\  Strictly speaking,\ Allen-Dynes formula has
been derived
for the electron---phonon model,\ with certain assumptions about the phonon
spectrum.\    Its use for general Boson---exchange model here serves only for
illustrative purposes.\
At relatively weak coupling $\lambda\leq 1.5$ Allen---Dynes expression
effectively
reduces to McMillan formula\cite{McM}:
\begin{equation}
T_{c}=\frac{\omega_{log}}{1.20}exp\left\{-\frac{1.04(1+\lambda)}{\lambda-\mu^{\star}(1+0.62\lambda)}\right\}
\label{McMTc}
\end{equation}
which for the weak coupling limit gives the usual BCS
result $T_{c}\sim <\omega>exp(-1/\lambda-\mu^{\star})$.\
For very strong pairing interaction Eq. (\ref{ADTc}) gives the asymptotic
behavior $T_{c}\approx 0.18\sqrt{\lambda <\omega^{2}>}$.\ In most parts of this
review we shall limit ourselves to weak coupling approximation.\ Coulomb
pseudopotential $\mu^{\star}$ in
the ``pure'' system is given by:
\begin{equation}
\mu^{\star}=\frac{\mu}{1+\mu ln \frac{E_{F}}{<\omega>}}
\label{mustar}
\end{equation}
where $\mu$ is the dimensionless Coulomb constant.\ The
mechanism of $T_{c}$ degradation under disordering due to
the growth of Coulomb repulsion is reflected in the appropriate growth of
$\mu^{\star}$ \cite{AMR,Bul2}.

The singlet gap function with a simple $s$-wave symmetry which we have
discussed
above has a non-zero amplitude at zero separation of the two electrons in the
pair. Thus it must pay the energy price of short-range repulsion due to a
finite $\mu$. In recent years a number of new mechanisms of superconducting
pairing were proposed  which try to eliminate the effect of repulsion assuming
a pair wavefunction which vanishes at zero separation. This is equivalent to
the requirement that the sum over all momentum of the BCS gap function
$\Delta$ must vanish\cite{Abrh92}:
\begin{equation}
\Delta({\bf r})=<\psi\uparrow({\bf r})\psi\downarrow({\bf r})>=
\sum_{{\bf k}}\Delta({\bf k})=0
\label{BCSgapzer}
\end{equation}
A number of rather exotic schemes for this were proposed\cite{Abrh92},\ but
probably the simplest way of satisfying this requirement is by means of higher
angular momentum pairing,\ e.g.\ $d$-wave which became rather popular as a
possible explanation of high-$T_{c}$ superconductivity within the
spin-fluctuation
exchange mechanism\cite{Scal,Pin92,Pin93,Pin94}. The sum in
Eq.(\ref{BCSgapzer}) is
then zero because the gap changes sign as ${\bf k}$ goes around the Fermi
surface. This leads to a large extent to cancellation of Coulomb
pseudopotential
effects. However,\ this type of pairing is extremely sensitive to any kind of
disordering (Cf.\ Appendix C) and superconductivity is destroyed long before
localization transition. For these reasons we shall not discuss disorder
effects
in such superconductors in this review. The same applies to more exotic
pairing schemes such as the odd-gap pairing\cite{KuchSadov93},\ where the usual
scattering suppression of $T_{c}$ is also very strong.

Among mechanisms discussed for high-$T_{c}$ superconductors we should also
mention different types of so-called van-Hove
scenarios\cite{Friedel,Tsuei,Campuz,Dagot},\
which are based upon the idea of $T_{c}$-enhancement due to some kind of the
density of states singularity close to the Fermi level. For all such mechanisms
rather strong $T_{c}$ suppression may be due to the potential scattering
smoothing out these singularities. Again we shall not discuss these mechanisms
in our review as having nothing to do with localization effects.

\subsubsection{Coulomb Kernel}

Let us use again the exact eigenstate $\phi_{\nu}({\bf r})$ representation for
an electron in random system,\ with exact energy levels $\varepsilon_{\nu}$.\
These functions and energies may  correspond either to extended or to localized
states.\ Consider the one---electron Green's
function in this representation and take its diagonal element
$G_{\nu\nu}(\varepsilon)$.\ The
influence of interaction is described by the appropriate irreducible
self---energy $\Sigma_{\nu}(\varepsilon)$ \cite{Migd,AAL81}:
\begin{equation}
G_{\nu\nu}(\varepsilon)=\frac{1}{\varepsilon-\varepsilon_{\nu}-\Sigma_{\nu}(\varepsilon)}
\label{Gnunu}
\end{equation}
Here energy zero is at the Fermi level.\ Let us introduce
a ``self---energy'' $\Sigma_{E}(\varepsilon)$ averaged
over some surface of constant energy $E=\varepsilon_{\nu}$ and over random
field configurations \cite{AAL81}:
\begin{equation}
\Sigma_{E}(\varepsilon)=\frac{1}{N(E)}<\sum_{\nu}\delta(E-\varepsilon_{\nu})\Sigma_{\nu}(\varepsilon)>
\label{SigmaE}
\end{equation}
Consider model with short---range static interelectron interaction $v({\bf
r-r'})$.\ Then for the simplest Fock
diagram shown in Fig. \ref{fig7} we find:
\begin{equation}
\Sigma^{F}_{\mu}=-\int d{\bf r}\int d{\bf r'} v({\bf
r-r'})\sum_{\nu}f_{\nu}\phi^{\star}_{\mu}({\bf r'})\phi^{\star}_{\nu}({\bf
r})\phi_{\mu}({\bf r})\phi_{\nu}({\bf r'})
\label{Fok}
\end{equation}
where $f_{\nu}=f(\varepsilon_{\nu})$ is Fermi distribution function.\
Accordingly from Eq. (\ref{SigmaE}) we get \cite{KazSad84}:
\begin{equation}
\Sigma^{F}_{E}=-\int_{-\infty}^{\infty} d\omega f(E+\omega)\int d{\bf r}\int
d{\bf r'} v({\bf r-r'})\ll\rho_{E}({\bf r})\rho_{E+\omega}({\bf r'})\gg^{F}
\label{SigmaF}
\end{equation}
where we again introduced Berezinskii---Gorkov spectral density defined in Eq.
(A2) or Eq. (\ref{BGsd}).

Let us define the Coulomb kernel by the following functional derivative:
\begin{equation}
K_{c}(E-E')=-\frac{\delta\Sigma^{F}_{E}}{\delta f(E')}
\label{Kdef}
\end{equation}
which characterize the change of electron energy due to a variation of its
distribution function.\ It
is easy to see that:
\begin{eqnarray}
K_{c}(\omega)=\frac{1}{N(E)}<\sum_{\mu\nu}<\mu\nu|v({\bf
r-r'})|\nu\mu>\delta(E-\varepsilon_{\nu})\delta(E+\omega-\varepsilon_{\mu})>=
\nonumber \\
=\int d{\bf r}\int d{\bf r'} v({\bf r-r'})\ll\rho_{E}({\bf
r})\rho_{E+\omega}({\bf r'})\gg^{F}
\label{Kcsd}
\end{eqnarray}
is actually Fock---type matrix element of interaction averaged over two
surfaces of constant energy
$E$ and $E'=E+\omega$ and over disorder.\ We can use $K_{c}(\omega)$ as a
kernel in the
linearized gap equation (Cf.\ Appendix B) determining $T_{c}$ which is a
reasonable generalization of  a Coulomb kernel used in the theory of ordered
superconductors \cite{VIK}.\  In momentum representation:
\begin{equation}
K_{c}(\omega)=\int \frac{d^{3}{\bf q}}{(2\pi)^{3}} v({\bf q})
\ll\rho_{E}\rho_{E+\omega}\gg^{F}_{{\bf q}}
\label{Kc}
\end{equation}
In the weak coupling approximation over pairing interaction it is the only
relevant Coulomb contribution in the gap equation  (Cf.\ Appendix B),\ in case
of strong coupling there are additional contributions,\ e.g.\ connected with
diffusional renormalization of the density of
states Eq. (\ref{AAcusp})\cite{Bel87a,Bel87b,Bel89,Bel85,MEF85}.\  We refer to
these papers
for the detailed analysis of the density of states effects upon $T_{c}$.

In the following we assume point---like interaction: $v({\bf q})=v_{0}$.\
During our discussion of
localization we have discovered that for small $\omega\ll \gamma$ and $q\ll
l^{-1}$ Gorkov---Berezinskii spectral density acquires a diffusional
contribution:
\begin{equation}
\ll\rho_{E}\rho_{E+\omega}\gg^{F diff}_{{\bf q}}=\frac{1}{\pi N(E)}Im
\Phi^{RA}_{E}({\bf q}\omega)
\label{rrdif}
\end{equation}
where
\begin{equation}
\Phi^{RA}_{E}({\bf q}\omega)=-\frac{N(E)}{\omega+iD_{E}(\omega)q^{2}}
\label{Fidif}
\end{equation}
and the generalized diffusion coefficient in metallic phase is given by:
\begin{equation}
D_{E}(\omega)\approx
\left\{ \begin{array}{l}
D_{E} \qquad |\omega|\ll \omega_{c}\approx 2\gamma(\sigma/\sigma_{c})^{3} \\
D_{0}\left(-\frac{i\omega}{2\gamma}\right)^{1/3} \qquad |\omega|\gg \omega_{c}
\end{array} \right.
\label{Diffg}
\end{equation}
In the absence of disorder this diffusional contribution disappears and the
kernel $K_{c}(\omega)$
for $|\omega|<E_{F}$ reduces to usual Coulomb potential
$\mu=N(E)v_{o}$.\cite{Genn,VIK} Accordingly we can use the following
approximation \cite{Bul2}:
\begin{equation}
K_{c}(\omega)\approx \mu\theta(E_{F}-|\omega|)+K_{c}^{diff}(\omega)
\label{Kcbs}
\end{equation}
where
\begin{equation}
K_{c}^{diff}(\omega)=\int\frac{d^{3}{\bf q}}{(2\pi)^{3}}
v_{0}\ll\rho_{E}\rho_{E+\omega}\gg^{F diff}_{{\bf q}}
\label{Kcdiff}
\end{equation}
This form of the Coulomb kernel gives correct interpolation between the strong
disorder limit and
``pure'' case.\ Note,\ that in case of disordered system besides diffusional
contribution which contains singularities associated with   Anderson transition
there also appear ``regular'' contributions to $K_{c}(\omega)$ which may be
modelled by $\mu$,\ making it different from its value in ``pure'' system.\
Diffusional term in $K_{c}(\omega)$ is connected with diffusion renormalization
of electron---electron interaction vertex
\cite{AltAr79,AltAr82,AAL81,Lee82,KazSad83,KazSad84}.\
Fig. \ref{fig7} shows diagrams of standard perturbation
theory responsible for this renormalization.\ In case of
the approach based upon self---consistent theory of localization ``triangular''
vertex defined by Fig. \ref{fig7} (c) is given by \cite{KazSad83,KotSad85}:
\begin{equation}
\gamma^{RA}({\bf q}\omega)\approx \frac{2\gamma}{-i\omega+D_{E}(\omega)q^{2}}
\qquad \omega\ll \gamma \quad q\ll l^{-1}
\label{gammaRA}
\end{equation}
Singularity of Eq. (\ref{gammaRA}) for small $\omega$ and
$q$ leads to significant growth of interaction in disordered system.\ Actually
this expression is the same
as in a ``dirty'' metal \cite{AltAr79} but with the replacement of Drude
diffusion coefficient by the generalized one.

\subsubsection{Electron---Phonon Interaction}

The case of electron---phonon interaction is different.\
Diffusion renormalization of electron---phonon vertex is unimportant because
the relevant corrections compensate each other if we take into account impurity
vibrations \cite{Pipp,KS76,FKK80}.\ Surely the value of electron---phonon
contribution to pairing interaction do change in a
disordered system in comparison with ``pure'' case \cite{KS76}.\ However,\
these changes are relatively insignificant in the sense of absence of drastic
changes
at the Anderson transition.\ We shall demonstrate the
absence of diffusion renormalization of electron---phonon
vertex using the lowest order diagrams of perturbation theory following the
approach of Ref.\cite{FKK80}.

Let us limit our analysis to homogeneous continuous medium.\ The appearance of
deformation ${\bf u}$ leads to
the variation of density of the medium given by $\delta\rho=-\rho div{\bf u}$.\
Accordingly,\ taking into account the electroneutrality condition we get the
variation of electron density as $\delta n = -n div{\bf u}$.\ This leads to the
following change of the free electron Green's function:
\begin{eqnarray}
\delta G^{-1}(E{\bf p})=-n div{\bf u}\frac{d}{dn}[E-v_{F}(|{\bf p}|-p_{F})]=
\nonumber \\
=-nv_{F} div{\bf u} \frac{dp_{F}}{dn}=-\frac{1}{3}v_{F}p_{F} div{\bf u}
\label{dG}
\end{eqnarray}
where we have used $n=p^{3}_{F}/(3\pi^{2})$.\ Let us define electron---phonon
vertex ${\bf\Lambda}$ by:
\begin{equation}
\frac{\delta G}{\delta {\bf u}}=G{\bf\Lambda}G=-G\frac{\delta G^{-1}}{\delta
{\bf u}}G; \qquad
{\bf\Lambda}=-\frac{\delta G^{-1}}{\delta {\bf u}}
\label{Lambvert}
\end{equation}
For ${\bf u}({\bf r},t)={\bf u}exp(i{\bf qr}-i\omega t)$ we get from Eq.
(\ref{dG}):
\begin{equation}
\delta G^{-1}(E{\bf p})=-i{\bf qu}\frac{v_{F}p_{F}}{3}
\label{dGq}
\end{equation}
so that the ``bare'' electron---phonon vertex ($i$ is
vector index):
\begin{equation}
\Lambda_{1i}^{(0)}=iq_{i}\frac{v_{F}p_{F}}{3}
\label{Lamb}
\end{equation}
Consider the system with impurities randomly placed at
points ${\bf R}_{n}$ which create the potential:
\begin{equation}
U({\bf r})=\sum_{n}V({\bf r-R}_{n})
\label{impp}
\end{equation}
Vibrations of the medium lead to vibrations of impurity
atoms,\ so that ${\bf R}_{n}\rightarrow {\bf R}_{0n}+{\bf u}_{n}(t)$ with ${\bf
u}_{n}(t)={\bf u}exp(i{\bf qR}_{0n}-i\omega t)$.\ Random field of static
impurities
leads to a simplest self---energy correction given by
Fig. \ref{fig8} (a) \cite{Edw58,AGD}.\ Impurity vibrations can be accounted for
by the additional interaction term:
\begin{displaymath}
\delta V({\bf r-R}_{n})=\frac{\partial V({\bf r-R}_{n0})}
{\partial {\bf R}_{n0}}{\bf u}exp(i{\bf qR}_{n0}-i\omega t)
\end{displaymath}
so that
\begin{eqnarray}
\Lambda_{2i} u_{i}=\frac{\delta\Sigma}{\delta u_{i}}u_{i}= \nonumber \\
=<\sum_{n}\left\{\frac{\partial V({\bf r-R}_{n0})}{\partial R_{n0}^{i}}G({\bf
r}t,{\bf r'}t')V({\bf r'-R}_{n0})u_{in}+ \right.  \nonumber \\
\left.+V({\bf r-R}_{n0})G({\bf r}t,{\bf r'}t')\frac{\partial V({\bf
r'-R}_{n0})}{\partial R_{n0}^{i}} u_{in} \right\}>
\label{Lamb2}
\end{eqnarray}
where the angular brackets define as usual the averaging
over random impurity positions.\ In momentum representation and for
point---like impurities we get in
lowest order over $\omega/E_{F}$ and $q/p_{F}$:
\begin{eqnarray}
\Lambda_{2i}({\bf p,q})=\rho V^{2}\int\frac{d^{3}{\bf
p'}}{(2\pi)^{3}}[-i(p_{i}-p'_{i})G(E{\bf p'})+i(p'_{i}-p_{i})G(E{\bf p'})]=
\nonumber \\
=2\rho V^{2}\int\frac{d^{3}{\bf p'}}{(2\pi)^{3}}[-i(p_{i}-p'_{i})G(E{\bf p'})]=
\nonumber \\
=2\pi \rho V^{2} N(E)p_{i} =2\gamma p_{i}
\label{Lambd2}
\end{eqnarray}
The relevant diagrams are shown in Fig. \ref{fig8} (b) \cite{NT74}.\ ``Bare''
electron---phonon vertex is thus
given by the sum of three diagrams shown in Fig. \ref{fig8} (b) and reduces to:
\begin{equation}
\Lambda_{i}^{(0)}=\Lambda_{1i}^{(0)}+\Lambda_{2i}^{(0)}=
iq_{i}\frac{v_{F}p_{F}}{3}+2\gamma p_{i}
\label{barevert}
\end{equation}
Diffusion renormalization of electron---phonon vertex can
appear due to impurity scattering ladder corrections as shown in Fig.
\ref{fig9} (a).\ Similar diagrams shown in
Fig. \ref{fig7} (c) lead to diffusion renormalization of
Coulomb vertex.\ However,\ in case of electron---phonon
interaction we have to make the same renormalization of
three diagrams of Fig. \ref{fig7} (c).\ Let us consider
simplest corrections shown in Fig. \ref{fig9} (b,c,d).\
For the contribution of graph of Fig. \ref{fig9} (b) we
have:
\begin{eqnarray}
\Lambda_{1i}^{(1)}=\rho V^{2}iq_{i}\frac{v_{F}p_{F}}{3}
\int \frac{d^{3}{\bf p'}}{(2\pi)^{3}}G(E{\bf p'})G(E+\omega{\bf p'+q})\approx
\nonumber \\
\approx iq_{i}\frac{v_{F}p_{F}}{3}[1+i\omega/2\gamma-D_{0}q^{2}/2\gamma]\approx
iq_{i}\frac{v_{F}p_{F}}{3} \qquad \omega,q\rightarrow 0
\label{La}
\end{eqnarray}
and for the sum of graphs of Fig. \ref{fig9} (c,d):
\begin{eqnarray}
\Lambda_{2i}^{(1)}=2\rho V^{2}\gamma \int \frac{d^{3}{\bf p'}}{(2\pi)^{3}}
G(E{\bf p'})G(E+\omega {\bf p'+q})p_{i}'\approx \nonumber \\
\approx 2\rho V^{2}\gamma q_{i} \int \frac{d^{3}{\bf p'}}{(2\pi)^{3}}
p'_{i}G(E{\bf p'})\frac{\partial}{\partial p_{i}}G(E+\omega{\bf p'})\approx
\nonumber \\
\approx 2\gamma\rho V^{2}q_{i}p_{F}\int \frac{d^{3}{\bf p'}}{(2\pi)^{3}}
\frac{v_{F}}{3}G(E{\bf p'})G^{2}(E{\bf p'})=-iq_{i}\frac{v_{F}p_{F}}{3}
\label{Lbc}
\end{eqnarray}
Thus for $\omega\rightarrow 0$,\ $q\rightarrow 0$ we obtain:
\begin{equation}
\Lambda_{1i}^{(1)}+\Lambda_{2i}^{(1)}=0
\label{cancel}
\end{equation}
and we have total cancellation of initial diagrams contributing to diffusion
ladder.\ Apparently there is no diffusion renormalization of electron---phonon
vertex
(for $\omega,q\rightarrow 0$):\ this cancellation is valid for any graph
obtained from the simplest corrections by adding further impurity lines to the
ladder.\ Similar cancellation takes place in case of adding to diagrams of Fig.
\ref{fig9} (b,c,d) corrections due to maximally crossed impurity lines (Cooper
channel).\ Thus there is no significant change of electron---phonon vertex due
to Cooperon and the
only relevant contribution to electron---phonon vertex in
impure system is defined by the sum of diagrams of Fig. \ref{fig8} (b) leading
to Eq. (\ref{barevert}) which does
not contain diffusion type renormalization.\ Localization
appears via generalized diffusion coefficient which replaces the Drude one.\
Thus localization singularities
does not appear in electron---phonon vertex,\ though surely this interaction is
really changed by disorder
scattering in comparison with ``pure'' case.\ Of course,\ the question of
whether localization effects contribute to renormalization of electron---phonon
coupling is still open to discussion\cite{Bel87a}.\ Probably more important
aspect of this problem is reflected by the fact that superconductivity is
actually
determined not by electron---phonon vertex itself,\ but by the famous integral
expression over
the phonon spectrum of Eliashberg function $\alpha^{2}(\omega)F(\omega)$  which
defines the pairing constant $\lambda$.\cite{AD} This integration will
apparently smooth out all possible singularities.

In the following we shall model pairing interaction due
to phonon exchange by some constant $\lambda$ as in BCS
model.\ Of course we must stress that this constant is
different from that in regular metal.\ It is constant in a sense that it does
not contain singularities due to
metal---insulator transition.\ Electron---phonon kernel
in the linearized gap equation (Cf.\ Appendix B) can be
taken in the simplest form:
\begin{equation}
K_{ph}(E,E')=
\left\{ \begin{array}{l}
-\lambda \qquad |E|,|E'|<\omega_{D} \\
0 \qquad |E|,|E'|>\omega_{D}
\end{array}
\right.
\label{Kph}
\end{equation}
and consider $\lambda$ as relatively weakly dependent on
disordering.\ More detailed discussion of electron---phonon pairing in
disordered systems can be found in Refs.\cite{KS76,Bel87a,Bel87b}.

As we mentioned above it is quite difficult to speculate
on disorder dependence of pairing interaction in high---temperature
superconductors.\ In case of the ``marginal''
Fermi---liquid approach \cite{Varma89,Varma90} pairing
interaction can be modelled as in Eq. (\ref{Kph}) with
the replacement of Debye frequency $\omega_{D}$ by some
phenomenological {\em electronic} frequency $\tilde\omega_{c}$ which we briefly
mentioned above while discussing localization in ``marginal'' Fermi---liquid.\
In the following we shall just assume that this pairing
interaction is weakly dependent on disorder as in the case
of phonon mechanism of pairing.

\subsubsection{Metallic Region}

In metallic region we can use Eqs. (\ref{Kc}---\ref{Fidif}) and Eq.
(\ref{Kcbs}) and find the diffusional contribution to Coulomb kernel:
\begin{eqnarray}
K_{c}^{diff}(\omega)=-\int \frac{d^{3}{\bf q}}{(2\pi)^{3}}
v_{0}Im\frac{1}{\omega+iD_{E}q^{2}}\approx \nonumber \\
\approx\frac{v_{0}}{2\pi ^{3}}\left
[\frac{1}{|D_{E}(\omega)l|}-\frac{|\omega|^{1/2}}{|D_{E}^{3/2}(\omega)|}\right]\approx \nonumber \\
\approx \frac{v_{0}}{2\pi^{3}} \left\{\begin{array}{l}
\frac{1}{D_{E}l}-\frac{|\omega|^{1/2}}{D_{E}^{3/2}} \qquad |\omega|\ll
\omega_{c} \\
\frac{1}{D_{0}l}\left(\frac{\omega}{2\gamma}\right)^{-1/3} \qquad |\omega|\gg
\omega_{c}
\end{array} \right.
\label{difKc}
\end{eqnarray}
Accordingly for the Coulomb kernel defined by Eq. (\ref{Kcbs}) we get
\cite{Bul2}:
\begin{equation}
K_{c}(\omega)=\mu\theta(E_{F}-|\omega|)+\frac{\mu}{p_{F}l}\left\{\begin{array}{l}
\frac{\sigma_{c}}{\sigma} \qquad |\omega|<\omega_{c} \\
\frac{1}{p_{F}l}\left(\frac{\omega}{2\gamma}\right)^{-1/3} \qquad
\omega_{c}<\omega<\gamma\sim E_{F}
\end{array}
\right.
\label{Kccc}
\end{equation}
Upper limit cut---off in the integral in Eq. (\ref{difKc}) was taken $\sim
l^{-1}$.\ Rough estimate of
contribution of higher momenta can be achieved introducing cut---off $\sim
p_{F}$ (Cf.\ Ref.\cite{Bel85}).\ This will cancel $(p_{F}l)^{-1}$ in
Eq. (\ref{Kccc}).\ Close to Anderson transition $l^{-1}\sim p_{F}$ and this
correction is irrelevant.\ We shall assume that far from transition these
higher momenta corrections can be included in the definition of $\mu$.
{}From Eq. (\ref{Kccc}) we can see that diffusion renormalization of Coulomb
kernel leads to substantial
growth of Coulomb repulsion close to Anderson transition
(i.e.\ when conductivity drops below $\sigma_{c}$---``minimal metallic
conductivity'').

Superconducting transition temperature $T_{c}$ is determined by the linearized
gap equation \cite{VIK} which in the weak coupling approximation can be written
as (Cf.\ Appendix B)\cite{Dolg77,DolgSad}:
\begin{eqnarray}
\Delta(\omega)=\lambda\theta(<\omega>-\omega)\int_{0}^{<\omega>}\frac{d\omega'}{\omega'}\Delta(\omega')th\frac{\omega'}{2T_{c}}- \nonumber \\
-\theta(E_{F}-\omega)\int_{0}^{E_{F}}\frac{d\omega'}{\omega'}K_{c}(\omega-\omega')\Delta(\omega')th\frac{\omega'}{2T_{c}}
\label{GapDS}
\end{eqnarray}
Consider metallic region and take $\omega_{c}\gg <\omega>$ which in accordance
with $\omega_{c}$ estimate given in Eq. (\ref{Diffg}) roughly corresponds to
$\sigma\geq \sigma_{c}$ for typical $E_{F}/<\omega>\sim 10^{2}$,\ so that the
system is not very close to Anderson transition.\ The change of $T_{c}$ due to
diffusion contribution in Coulomb kernel
Eq. (\ref{Kccc}) can be determined by perturbation theory over
$K_{c}^{diff}(\omega)$ in gap
equation.\ First iteration of Eq. (\ref{GapDS}) gives:
\begin{equation}
\frac{\delta T_{c}}{T_{co}}\approx
\frac{\int\limits_{0}^{\infty}\frac{d\omega}{\omega}\int\limits_{0}^{\infty}\frac{d\omega'}{\omega'}\Delta_{0}(\omega)th\frac{\omega}{2T_{co}}K_{c}^{diff}(\omega-\omega')\Delta_{0}(\omega')th\frac{\omega'}{2T_{co}}}{\frac{1}{2T_{co}}\int\limits_{0}^{\infty}d\omega [\Delta_{0}(\omega)]^{2}[ch\frac{\omega}{2T_{co}}]^{-2}}
\label{delTc}
\end{equation}
where $\Delta_{0}(\omega)$ is the usual ``two---step'' solution of Eg.
(\ref{GapDS})\cite{Genn,VIK}
which is valid for standard form of Coulomb kernel
$K_{c}(\omega)=\mu\theta(E_{F}-|\omega|)$,
\begin{equation}
T_{co}=1.13 <\omega>exp\left(-\frac{1}{\lambda-\mu^{\star}_{0}}\right)
\label{Tco}
\end{equation}
is a critical temperature in regular superconductor when the Coulomb
pseudopotential is given
by :
\begin{equation}
\mu_{0}^{\star}=\frac{\mu}{1+\mu ln\frac{E_{F}}{<\omega>}}
\label{mupure}
\end{equation}
Using the first relation in Eq. (\ref{Kccc}) we get from Eq. (\ref{delTc}):
\begin{equation}
\frac{\delta T_{c}}{T_{c0}}\approx
-\frac{\mu}{(\lambda-\mu_{0}^{\star})^{2}}\frac{1}{p_{F}l}\frac{\sigma_{c}}{\sigma}
\label{delTco}
\end{equation}
This change of $T_{c}$ is equivalent to the following change of Coulomb
pseudopotential \cite{Bul2}:
 \begin{equation}
\delta\mu^{\star}\approx \mu\frac{\sigma^{2}_{c}}{\sigma(\sigma+\sigma_{c})}
\label{deltmu}
\end{equation}
where we have used  Eq. (\ref{sigmaBS}) and $p_{F}l\approx
\sigma_{0}/\sigma_{c}=(\sigma+\sigma_{c})/\sigma_{c}$ to replace $p_{F}l$ in
Eq. (\ref{delTco}).\ As we noted above this later factor disappears from Eq.
(\ref{delTco}) if we use
cut---off at $q\sim p_{F}$ in Eq. (\ref{difKc}).\  According to Eq.
(\ref{deltmu}) Coulomb pseudopotential $\mu^{\star}$ grows as $\sigma$ drops
and this dependence is more strong
than a similar one obtained in Ref.\cite{AMR},\ which is connected with our use
of the results of
self---consistent theory of localization.\ Method of Ref.\cite{AMR} is based
upon the use of $q$---dependence   of  diffusion coefficient as given by Eq.
(\ref{Dq}) .\ Our expression for $\delta\mu^{\star}$ leads to a significant
growth of $\mu^{\star}$ for conductivities $\sigma\leq 10^{3}
Ohm^{-1}cm^{-1}$.\ This growth can easily explain the typical $T_{c}$
degradation in ``very dirty'' superconductors as their conductivity in normal
state drops approaching the Ioffe---Regel limit \cite{IR}.\ At the same time
expressions for $\mu^{\star}$ proposed in Ref.\cite{AMR} can explain
experimental data only under the assumption that a characteristic conductivity
scale determining $\mu^{\star}$ is an order of magnitude larger than
Ioffe---Regel limit.,\ for which we see no serious grounds.\ More extensive
discussion can be found in Ref.\cite{Bel87a}.

Let us consider now the situation at the mobility edge itself,\ when $\sigma=0$
and $\omega_{c}=0$ so that $K_{c}(\omega)$ is determined by the second
expression in Eq. (\ref{Kccc}) for all frequencies below $\gamma\sim E_{F}$.\
In this case we can show \cite{Bul2} that the influence of Coulomb repulsion on
$T_{c}$ is again described by effective
pseudopotential $\mu^{\star}$ which can be estimated as:
\begin{equation}
\mu^{\star}\sim \alpha\mu \left(\frac{<\omega>}{2\gamma}\right)^{-1/3} \qquad
\alpha\sim 1
\label{muedge}
\end{equation}
In this case $T_{c}$ may remain finite at the mobility edge only under very
strict conditions:\ both
$E_{F}\sim \gamma$ and $\mu$ must be very small,\ while $\lambda$ must be at
least close to unity.\ As a crude estimate we can demand something like
$\lambda\sim 1$,\ $\mu\leq 0.2$ and
$E_{F}\leq 10^{3}T_{c0}$.\ Obviously only some narrow band superconductors like
Chevrel phases can satisfy these conditions among traditional systems.\
High---$T_{c}$ superconductors are especially promising.\ Experimental
situation will be discussed later.

Using Eq. (\ref{deltmu}) and Eq. (\ref{muedge}) we can write down a simple
interpolation formula for the conductivity dependence of
$\mu^{\star}$:\cite{Bul2}
\begin{equation}
\mu^{\star}\approx \mu^{\star}_{0}+\frac{\alpha\mu
(<\omega>/2\gamma)^{-1/3}-\mu_{0}^{\star}}{1+(<\omega>/2\gamma)^{-1/3}\sigma(\sigma+\sigma_{c})/\sigma_{c}^{2}}
\label{muinter}
\end{equation}
To get an expression via observable parameters take into account
$<\omega>/\gamma\approx
(<\omega>/E_{F})(1+\sigma/\sigma_{c})$.\ These expressions describe continuous
crossover from the region of weak localization corrections to the vicinity of
Anderson transition where its
influence upon $T_{c}$ becomes very strong.\ This crossover takes place at
$\omega_{c}\sim <\omega>$.

\subsubsection{Localization Region}

Let us now consider Anderson insulator.\ According to
Eq. (\ref{Kc}) and Eq. (A9) Coulomb kernel acquires in
this case $\delta(\omega)$---contribution:
\begin{eqnarray}
K_{c}^{loc}(\omega)=v_{0}A_{E}\delta(\omega)=v_{0}\frac{1}{N(E)}<\sum_{\nu}\delta(E-\varepsilon_{\nu})|\phi({\bf r})|^{2}|\phi({\bf r})|^{2}> \\
A_{E}=A_{E}({\bf r-r'})|_{{\bf r=r'}}\sim R_{loc}^{-3}
\label{Kcloc}
\end{eqnarray}
which is actually connected with ``Hubbard---like'' repulsion of electrons in a
single quantum state becoming
nonzero in localization region \cite{Mott71,KazSad83,Sad86}.\ This mechanism
acts in
addition to diffusion contributions in Coulomb pseudopotential $\mu^{\star}$
considered above,\ which
are due to ``regular'' part of Gorkov---Berezinskii
spectral density.\ Using Eq. (\ref{Kcloc}) as a full
Coulomb vertex in linearized gap equation (\ref{GapDS})
we can solve it exactly \cite{Bul2} and find:
\begin{equation}
\Delta(\omega)=\frac{\theta(<\omega>-|\omega|)\Delta_{1}}
{1+\frac{\mu A_{E}}{2N(E)}\frac{1}{\omega}th\frac{\omega}{2T_{c}}}
\label{Deltaloc}
\end{equation}
where
\begin{equation}
\Delta_{1}=\lambda\int_{0}^{<\omega>}d\omega\Delta(\omega)\frac{1}{\omega}th\frac{\omega}{2T_{c}}
\label{Delta1}
\end{equation}
and equation for $T_{c}$ takes the form:
\begin{equation}
1=\lambda\int_{0}^{<\omega>}d\omega\frac{th\frac{\omega}{2T_{c}}}{\omega+\frac{\mu A_{E}}{2N(E)}th\frac{\omega}{2T_{c}}}
\label{Tclocl}
\end{equation}
To account for ``regular'' diffusion contributions to
$\mu^{\star}$ we can just replace here
$\lambda\rightarrow\lambda^{\star}=\lambda-\mu^{\star}$,\
where $\mu^{\star}$ is given by Eq. (\ref{muedge}).\
Then our equation for $T_{c}$ can be approximately
represented by\cite{Bul2}:
\begin{equation}
ln\frac{T^{\star}}{T_{c}}\approx \psi\left(\frac{1}{2}+\frac{\mu
A_{E}}{4T_{c}N(E)}\right)-\psi\left(\frac{1}{2}\right)
\label{Tclocliz}
\end{equation}
where $\psi (x)$ is digamma function,\ and $T^{\star}$ is
taken to be equal to $T_{c}$ of the system at the mobility edge which is given
by Eq. (\ref{Tco}) with $\mu^{\star}_{0}$ replaced by $\mu^{\star}$ from Eq.
(\ref{muedge}).\ Here we slightly overestimate the role
of Coulomb repulsion in localization region.\ We can see that this additional
``Hubbard---like'' repulsion acts
upon $T_{c}$ as magnetic impurities \cite{Genn,VIK} with effective spin---flip
scattering rate:
\begin{equation}
\frac{1}{\tau_{sf}}=\pi\frac{\mu A_{E}}{N(E)}\sim
\frac{\mu}{N(E)R^{3}_{loc}(E)}
\label{tausf}
\end{equation}
Obviously this result is connected with the appearance below the mobility edge
of the ``band'' of singly occupied electron states of the width
\cite{Mott71,KazSad83,Sad86,Kam80} $v_{0}R_{loc}^{-3}$.\ 
Superconductivity persists until
$\tau_{sf}^{-1}<0.57T^{\star}_{c}$,\ i.e. until
\begin{equation}
R_{loc}(E)>\left[\frac{\mu}{N(E)T^{\star}_{c}}\right]^{1/3}\sim
(\xi_{0}p_{F}^{-2})^{1/3}\sim (\xi_{0}l^{2})^{1/3}
\label{BScritr}
\end{equation}
where the last estimates are valid for typical values of
parameters and correspond to the simple estimate of Eq. (\ref{BScrit}).\ Thus
the Coulomb repulsion in a single
(localized) quantum state leads to a sharp reduction of $T_{c}$ below the
mobility edge even if superconductivity
survived up to the Anderson transition.\ Another interpretation of this effect
is the influence of ``free'' spins of Mott's band of singly occupied states
below the Fermi level of Anderson insulator.

Coulomb gap \cite{EfrShkl75,Efr76,ShklEfr79,ShklEfr79,EfrShkl85} effects can be
neglected here \cite{Bul2} because according to the estimates given in Eq.
(\ref{cgap}) and Eq. (\ref{gapest}) the Coulomb gap width:
\begin{equation}
\Delta_{c}\sim [N(E)R_{loc}^{3}(E)]^{-1}\ll T_{c}\sim\Delta
\label{coulgapineq}
\end{equation}
i.e.\ is small in comparison to superconducting gap $\Delta$ (or $T_{c}$) under
conditions given by Eq. (\ref{BScrit}) which is necessary for the observation
of
superconductivity in localization region.

\subsubsection{Spin Fluctuations}

As we mentioned during our discussion of interaction
effects upon Anderson transition the role of magnetic
fluctuations (spin effects) in general becomes stronger
as we approach metal---insulator transition.\ The band of
single---occupied states is being formed below the Fermi
level of Anderson insulator,\ which is equivalent to the
appearance of localized moments \cite{Mott71,Sad86,Kam80}.\ These effects
actually may
become important already before metal---insulator
transition\cite{Fin84b,KirkBel90,BelKirk90,BelKirk91,AltAr83a,AltAr83b,AltAr85},\ and lead to additional mechanism
of $T_{c}$ degradation.\ Unfortunately there is no complete theoretical
understanding of these effects and
accordingly only few estimates can be done concerning
superconductivity.\ Here we shall mention only some of
these crude estimates following Refs.\cite{Fuk85a,Fuk85b,EFM85}.

In the framework of Hubbard model with weak disorder it
can be shown \cite{Fuk85a} that the spin susceptibility
is represented by:
\begin{equation}
\chi_{s}=\frac{\chi_{0}}{1-UN(E)+\gamma_{0}-\gamma'}=
\frac{\chi_{0}}{\eta_{0}-\gamma'}\equiv \frac{\chi_{0}}{\eta}
\label{chiHb}
\end{equation}
where $\chi_{0}$ is spin susceptibility of free electrons,\
$\eta_{0}=1-UN(E)+\gamma_{0}$ is enhancement factor for the ordered case ($U$
is Hubbard interaction,\
$\gamma_{0}$ is correlation correction to RPA approximation),\ $\gamma'$ is the
correction due to the
interference of Hubbard interaction and disorder scattering:
\begin{equation}
\gamma'=B\lambda^{2} \qquad
B=6\sqrt{3}\pi^{2}[N(E)U]^{2}\left\{1-\frac{1}{2}UN(E)\right\}
\label{gamcor}
\end{equation}
Here $\lambda=1/(2\pi E\tau)=1/(p_{F}l)$ is the usual
perturbation theory parameter for disorder scattering.\
As $\gamma'>0$ we can see from Eq. (\ref{chiHb}) that
disordering leads to diminishing denominator $\eta=\eta_{0}-\gamma'$.\ If we
reach a critical disorder
defined by:
\begin{equation}
\lambda_{c}=\sqrt{\frac{\eta_{0}}{B}} \qquad
p_{F}l|_{c}=2\sqrt{\frac{B}{\eta_{0}}}
\label{chiinf}
\end{equation}
we get $\chi_{s}\rightarrow \infty$.\ It should be stressed
that this divergence of $\chi_{s}$ in a disordered system
must not be identified with any kind of ferromagnetic instability but may
signify something like the appearance of a spin---glass state or just of
localized moments.\
In any case it means the growth of spin dependent effects
under disordering.

If the initial enhancement of spin susceptibility is strong enough (e.g.\ due
to a large $U$),\ i.e.\ $\eta_{0}\ll 1$,\ the critical disorder defined by Eq.
(\ref{chiinf}) may be lower than the critical disorder
for Anderson localization,\ appearing at $p_{F}l\sim 1$.\
Then these spin dependent effects may become important well before Anderson
transition.\ In the opposite case these effects will appear only very close to
metal---insulator
transition.\ In general case the relation between these
two transitions depends on parameters.

If spin fluctuations are strong enough ($\eta\ll 1$) a
strong mechanism for $T_{c}$ degradation in superconducting state
appears\cite{Fuk85b} analogous to similar effect due to magnetic impurities
\cite{Genn,VIK}:
\begin{equation}
ln\frac{T_{c0}}{T_{c}}=\psi\left(\frac{1}{2}+\rho\right)-
\psi\left(\frac{1}{2}\right)
\label{magim}
\end{equation}
where \cite{Fuk85b}:
\begin{equation}
\rho =\frac{9\sqrt{3}\pi}{2}\lambda^{2}\frac{UN(E)}{\eta}
=\frac{9\sqrt{3}}{2}\left[\frac{UN(E)}{B}\right]\frac{\lambda^{2}}{\lambda_{c}^{2}-\lambda^{2}}
\label{rhospin}
\end{equation}
As $\rho$ from Eq. (\ref{rhospin}) diverges as $(\lambda_{c}-\lambda)^{-1}$ for
$\lambda\rightarrow\lambda_{c}$ superconducting transition temperature $T_{c}$
drops to zero.

If $\lambda_{c}\ll 1$,\ which is possible for $\eta_{0}\ll 1$,\
superconductivity will be destroyed long before metal---insulator transition.\
In the opposite case this mechanism may lead to its destruction on the either
side of metal---insulator transition depending on the parameters of the
system,\ such as $U$.\
In general we need a more accurate analysis which must
include the mutual interplay of magnetic fluctuations and disorder scattering
leading to metal---insulator transition.\ In any case magnetic mechanisms of
$T_{c}$ degradation close to metal---insulator transition may be
as important as Coulomb effects considered above.



\subsection{Ginzburg-Landau Theory and Anderson Transition}

\subsubsection{General Analysis}
The main result of the previous analysis may be formulated as follows.\ Despite
many mechanisms
leading to $T_{c}$ degradation and destruction of superconductivity in strongly
disordered
systems there seems to be no general rule prohibiting a possibility of a
superconducting state in
Anderson insulator.\ Of course we must meet very rigid conditions if we hope to
observe this
rather exotic state.\ There is almost no chance to observe it in traditional
superconductors but
high---$T_{c}$ systems seem promising.\ The following analysis will be based on
the general
assumption that $T_{c}$ survives in a strongly disordered system or even in
Anderson insulator,\
i.e.\ that these strict conditions are met.\ Our aim is to study
superconducting properties of such a
strongly disordered system to determine specific characteristics which will
make this case different
from the usual case of ``dirty'' superconductors.\ We shall see that even
before Anderson transition
there are significant deviations from the predictions of standard theory which
make strongly
disordered system different.\ So on the practical side our aim is simply to
generalize the usual
theory of ``dirty'' superconductors for the case of strong disorder in a sense
of the mean free path
becoming of the order of interatomic spacing or $l\sim p_{F}^{-1}$.

To claim that  superconductivity is possible close to disorder---induced
metal---insulator transition
it is not sufficient just to demonstrate the finite values of $T_{c}$.\ Even
more important is to show
the existence of superconducting response to an external electromagnetic
potential ${\bf A}$.\
In general case the analysis of response functions of a superconductor with
strong disorder seems
to be a difficult task.\ However,\ close to $T_{c}$ significant simplifications
take place and actually we only have to show that the free---energy density of
the system can be expressed in the
standard Ginzburg---Landau form \cite{GL,Gor59,Genn}:
\begin{equation}
F=F_{0}+A|\Delta|^{2}+\frac{1}{2}B|\Delta|^{4}+C|(\nabla -\frac{2ie}{\hbar
c}{\bf A})\Delta|^{2}
\label{FGL}
\end{equation}
where $F_{n}$ is free energy density of the normal state.\ Our problem is thus
reduced to a microscopic derivation of expressions for the coefficients $A$,\
$B$,\ and $C$ of Ginzburg---Landau expansion Eq. (\ref{FGL}) taking into
account the possibility of electron localization.\
This will be the generalization of the famous Gorkov's derivation \cite{Gor59}
of similar expressions for the case of ``dirty'' superconductors.\ Such
analysis was first done by  Bulaevskii and Sadovskii \cite{Bul1,Bul2} and later
by Kotliar and Kapitulnik \cite{KK1,KK2}.\
Recently the same results were obtained by Kravtsov \cite{Krav91}.

Within the BCS model coefficients $A$ and $B$ which determine the transition
temperature and
the equilibrium value of the order---parameter $\Delta$ do not change in
comparison with their
values found in the theory of ``dirty'' superconductors,\ even if the system is
close to Anderson
transition.\ This corresponds to the main statement of Anderson theorem.\ Less
trivial is the
behavior of the coefficient $C$,\ which in fact defines the superconducting
response.\ In the usual
theory of ``dirty'' superconductors \cite{Gor59} this coefficient is
proportional to diffusion coefficient of electrons,\ i.e.\ to conductivity (at
$T=0$).\ As the Fermi level approaches the mobility edge conductivity drops to
zero.\ However,\ we shall see that the coefficient $C$ remains
finite in the vicinity of Anderson transition,\ even in the region of localized
states.

To derive Ginzburg---Landau coefficients we must know the two---electron
Green's function in the
normal state \cite{Gor59}.\ Let us introduce the following two-particle
Matsubara Green's functions in momentum representation \cite{Bul2}:
\begin{equation}
\Psi_{E}({\bf q},\omega_{m},\varepsilon_{n})=-\frac{1}{2\pi i}\sum_{{\bf
p_{+}p'_{-}}}<G({\bf p_{+},p'_{+}},-\varepsilon_{n}+\omega_{m})G({\bf
-p'_{-},-p_{-}},-\varepsilon_{n})>
\label{Psim}
\end{equation}
\begin{equation}
\Phi_{E}({\bf q},\omega_{m},\varepsilon_{n})=-\frac{1}{2\pi i}\sum_{{\bf
p_{+}p'_{-}}}<G({\bf p_{+},p'_{+}},-\varepsilon_{n}+\omega_{m})G({\bf
p'_{-},p_{-}},-\varepsilon_{n})>
\label{Phim}
\end{equation}
where ${\bf p_{+-}=p^{+}_{-}q}/2$ and $\omega_{m}=2\pi mT$.
Graphically these functions are represented in Fig. \ref{fig10}.\ Then
Ginzburg---Landau coefficients are defined by \cite{Gor59,TSK82}:
\begin{equation}
A=\frac{1}{g}+2\pi iT\sum_{\varepsilon_{n}}\Psi_{E}({\bf
q}=0\omega_{m}=2\varepsilon_{n})
\label{A}
\end{equation}
\begin{equation}
C=i\pi T \sum_{\varepsilon_{n}}\frac{\partial^{2}}{\partial q^{2}}\Psi_{E}({\bf
q}\omega_{m}=2\varepsilon_{n})|_{q=0}
\label{C}
\end{equation}
Thus the superconducting properties are determined by the Green's function
$\Psi_{E}$ describing the propagation of electronic (Cooper) pair.\ At the same
time we have seen that the Green's function $\Phi_{E}$ determines transport
properties of a normal metal and Anderson transition.\ In case of
time---invariance (i.e.\ in the absence of external magnetic field or magnetic
impurities) we have \cite{YOF81}:
\begin{equation}
\Psi_{E}({\bf q}\omega_{m}\varepsilon_{n})=\Phi_{E}({\bf
q}\omega_{m}\varepsilon_{n})
\label{timeinv}
\end{equation}
and it is sufficient to know only $\Phi_{E}({\bf
q}\omega_{m}=2\varepsilon_{n})$ to determine Ginzburg---Landau coefficients.

As a one---electron model of Anderson transition we can take the
self---consistent theory of
localization which will allow  us to perform all calculations explicitly.\ We
only have to formulate the
main equations of this theory in Matsubara formalism (finite $T$) \cite{Bul2}.\
For small $q$ and
$\omega_{m}$,\ analogously to Eq. (\ref{phidiff}),\ we have:
\begin{equation}
\Phi_{E}({\bf
q}\omega_{m})=-\frac{N(E)}{i|\omega_{m}|+D_{E}(|\omega_{m}|)q^{2}}
\qquad \omega_{m}=2\pi mT
\label{phidiffm}
\end{equation}
where the generalized diffusion coefficient $D_{E}(\omega_{m})$ is determined
by the self---consistency equation analogous to Eq. (\ref{D}):
\begin{equation}
\frac{D_{0}}{D_{E}(\omega_{m})}=1-\frac{i}{\pi N^{2}(E)}\sum_{|{\bf q}|<k_{0}}
\Phi_{E}({\bf q}\omega_{m})
\label{Dm}
\end{equation}
In three-dimensional case Eq. (\ref{Dm}) reduces to (Cf.\ Eq. (\ref{D3fr})):
\begin{equation}
\frac{D_{E}(\omega_{m})}{D_{0}}=1-\frac{\lambda}{\lambda_{c}}+\frac{\pi}{2}\frac{\lambda}{\lambda_{c}}\left[\frac{D_{0}}{D_{E}(\omega_{m})}\frac{\omega_{m}}{2\gamma}\right]^{1/2}
\label{D3frm}
\end{equation}
where we have used the same notations as in our discussion of self---consistent
theory of localization.
Analogously to Eq. (\ref{difffreq}) and with accuracy sufficient for our aims
we can write down the solution
of Eq. (\ref{D3frm}) as:
\begin{equation}
D_{E}(\omega_{m})\approx Max
\left\{D_{E}\frac{\omega_{m}}{\omega_{m}+3D_{E}\omega_{0}^{2}(E)/v_{F}^{2}};\quad D_{0}\left(\frac{\omega_{m}}{2\gamma}\right)^{1/3}\right\}
\label{difffrm}
\end{equation}
where $D_{E}$ is the renormalized diffusion coefficient  defined in Eq.
(\ref{D3dgen}) and $\omega_{0}$ is the fundamental frequency defined  by Eq.
(\ref{omega}),\ which signals a
transition to insulator.

As we have already noted Ginzburg---Landau coefficients $A$ and $B$ are given
by the usual
expressions valid also for ``dirty'' superconductors \cite{Gor59,Bul2}:
\begin{equation}
A=N(E_{F})ln\frac{T}{T_{c}}\approx N(E_{F})\frac{T-T_{c}}{T_{c}}
\label{Adirt}
\end{equation}
where $T_{c}$ is given by the usual BCS relation of Eq. (\ref{BCSTc}),\ and
\begin{equation}
B=\frac{7\zeta(3)}{8\pi^{2}T_{c}^{2}}N(E_{F})
\label{Bdirt}
\end{equation}
where $\zeta(x)$ is Riemann zeta---function ($\zeta(3)=1.202...$).\ These
coefficients depend on disorder only through the appropriate disorder
dependence of $N(E_{F})$ and are valid even in localized phase.\ This is
equivalent to the main statement of Anderson theorem.

Significant changes appear in the gradient term coefficient $C$.\ Using Eqs.
(\ref{C})---(\ref{phidiffm}) with Eq. (\ref{difffrm}) we can find that in
different limiting cases this coefficient
can be expressed as \cite{Bul1,Bul2}:
\begin{equation}
C\equiv N(E_{F})\xi^{2}\approx N(E_{F})\left\{
\begin{array}{l}
\frac{\pi}{8T_{c}}D_{E_{F}} \qquad \xi_{loc}(E_{F})
<(\xi_{0}l^{2})^{1/3};\quad E_{F}>E_{c} \\
\left(\frac{D_{0}l}{T_{c}}\right)^{2/3}\approx (\xi_{0}l^{2})^{2/3} \qquad
\xi_{loc}(E_{F})>(\xi_{0}l^{2})^{1/3} \quad E_{F}\sim E_{c} \\
R^{2}_{loc}(E_{F})ln\frac{1.78 D_{E_{F}}}{\pi T_{c}R^{2}_{loc}(E_{F})} \qquad
R_{loc}(E_{F})<(\xi_{0}l^{2})^{1/3}; \quad E_{F}<E_{c}
\end{array}
\right.
\label{CBS}
\end{equation}
where we have defined the coherence length $\xi$,\ and $\xi_{0}=0.18
v_{F}/T_{c}$ is BCS
coherence length,\ $l$ as usual is the mean free path.
Practically the same results were obtained in Refs.\cite{KK1,KK2}using the
approach based upon elementary scaling theory of localization,\ which is as
we already noted is equivalent to our use of self---consistent theory of
localization.\ In Ref.\cite{Krav91}
the same results were confirmed using the $\sigma$---model approach to
localization.

In metallic state,\ as Fermi level $E_{F}$ moves towards the mobility edge
$E_{c}$ localization correlation length
$\xi_{loc}$ grows and the coefficient $C$ initially drops as the generalized
diffusion coefficient $D_{E_{F}}$,\ i.e.\ as conductivity of a system in the
normal state.\
However,\ in the vicinity of Anderson transition while
$\sigma\rightarrow 0$ the drop of $C$ coefficient saturates and it remains
finite even for $E_{F}<E_{c}$,\
i.e.\ in Anderson insulator.\ With further lowering of
$E_{F}$ into localization region (or with $E_{c}$ growth
with disorder) the $C$ coefficient is being determined by
localization radius $R_{loc}$ which diminishes as $E_{F}$
moves deep into insulating state.\ However,\ remembering Eq. (\ref{BScrit}) and
Eq. (\ref{critBS}) we recognize
that our analysis is valid only for large enough values
of localization length,\ which satisfy Eq. (\ref{critBS}).\ In this sense the
last asymptotics in
Eq. (\ref{CBS}) is actually outside these limits of
applicability.

The finite value of the coefficient $C$ in Ginzburg---Landau expansion in the
vicinity of Anderson transition
signifies the existence of superconducting (Meissner)
response to an external magnetic field.\ Accordingly,\
for $T<T_{c}$ the system can undergo a transition from
Anderson insulator to superconductor.\ The physical meaning of this result can
be understood from the following qualitative picture (Cf.\ Ref.\cite{IS81}
where
the similar estimates were used for the granular metal).\
In Anderson insulator all electrons with energies $E$
close to Fermi level are localized in spatial regions
of the size of $\sim R_{loc}(E)$.\ Nearby regions are connected by some
tunneling amplitude ${\cal V}$ which determines the probability of electron
transition between such regions as:
\begin{equation}
P_{T}\approx 2\pi |{\cal V}|^{2}N(E)R_{loc}^{3}(E)
\label{tunnpr}
\end{equation}
However,\ Anderson localization means that
\begin{equation}
|{\cal V}|<\frac{1}{N(E)R_{loc}^{3}(E)}
\label{Vtunins}
\end{equation}
and coherent tunneling between states localized in these
regions is impossible,\ and we have $P_{T}<2\pi N^{-1}(E)R_{loc}^{-3}$.\ At the
same time if conditions given
by Eq. (\ref{BScrit}) or Eq. (\ref{critBS}) are satisfied
inside each region $\sim R_{loc}$ Cooper pairs may form
and superconducting gap $\Delta$ appears in the spectrum.\ Then a kind of
``Josephson'' coupling appears
between regions of localized states which determines the
possibility of {\em pairs} tunneling:
\begin{equation}
E_{J}\approx \pi^{2}[N(E)R^{3}_{loc}(E)]^{2}|{\cal V}|^{2}\Delta
\label{Jostun}
\end{equation}
It is easy to see that for
\begin{equation}
\Delta > \frac{2}{\pi}\frac{1}{N(E)R_{loc}^{3}(E)}
\label{DelJ}
\end{equation}
we have $E_{J}>P_{T}$,\ so that if Eq. (\ref{BScrit}) is
satisfied we can get $E_{J}\gg N^{-1}(E)R^{-3}(E)$ despite of Eq.
(\ref{Vtunins}) and tunneling of pairs between nearby regions of localized
states is possible,\
even in the absence of single---particle tunneling.

It is convenient to rewrite Eq. (\ref{CBS}) using the
the relation between generalized diffusion coefficient
and conductivity like Eq. (\ref{condM}) as well as Eqs. (\ref{sig3d}),\
(\ref{sigmaBS}).\ Then using the Ginzburg---Landau expansion and the
expressions for its
coefficients we can easily find the temperature dependent
coherence length $\xi(T)$ \cite{Genn,Bul1,Bul2}:
\begin{equation}
\xi^{2}(T)=\frac{T_{c}}{T_{c}-T}\left\{
\begin{array}{l}
\xi_{0}l\frac{\sigma}{\sigma+\sigma_{c}} \qquad \sigma > \sigma^{\star} \quad
(E_{F}>E_{c}) \\
(\xi_{0}l^{2})^{2/3} \qquad \sigma < \sigma^{\star} \quad
(E_{F}\sim E_{c})
\end{array}
\right.
\label{xiT}
\end{equation}
where $\sigma_{c}=e^{2}p_{F}/(\pi^{3}\hbar^{2})$ and characteristic
conductivity scale $\sigma^{\star}$ is
given by
\begin{equation}
\sigma^{\star}\approx \sigma_{c}(p_{F}\xi_{0})^{-1/3}
\approx \sigma_{c}\left(\frac{T_{c}}{E_{F}}\right)^{1/3}
\label{sigstar}
\end{equation}
Thus in the region of very small conductivities $\sigma < \sigma^{\star}$ the
scale of $\xi(T)$ is defined
not by $\xi\sim \sqrt{\xi_{0}l}$ as in the usual theory of
``dirty'' superconductors \cite{Gor59,Genn} but by the
new length $\xi\sim (\xi_{0}l^{2})^{1/3}\sim (\xi_{0}/p_{F}^{2})^{1/3}$,\ which
now is the characteristic size of Cooper pair close to Anderson
transition.

In a case if $\omega^{1/3}$---law for a diffusion coefficient at the mobility
edge is invalid and we have $\omega^{\delta}$---behavior,\ with some unknown
critical
exponent $\delta$ (which is possible because the modern theory actually cannot
guarantee precise values of critical exponents at Anderson transition
\cite{Weg76,Sad86}) we can easily show in a similar way
that for conductivities $\sigma < \sigma^{\star}\approx
\sigma_{c}(p_{F}\xi_{0})^{-\delta}$
the coherence length is defined by $\xi\sim
\xi_{0}^{\frac{1-\delta}{2}}l^{\frac{1+\delta}{2}}$.\ Qualitatively this leads
to the same type of behavior as above.

{}From Eq. (\ref{xiT}) we can see that $\xi^{2}(T)$ initially diminishes as we
approach metal---insulator transition proportionally to $\sigma$ as in the case
of
a ``dirty'' superconductor.\ However,\ already in metallic region for
$\sigma<\sigma^{\star}$ it diminishes
much slower remaining finite both at the transition itself and below.

The superconducting electron density $n_{s}$ can be defined as \cite{Genn}:
\begin{equation}
n_{s}(T)=8mC\Delta^{2}(T)=8mC(-A)/B
\label{nsGL}
\end{equation}
Close to Anderson transition we can estimate:
\begin{equation}
n_{s}\sim mN(E_{F})\xi^{2}\Delta^{2}\sim
mp_{F}(\xi_{0}/p_{F}^{2})^{2/3}\Delta^{2}\sim
n(T_{c}^{1/2}/E^{2}_{F})^{2/3}(T_{c}-T)
\label{nsloc}
\end{equation}
where $n\sim p_{F}^{3}$ is total electron density.\ If we take here $T\sim
0.5T_{c}$
i.e.\ more or less low temperatures we get a simpler estimate:
\begin{equation}
n_{s}\sim n\left(\frac{T_{c}}{E_{F}}\right)^{4/3}
\label{nslocl}
\end{equation}
which is actually valid up to $T=0$,\ as we shall see below.\ From these
estimates
we can see that only a small fraction of electrons are superconducting in a
strongly disordered case.\ However this confirms a possibility of
superconducting
response of Anderson insulator.

Characteristic conductivity $\sigma^{\star}$ defined in
Eq. (\ref{sigstar}) gives an important conductivity scale
at which significant influence of localization effects
upon superconducting properties appear \cite{Bul2}.\ While $\sigma_{c}$ is of
the order of Mott's ``minimal
metallic conductivity'' \cite{Mott74,MottDav} $\sigma^{\star}$ is in general
even lower.\ However,\ for
small enough Cooper pairs (i.e.\ small $\xi_{0}$ which is
characteristic of strong coupling and high---$T_{c}$ superconductors) it is
more or less of the order of $\sigma_{c}$.\ Experimentally it can be defined as
a
conductivity scale at which significant deviations from
predictions of the standard theory of ``dirty'' superconductors appear under
disordering.

We must stress that these results show the possibility of Cooper pairs being
delocalized in Anderson
insulator,\ while single---particle excitations of such
superconductor are apparently localized,\ which may lead
to some peculiar transport properties of ``normal''
electrons for $T<T_{c}$.\ First attempts to explore this peculiar situation
were undertaken in Refs.\cite{Opp87,Opp88,Opp90,Opp91}.

These results are easily generalized for the case of strongly anisotropic
quasi---two---dimensional systems
such as high---$T_{c}$ superconducting oxides.\ Using the
analysis of such systems within the self---consistent
theory of localization \cite{PrFir84} we can write down
the following Matsubara generalization of Eq. (\ref{alphafreq}):
\begin{equation}
\frac{D_{j}(\omega_{m})}{D^{0}_{j}}\approx \left\{
\begin{array}{l}
Max\left[\frac{E_{F}-E_{c}}{E_{c}};(2\pi
E_{F}w\tau^{2})^{-2/3}(\omega_{m}\tau)^{1/3}\right] \qquad \omega_{m}\ll
w^{2}\tau \\
1-\frac{1}{2\pi E_{F}\tau}ln\left(\frac{1}{\omega_{m}\tau}\right) \qquad
\omega_{m}\gg w^{2}\tau
\end{array}\right.
\label{difManis}
\end{equation}
where $j=\|,\bot$.\ Now carrying out calculations similar
to that of Ref.\cite{Bul2} we obtain for the coefficients of gradient terms in
Ginzburg---Landau expansion \cite{Sad89,GoSad89}:
\begin{equation}
C_{\|,\bot}=N(E_{F})\xi^{2}_{\|,\bot}
\label{Canis}
\end{equation}
where for the coherence lengths $\xi_{\|,\bot}$ we obtain
a number of different expressions,\ depending on the value of the ratio
$w^{2}\tau/2\pi T_{c}\hbar$ which
determines as we shall see the ``degree of two---dimensionality'' of the
problem under study.\ For the case of $w^{2}\tau/2\pi T_{c}\hbar\gg 1$,\
corresponding
to an anisotropic but three---dimensional system,\ we have:
\begin{equation}
\xi^{2}_{\|,\bot}=\frac{\pi}{8T_{c}}D^{0}_{\|,\bot}\left(\frac{E_{F}-E_{c}}{E_{c}}\right)\approx \xi^{0}_{\|,\bot}l_{\|,\bot}\left(\frac{E_{F}-E_{c}}{E_{c}}\right)
\label{xianismet}
\end{equation}
where $\xi^{0}_{\|}\sim \hbar v_{F}/T_{c}$,\ $\xi^{0}_{\bot}\sim
wa_{\bot}/T_{c}$,\ $l_{\|}=v_{F}\tau$ and $l_{\bot}=wa_{\bot}\tau/\hbar$ are
the longitudinal and transverse BCS coherence lengths and mean free paths.\ The
above expressions are valid in the conductivity region
$\sigma_{\|}>\sigma^{\star}$,\ where
\begin{equation}
\sigma^{\star}\sim
\sigma_{\|}^{c}\frac{\xi^{0}_{\|}}{l_{\|}}\left(\frac{T_{c}^{2}}{E_{F}w}\right)^{2/3}
\label{sigstaranis}
\end{equation}
Here $\sigma^{c}_{\|}$ was defined in Eq. (\ref{mmcq2d}).\
The condition of $w^{2}\tau/2\pi T_{c}\hbar\gg 1$ is
equivalent to the requirement:
\begin{equation}
\xi_{\bot}\sim \sqrt{\xi^{0}_{\bot}l_{\bot}}\gg a_{\bot}
\label{3dpair}
\end{equation}
which clarifies its physical meaning:\ the transverse size of a Cooper pair
must be much greater than interplane lattice spacing.\ In this case we have
just
anisotropic three-dimensional superconductivity.

In the immediate vicinity of the Anderson transition,\
for $\sigma_{\|}<\sigma^{\star}$ we have:
\begin{equation}
\xi^{2}_{\|,\bot}\approx (1-2^{-5/3})(16\pi^{4})^{-1/3}
\zeta(5/3)\frac{D_{\|,\bot}}{(E_{F}T_{c}w)^{2/3}\tau}\approx
(\xi^{0}_{\|,\bot})^{2}\left(\frac{T_{c}^{2}}{E_{F}w}\right)^{2/3}
\label{xianismobed}
\end{equation}
It is easy to see that for $w\sim E_{F}$ all these
expressions naturally go over to those derived above for the
three---dimensional case.

For the case of $w^{2}\tau/2\pi T_{c}\hbar < 1$ which
corresponds to ``almost two---dimensional'' case of
\begin{equation}
\xi_{\bot}\sim \sqrt{\xi^{0}_{\bot}l_{\bot}}\leq a_{\bot}
\label{2dpair}
\end{equation}
i.e.\ of transverse size of Cooper pairs smaller than
interplane spacing,\ we have
\begin{equation}
\xi^{2}_{\|,\bot}\approx \left\{
\begin{array}{l}
\frac{D^{0}_{\|,\bot}}{\pi T_{c}}\frac{E_{F}-E_{c}}{E_{c}} \qquad (\sigma_{\|}
> \sigma^{\star}) \\
\frac{D^{0}_{\|,\bot}}{(4\pi^{2}E_{F}T_{c}w)^{2/3}\tau}
\qquad (\sigma_{\|} < \sigma^{\star}) \end{array}\right\}+
(\pi^{2}/8-1)\frac{D^{0}_{\|,\bot}}{\pi T_{c}}\left(1-\frac{1}{2\pi
E_{F}\tau}ln\frac{1}{2\pi T_{c}\tau}\right)
\label{xiq2d}
\end{equation}
Essential difference from just anisotropic case of Eq.
(\ref{xianismet}) and Eq.(\ref{xianismobed}) is the
appearance here of a second term of ``two-dimensional''
type.\ In purely two---dimensional problem ($w=0$) we
have \cite{TSK82}:
\begin{equation}
\xi^{2}_{\|}=\frac{\pi D^{0}_{\|}}{8T_{c}}\left(1-\frac{1}{2\pi
E_{F}\tau}ln\frac{1}{2\pi T_{c}\tau}\right)
\label{xi2d}
\end{equation}
For high---$T_{c}$ oxides it is reasonable to estimate
$\xi^{0}_{\|}\sim l_{\|}$,\ $T_{c}\sim w$,\ $T_{c}\sim 0.1E_{F}$,\ so that
$\sigma^{\star}\sim \sigma^{c}_{\|}$,\ i.e.\ these systems are always more or
less close to the Anderson transition.\ For $T_{c}\sim w$ and $\hbar/\tau\sim
E_{F}$ which is characteristic of
rather strongly disordered case,\ we have $w^{2}\tau/2\pi T_{c}\hbar<1$,\ so
that for these systems we can realize
almost two---dimensional behavior,\ though in general
high---$T_{c}$ oxides are apparently an intermediate case
between strongly anisotropic three---dimensional and
nearly two---dimensional superconductors.

The significant change of Ginzburg---Landau coefficients and the new
scale of coherence length close to the Anderson transition lead to an increased
width of critical region
of thermodynamic fluctuations near $T_{c}$\cite{KK1,KK2}.\ These are well known
to be important for any second---order phase transition.\ The
width of the critical region is defined by the so called
Ginzburg criterion \cite{Kad,PP} which may be expressed
via the coefficients of Landau expansion.\ Mean---field
approximation for the order parameter in Landau theory is
valid (for $d=3$) for\cite{Kad,PP}
\begin{equation}
1\gg\left|\frac{T-T_{c}}{T_{c}}\right|\gg\frac{B^{2}T^{2}_{c}}{\alpha
C^{3}}\equiv \tau_{G}
\label{Gi}
\end{equation}
where $\alpha$ is defined by $A=\alpha (T-T_{c})/T_{c}$.\
In case of superconducting transition we have: $\alpha=N(E_{F})$,\ $B\sim
N(E_{F})/T^{2}_{c}$ and $C=N(E_{F})\xi^{2}$.\ Accordingly,\ from Eq. (\ref{Gi})
we get the following estimate for the critical region:
\begin{equation}
\tau_{G}\sim \frac{1}{N^{2}(E_{F})\xi^{6}T^{2}_{c}}\sim
\left(\frac{E_{F}}{T_{c}}\right)^{2}\frac{1}{\xi^{6}p^{6}_{F}}
\label{Gisc}
\end{equation}
In the ``pure'' limit $\xi =\xi_{0}\sim v_{F}/T_{c}$ and we get $\tau_{G}\sim
(T_{c}/E_{F})^{4}$,\ so that critical region is practically unobservable.\ In a
``dirty'' superconductor $\xi\sim \sqrt{\xi_{0}l}$ and
\begin{equation}
\tau_{G}\sim \left(\frac{T_{c}}{E_{F}}\right)\frac{1}{(p_{F}l)^{3}}
\label{Gidirt}
\end{equation}
and again we have $\tau_{G}\ll 1$.\ However,\ for a superconductor close to
mobility edge $\xi\sim
(\xi_{0}/p_{F}^{2})^{1/3}$ and from Eq. (\ref{Gisc}) we
get:\cite{KK1,KK2}
\begin{equation}
\tau_{G}\sim 1
\label{Gime}
\end{equation}
Note that in fact $\tau_{G}$ may still be small because of numerical constants
which we have dropped in our estimates.\ Anyhow,\ the critical region in this
case
becomes unusually wide and superconducting transition
becomes similar in this respect to superfluid transition
in Helium.\ Fluctuation effects may thus become observable
even in bulk three---dimensional superconductor.\ Note
that in localized phase $\xi\sim R_{loc}$ and $\tau_{G}\sim
[N^{2}(E_{F})R^{6}_{loc}T^{2}_{c}]^{-1}>1$
if the condition given by Eq. (\ref{BScrit}) is violated.

Finally we should like to mention that thermodynamic fluctuations lead
\cite{KK1,KK2} to an
additional mechanism of $T_{c}$ degradation for a system which is close to
Anderson transition.\ This follows from the general result on thr reduction of
mean---field transition temperature due to critical fluctuations.\ If these
fluctuations are small (and we can use the so called one---loop approximation)
for a three-dimensional system it can be shown that \cite{KK1,KK2}:
\begin{equation}
T_{c}=T_{c0}-\frac{7\zeta(3)}{16\pi^{4}\xi^{3}N(E_{F})}
\label{Tcfl}
\end{equation}
where $T_{c0}$ is the mean---field transition temperature.\ If we use here our
expressions for $\xi$ valid close to metal---insulator transition we easily
find for $\sigma>\sigma^{\star}$\cite{KK2}:
\begin{equation}
T_{c}\approx
T_{co}\left[1-0.5\left(\frac{\sigma_{c}}{\sigma}\right)^{3/2}\left(\frac{T_{c0}}{E_{F}}\right)^{1/2}\right]
\label{TcKK}
\end{equation}
For $\sigma<\sigma^{\star}$ this fluctuation correction saturates as the
further drop of coherence length stops there.\ Obviously higher---order
corrections are important here,\ but unfortunately little is known on the
importance of this mechanism of $T_{c}$ degradation outside the limits of
one---loop approximation.

\subsubsection{Upper critical field}

Direct information on the value of $\xi^{2}(T)$ can be
obtained from the measurements of the upper critical
field $H_{c2}$ \cite{Genn}:
\begin{equation}
H_{c2}=\frac{\phi_{0}}{2\pi\xi^{2}(T)}
\label{Hc2}
\end{equation}
where $\phi_{0}=\pi c\hbar/e$ is superconducting magnetic flux quantum.\ Using
Eq. (\ref{xiT}) we obtain the following relation between normal state
conductivity $\sigma$,\ the slope of the upper critical field at $T=T_{c}$
given by $(dH_{c2}/dT)_{T_{c}}$ and the value of
electronic density of states at the Fermi level (per one
spin direction) $N(E_{F})$\cite{Bul1,Bul2}:
\begin{equation}
-\frac{\sigma}{N(E_{F})}\left(\frac{dH_{c2}}{dT}\right)_{T_{c}}\approx \left\{
\begin{array}{l}
\frac{8e^{2}}{\pi^{2}\hbar}\phi_{0} \qquad \sigma > \sigma^{\star} \\
\phi_{0}\frac{\sigma}{N(E_{F})(\xi_{0}l^{2})^{2/3}T_{c}}\approx
\phi_{0}\frac{\sigma}{[N(E_{F})T_{c}]^{1/3}}
\qquad \sigma < \sigma^{\star}
\end{array}
\right.
\label{Gorrel}
\end{equation}
For $\sigma > \sigma^{\star}$ the r.h.s.\ of Eq. (\ref{Gorrel}) contains only
the fundamental constants.\
This so called Gorkov's relation \cite{Gor59} is often
used to interpret experimental data in ``dirty'' superconductors.\ Using it we
may find $N(E_{F})$ for
different degrees of disorder from measurements of
$(dH_{c2}/dT)_{T_{c}}$ and conductivity $\sigma$.\ On the
other hand $N(E_{F})$ can in principle be determined from
independent measurements e.g.\ of electronic contribution to specific heat.\
However,\ our expression for $\sigma < \sigma^{\star}$ which is valid close to
metal---insulator transition shows that in this region Gorkov's relation
becomes invalid and its use can ``simulate'' the drop of
$N(E_{F})$ with the growth of resistivity (disorder).\
Roughly speaking Eq. (\ref{Gorrel}) shows that under the
assumption of relatively smooth change of $N(E_{F})$ and
$T_{c}$ with disorder the usual growth of $(dH_{c2}/dT)_{T_{c}}$ with disorder
saturates in conductivity region of $\sigma < \sigma^{\star}$  close
to the Anderson transition and the slope of the upper
critical field becomes independent of resistivity.\
This stresses the importance of independent measurements of $N(E_{F})$.

Note that the qualitative behavior given by Eq. (\ref{Gorrel}) is retained also
in the case when $\omega^{\delta}$---dependence of diffusion coefficient
at the mobility edge (with some arbitrary critical exponent $\delta$),\ only
the expression for $\sigma^{\star}$ is changed as noted above.\ Thus this
behavior is not related to any specific approximations of
self---consistent theory of localization,\ except the
general concept of continuous transition.

For an anisotropic (quasi---two---dimensional) system we
have similar relations:
\begin{equation}
\left(\frac{dH_{c2}^{\bot}}{dT}\right)_{T_{c}}=-\frac{\phi_{0}}
{2\pi\xi^{2}_{\|}T_{c}}
\label{dHc2perp}
\end{equation}
\begin{equation}
\left(\frac{dH_{c2}^{\|}}{dT}\right)_{T_{c}}=-
\frac{\phi_{0}}{2\pi\xi_{\|}\xi_{\bot}T_{c}}
\label{dHc2paral}
\end{equation}
with $\xi_{\|,\bot}$ given above during our discussion after Eq.
(\ref{Canis}).\ This leads to relations and qualitative behavior similar to Eq.
(\ref{Gorrel}).\
However,\ we should like to note an especially interesting relation for the
anisotropy of the slopes of
the upper critical field \cite{Sad89}:
\begin{equation}
\frac{(dH_{c2}^{\|}/dT)_{T{c}}}{(dH_{c2}^{\bot}/dT)_{T{c}}}=\frac{\xi_{\|}}{\xi_{\bot}}=\frac{v_{F}}{wa/\hbar}
\label{anislop}
\end{equation}
We see that the anisotropy of $(dH_{c2}/dT)_{T_{c}}$ is
actually determined by the anisotropy of the Fermi velocity irrespective of the
regime of superconductivity:
from the ``pure'' limit,\ through the usual ``dirty'' case,\ up to the vicinity
of the Anderson transition.

The above derivation of $C$ coefficient of Ginzburg---Landau expansion
explicitly used the time---reversal invariance expressed 
by Eq. (\ref{timeinv}).\ This is valid in the absence of the external
magnetic field and magnetic impurities.\ Accordingly the previous results for
the upper critical field are formally valid in the limit of infinitesimal
external field and this is sufficient for the demonstration of superconducting
(Meissner) response  and for the determination of
$(dH_{c2}/dT)_{T_{c}}$,\ because $H_{c2}\rightarrow 0$ as $T\rightarrow
T_{c}$.\ In a finite
external field we must take into account its influence upon localization.\ The
appropriate analysis
was performed in Refs.\cite{KS91a,KS91b} and with a slightly different method
in Ref.\cite{Alba91}.\ The results are essentially similar and below we shall
follow Ref.\cite{KS91a}.\ The standard scheme for the analysis of
superconducting transition in an
external magnetic field \cite{Gor59,Genn,WHH,SJS} gives the following equation
determining the
temperature dependence of $H_{c2}(T)$:
\begin{equation}
ln\frac{T}{T_{c}}=2\pi
T\sum_{\varepsilon_{n}}\left\{\frac{1}{2|\varepsilon_{n}|+2\pi
D_{2}(2|\varepsilon_{n}|)H/\phi_{0}}-\frac{1}{2|\varepsilon_{n}|}\right\}
\label{Hc2T}
\end{equation}
where $D_{2}(2|\varepsilon_{n}|)$ is the generalized diffusion coefficient in
the Cooper channel as defined after Eqs. (\ref{M1}) and (\ref{M2}).\ Eq.
(\ref{Hc2T}) is valid \cite{Genn} for
\begin{equation}
R_{H}=\frac{mcv_{F}}{eH}\gg \xi
\label{Rh}
\end{equation}
 $R_{H}$ is Larmor radius of an electron in magnetic field,\ $\xi$ is the
coherence length.\ Note that Eq. (\ref{Hc2T}) describes only the orbital motion
contribution to $H_{c2}$.\ In fact $H_{c2}$ is also limited by the paramagnetic
limit \cite{Genn,SJS}:
\begin{equation}
\frac{1}{2}g_{0}\mu_{B}H<\Delta
\label{pmlim}
\end{equation}
where $g_{0}$ is the usual $g$---factor of an electron,\ $\mu_{B}$ is Bohr
magneton.

Standard approach of the theory of ``dirty'' superconductors is based upon the
replacement of
$D_{2}(2|\varepsilon_{n}|)$ in Eq. (\ref{Hc2T}) by Drude diffusion coefficient
$D_{0}$ which is
valid for a metal with $l\gg p_{F}^{-1}$.\ For a system which is close to the
Anderson transition we
must take into account both the frequency dependence of diffusion coefficient
and the fact that in
magnetic field $D_{2}$ is not equal to $D_{1}$ --- the usual diffusion
coefficient determining electronic transport.\ Actually we shall see that the
external magnetic field influence upon localization leads to rather small
corrections to $H_{c2}(T)$ practically everywhere except the
region of localized states \cite{KS91a}.\ Thus we may really neglect this
influence as a first
approximation as that was done in Refs.\cite{Bul1,Bul2} and start with the
replacement of
$D_{2}$ in Eq. (\ref{Hc2T}) by $D_{1}=D_{E}$,\ where $D_{E}$ is the frequency
dependent generalized diffusion coefficient in the absence of magnetic field.\
Detailed analysis of Eq. (\ref{Hc2T}) can be found in Ref.\cite{KS91a}.

Summation over Matsubara frequencies in Eq. (\ref{Hc2T}) must be cut---off at
some frequency of
the order of $<\omega>$ --- the characteristic frequency of Bose excitations
responsible for
pairing interaction.\ It is convenient here to measure the distance from
Anderson transition (degree
of disorder) via frequency $\omega_{c}$ defined in Eqs.
(\ref{omegc}),(\ref{omegac}) or Eq. (\ref{omegacr}) .\ If a system is far from
Anderson transition,\ so that $\omega_{c}\gg <\omega>$  we can completely
neglect the frequency dependence of diffusion coefficient and
find the usual results of the theory of ``dirty'' superconductors:
\begin{equation}
H_{c2}(T)= \frac{4}{\pi^{2}}\frac{\phi_{0}T_{c}}{D_{0}}ln\frac{T_{c}}{T} \qquad
T\sim T_{c}
\label{Hc2Tcdirt}
\end{equation}
\begin{equation}
H_{c2}=\frac{1}{2\gamma}\frac{\phi_{0}T_{c}}{D_{0}}\left[1-\frac{1}{24}\left(\frac{4\gamma T}{T_{c}}\right)^{2}\right] \qquad T\ll T_{c}
\label{Hc2zerodirt}
\end{equation}
where $\gamma=1.781...$.\ For the $H_{c2}$ derivative at $T=T_{c}$ we find from
here the first
relation of Eq. (\ref{Gorrel}),\ and $H_{c2}(T=0)$ is conveniently expressed
as\cite{Gor59,WHH}:
\begin{equation}
-\frac{H_{c2}(0)}{T_{c}(dH_{c2}/dT)_{T_{c}}}=\frac{\pi^{2}}{8\gamma}\approx
0.69
\label{Hc2zerslop}
\end{equation}
In this case $H_{c2}(T)$ curve is convex at all temperatures below
$T_{c}$.\cite{Gor59,WHH,Genn,SJS} Very close to the Anderson transition,\ when
$\omega_{c}\ll 2\pi T$ ,\ only $\omega^{1/3}$ behavior of diffusion coefficient
is important in
Eq. (\ref{Hc2T}) and it takes the following form\cite{KS91a}:
\begin{equation}
ln\frac{T}{T_{c}}=\sum_{n=0}^{\infty}\left\{[(n+1/2)+(n+1/2)^{1/3}(E/4\pi
T)^{2/3}(\omega_{H}/E)]^{-1}-[n+1/2]^{-1}\right\}
\label{hc2tme}
\end{equation}
where $\omega_{H}=eH/mc$.\ From here we get:
\begin{equation}
H_{c2}(T)=m\frac{\phi_{0}}{\pi}\frac{(4\pi)^{2/3}}{c_{1}}T^{2/3}E^{1/3}ln\frac{T_{c}}{T} \qquad T\sim T_{c}
\label{hc2tcme}
\end{equation}
\begin{equation}
H_{c2}(T)=m\frac{\phi_{0}}{\pi}(\pi/\gamma)^{2/3}T_{c}^{2/3}E^{1/3}\left[1-\frac{2}{3}c_{2}\left(\frac{4\gamma T}{T_{c}}\right)^{2/3}\right]  \qquad T\ll T_{c}
\label{hc2zerme}
\end{equation}
where $c_{1}=\sum_{n=0}^{\infty}(n+1/2)^{-5/3}\approx 4.615$ and $c_{2}\approx
0.259$.\ From these expressions we get:
\begin{equation}
-\frac{1}{N(E)}\left(\frac{dH_{c2}}{dT}\right)_{T_{c}}=\frac{(4\pi)^{2/3}}{\pi
c_{1}}m\phi_{0}(E/T_{c})^{1/3}=\frac{2\pi}{c_{1}}\frac{\phi_{0}}{[N(E)T_{c}]^{1/3}}
\label{slopeme}
\end{equation}
which makes precise the second relation in Eq. (\ref{Gorrel}),\ while for
$H_{c2}(T=0)$ we obtain:
\begin{equation}
-\frac{H_{c2}(0)}{T_{c}(dH_{c2}/dT)_{T_{c}}}=\frac{c_{1}}{(4\gamma)^{2/3}}\approx 1.24
\label{hc2zerome}
\end{equation}
As was first noted in Refs.\cite{Bul1,Bul2} this ratio for the system at the
mobility edge is significantly larger than its classical value 0.69.\ In this
case $H_{c2}(T)$ curve is concave for
all temperatures below $T_{c}$.\cite{Bul2}Detailed expressions for the
intermediate disorder
when $2\pi T\ll \omega_{c}\ll <\omega>$ can be found in Ref.\cite{KS91a}.

On Fig.\ref{fig11} we present the results of numerical solution of Eq.
(\ref{Hc2T}) for the different
values of characteristic frequency $\omega_{c}$,\ i.e. for the different
disorder.\ A smooth crossover from the classical behavior of the theory of
``dirty'' superconductors \cite{WHH,Genn,SJS} to anomalous temperature
dependence close to the Anderson transition
\cite{Bul2} is clearly seen.

Below the mobility edge (i.e.\ in Anderson insulator) and
for $\omega_{c}=1/(2\pi^{2}N(E)R^{3}_{loc})\ll 2\pi T$,\ i.e.\ very close to
mobility edge we can again use $\omega^{1/3}$---behavior of diffusion
coefficient and find the same temperature dependence of $H_{c2}$ as at the
mobility edge itself or just above it.\ For $2\pi T\ll\omega_{c}\ll2\pi T_{c}$
Eq. (\ref{Hc2T}) takes the
form \cite{KS91a}:
\begin{eqnarray}
ln\frac{T}{T_{c}}=\sum_{n=0}^{n_{0}-1}\{(n+1/2)[1+(E/\omega_{c})^{2/3}(\omega_{H}/E)]\}^{-1}+
\nonumber \\
+\sum_{n=n_{0}}^{\infty}\{(n+1/2)+(n+1/2)^{1/3}(E/4\pi
T)^{2/3}(\omega_{H}/E)\}^{-1}-\sum_{n=0}^{\infty}(n+1/2)^{-1}
\label{Hc2Tandins}
\end{eqnarray}
where $n_{0}=\omega_{c}/4\pi T_{c}$ corresponds to a change of frequency
behavior of diffusion coefficient.\
Defining $x=\omega_{H}/\omega_{c}^{2/3}E^{1/3}$ we can reduce Eq.
(\ref{Hc2Tandins}) to:
\begin{equation}
ln(T/T_{c})= x ln(\gamma\omega_{c}/\pi T_{c})+\frac{3}{2}(1+x)ln(1+x)
\label{Hc2andins}
\end{equation}
which implicitly defines $H_{c2}(T)$ and shows \cite{KS91a}
that now $H_{c2}(T)\rightarrow \infty$ for $T\rightarrow 0$
(logarithmic divergence).\ Numerical solution of Eq. (\ref{Hc2Tandins}) is
shown at the insert in Fig.\ref{fig11}.\ Below we shall see however,\ that this
divergence of $H_{c2}$ is lifted by the inverse influence of magnetic field
upon diffusion.

Let us now turn to the problem of magnetic field influence upon diffusion and
its consequences for $H_{c2}$ temperature behavior.\ If we are far from the
Anderson transition magnetic field influence is small on
parameter $\sim \sqrt{\omega_{H}/E}$ and its influence
upon $H_{c2}$ is insignificant.\ Close to the transition
magnetic field correction may overcome the value of $D(H=0)$ and we have to
consider its influence in detail
\cite{KS91a}.\ Accordingly we shall limit ourselves with
the case of $\omega_{c}/E\ll (\omega_{H}/E)^{2/3}$ for
which we have already discussed the magnetic field behavior of generalized
diffusion coefficient in Cooper
channel.\ It was given in Eq. (\ref{D22}) and Eq. (\ref{D222}).\ In this case
we have seen that characteristic frequency $\omega_{c}$ is replaced by:
\begin{equation}
\omega'_{c}=(\varphi\omega_{H}/E)^{3/2}E
\label{omegcprim}
\end{equation}
where $\varphi=W^{2}/2\approx 0.18$.($W$ was defined
during our discussion of localization in magnetic field).\  For $T\sim T_{c}$
there is no change in the slope of $H_{c2}$  given by Eq. (\ref{slopeme}) as
was noted already in Ref.\cite{Bul2}.\ Here we shall consider the case of $T\ll
T_{c}$.

For $2\pi T>\omega'_{c}$ in all sums over Matsubara frequencies we can take
$D(\omega)\sim \omega^{1/3}$ and actually we can neglect magnetic field
influence upon diffusion.\ In this case $H_{c2}(T)$ behaves like in Eq.
(\ref{hc2zerme}) i.e.\ as at the mobility edge in the absence of magnetic field
effects.\ For
$2\pi T<\omega'_{c}$ equation for $H_{c2}(T)$ takes the
form \cite{KS91a}:
\begin{eqnarray}
ln\frac{T}{T_{c}}=\sum_{n=0}^{n_{0}-1}[(n+1/2)+(\omega_{c}/E)^{1/3}(\omega_{H}/4\pi T)]^{-1}+
\nonumber \\
+\sum_{n=n_{0}}^{\infty}\{(n+1/2)+(n+1/2)^{1/3}(E/4\pi
T)^{2/3}(\omega_{H}/E)\}^{-1}-\sum_{n=0}^{\infty}(n+1/2)^{-1}
\label{Hc2TH}
\end{eqnarray}
where $n_{0}=\omega'_{c}/4\pi T$.\ In this case we find
\begin{equation}
H_{c2}(T)=m\frac{\phi_{0}}{\pi}(1+\varphi)^{-1/3}(\pi/\gamma)^{2/3}T_{c}^{2/3}E^{1/3}\left[1-\frac{4\gamma}{3\varphi^{1/3}(1+\varphi)}\frac{T}{T_{c}}\right]
\label{Hc2HlowT}
\end{equation}
Accordingly we have
\begin{equation}
-\frac{H_{c2}(0)}{T_{c}(dH_{c2}/dT)_{T_{c}}}=(1+\varphi)^{-1/3}\frac{c_{1}}{(4\gamma)^{2/3}}\approx 1.18
\label{Hc2THzerome}
\end{equation}
and the change in comparison with Eq. (\ref{hc2zerome}) is actually small.\
However,\ for $2\pi T<\omega'_{c}$
the $H_{c2}(T)$ curve becomes convex.\ The inflexion point can be estimated as
$T^{\star}=\omega'_{c}/2\pi\approx 0.02T_{c}$.\ This behavior is shown in the
insert on Fig.\ref{fig11}.

Consider now insulating region.\ We shall see that the
magnetic field effects on diffusion lead to the effective
cut---off of the weak divergence of $H_{c2}$ as $T\rightarrow 0$ noted above.\
Generalized diffusion
coefficient $D_{2}$ in insulating phase and at low enough frequencies is
determined by the following equation\cite{KS91a}:
\begin{equation}
2mD_{2}=-(\omega_{c}/E)^{1/3}+(-i\omega/E)^{1/2}(2mD_{2})^{-1/2}+
\frac{1}{2}W(2\omega_{H}/E)^{1/2}
\label{D2Hins}
\end{equation}
Now we can see that the external field defined by
\begin{equation}
\frac{W}{2}\sqrt{2\omega_{H}/E}>(\omega_{c}/E)^{1/3}
\label{Hmi}
\end{equation}
transfers the system from insulating to metallic state.\
If the system remains close to mobility edge we can estimate the upper critical
field as above by $\omega_{H}\approx (\pi/\gamma)^{2/3}T_{c}^{2/3}E^{1/3}$
and Eq. (\ref{Hmi}) reduces to:
\begin{equation}
\omega_{c}\approx\frac{1}{2\pi^{2}N(E)R_{loc}^{3}}<\frac{\pi}{\gamma}(W/\sqrt{2})^{3}T_{c}\approx 0.14T_{c}
\label{omegacnosc}
\end{equation}
and practically in all the interval of localization lengths where according to
our main criterion of Eq. (\ref{BScrit}) we can have superconductivity in
Anderson
insulator {\em the upper critical field in fact destroys
localization} and the system becomes metallic.\ Accordingly there is no way to
observe the divergence of the upper critical field as $T\rightarrow 0$ and the
$H_{c2}(T)$ curves in ``insulating'' phase all belong to the region between the
curves of $H_{c2}(T)$ at the mobility edge defined in the absence of magnetic
field
(curve $3$ in the insert on Fig.\ref{fig11}) and at the
mobility edge defined in magnetic field (curve $1$ in the
insert).\ This result actually shows that it may be difficult to confirm the
insulating ground
state of strongly disordered superconducting system just applying strong enough
magnetic field to destroy superconductivity and perform usual transport
measurements at low temperatures.

Note that another mechanism for the change of  $H_{c2}(T)$ at low temperatures 
was proposed by Coffey,\ Levin and Muttalib
\cite{CLM85}.\ They have found the enhancement of $H_{c2}$
at low temperatures due to the magnetic field dependence
of the Coulomb pseudopotential $\mu^{\star}$ which appears
via the magnetic field dependence of diffusion coefficient.\
Magnetic field suppression of localization effects leads to
the reduction of Coulomb pseudopotential enhancement due to
these effects \cite{AMR}.\ Accordingly we get the enhancement of $H_{c2}$ at
low temperatures.\ Unfortunately
the apparently more important effects of the frequency dependence of
generalized diffusion coefficient were dropped.

Returning to general criteria of validity of Eq. (\ref{Hc2T}) we note that the
condition of $R_{H}\gg\xi$
is reduced to $\omega_{H}\ll T_{c}^{1/3}E_{F}^{2/3}$ which is obviously
satisfied in any practical case.\ Note,\ however,\ that our estimates for
$H_{c2}$ at low temperatures lead to
$\omega_{H}=\Delta_{0}(E_{F}/\Delta_{0})^{1/3}>\Delta_{0}$ which can easily
overcome paramagnetic limit.\ In this case the experimentally observed $H_{c2}$
of course will
be determined by paramagnetic limit and anomalous behavior due to localization
will be unobservable at low
temperatures.\ At the same time in case of $H_{c2}$ being determined by
paramagnetic limit it may become possible to obtain insulating ground state
of the system applying the strong enough magnetic field.\ Note that the
effective masses entering to
cyclotron frequency and paramagnetic splitting may be
actually very different and there may be realistic cases
when orbital critical field may dominate at low $T$.\ For
$T\sim T_{c}$ $H_{c2}$ is always determined by orbital contribution.

Similar analysis can be performed for the two-dimensional and
quasi-two-dimensional cases\cite{KS93},\ which are important mainly due
to quasi-two-dimensional nature of high-temperature superconductors.\ We
shall limit ourselves only to the case of magnetic field perpendicular to
highly conducting planes,\ when the temperature dependence of $H_{c2}(T)$
is again determined by Eq.(\ref{Hc2T}) with $D_{2}(\omega)$ having the
meaning of diffusion coefficient in Cooper channel along the plane.

If we neglect the magnetic field influence upon diffusion the frequency
dependence of diffusion coefficient in purely two-dimensional case is
determined by Eq.(\ref{d2litl}).\ It is easy to see that the possible anomalies
in the temperature behavior of the upper critical field due to the frequency
dependence of diffusion coefficient will appear only at temperatures
$T\ll \frac{e^{-1/\lambda}}{\tau}$.\ At higher temperatures we obtain the
usual dependence of the "dirty" limit.\
Accordingly,\ from Eq.(\ref{Hc2T}) we
obtain two different types of behavior of $H_{c2}(T)$:
\begin{enumerate}
\item For $T_{c}\gg\frac{e^{-1/\lambda}}{\tau}$
\begin{equation}
H_{c2}(T)=\frac{4}{\pi^{2}}\frac{\phi_{0}}{D_{0}}Tln\left(\frac{T_{c}}{T}\right)\mbox{ for }T\sim T_{c}
\label{Hc11}
\end{equation}
\begin{equation}
H_{c2}(T)=\frac{1}{2\gamma}\frac{\phi_{0}T_{c}}{D_{0}}(1-2.12{\left(\frac{T}{T_{c}}\right)}^{2})\mbox{ for }\frac{e^{-1/\lambda}}{\tau}\ll T\ll T_{c}
\label{Hc22}
\end{equation}
For $T\ll\frac{e^{-1/\lambda}}{\tau}$ the upper critical field
is defined by the equation:
\begin{equation}
ln\left(\frac{\gamma}{2\pi}\frac{e^{-1/\lambda}}{\tau
T}\right)=(1+4\pi\frac{D_{0}}{\phi_{0}}\frac{\tau
H_{c2}}{e^{-1/\lambda}})ln\left(\frac{\gamma}{2\pi}\frac{e^{-1/\lambda}}{\tau
T_{c}}(1+4\pi\frac{D_{0}}{\phi_{0}}\frac{\tau H_{c2}}{e^{-1/\lambda}})\right)
\label{Hc33}
\end{equation}
from which we can explicitly obtain the dependence of $T(H_{c2})$.

Thus,\ up to very low temperatures of the order of
$\sim\frac{e^{-1/\lambda}}{\tau}$
the upper critical field is determined by Drude diffusion coefficient and we
obtain the standard $H_{c2}(T)$ dependence of a "dirty" superconductor.
The ratio $-\frac{H_{c2}(T)}{T_{c}{(\frac{dH_{c2}}{dT})}_{T_{c}}}$ for
$\frac{e^{-1/\lambda}}{\tau}\ll T\ll T_{c}$
is equal to the usual value of 0.69.
For low temperatures $T\ll \frac{e^{-1/\lambda}}{\tau}$ we obtain significant
deviations from the predictions of the usual theory of "dirty" superconductors.
$H_{c2}(T)$ dependence acquires the positive curvature and the upper critical
field diverges as $T\rightarrow 0$.
The behavior of the upper critical field for the case of $T_{c}\gg
\frac{e^{-1/\lambda}}{\tau}$
is shown in Fig.\ref{fig12a},\ curve $1$.

\item
For $T_{c}\ll \frac{e^{-1/\lambda}}{\tau}$ the upper critical
field behavior for any temperature is defined by Eq.(\ref{Hc33}).
$H_{c2}(T)$---dependence acquires positive curvature and $H_{c2}$
diverges for $T\rightarrow 0$.  For small fields
$H_{c2}\ll\frac{\phi_{0}}{D_{0}}\frac{e^{-1/\lambda}}{\tau}$,\ i.e. for
$T\sim T_{c}$,\ Eq.(\ref{Hc33}),\ gives the explicit expression for $H_{c2}$:
\begin{equation}
H_{c2}=\frac{1}{4\pi}\frac{\phi_{0}}{D_{0}}\frac{e^{-1/\lambda}}{\tau}\frac{ln\left(\frac{T_{c}}{T}\right)}{ln\left(\frac{\gamma}{2\pi}\frac{e^{-1/\lambda}}{\tau T}\right)}
\label{Hc3n}
\end{equation}
The slope of $H_{c2}(T)$ at superconducting transition is determined by:
\begin{equation}
-\frac{\sigma_{0}}{N(E)}
{\left(\frac{dH_{c2}}{dT}\right)}_{T_{c}}=\frac{e^{2}}{2\pi}\phi_{0}\frac{e^{-1/\lambda}}{\tau T_{c}ln\left(\frac{\gamma}{2\pi}\frac{e^{-1/\lambda}}{\tau T_{c}}\right)}
\label{Gorkov1}
\end{equation}

The behavior of the upper critical field for the case of
$T_{c}\ll\frac{e^{-1/\lambda}}{\tau}$ is shown in Fig.\ref{fig12b},\
curve $1$.
\end{enumerate}

It is clearly seen from Eqs.(\ref{d2litl}) and
(\ref{d2full}) that magnetic field influence upon diffusion becomes
relevant only for high enough magnetic fields
$H_{c2}\gg\frac{\phi_{0}}{D_{0}}\frac{e^{-1/\lambda}}{\tau}$,\ i.e.
for very low temperatures $T\ll T_{c}$.\ If we use Eq.(\ref{d2full})
in the main equation (\ref{Hc2T}),\ we obtain the following results:
\begin{enumerate}
\item
The case of
$\frac{e^{-1/\lambda}}{\tau}\ll T_{c}\ll \frac{1}{\tau}$.

For high enough temperatures
$T\gg\frac{e^{-1/\lambda}}{\tau}$ the diffusion coefficient
entering Eq.(\ref{Hc2T}) coincides with Drude's
$D_{0}$ and the upper critical field is determined by Eqs.(\ref{Hc11}) and
(\ref{Hc22}).

For $\frac{e^{-1/{\lambda}^{2}ln\left(\frac{\gamma^{2}}{\pi}\frac{1}{\tau
T_{c}}\right)}}{\tau}\ll T\ll\frac{e^{-1/\lambda}}{\tau}$ we obtain:
\begin{equation}
H_{c2}(T)=\frac{1}{2\gamma}\frac{\phi_{0}T_{c}}{D_{0}}(1-3.56\frac{T}{T_{c}})
\label{Hc2H}
\end{equation}
Eq.(\ref{Hc2H}) differs from Eq.(\ref{Hc22}) only by temperature
dependent corrections and we can say that the magnetic field influence
upon diffusion in this case leads to the widening of the temperature
region where we can formally apply the usual theory of "dirty"
superconductors.

For
$T\ll\frac{e^{-1/{\lambda}^{2}ln\left(\frac{\gamma^{2}}{\pi}\frac{1}{\tau
T_{c}}\right)}}{\tau}$ the upper critical field is defined by:
\begin{equation}
ln\left(\frac{\gamma}{2\pi e}\frac{e^{-1/\lambda^{2}lnQ}}{\tau T}\right) =
\frac{2\gamma}{Q}\frac{\lambda
lnQ}{e^{-1/\lambda^{2}lnQ}}ln\left(\frac{\gamma^{2}}{\pi Q}\frac{1}{\tau
T_{c}}\right) \label{Hc3H}
\end{equation} where
$Q=\frac{\gamma}{2\pi}\frac{\phi_{0}}{D_{0}H_{c2}} \frac{1}{\tau}$.
From Eq.(\ref{Hc3H}) we can obtain the explicit dependence $T(H_{c2})$.
The upper critical field in this case is slightly concave as in case of
Eq.(\ref{Hc33}) where we have neglected the magnetic field influence
upon diffusion.\ However,\ now we have no divergence of
$H_{c2}$ for $T\rightarrow 0$ and
\begin{equation}
H_{c2}(T=0)=\frac{\gamma}{2\pi}\frac{\phi_{0}}{D_{0}} \frac{1}{\tau}
\label{Hc(0)H}
\end{equation}
In fact the value of $H_{c2}(T=0)$ will be even smaller,\ because for
these values of the field the number of Landau levels below the cutoff
will be of the order of unity and we are now outside the limits of
applicability of Eqs.(\ref{sisdif1}).\ However,\ the order of
magnitude of $H_{c2}(T=0)$ given by Eq.(\ref{Hc(0)H}) is correct.
$H_{c2}(T)$ behavior with the account of magnetic field influence upon
diffusion is shown in Fig.\ref{fig12a},\ curve $2$.

\item
The case of $T_{c}\ll\frac{e^{-1/\lambda}}{\tau}$.

For small fields
$H_{c2}\ll\frac{\phi_{0}}{D_{0}}\frac{e^{-1/\lambda}}{\tau}$,\ i.e. for
$T\sim T_{c}$,\ magnetic field influence upon diffusion is irrelevant
and the upper critical field is determined by Eq.(\ref{Hc3n}).
For low temperatures $H_{c2}(T)$ is determined by Eq.(\ref{Hc3H}),\ i.e.
magnetic field influence upon diffusion liquidates the divergence of
the upper critical field as $T\rightarrow 0$.
The behavior of $H_{c2}(T)$ for
$T_{c}\ll\frac{e^{-1/\lambda}}{\tau}$ is shown in Fig.\ref{fig12b},\
curve $2$.

\end{enumerate}

It should be noted that the case of
$T_{c}\ll\frac{e^{-1/\lambda}}{\tau}$ is possible only for
sufficiently strong disorder.
For typical $T_{c}\sim {10}^{-4}E_{F}$ this case can occur only for
$\lambda >0.2$.\ Superconducting pairing can exist only in case the
condition similar to Eq.(\ref{BScrit}) is satisfied.\ In two-dimensional case
this condition leads to
inequality $T_{c}\gg\lambda\frac{e^{-1/\lambda}}{\tau}$ which makes the
region under discussion rather narrow.

Quasi-two-dimensional case was extensively discussed in Ref.\cite{KS93}.
Situation here is in many respects similar to that of two-dimensions,\ e.g.
the anomalies in the upper critical field behavior due to the frequency
dependence of diffusion coefficient appear only for temperatures
$T\ll\frac{e^{-1/\lambda}}{\tau}$,\ while at higher temperatures
$H_{c2}(T)$ is well described by the usual theory of "dirty"
superconductors.\ As the interplane transfer integral $w$ grows the
smooth transition from purely two-dimensional behavior to that of
three-dimensional isotropic system can be demonstrated.In case of
$T_{c}\gg\frac{e^{-1/\lambda}}{\tau}$ deviations from the usual
temperature behavior of $H_{c2}$ is observed only for very low
temperatures
$T\ll\frac{e^{-1/\lambda}}{\tau}$,\ while close to $T_{c}$ there are
no significant changes from the standard dependence of $H_{c2}(T)$.
For $T_{c}\ll\frac{e^{-1/\lambda}}{\tau}$ as interplane transfer integral
$w$ grows the temperature dependence of $H_{c2}(T)$ changes from
purely two-dimensional concave behavior for all temperatures to
convex three-dimensional like dependence.\ In Fig.\ref{fig12c} we show the
typical transformations of $H_{c2}(T)$ behavior as transfer integral $w$
changes driving the system through metal-insulator transiton\cite{KS93}.
This clearly demonstrates the sharp anomalies in $H_{c2}$ behavior which
can appear due to localization effects.



\subsection{Fluctuation Conductivity Near Anderson Transition}

Fluctuation conductivity of Cooper pairs (above $T_{c}$) is especially
interesting in strongly disordered system because the usual single---particle
contribution to conductivity drops to
zero as the system moves towards Anderson transition.\ We shall use the
standard approach
\cite{AL68} which takes into account fluctuational Cooper pairs formation above
$T_{c}$.\ We
assume that it is possible to neglect the so called Maki---Thompson correction
which describes
the increased one---particle contribution to conductivity due to
superconducting fluctuations \cite{KelKor72}.\ We  expect that these estimates
\cite{Bul3} will enable us to find a correct scale
of fluctuation conductivity close to mobility edge.

Consider first the averaged fluctuation propagator:
\begin{equation}
L^{-1}({\bf q},\Omega_{k})=\lambda^{-1}-\Pi({\bf q},\Omega_{k})
\label{Lpr}
\end{equation}
where the polarization operator
\begin{eqnarray}
\Pi({\bf q},\Omega_{k})=\sum_{\varepsilon_{n}}\sum_{{\bf pp'}}<G({\bf
p_{+}p'_{+}}\varepsilon_{n}+\Omega_{k})G({\bf p_{-}p'_{-}}-\varepsilon_{n})>=
\nonumber \\
=-2i\pi T \sum_{\varepsilon_{n}}\Phi_{E}({\bf
q},\omega_{m}=-2\varepsilon_{n}+\Omega_{k}) \qquad \omega_{m}=2\pi mT
\label{Pifl}
\end{eqnarray}
During our analysis of Ginzburg---Landau coefficients we were interested in
$\omega_{m}=2\varepsilon_{n}$,\ so that one of the Green's functions in
$\Phi_{E}$ was automatically retarded,\ while the other was advanced.\ Now we
need a more general expression of Eq. (\ref{Pifl}) with
$\omega_{m}=2\varepsilon_{n}+\Omega_{k}$.\ Accordingly,\ instead of
Eq. (\ref{phidiffm}) we must use the following expression with additional
$\theta$-function:
\begin{equation}
\Phi_{E}({\bf q},\omega_{m}=2\varepsilon_{n}+\Omega_{k})=-\frac{N(E)\theta
[\varepsilon_{n}(\varepsilon_{n}+\Omega_{k})]}{i|2\varepsilon_{n}+\Omega_{k}|+iD_{E}(|2\varepsilon_{n}+\Omega_{k}|)q^{2}}
\label{Phitheta}
\end{equation}
where the generalized diffusion coefficient is again determined by Eq.
(\ref{D3frm}) and Eq. (\ref{difffrm}).\ From Eqs.
(\ref{Lpr})---(\ref{Phitheta}),\ performing summation over $\varepsilon_{n}$ we
get the following form of fluctuation propagator for small ${\bf q}$
($D_{E}q^{2}<T$):
\begin{equation}
L^{-1}({\bf q},\Omega_{k})=-N(E)\left\{ln\frac{T}{T_{c}}+
\psi\left(\frac{1}{2}+\frac{|\Omega_{k}|}{4\pi T}\right)-
\psi\left(\frac{1}{2}\right)+\eta(|\Omega_{k}|)q^{2}\right\}
\label{LqT}
\end{equation}
where
\begin{eqnarray}
\eta(|\Omega_{k}|)=4\pi
T\sum_{n=0}^{\infty}\frac{D_{E}(2\varepsilon_{n}+|\Omega_{k}|)}{(2\varepsilon_{n}+|\Omega_{k}|)^{2}}= \nonumber \\
=\left\{\begin{array}{l}
\frac{D_{E}}{4\pi T}\psi'(\frac{1}{2}+\frac{|\Omega_{k}|}{4\pi T}) \qquad
\xi_{loc}< (\xi_{0}l^{2})^{1/3} \quad E>E_{c} \\
\frac{D_{0}}{(4\pi
T)^{2/3}(2\gamma)^{1/3}}\zeta(\frac{5}{3};\frac{1}{2}+\frac{|\Omega_{k}|}{4\pi
T}) \qquad \xi_{loc}> (\xi_{0}l^{2})^{1/3}
\end{array}\right.
\label{etaa}
\end{eqnarray}
It is also useful to know the form of fluctuation propagator  for
$|\Omega_{k}|\gg T$.\ In this case,\ close
to the Anderson transition,\ we may replace the sum over
$\varepsilon_{n}$ in Eq. (\ref{Pifl}) by integral,\ while
far from the transition it can be calculated exactly.\ As
a result we get:
\begin{equation}
L^{-1}({\bf q},\Omega_{k})=-N(E)\left\{
\begin{array}{l}
ln\frac{T}{T_{c}}+\psi\left(\frac{1}{2}+\frac{|\Omega_{k}|}{4\pi
T}+\frac{D_{E}q^{2}}{4\pi T}\right)-\psi(\frac{1}{2}) \qquad
\xi_{loc}<(\xi_{0}l^{2})^{1/3} \quad E>E_{c} \\
ln\frac{T}{T_{c}}+\frac{3}{2}ln\left[\left(\frac{|\Omega|}{4\pi
T}\right)^{2/3}+\frac{D_{0}q^{2}}{(4\pi T)^{2/3}(2\gamma)^{1/3}}\right] \qquad
\xi_{loc}> (\xi_{0}l^{2})^{1/3}
\end{array}\right.
\label{LqTT}
\end{equation}
Diagrams determining fluctuation conductivity are shown in Fig.\ref{fig12}.\
Contributions of graphs Fig.\ref{fig12} (a) and (b) are nonsingular close to
$T_{c}$ because at least one of fluctuation propagators
transfers a large momentum of the order of $p_{F}$.\ Thus
we have to consider independent contributions ${\bf B}$ formed by three Green's
functions.\ We can calculate these contributions using the usual approximations
of self---consistent theory of localization taking into account the
renormalization of triangular vertices by maximally---crossed graphs
\cite{KazSad83,KotSad85} (Cf.\ Eq. (\ref{gammaRA}) as in Fig.\ref{fig12} (c).\
We shall
neglect graphs like in Fig.\ref{fig12} (d) where the topology of disorder
scattering lines is not reduced to
the renormalization of triangular vertices.\ We assume
that these approximations are sufficient at least for a
qualitative inclusion of localization effects.\ Note that
it is sufficient to calculate the contribution of three
Green's functions ${\bf B}({\bf q},\Omega_{k},\omega_{m})$ for small ${\bf q}$
and zero
external frequency $\omega_{m}=0$.\ It can be easily found differentiating the
polarization operator of Eq. (\ref{Pifl}):
\begin{equation}
{\bf B}({\bf q},\Omega_{k},0)={\bf q}C=-\frac{\partial}{\partial {\bf
q}}\Pi({\bf q},\Omega_{k})
\label{qC}
\end{equation}
The contribution of diagram of Fig.\ref{fig12} (c) to the
operator of electromagnetic response  \cite{AGD} is determined by the following
expression:
\begin{equation}
Q_{\alpha\beta}=-\frac{4e^{2}T}{m^{2}}\sum_{\Omega_{k}}
\int\frac{d^{3}{\bf q}}{(2\pi)^{3}}(Cq_{\alpha})(Cq_{\beta})L({\bf
q},\Omega_{k})L({\bf q},\Omega_{k}+\omega_{m})
\label{Qab}
\end{equation}
Close to $T_{c}$ we can also neglect the dependence of $C$ on $\Omega_{k}$.\
Then $C$ reduces to Eq. (\ref{CBS}) and we have $C=N(E)\xi^{2}$.\ Fluctuation
propagator analytically continued to the upper halfplane of complex $\omega$
takes the usual form:
\begin{equation}
L({\bf
q},\omega)=-\frac{1}{N(E)}\frac{1}{\frac{T-T_{c}}{T_{c}}-\frac{i\pi\omega}{8T_{c}}+\xi^{2}q^{2}}
\label{Lqomega}
\end{equation}
Further calculations can be performed in a standard way
and for fluctuation conductivity for $(T-T_{c})/T_{c}\ll 1$ we get the usual
result \cite{AL68}:
\begin{equation}
\sigma_{AL}=\frac{e^{2}}{32\xi\hbar}\left(\frac{T_{c}}{T-T_{c}}\right)^{1/2}
\label{sigAL}
\end{equation}
but with the coherence length $\xi$ being defined as (Cf.
Eq.(\ref{CBS})):
\begin{equation}
\xi=\left\{\begin{array}{l}
\left(\frac{\xi_{0}l}{p_{F}\xi_{loc}}\right)^{1/2} \qquad
\xi_{loc}<(\xi_{0}l^{2})^{1/3} \quad E>E_{c} \\
(\xi_{0}l^{2})^{1/3}\sim (\xi_{0}/p_{F}^{2})^{1/3} \qquad
\xi_{loc}>(\xi_{0}l^{2})^{1/3} \quad E\sim E_{c}
\end{array}\right.
\label{xiAL}
\end{equation}
{}From these estimates we can see that as the system approaches the Anderson
transition a temperature interval
where the fluctuation contribution to conductivity is
important widens.\ Fluctuation Cooper pair conductivity
becomes comparable with a single---particle one for
$\sigma<\sigma^{\star}\approx \sigma_{c}(p_{F}\xi_{0})^{-1/3}\approx
\sigma_{c}(T_{c}/E_{F})^{1/3}$,\ i.e.\ close
enough to mobility edge.\ In fact this confirms the above
picture of Cooper pairs remaining delocalized while single---particle
excitations localize as the system undergoes metal---insulator transition.

It is not difficult to find also the fluctuation contribution to diamagnetic
susceptibility \cite{Bul3}.\
Close to $T_{c}$ it is determined by a standard expression:
\begin{equation}
\chi_{fl}=-\frac{e^{2}T_{c}}{6\pi
c^{2}}\xi\left(\frac{T_{c}}{T-T_{c}}\right)^{1/2}
\label{chifl}
\end{equation}
where the coherence length is again defined as in Eq. (\ref{xiAL}).

Thus our expressions for fluctuation effects follow more or less obviously from
our general picture of Ginzburg---Landau expansion:\ for the system close to
Anderson transition we have only to replace the usual coherence
length $\sqrt{\xi_{0}l}$ of a ``dirty'' superconductor
by $\xi\sim(\xi_{0}l^{2})^{1/3}\sim(\xi_{0}/p_{F})^{1/3}$.



\section{Superconductivity in Anderson Insulator at $T=0$}

We have already considered the superconducting response
of a system which is close to Anderson transition within
Ginzburg---Landau approximation,\ i.e.\ for temperatures
$T\sim T_{c}$.\ In fact it is not difficult to obtain similar results also for
$T=0$ \cite{LeeMa}.

Superconducting current density at $T=0$ is given by \cite{Genn}:
\begin{equation}
{\bf j}_{s}=-\frac{n_{s}e^{2}}{mc}{\bf A}
\label{Lndrsp}
\end{equation}
where $n_{s}$ is superconducting electron density,\ {\bf A} is vector potential
of an external magnetic field.\ On the other hand,\ using exact eigenstates
representation
DeGennes has obtained the following beautiful relation between superconducting
response at $T=0$ and conductivity of a system in the normal state
\cite{Genn,LeeMa}:
\begin{equation}
{\bf j}_{s}=\left\{\frac{1}{2\pi c}\int d\xi\int d\xi'
L(\xi,\xi')Re\sigma(\xi-\xi')-\frac{ne^{2}}{mc}\right\}
{\bf A}
\label{Genrel}
\end{equation}
All characteristics of a superconducting state are contained here in the
kernel:
\begin{equation}
L(\xi,\xi')=\frac{1}{2}\frac{EE'-\xi\xi'-\Delta_{0}^{2}}
{EE'(E+E')}
\label{Genkern}
\end{equation}
where $E=\sqrt{\xi^{2}+\Delta^{2}_{0}}$ and $\Delta_{0}$
is superconducting gap at $T=0$.\ Note that in normal state ${\bf j}_{s}=0$ and
we can rewrite Eq. (\ref{Genrel}) as:
\begin{equation}
{\bf j}_{s}=\frac{1}{2\pi c}\int d\xi\int d\xi'
[L(\xi,\xi')|_{\Delta=\Delta_{0}}-L(\xi,\xi')|_{\Delta=0}]{\bf A}
\label{Genrsp}
\end{equation}
Taking into account that
$L(\xi,\xi')|_{\Delta=\Delta_{0}}-L(\xi,\xi')|_{\Delta=0}$ for large
$|\xi-\xi'|$ drops as
$|\xi-\xi'|^{-3}$ it is sufficient to know only the low---frequency response of
a system in normal state.\ In particular,\ for ``pure'' system (with no
scattering) we
have $Re\sigma(\omega)=(ne^{2}/m)\pi^{-1}\delta(\omega)$
and comparing Eq. (\ref{Lndrsp}) with Eq. (\ref{Genrsp})
it is immediately clear that at $T=0$ we have $n_{s}=n$,\
i.e.\ in an ideal system all electrons are superconducting.

Close to the Anderson transition we can use the results
of elementary scaling theory of localization,\ e.g.\
Eq. (\ref{weg}) and Eq. (\ref{3dcond}) to write
\begin{equation}
\sigma(\omega)\approx
\left\{ \begin{array}{l}
A\frac{g_{c}}{\xi_{loc}}  \qquad \omega<\omega_{c} \\
A\frac{g_{c}}{\xi_{loc}}\left(\frac{\omega}{\omega_{c}}\right)^{1/3} \qquad
\omega > \omega_{c}
\end{array} \right.
\label{sclcnd}
\end{equation}
where $\omega_{c}\sim [N(E)\xi_{loc}^{3}]^{-1}$ is defined in Eq.
(\ref{omegc}),\ $g_{c}$ is the critical
conductance of scaling theory ($g_{c}\sim 1$),\ $A\sim 1$.\ From Eq.
(\ref{Genkern})
and Eq. (\ref{Genrsp}) it is clear that the main contribution into integral in
Eq. (\ref{Genrsp}) comes from $|\xi-\xi'|\sim \Delta_{0}$,\ so that the value
of
$n_{s}$ depends on the relation between $\Delta_{0}$ and $\omega_{c}$.\ For
$\Delta_{0} < \omega_{c}$ we  have $\sigma(\Delta_{0})=Ag_{c}/\xi_{loc}$ and
\begin{equation}
n_{s}=A\frac{m}{e^{2}}\Delta_{0}\frac{g_{c}}{\xi_{loc}}
\label{nscm}
\end{equation}
For $\Delta_{0} > \omega_{c}$ we have
$\sigma(\Delta_{0})=Ag_{c}[N(E)\Delta_{0}]^{1/3}$ and
it becomes independent on the further growth of $\xi_{loc}$ in the region of
$\xi_{loc}>[N(E)\Delta_{0}]^{-1/3}$.\ Accordingly $n_{s}$ does not vanish at the
mobility edge but saturate at
\begin{equation}
n_{s}=A\frac{m}{e^{2}}g_{c}[N(E)\Delta_{0}]^{1/3}
\label{nsci}
\end{equation}
In localization region we can write instead of Eq. (\ref{sclcnd})
\begin{equation}
\sigma(\omega)\approx
\left\{ \begin{array}{l}
0 \qquad \omega<\omega_{c} \\
Ag_{c}[N(E)\omega]^{1/3} \qquad \omega>\omega_{c}
\end{array} \right.
\label{sccndlc}
\end{equation}
which again leads to $\sigma(\Delta_{0})\approx Ag_{c}[N(E)\Delta_{0}]^{1/3}$
and Eq. (\ref{nsci}) remains valid until $R_{loc}>[N(E)\Delta_{0}]^{-1/3}$.
Thus the density of superconducting electrons $n_{s}$ remains finite close to
Anderson transition both in metallic and insulating states.
From Eq. (\ref{nsci}) it is easy to see that close to Anderson
transition
\begin{equation}
\frac{n_{s}}{n}\sim \left(\frac{\Delta_{0}}{E_{F}}\right)^{4/3}
\label{ns}
\end{equation}
This coincide with an estimate of Eq. (\ref{nslocl})
based upon Ginzburg---Landau expansion.\
For typical $\Delta_{0}$ and $E_{F}$ only small part ($\sim 10^{-4}$ in
traditional
superconductors) of conduction electrons form Cooper pairs.\ The condition of
$R_{loc}>[N(E)\Delta_{0}]^{-1/3}\sim a(E_{F}/\Delta_{0})^{1/3}$ as discussed
above defines the size of possible superconducting region in Anderson insulator.
This region is of course quite narrow,\ e.g.\ if metal---insulator transition 
takes place with a change of some external parameter $x$ 
(impurity concentration,\ pressure,\ fluence of fast neutrons etc.),\
so that $R_{loc}\sim a|(x-x_{c})/x_{c}|^{-\nu}$,\ then for $\nu\approx 1$ and
typical $E_{F}/\Delta_{0}\sim 10^{4}$ we get $|x-x_{c}|< 0.1x_{c}$.

These estimates are in complete accordance with the results of our discussion
of Ginzburg---Landau approximation \cite{Bul1,Bul2} and in fact we now have 
the complete qualitative picture of superconductivity in Anderson 
insulator both for $T\sim T_{c}$ and $T\rightarrow 0$,\ i.e.\ in the ground 
state.


\newpage
\section{STATISTICAL FLUCTUATIONS OF SUPERCONDUCTING  ORDER PARAMETER}


The previous discussion of superconductivity in a strongly disordered system is
based upon important
assumption of the existence of self---averaging
superconducting order---parameter $\Delta$.\ This assumption was first used in
the theory of ``dirty''
superconductors \cite{AG58,AG59,And59,Genn} and also in
all early papers on the interplay of localization and
superconductivity.\ It was expected that spatial fluctuations of this order
parameter $\Delta({\bf r})$
are actually small and we can always use some disorder
averaged parameter $<\Delta({\bf r})>$.\ It seems natural
for $\sigma\gg\sigma_{c}$ and it really can be justified
in this region as we shall see below.\ However,\ close to
the mobility edge there are no special reasons to believe
in correctness of this assumption.\ In this case electronic characteristics of
the system become strongly
fluctuating and we shall see that these lead to the strong spatial ({\em
statistical}) fluctuations of superconducting order parameter,\ or even to the
regime inhomogeneous superconductivity.\ At the same time we must stress that
these fluctuations are in some sense similar to the usual thermodynamic
critical fluctuations of the order parameter and become important in some new
critical region (we call it statistical critical region) close to $T_{c}$.\ In
this sense all the previous analysis is just a kind of statistical mean---field
approximation and of course it is a necessary step for further studies taking
into account the statistical fluctuations.\ The importance of these
fluctuations is
stressed by the fact that the statistical critical region widens (similarly to
the usual critical region) as
the system goes to the Anderson transition and apparently
the role of fluctuations becomes decisive for the physics
of the interplay of localization and superconductivity.

\subsection{Statistical Critical Region}

Here we shall start by a demonstration of the appearance
of the new type of fluctuations which are at least of the
same importance as the usual critical fluctuations of
superconducting order---parameter.\ We call them statistical fluctuations
\cite{BulSad86} and their nature is
closely connected to the problem of self---averaging
properties of this order parameter (i.e.\ with a possibility of decoupling
transforming Eq. (\ref{Gap})
into Eq. (\ref{GAP})).\ We shall more or less follow
Ref.\cite{BulSad86},\ equivalent results were recently obtained in
Ref.\cite{Ng91}.

Let us return to the Eq. (\ref{lingap}) and analyze
the situation in more details.\ We shall use a simple
iteration procedure assuming that fluctuations of the
kernel $K({\bf rr'})$ due to disorder are small.\ Similar
approach was first used in Ref.\cite{Genn62}.\ In this case we can represent
$K({\bf rr'})$ and $\Delta({\bf r})$ as
\begin{eqnarray}
K({\bf rr'})=K_{0}({\bf r-r'})+K_{1}({\bf rr'}); \qquad
K_{0}({\bf r-r'})=<K({\bf rr'})> \nonumber \\
\Delta({\bf r})=<\Delta> + \Delta_{1}({\bf r})
\label{iter}
\end{eqnarray}
where $<\Delta>$ is the solution of linearized gap equation
with averaged kernel $K_{0}({\bf r-r'})$ while $\Delta_{1}({\bf r})$ is the
first order correction over
the perturbation defined by $K_{1}({\bf rr'})$.\ We have
seen that the linearized gap equation Eq. (\ref{GAP})
with the averaged kernel $K_{0}({\bf r-r'})$ determines
the standard transition temperature of BCS theory given by Eq. (\ref{BCSTc})
which we shall now denote as $T_{c0}$.\ In the first order over $K_{1}$ there
is no
correction to $T_{c0}$:\ $<K_{1}>$=0.\ In the second order of this perturbation
theory we obtain the following
change of transition temperature,\ defined as the temperature of appearance of
homogeneous order---parameter:
\begin{eqnarray}
\frac{T_{c}-T_{c0}}{T_{c0}}=\frac{1}{\lambda_{p}}
\int\frac{d^{3}{\bf q}}{(2\pi)^{3}}\frac{K_{1}({\bf q}0)K_{1}(0{\bf q})}
{1-K_{0}({\bf q},T_{c})} \nonumber \\
K_{0}=\int d{\bf r}e^{i{\bf qr}}K({\bf r},T_{c})
\label{Tcchange}
\end{eqnarray}
where
\begin{eqnarray}
K_{1}(0{\bf q})=K_{1}(-{\bf q}0)=\int d{\bf r}\int d{\bf r'}e^{i{\bf
qr}}[K({\bf rr'})-K_{0}({\bf r-r'})]= \nonumber \\
=\lambda_{p}\int\limits_{0}^{<\omega>}\frac{dE}{E}th\frac{E}{2T_{c}}\int d{\bf
r}e^{i{\bf qr}}\left[\frac{1}{N(E)}
\sum_{\mu}|\phi_{\mu}({\bf r})|^{2}\delta(E-\varepsilon_{\mu})-1\right]
\label{K1}
\end{eqnarray}
Here $\lambda_{p}=gN(E_{F})$ and we have used the completeness and
orthonormality of exact eigenfunctions
$\phi_{\mu}({\bf r})$.\ It is obvious that correction to
$T_{c0}$ given by Eq. (\ref{Tcchange}) is always positive.\ After averaging Eq.
(\ref{Tcchange}) over disorder we get the relative change of transition
temperature due to fluctuations as
\begin{eqnarray}
\frac{\delta T_{c}}{T_{c0}}=\left\langle\frac{T_{c}-T_{c0}}{T_{c0}}
\right\rangle=\lambda_{p}\int \frac{d^{3}{\bf q}}{(2\pi)^{3}}
\frac{\varphi({\bf q})}{1-K_{0}({\bf q},T_{c})} \nonumber \\
\varphi({\bf q})=\int d{\bf r} e^{i{\bf qr}}\varphi({\bf r})
\label{delTcc}
\end{eqnarray}
where
\begin{equation}
\varphi({\bf
r})=\int\limits_{0}^{<\omega>}\frac{dE}{E}th\frac{E}{2T_{c}}\int\limits_{0}^{<\omega>}\frac{dE'}{E'}th\frac{E'}{2T_{c}}\left\{\frac{1}{N(E)}\ll\rho_{E}({\bf r})\rho_{E'}(0)\gg^{H}-1\right\}
\label{fir}
\end{equation}
and we have introduced the spectral density of Eq. (A3)
\begin{equation}
\ll\rho_{E}({\bf r})\rho_{E'}({\bf
r'})\gg^{H}=\frac{1}{N(E)}<\sum_{\mu\nu}|\phi_{\mu}({\bf
r})|^{2}|\phi_{\nu}({\bf
r'})|^{2}\delta(E-\varepsilon_{\mu})\delta(E'-\varepsilon_{\nu})>
\label{rhoH}
\end{equation}
which is actually a correlation function of local densities of states.

Remember now that in a ``dirty'' system \cite{SJS}:
\begin{eqnarray}
1-K_{0}({\bf q},T)=1-2\pi
T\lambda_{p}\sum_{\varepsilon_{n}}\frac{1}{2|\varepsilon_{n}|+D_{E}(2|\varepsilon_{n}|)q^{2}}\approx \nonumber \\
\approx\lambda_{p}\left[\frac{T-T_{c0}}{T_{c0}}+
\xi^{2}q^{2}\right] \qquad \varepsilon_{n}=(2n+1)\pi T
\label{denom1K}
\end{eqnarray}
where $\xi$ is the coherence length defined previously
e.g.\ in Eq. (\ref{CBS}).\  The approximate equality here is valid for
$|T-T_{c0}|/T_{c}\ll 1, \quad \xi^{2}q^{2}\ll 1$.\ From Eq. (\ref{delTcc}) and
Eq. (\ref{denom1K}) we get the change of transition temperature in the
following form:
\begin{equation}
\frac{\delta T_{c}}{T_{c0}}=\int\frac{d^{3}{\bf
q}}{(2\pi)^{3}}\frac{\varphi({\bf q})}{\xi^{2}q^{2}}
\label{deltaTc}
\end{equation}
Here we must cut---off integration at $q\sim \xi^{-1}$ in accordance with
limits of applicability of the last expression in Eq. (\ref{denom1K}).\
However,\ the contribution of short---wave fluctuations here may be
also important.

The Ginzburg---Landau functional expressed via non---averaged order parameter
$\Delta({\bf r})$
has the following form \cite{Genn}:
\begin{equation}
F\{\Delta\}=\int d{\bf r} \left\{\frac{N(E_{F})}{\lambda_{p}}|\Delta({\bf
r})|^{2}-N(E_{F})\int d{\bf r'} K({\bf rr'})\Delta({\bf r'})\Delta({\bf
r})+\frac{1}{2}B|\Delta({\bf r})|^{4}\right\}
\label{GLnonav}
\end{equation}
where we have neglected the fluctuations of pairing interaction $\lambda_{p}$
and of the coefficient $B$,\ which is defined by the standard expression given
in Eq. (\ref{Bdirt}).\ Using Eqs.
(\ref{iter})---(\ref{K1}) we can find Ginzburg---Landau equations which
describe the slow changes of $\Delta({\bf r})$:
\begin{equation}
\left\{N(E_{F})\frac{T_{c0}-T}{T_{c0}}+\delta A({\bf r})-B|\Delta({\bf
r})|^{2}+C\frac{\partial^{2}}{\partial r^2}\right\}\Delta({\bf r})=0
\label{GLfl}
\end{equation}
where
\begin{equation}
\delta A({\bf
r})=N(E_{F})\int\limits_{0}^{<\omega>}\frac{dE}{E}th\frac{E}{2T_{co}}\left\{\frac{1}{N(E_{F})}\sum_{\nu}|\phi_{\nu}({\bf r})|^{2}\delta(E-\varepsilon_{\nu})-1\right\}
\label{delA}
\end{equation}
describes the fluctuations of the coefficient $A$ of Ginzburg---Landau
expansion and we have
neglected the fluctuations of the $C$ coefficient .

Ginzburg---Landau equations with fluctuating coefficients were analyzed for the
first time by Larkin
and Ovchinnikov\cite{LO71}.\  It was shown that $\delta A({\bf
r})$---fluctuations lead to a shift of transition temperature given by Eq.
(\ref{deltaTc}) and the solution of Eq. (\ref{GLfl}) for the order parameter in
the first order over fluctuations has the form of Eq. (\ref{iter}) with:
\begin{eqnarray}
\Delta_{1}({\bf r})=\int\frac{d^{3}{\bf q}}{(2\pi )^{3}}\Delta_{1}({\bf
q})e^{i{\bf qr}} \nonumber \\
\Delta_{1}({\bf q})=-\frac{<\Delta>}{N(E_{F})}\frac{\delta A({\bf
q})}{\xi^{2}q^{2}+2\tau}
\label{delt1q}
\end{eqnarray}
where $\tau=(T_{c}-T)/T_{c}$ is temperature measured relative to the new
transition temperature.\ The mean---square fluctuation of the order---parameter
itself is determined from
Eq. (\ref{delt1q}) by:
\begin{equation}
\frac{<\Delta^{2}>}{<\Delta>^{2}}-1=\int\frac{d^{3}{\bf
q}}{(2\pi)^{3}}\frac{\varphi({\bf q})}{[\xi^{2}q^{2}+2\tau]^{2}}
\label{deltafl}
\end{equation}
where $\varphi({\bf q})$ was introduced in Eqs. (\ref{delTcc}),\ (\ref{fir}).\
It is important to note
that fluctuations of $\Delta({\bf r})$ as opposed to $T_{c}$---shift are
determined by small ${\bf q}$ behavior of $\varphi({\bf q})$.

We can see now that all the physics of statistical fluctuations is described by
the correlation
function of local densities of states (or spectral density of Eq.
(\ref{rhoH})).\ This function was determined above in Eqs. (\ref{locdencor}),\
(\ref{locdencor3d}) within self---consistent theory of
localization or by Eqs. (\ref{Ksclg}),\ (\ref{Dscl}) which follow from from
scaling approach close to the
mobility edge.

Using Eq. (\ref{locdencor3d}) for the metallic state not very close to the
mobility edge we can get from Eq. (\ref{fir}):
\begin{equation}
\varphi({\bf q}=0)\sim \frac{\xi}{N^{2}(E_{F})D_{0}^{2}}
\label{fimet}
\end{equation}
where $\xi=\sqrt{\xi_{0}l}$ and $D_{0}$ is the Drude diffusion coefficient.\
Estimating the $T_{c}$---shift
from Eq. (\ref{deltaTc}) we get:
\begin{equation}
\frac{\delta T_{c}}{T_{co}}\sim \frac{1}{N^{2}(E_{F})D_{0}^{2}\xi^{2}}\sim
\frac{T_{c}}{E_{F}}\frac{1}{(p_{F}l)^{3}}\sim \tau_{G}
\label{Tcshift}
\end{equation}
where $\tau_{G}$ is the size of Ginzburg critical region
defined by Eq. (\ref{Gi}).\ We have seen that in the
usual ``dirty'' superconductor $\tau_{G}\ll 1$.\ For the
order---parameter fluctuations from Eq. (\ref{deltafl})
we obtain:
\begin{equation}
\frac{<\Delta^{2}>}{<\Delta>^{2}}-1\approx \frac{1}{8\pi}\frac{\varphi({\bf
q}=0)}{\xi^{3}\sqrt{2|\tau|}}\approx \left(\frac{\tau_{D}}{|\tau|}\right)^{1/2}
\label{deltaflmet}
\end{equation}
{}From here we can see that the width of the temperature region where
statistical fluctuations are important is
given by:
\begin{equation}
\tau_{D}\sim \frac{\varphi^{2}(0)}{\xi^{6}}\sim
\frac{1}{N^{4}(E_{F})D_{0}^{4}\xi^{4}}\sim
\left(\frac{T_{c}}{E_{F}}\right)^{2}\frac{1}{(p_{F}l)^{6}}\sim \tau_{G}^{2}
\label{stcr}
\end{equation}
It is obvious that in a``dirty'' superconductor we have
$\tau_{D}\ll\tau_{G}\ll 1$ and statistical fluctuations are absolutely
unimportant.

Situation change for a system which is close to the mobility edge.\ Using Eq.
(\ref{locdencor3d}) with
$D_{0}$ replaced by $D_{0}(\omega/\gamma)^{1/3}$ or Eqs.
(\ref{Ksclg})---(\ref{Dscl}) we obtain:
\begin{equation}
\varphi({\bf q})\approx \frac{\gamma^{1/2}}{N^{2}(E_{F})D_{0}^{3}T_{c}}\int
\limits_{0}^{T_{c}}\frac{d\omega}{\omega^{1/2}}[\omega^{2}+D_{0}\gamma^{-2/3}\omega^{2/3}q^{4}]^{-1/4}\sim \xi^{3}ln\frac{1}{\xi q}
\label{fimobed}
\end{equation}
where $\xi\sim (\xi_{0}p_{F}^{-2})^{1/3}$.\ Similarly we
get:
\begin{equation}
\frac{<\Delta^{2}>}{<\Delta>^{2}}-1\approx
\int\limits_{0}^{\xi^{-1}}\frac{\xi^{3}q^{2}dq}{(\xi^{2}q^{2}+2\tau)^{2}}ln\frac{1}{q\xi}\sim \frac{1}{\sqrt{|\tau|}}ln\frac{1}{|\tau|}
\label{deltflmobed}
\end{equation}
{}From Eq. (\ref{deltflmobed}) it follows that close to the
mobility edge statistical fluctuations become important
and even overcome thermodynamic fluctuations due to the
logarithmic factor in $\varphi({\bf q})$.\ Thus in this
region we have $\tau_{D}>\tau_{G}\sim 1$.

The crossover from the regime of weak statistical fluctuations
($\tau_{D}\ll\tau_{G}$) to the strong
fluctuation regime occurs at the conductivity scale
$\sigma\sim\sigma^{\star}\approx\sigma_{c}(p_{F}\xi_{0})^{-1/3}$ which was
extensively discussed above.\ Thus close to the mobility edge the
superconducting order---parameter is no more a self---averaging quantity.\ Here
the mean---field theory approach becomes formally invalid
due to thermodynamic and also because of statistical fluctuations.\ Below we
shall analyze this situation in
more details.

Finally we shall briefly discuss the region of localized
localized states.\ The appearance here of a singular
$\delta(\omega)$---contribution to the correlator of
local densities of states given by Eqs. (A8)---(A10) leads to the additional
contribution to $\varphi({\bf q})$:
\begin{eqnarray}
\varphi({\bf
q})=\int\limits_{0}^{<\omega>}\frac{dE}{E^{2}}\left(th\frac{E}{2T_{c}}\right)\frac{A_{E}({\bf q})}{N(E_{F})}+\cdots\sim\frac{A_{E_{F}}}{N(E_{F})T_{c0}}+\cdots = \nonumber \\
=\frac{1}{N(E_{F})T_{c}(1+R_{loc}^{2}q^{2})}+\cdots
\label{filoc}
\end{eqnarray}
Accordingly a new contribution to $\Delta({\bf r})$ fluctuations is given by:
\begin{equation}
\frac{<\Delta^{2}>}{<\Delta>^{2}}-1\approx
\frac{1}{N(E_{F})T_{c}}\int\limits_{0}^{R_{loc}^{-1}}
\frac{q^{2}dq}{(\xi^{2}q^{2}+2|\tau|)^{2}(1+R_{loc}^{2}q^{2})}\sim
\frac{1}{N(E_{F})T_{c}R_{loc}^{3}\tau^{2}}
\label{deltafloc}
\end{equation}
and it grows fast as the localization length $R_{loc}$
diminishes.\ Using our main criterion of superconductivity in localized phase
given by Eq. (\ref{BScrit}) we can see that in all region of possible
superconductivity statistical fluctuations of $\Delta({\bf r})$ remain of the
order of unity and are
important in rather wide temperature interval around $T_{c}$.



\subsection{Superconducting Transition at Strong Disorder}
We consider now superconductivity in systems with strong statistical
fluctuations of the ``local transition temperature'' $T_{c}({\bf r})$ as
described by Eq. (\ref{GLfl}) and Eq. (\ref{delA}).\ In this analysis we shall
follow Refs.\cite{BPS87,BPS89}.\ For simplicity we assume Gaussian nature
of these fluctuations.\ Note,\ however,\ that close to the mobility edge the
fluctuations of local density of states become strongly non---Gaussian
\cite{Ler88} and this can complicate the situation.\ Unfortunately the
importance of this non---Gaussian behavior for superconductivity has
not been studied up to now.\ We shall see that in our model ,\ depending on the
degree of disorder,\ which we shall measure by the ratio $\tau_{D}/\tau_{G}$ ,\
two types of superconducting transition are possible.\ For $\tau_{D}$ smaller
than some critical value $\tau_{D}^{\star}$ the superconducting transition is
the usual second---order phase transition
at $T=T_{c}$.\ The superconducting order---parameter is in this case equal to
zero for $T>T_{c}$ and is spatially homogeneous over scales exceeding the
correlation length $\xi(T)$
below $T_{c}$.\ Statistical fluctuations lead only to a change of critical
exponents at the transition\cite{Khm75,Lub75}.\

At $\tau_{D}>\tau_{D}^{\star}$ the superconducting state appears in
inhomogeneous fashion
even if the correlation length of disorder induced fluctuations of $T_{c}({\bf
r})$ is small compared with the superconducting correlation length $\xi$
(microscopic disorder).\ This case
was first analyzed by Ioffe and Larkin\cite{IL81}.\ Investigating the case of
extremely strong disorder they have shown that as the temperature is lowered
the normal phase acquires localized superconducting regions (drops) with
characteristic size determined by $\xi(T)$.\ Far from $T_{c}$ their density is
low,\ but with further cooling the density and dimensions of the drops increase
and they begin to overlap leading to a kind of percolative superconducting
transition.

According to our previous estimates,\ if we take into account only the
fluctuations of local density
of states,\ the parameter $\tau_{D}/\tau_{G}$ increases from very small values
to a value greater than unity as the system moves towards the mobility edge.\
An onset of an inhomogeneous superconducting regime is therefore to be expected
as the localization transition
is approached.

Our  treatment of superconductors with large statistical fluctuations will be
based on the Ginzburg---Landau functional:
\begin{eqnarray}
F\{{\bf A}({\bf r}),\Delta({\bf r})\}=\int d{\bf r}\left\{\frac{{\bf
B}^{2}({\bf r})}{8\pi}+N(E_{F})\left[(\tau+t({\bf r}))|\Delta({\bf
r})|^{2}+\right.\right.\nonumber\\
\left.\left.+\xi^{2}\left|\left(\nabla-\frac{2ie}{\hbar c}{\bf A}({\bf
r}))\right)\Delta({\bf r})\right|^{2}+\frac{1}{2}\lambda|\Delta({\bf
r})|^{4}\right]\right\}
\label{GLAB}
\end{eqnarray}
where ${\bf B}=rot{\bf A}$ is magnetic field and we have redefined the
coefficient of quartic term
as $B=N(E_{F})\lambda$.\ Here $t({\bf r})$ is defined by Eq. (\ref{delA}) as
$\delta A({\bf r})=N(E_{F})t({\bf r})$ and plays the role of the fluctuation of
local ``critical temperature'',\ which
appears due to fluctuations of local density of states.\ In general case it
also can have contributions from local fluctuations of pairing interaction or
other types of microscopic inhomogeneities.\ As noted above we assume Gaussian
statistics of these fluctuations,\ though
real situation close to the mobility edge may be more complicated
\cite{Ler88}.\ Given the distribution of $t({\bf r})$,\ the free energy of the
system and the order---parameter correlator are
equal to:
\begin{equation}
{\cal F}\{t({\bf r})\}=-TlnZ, \quad Z=\int D\{{\bf A},\Delta\}exp[-F\{{\bf
A}({\bf r}),\Delta({\bf r})\}/T]
\label{freeen}
\end{equation}
\begin{equation}
<\Delta({\bf r})\Delta({\bf r'})>=Z^{-1}\int D\{{\bf A},\Delta\}\Delta({\bf
r})\Delta({\bf r'})exp[-F\{{\bf A}({\bf r}),\Delta({\bf r})\}/T]
\label{corfun}
\end{equation}
and must be averaged over the Gaussian distribution of $t({\bf r})$.\ From our
definition of $t({\bf r})$ and using the approach of the previous section,\
assuming the short---range of fluctuations of local density of states(on the
scale of $\xi$),\ it is easy to estimate the correlator of $t({\bf r})$ as:
\begin{equation}
<t({\bf r})t({\bf r'})>=\gamma\delta({\bf r-r'}), \qquad \gamma\approx
\tau_{D}^{1/2}\xi^{3}
\label{tcorr}
\end{equation}
Then  the probability of a configuration with a given $t({\bf r})$ is given by
\begin{equation}
{\cal P}\{t({\bf r})\}=exp\left[-\frac{1}{2\gamma}\int d{\bf r}t^{2}({\bf
r})\right]
\label{Gauss}
\end{equation}
The problem reduces thus to calculation of the functions ${\cal F}\{t({\bf
r})\}$ and $<\Delta({\bf r})\Delta({\bf r'})>$  and their subsequent averaging
over ${\cal P}\{t({\bf r})\}$.

We shall limit ourselves to consideration of noninteracting drops and no
vortices.\ Then we can consider the phase of the order---parameter $\Delta({\bf
r})$ as nonsingular.\ After the gauge
transformation
\begin{eqnarray}
{\bf A}({\bf r})\rightarrow {\bf A}({\bf r})+(c\hbar/2e)\nabla\phi({\bf r})
\nonumber \\
\Delta({\bf r})\rightarrow \Delta({\bf r})exp[-i\phi({\bf r})]
\label{gauge}
\end{eqnarray}
where $\phi({\bf r})$ is the {\em phase of the order parameter} we can use real
$\Delta({\bf r})$
and Ginzburg---Landau functional of Eq. (\ref{GLAB}) becomes:
\begin{eqnarray}
F\{{\bf A}({\bf r}),\Delta({\bf r})\}=\int d{\bf r}\left\{\frac{{\bf
B}^{2}({\bf r})}{8\pi}+N(E_{F})\left[(\tau+t({\bf r}))\Delta^{2}({\bf
r})+\frac{4e^{2}\xi^{2}}{c^{2}\hbar^{2}}{\bf A}^{2}({\bf r})\Delta^{2}({\bf
r})+\right.\right.\nonumber\\
\left.\left.+\xi^{2}(\nabla\Delta({\bf
r}))^{2}+\frac{1}{2}\lambda\Delta^{4}({\bf r})\right]\right\}
\label{GLreal}
\end{eqnarray}
Integration over phase in Eq. (\ref{freeen}) gives an inessential constant
factor to the partition
function which we disregard.

To average the logarithm of the partition function Eq. (\ref{freeen}) over
$t({\bf r})$ we can use the replica
trick\cite{GrLut76} which permits the averaging to be
carried out in explicit form.\ We express the average free energy Eq.
(\ref{freeen}) of the system in the form:
\begin{equation}
<{\cal F}>=-T \lim_{n\rightarrow 0}\frac{1}{n}[<Z^{n}>-1]
\label{avfreeen}
\end{equation}
To calculate $<Z^{n}>$ in accordance with the idea of the
replica method,\ we first assume $n$ to be an arbitrary integer.\ Expressing
$Z^{n}$ in terms of an $n$---fold
functional integral over the fields of the replicas ${\bf A}_{\alpha}$,\
$\Delta_{\alpha}({\bf r})$,\ $\alpha=1,...,n$ and carrying out exact Gaussian
averaging over $t({\bf r})$,\ we get
\begin{equation}
<Z^{n}>=\int D\{{\bf A},\Delta\}exp[-S_{n}\{{\bf A}_{\alpha},\Delta_{\alpha}\}]
\label{partitn}
\end{equation}
\begin{eqnarray}
S\{{\bf A}_{\alpha},\Delta_{\alpha}\}=\int d{\bf
r}\left\{\sum_{\alpha}^{n}\frac{{\bf B}^{2}({\bf r})}{8\pi
T}+\frac{N(E_{F})}{T}\sum_{\alpha}^{n}\left[\tau
\Delta_{\alpha}({\bf r})^{2}+\frac{4e^{2}\xi^{2}}{c^{2}\hbar^{2}}{\bf
A}^{2}_{\alpha}({\bf r})\Delta^{2}_{\alpha}({\bf r})+\right.\right.\nonumber\\
\left.\left.+\xi^{2}(\nabla\Delta_{\alpha}({\bf
r}))^{2}+\frac{1}{2}\lambda\Delta_{\alpha}^{4}({\bf
r})\right]-\frac{1}{2}\frac{N(E_{F})}{T}\tilde\gamma
\left[\sum_{\alpha=1}^{n}\Delta_{\alpha}^{2}({\bf r})\right]^{2}\right\}
\nonumber
\end{eqnarray}
The last expression here represents the ``effective action'' and
$\tilde\gamma=\gamma N(E_{F})/T_{c}\approx
\tau_{D}^{1/2}N(E_{F})/T_{c}$ grows with disorder .\
Note that the random quantities $t({\bf r})$ have already dropped out of these
expressions,\ and that the action
$S\{{\bf A}_{\alpha},\Delta_{\alpha}\}$ is translationally invariant.\ For the
correlator of Eq. (\ref{corfun}) we obtain:
\begin{equation}
<\Delta({\bf r})\Delta({\bf r'})>=\lim_{n\rightarrow 0}
\frac{1}{n}\int D\{{\bf A},\Delta\}exp[-S_{n}\{{\bf
A}_{\alpha},\Delta_{\alpha}\}]\sum_{\alpha=1}^{n}\Delta_{\alpha}({\bf
r})\Delta_{\alpha}({\bf r'})
\label{repcorr}
\end{equation}
where we have symmetrized over the replica indices.

Far from the region of strong fluctuations of the order
parameter $|\tau|\gg\tau_{D},\tau_{G}$ the functional
integrals in Eq. (\ref{partitn}) and Eq. (\ref{corfun})
can be calculated by the saddle---point method.\ The extrema of the action are
determined by classical equations:
\begin{equation}
\left[\tau-\xi^{2}\nabla^{2}+\lambda\Delta_{\alpha}^{2}-
\tilde\gamma\sum_{\beta=1}^{n}\Delta_{\beta}^{2}({\bf
r})\right]\Delta_{\alpha}({\bf r})=0 \qquad {\bf A}_{\alpha}=0
\label{extrsol}
\end{equation}
The nontrivial conclusion is that these equations for
$\Delta_{\alpha}({\bf r})$ besides spatially homogeneous
solutions do have localized solutions with finite action ({\em instantons}).\
These correspond at $\tau>0$ to
superconducting drops.\ We shall limit ourselves to a
picture of noninteracting drops and consider only instanton solutions above
$T_{c}$ (at $\tau>0$).\ We shall be interested only in those solutions that
admit
analytic continuation as $n\rightarrow 0$.\ We designate them
$\Delta^{(i)}_{\alpha}({\bf r})$,\ where the superscript $i$ labels the type of
solution.\ To find their contribution we must expand the action of Eq.
(\ref{partitn}) up to the terms quadratic in deviations
$\varphi_{\alpha}({\bf r})=\Delta_{\alpha}({\bf r})-\Delta_{\alpha}^{(i)}({\bf
r})$.\ It can be shown that
fluctuations of the fields ${\bf A}_{\alpha}({\bf r})$
can be neglected if we consider noninteracting drops \cite{BPS87,BPS89}.

For $\tau>0$ and for $\tilde\gamma > \lambda$ Eq. (\ref{extrsol}) possess
(besides the trivial solution $\Delta_{\alpha}=0$) the following nontrivial
solution with finite action (instanton)(Cf.\ Refs.\cite{Car78,Sad79,Sad86}):
\begin{equation}
\begin{array}{l}
\Delta_{\alpha}^{(i)}({\bf r})=\Delta_{0}({\bf r})\delta_{\alpha i},\qquad
i=1,...,n \\
\Delta_{0}({\bf
r})=\sqrt{\frac{\tau}{\tilde\gamma-\lambda}}\chi\left[\frac{r}{\xi(T)}\right],
\qquad \xi(T)=\frac{\xi}{\sqrt{\tau}}
\end{array}
\label{inst}
\end{equation}
where the dimensionless function $\chi(x)$ satisfies the
condition $d\chi(x)/dx|_{x=0}=0$ and its asymptotic form:\ $\chi(x)\sim
x^{-1}exp(-x)$ for $x\gg 1$ (for spatial dimension $d=3$).\ The qualitative
form of this solution is shown in Fig. \ref{fig13}.

{}From Eq. (\ref{inst}) it is seen that instantons are oriented along axes of
replica space (there are
$n$ types of instanton solutions) which is due to the ``cubic anisotropy'' term
$\lambda\Delta^{4}_{\alpha}$ in the effective action of Eq. (\ref{partitn}).\
Index $i$ characterizes the direction in replica space along which the symmetry
breaking takes place.\ For
$\lambda\rightarrow 0$ the action becomes $O(n)$---symmetric  and instantons
take the form :
\begin{equation}
\Delta_{\alpha}({\bf r})=\Delta_{0}({\bf r})e_{\alpha}, \qquad
\sum_{\alpha=1}^{n}e_{\alpha}^{2}=1
\label{instOn}
\end{equation}
i.e.\ are oriented along arbitrary unit vector $\vec e$ in replica space.\
Such instantons earlier
were studied in the theory of localization \cite{Car78,Sad79,Sad86}.

The quadratic expansion of the effective action near instanton solution takes
the form (Cf.\ analogous treatment in Refs.\cite{Car78,Sad79,Sad86}):
\begin{equation}
S\{\Delta_{\alpha}\}=S\{\Delta_{\alpha}^{(i)}\}+\frac{1}{2}\int d{\bf
r}\sum_{\alpha,\beta}(\varphi_{\alpha}\hat
M^{(i)}_{\alpha\beta}\varphi_{\beta})
\label{gausaction}
\end{equation}
where the operator $M^{(i)}_{\alpha\beta}$ on instanton solutions is equal to:
\begin{equation}
\hat M^{(i)}_{\alpha\beta}=[\hat M_{L}\delta_{\alpha i}+\hat
M_{T}(1-\delta_{\alpha i})]\delta_{\alpha\beta}
\label{Mab}
\end{equation}
with
\begin{equation}
M_{L,T}=\frac{2N(E_{F})}{T}[-\xi^{2}\nabla^{2}+\tau U_{L,T}({\bf r})]
\label{Mlt}
\end{equation}
where
\begin{equation}
\begin{array}{l}
U_{L}({\bf r})=1-3\chi^{2}[r/\xi(T)]  \\
U_{T}({\bf r})=1-(1-\lambda/\tilde\gamma)^{-1}\chi^{2}[r/\xi(T)]
\end{array}
\label{MLT}
\end{equation}
The value of Gaussian functional integral is determined by the spectra of
eigenstates of operators
$M_{L}$ and $M_{T}$ .\ Detailed analysis can be found in
Refs.\cite{BPS87,BPS89}.\ The qualitative form of these spectra is shown in
Fig. \ref{fig14}.\ Operator $M_{L}$ always possess an eigenvalue
$\varepsilon_{1}^{L}=0$---the so called translation zero---mode,\ connected
with translation symmetry:\  instanton center may be placed anywhere in space,\
the action does not change .\ However,\ this is not a lowest eigenvalue of
$M_{L}$,\ there is always a negative eigenvalue
$\varepsilon_{0}^{L}<\varepsilon_{1}^{L}=0$.\ It can be shown rigorously that
it is the only negative eigenstate of $M_{L}$\cite{Zinn}.\ Operator $M_{T}$
possess also a single
negative eigenvalue $\varepsilon_{0}^{T}<0$\cite{BPS87,BPS89},\ however this
eigenvalue
tends to zero for $\lambda\rightarrow 0$ becoming the ``rotation''
zero---mode,\ reflecting the
arbitrary ``direction'' of instanton in replica space in the absence of cubic
anisotropy in the action \cite{Car78,Sad79,Sad86}.\ For
$\lambda=\lambda^{\star}=2/3\tilde\gamma$ we have $M_{L}=M_{T}$ and the spectra
of both operators coincide.

Including the contributions of instantons oriented along all the axes in
replica space we obtain the
following one---instanton contribution to the partition function entering Eq.
(\ref{avfreeen})\cite{BPS87,BPS89}:
\begin{equation}
<Z^{n}>=n\Omega\left(\frac{J_{L}}{2\pi}\right)^{d/2}[Det'M_{L}]^{-\frac{1}{2}}[Det M_{T}]^{\frac{1-n}{2}}exp\{-S_{0}(\tau)\}
\label{Zn}
\end{equation}
where $\Omega$ is the system volume,
\begin{equation}
J_{L}=\frac{1}{d}\int dr\left(\frac{\partial\Delta_{0}}{\partial
r}\right)^{2}=\frac{T}{2N(E_{F})}\frac{S_{0}(\tau)}{\xi^{2}}
\label{Jl}
\end{equation}
and the action at the instanton is given by:
\begin{equation}
S_{0}(\tau)={\cal A}\frac{\xi^{3}\tau^{1/2}}{\gamma-\lambda T/N(E_{F})}
\label{Sinst}
\end{equation}
where ${\cal A}\approx 37.8$ is a numerical constant\cite{LGP}.\ The prime on
$Det M_{L}$ means that we must exclude the zero---eigenvalue
$\varepsilon_{1}^{L}=0$ from the product of
eigenvalues determining this determinant.\ The condition of applicability of
the saddle---point
approximation looks like $S_{0}(\tau)\gg 1$,\ and in fact all our analysis is
valid outside the
critical regions both for thermodynamic and statistical fluctuations.

In the limit of $n\rightarrow 0$ the total cancellation of imaginary
contributions appearing due to negative eigenvalues takes place in Eq.
(\ref{Zn}) and using Eq.
(\ref{avfreeen}) we get for $\tilde\gamma>3/2\lambda$ the
following {\em real} contribution to the free energy:
\begin{equation}
{\cal F}=-\rho_{s}(\tau)T\Omega
\label{Finst}
\end{equation}
where the density of superconducting ``drops''
\begin{equation}
\rho_{s}(\tau)=\left[\frac{T}{4\pi
N(E_{F})}S_{0}(\tau)\right]^{3/2}\xi^{-3}\left[\frac{Det M_{T}}{Det'
M_{L}}\right]^{1/2}exp\{-S_{0}(\tau)\}
\label{rhosup}
\end{equation}
Thus for $\tilde\gamma>3/2\lambda$ even for $T>T_{c}$ the
superconducting ``drops'' (instantons) appear in the system which directly
contribute to the equilibrium free energy.\ This contribution given by Eqs.
(\ref{Finst})---
(\ref{rhosup}) exists along the usual thermodynamic fluctuations.\ The
condition of $\tilde\gamma>3/2\lambda$
defines critical disorder $\tau_{D}>\tau_{D}^{\star}>\tau_{G}$,\ and this
inhomogeneous picture of superconducting transition appears only for the case
of sufficiently strong statistical fluctuations.\ The knowledge of qualitative
structure of spectra of eigenvalues of $M_{L}$ and $M_{T}$ allows to analyze
different asymptotics of Eq. (\ref{Finst})\cite{BPS87,BPS89}.\ For
$\tilde\gamma S_{0}(\tau)\ll\lambda\ll\lambda^{\star}$ we get:
\begin{equation}
\rho_{s}(\tau)\approx\xi^{-3}(T)\left(\frac{\lambda}{\tilde\gamma}\right)^{1/2}S_{0}^{3/2}(\tau)exp[-S_{0}(\tau)]
\label{rhosupl}
\end{equation}
For $\lambda\rightarrow\lambda^{\star}$ we obtain:
\begin{equation}
\rho_{s}(\tau)\approx\xi^{-3}(T)\left(\frac{\lambda^{\star}}{\lambda}-1\right)^{3/2}S_{0}^{3/2}(\tau)exp[-S_{0}(\tau)]
\label{rhosuph}
\end{equation}
Thus the density of superconducting ``drops'' $\rho_{s}(\tau)$ vanishes as
$\lambda\rightarrow\lambda^{\star}$,\ they are destroyed
by thermodynamic fluctuations.

For the order---parameter correlator of Eq. (\ref{corfun}) we get the following
result:
\begin{equation}
<\Delta({\bf r})\Delta({\bf r'})>\approx\rho_{s}(\tau)\int d{\bf
R_{0}}\Delta_{0}({\bf r+R_{0}})\Delta_{0}({\bf r'+R_{0}})
\label{corfuninst}
\end{equation}
The integration over instanton center ${\bf R_{0}}$ here
means in fact averaging over different positions of ``drops''.\ Note that over
large distances this correlator decreases like $exp[-|{\bf r-r'}|/\xi(T)]$ and
does not contain the usual Ornstein---Zernike factor
$|{\bf r-r'}|^{-1}$.

We have found the free---energy of inhomogeneous superconducting state in the
temperature region $\tau\gg\tau_{D}$,\ where the ``drop'' concentration is
exponentially small and the picture of noninteracting
``drops'' is valid.\ They give exponentially small contribution to the specific
heat and diamagnetic susceptibility.\ The characteristic size of ``drops''
is determined by $\xi(T)$ and as $T\rightarrow T_{c}$ the ``drops'' grow and
begin to overlap leading to a
percolative superconducting transition.\ Thus for
$\tau_{D}>\tau_{D}^{\star}>\tau_{G}$ superconductivity first appears in
isolated ``drops''.\ This is similar
to the picture of decay of a metastable state in case of
the first---order phase transitions\cite{Lang67}.\ However,\ in this latter
case instantons give imaginary
contribution to the free energy determining the decay rate of a ``false''
equilibrium state (critical bubble formation).\ Here instanton contributions
lead as was noted above to real free energy and ``drops'' appear in
the true equilibrium state.

It is more or less obvious that between isolated ``drops'' a kind of Josephson
coupling may appear and lead to rather complicated phase diagram of the system
in
external magnetic field,\ e.g.\ including the ``superconducting glass''
phase\cite{JL85,JL86}.\
The existence of inhomogeneous regime of superconductivity
will obviously lead to the rounding of BCS---like singularities of
the density of states and superconductivity may become gapless.\
Note that diffusion---enhanced Coulomb interactions can also lead to the
gaplessness of strongly disordered superconductors via Coulomb---induced
inelastic
scattering \cite{BLM87}.\
Fluctuation conductivity in a similar inhomogeneous superconducting state was
studied in Ref.\cite{ChrK88}.\ Note the closely related problem of strongly
disordered superfluids \cite{MHL86,FF88}.\ Some results here may be quite
useful for the case of strongly disordered superconductors,\ though
the limitations of this analogy are also important.

A major unsolved problem here is the possible influence
of statistical fluctuations of gradient term coefficient
in Ginzburg---Landau expansion which has been neglected
above,\ or the equivalent problem (Cf.\ Eq.(\ref{nsGL}))
for superconducting electron density $n_{s}$.\ This problem was briefly
considered
for the case of weak disorder in Ref.\cite{SpZuz88}.\ It was shown that:
\begin{equation}
<(\delta n_{s}/n_{s})^{2}>\sim (\xi_{0}p_{F}^{2}l)^{-1}\sim
\frac{e^{4}}{g(\xi)}
\label{stns}
\end{equation}
where $g(\xi)=\sigma\xi$ is the conductunce of metallic sample with the size of
the order of superconducting coherence length $\xi = \sqrt{\xi_{0}l}$.\
Extrapolating this estimate up to the Anderson transition  using
$\xi=(\xi_{0}/p_{F}^{2})^{1/3}$ we get:
\begin{equation}
<(\delta n_{s}/n_{s})^{2}>\sim
\frac{e^{4}}{\sigma^{2}(\xi_{0}/p_{F}^{2})^{2/3}}
\label{stnsmi}
\end{equation}
Obviously we get $<(\delta n_{s}/n_{s})^{2}>\geq 1$ for
$\sigma\leq \sigma^{\star}$ so that statistical fluctuations of $n_{s}$
become important close to the
Anderson transition in the same region we have discussed
above.\ This further complicates the picture of superconducting transition and
can also be very important for the possible anomalous behavior of $H_{c2}$
which was recently studied on the similar lines in Ref.\cite{Sp95}.\
Some qualitative conjectures for the case of $<(\delta n_{s}/n_{s})^{2}>\geq 1$
were
formulated in Ref.\cite{SpKiv91},\ where it was argued that in this case there
will occur regions in the sample
with locally {\em negative} values of superfluid density.\ This is equivalent
to a negative sign of a Josephson coupling betweenthe ``drops''.\ In this
sense,\
the disordered superconductor is unlike a Bose liquid.\ This leads to an
important
prediction that in a small superconducting ring,\ if there is a segment with
negative $n_{s}$,\ the ground state of the ring will spontaneously break the
time---reversal invariance.\ The ground state will have nonzero supercurrent
and magnetic flux (or rather random,\ trapped fluxes in the ground state) and
will be two---fold degenerate.\ At longer times the symmetry will be restored
due to thermal activation of macroscopic quantum tunneling between the two
states,\
but according to Ref.\cite{SpKiv91} it can be expected that for dirty metal
rings
with conductance of the order of $e^{2}/\hbar$ there will be
``roughly 50\% chance that the ground state will break time---reversal
symmetry''.\
By the way this means that in the presence of disorder there may be no way to
distinguish between an anyon superconductor\cite{Laugh88} and a conventional
superconductor.\ Of course we must stress that these speculations are entirely
based upon a simple extrapolation of Eq.(\ref{stns}) to the vicinity of
metal---insulator transition and there is no complete theory of statistical
fluctuations of gradient term in this region at the moment.


\newpage
\section{SUPERCONDUCTIVITY IN STRONGLY DISORDERED METALS: EXPERIMENT}


Our review of experiments on strongly disordered superconductors will be in no
sense exhaustive.\ This is mainly a theoretical review and the author is in no
way an expert on experiment.\ However,\ we shall try to illustrate the
situation with the interplay of Anderson
localization and superconductivity in {\em bulk} (three---dimensional)
superconductors,\ both
traditional and high---temperature.\ Again we must stress
that we exclude any discussion of numerous data on thin
films which are to be described by two---dimensional
theories.\ In this case we just refer to existing
reviews\cite{Ramak,Belev,LiuGold}.\
Here we shall confine ourselves to a limited number of the experiments ,\ which
we consider most interesting from the point of view of illustration of some of
the ideas expressed above,\ just to convince the reader,\ that previous
discussion,\ while purely theoretic,\ has something to do with the real life.\
More than anywhere else in this review our choice of material is
based on personal interests of the author,\ or our direct involvement in the
discussion of experiments.\ We shall not deal with the general problem of
disorder influence upon superconductivity,\ but shall consider only the systems
which remain superconducting close to the disorder---induced metal---insulator
transition.

\subsection{Traditional Superconductors}

There exists a number of strongly disordered systems which remain
superconducting  close
to the metal---insulator transition induced by disorder.

The drop of $T_{c}$ with conductivity decrease from the value of the order of
$10^{4} Ohm^{-1}cm^{-1}$ was observed in amorphous
alloys of $GeAl$\cite{Nich82},\ $SiAu$\cite{Fur85} and $MoRe$\cite{Tenh81},\ in
Chevrel phase superconductors disordered by fast neutron irradiation,\ such as
$Pb_{1-x}U_{x}Mo_{6}S_{8}$\cite{Aleks83},\
$SnMo_{6}S_{8}$\cite{Dav83},\ $Mo_{6}Se_{8}$\cite{Dav86},\ in amorphous
$InO_{x}$\cite{Fior84},\ in $BaPb_{1-x}Bi_{x}O_{3}$ in the concentration
interval $0.25<x<0.30$\cite{Batl84} and in metallic glass
$Zr_{0.7}Ir_{0.3}$\cite{Poon85}.\ In all of these systems superconducting
transition is observed apparently   not  very far from the metal---insulator
transition.\ For a number of these systems,\  such as
$Pb_{1-x}U_{x}Mo_{6}S_{8}$,\
$SnMo_{6}S_{8}$,\ $Mo_{6}Se_{8}$,\  $Zr_{0.7}Ir_{0.3}$ and
$BaPb_{0.75}Bi_{0.25}O_{3}$\cite{Batl84} and some others a characteristic
strongly negative temperature
resistivity coefficient has been observed.\ Note,\ however,\ that this fact
alone in no way indicates that a
specimen is on one side or the other of the metal---insulator transition.\ 
The drop of $T_{c}$ close to the mobility edge apparently was also observed in
$As_{2}Te_{3}$\cite{Ber86}.\ However,\ in all of these systems
$T_{c}$ apparently vanishes before metal---insulator transition.\ Below we
present some of the data on these
and other similar systems.

On Fig.\ref{fig15} we show the dependence of $T_{c}$ and $|dH_{c2}/dT|_{T_{c}}$
in $SnMo_{5}S_{6}$ (Chevrel phase superconductor) on the fluence of fast
neutron irradiation (the number of neutrons which passed through a crossection
of a sample during irradiation)\cite{Dav83}.\
In the region of large fluences (large disorder),\ when the system becomes
amorphous,\ characteristic values of
conductivity in the normal state are of the order of
$\sim 10^{3}Ohm^{-1}cm^{-1}$,\ which is not far from the
values of ``minimal metallic conductivity'' $\sigma_{c}\sim 
5\ 10^{2}\ Ohm^{-1}cm^{-1}$,\ which define
the conductivity scale of disorder induced metal---insulator transition.\ The
negative temperature coefficient of resistivity was observed in this
conductivity range.\ The experimental data on $T_{c}$ decrease with the growth
of resistivity in this system were rather well fitted in Ref.\cite{Bul2} using
the $\mu^{\star}$ dependence on resistivity given by Eq. (\ref{muinter}).\ A
clear tendency for $|dH_{c2}/dT|_{T_{c}}$ saturation with disorder is also
observed.\
Analogous dependence of $T_{c}$ and $|dH_{c2}/dT|_{T_{c}}$ on the resistivity
in the normal
state for $Mo_{6}Se_{8}$ disordered by fast neutrons is shown in
Fig.\ref{fig16}\cite{Dav86}.\ Here superconductivity exists up to
conductivities $\sigma\sim
250\ Ohm^{-1}cm^{-1}$.\ Further disordering (irradiation) leads to the
destruction of superconducting state and
metal---insulator transition (an unlimited growth of
resistivity with decrease of $T$,\ with
variable---range hopping conduction\cite{Mott74,MottDav}
is observed).\ The slope of the upper critical field
$|dH_{c2}/dT|_{T_{c}}$ also has a tendency to saturate
with the growth of resistivity.\ Standard interpretation
of such behavior of $|dH_{c2}/dT|_{T_{c}}$ was based
upon the use of Gorkov's relation (Cf.\ first relation in Eq. (\ref{Gorrel}))
and lead to the conclusion of $N(E_{F})$ decrease under disordering.\ In fact,\
we have seen that no such conclusion can be reached for systems with
conductivities $\sigma<10^{3}Ohm^{-1}cm^{-1}$,\ because such saturation
behavior may be a natural manifestation of the approaching metal---insulator
transition.\ Similar dependences were observed in other Chevrel phase
superconductors \cite{Aleks83,GAC87,AAGE89}.\

In Fig.\ref{fig17} we show the dependence of conductivity
and $T_{c}$ on the parameter $p_{F}l/\hbar$ in amorphous  $InO_{x}$
alloy\cite{Fior84}.\ In Fig.\ref{fig18} from
Ref.\cite{Heb84} the data on the temperature dependence
of $H_{c2}$ in amorphous $In/InO_{x}$ (bulk) films are
presented for different degrees of disorder.\ We can see
that in the low temperature region $H_{c2}(T)$ deviates
from the standard temperature dependence,\ but apparently
confirm the qualitative form predicted above for systems
which are close to Anderson transition.\ The same system was also studied
in Ref.\cite{Ovadyahu}.\ In Fig.\ref{fig18a} we show the dependence  of
two characteristic energies on disorder which in the opinion of the
authors of Ref.\cite{Ovadyahu} demonstrate the narrow region of
coexistense of superconductivity and insulating state.\ In Fig.\ref{fig18b}
we show the dependencies of localization length and superconducting coherence
length on disorder according to Ref.\cite{Ovadyahu}.\ It demonstrates the
qualitative agreement with our general criterion of coexistense of
superconductivity and localization---localization length must be larger
or at least of the order of the size of the Cooper pair.

Very impressive are the data for amorphous
$Si_{1-x}Au_{x}$ alloy \cite{Nich82,Fur85,Fur86}.\ In Fig.\ref{fig19}
\cite{Fur85} the data on $T_{c}$ and
conductivity dependence on the gold concentration $x$
are shown.\ In Fig.\ref{fig20} $H_{c2}(T)$ dependence for
this system is shown for different alloy compositions \cite{Fur85}.\ From these
data it is clearly seen that
$T_{c}$ vanishes {\em before} metal---insulator transition.\ The
metal---insulator transition itself is {\em continuous},\ conductivity vanishes
linearly with the
decrease of gold concentration and the values of conductivity significantly
less than the estimated ``minimal metallic conductivity'' are definitely
observed.\ The system remains superconducting even for
such low conductivity values.\ The slope of $H_{c2}(T)$
at $T=T_{c}$ is practically constant ,\ despite the change of conductivity
(disorder) in rather wide range.\
This behavior apparently cannot be explained only by the
appearance of correlation pseudogap in the density of
states observed in Ref.\cite{Fur86},\ which becomes significant only very close
to metal---insulator transition.\ Low temperature deviation from standard
convex dependence on $T$ is also clearly seen.\ In Fig.\ref{fig21} from
Ref.\cite{Fur86} we show the temperature dependences of resistivity and
superconducting energy gap (determined by tunneling) of a sample with
$x=0.21$.\ It nicely demonstrates superconducting transition in the system
which is very close to disorder induced metal---insulator transition.\ Note,\
that according to Ref.\cite{Fur86} the superconducting energy gap in this
sample is substantially broadened which may indicate the
growth of statistical gap fluctuations due to the same
fluctuations of the local density of states.\ These data
are in obvious qualitative correspondence with the general theoretical picture
described throughout this review.

These data show that in systems which are superconducting
close to the disorder induced metal---insulator (Anderson) transition $T_{c}$
decreases rather fast and practically
in all reliable cases vanishes before the transition to
insulating state.\ At the same time the temperature dependence of $H_{c2}$ is
not described by the standard
theory of ``dirty'' superconductors both with respect to
$(dH_{c2}/dT)_{T_{c}}$ behavior and at low temperatures,\ where the upward
deviations from the standard dependence
are readily observed.\ This confirms most of our theoretical conclusions.

Some indications of a possible superconducting state in
the insulating phase of granular $Al$ and $Al-Ge$ were observed in
Refs.\cite{Deut81,Deut83}.\ Obviously,\ the granular systems are more or less
outside the scope of our review.\ However,\ we should like to mention that the
strong smearing of BCS---like density of states and the
gapless regime of superconductivity was observed (via tunneling measurements)
in Refs.\cite{Dyn84,Dyn86},\ close to the metal---insulator transition in these
systems.\ This may confirm our picture of statistical
fluctuation smearing of the density of states.\ Note,\
that more recent work on granular $Al$ \cite{Mill88}
apparently exclude the possibility of superconductivity
in the insulating phase.\ In this work a small amount of
$Bi$ was added to granular $Al$ in order to enhance spin-orbit scattering,\ which
leads to {\em antilocalization}
effect\cite{AltAr82}.\ This shifts both metal---insulator
and to the {\em same extent} the superconducting transition,\ with the
preservation of a narrow range of concentration on the metallic side where the
material is
not fully superconducting.\ The fact that the superconducting transition shifts
with metal---insulator
transition demonstrates that its position is determined by the vicinity of the
metal---insulator transition,\ and
that it is the impending transition to the insulating state which inhibits
superconductivity.\   Similar conclusions on superconductivity vanishing at the
point of metal---insulator transition were reached for amorphous
$Al_{x}Ge_{1-x}$\cite{Ned88} and amorphous $Ga-Ar$ mixtures\cite{Zint90}.\ This
later case is particularly interesting because it has been shown that
conductivity exponent at metal---insulator transition
here is $\nu\approx 0.5$ which places this system to a
different universality class than those discussed above
and similar to that observed in some doped uncompensated
semiconductors like $Si:P$\cite{Mill85}.\ Usual interpretation of this
difference is based upon the importance of interaction effects in these
systems\cite{KotLee87}.\ Starting with the value of
$T_{c}$ of amorphous $Ga$ ($T_{c}=7.6K$),\ $T_{c}$
decreases rather slowly with decreasing $Ga$ volume fraction $v$,\ until one
enters the critical region near
$v_{c}\approx 0.145$.\ Further approach to $v_{c}$ leads
to a rapid decrease of $T_{c}$.\ Taking McMillan formula
Eq.(\ref{McMTc}) for $T_{c}$ (with $\omega_{log}/1.20=320K$ and $\lambda=0.45$)
and assuming
negligible Coulomb repulsion $\mu^{\star}$ for pure amorphous $Ga$ the increase
of $\mu^{\star}$ on the
approach of metal---insulator transition can be determined from the
experimental data for $T_{c}$.\ This
increase is approximately given by $\mu^{\star}\sim (v-v_{c})^{-0.33}$.\ From
this it is easy to see that $T_{c}\rightarrow 0$ for $v\rightarrow v_{c}$,\ so
that
these data does not indicate the survival of superconductivity beyond
metal---insulator transition.\
These results are not surprising since we have seen the
existence of strong mechanisms of $T_{c}$ degradation
close to disorder induced metal---insulator transition.

The interesting new high-pressure metastable metallic phase of an amorphous
alloy $Cd_{43}Sb_{57}$ exhibiting
the gradual metal---insulator transition during the slow
decay at room temperature and atmospheric pressure has
been studied in Refs.\cite{Gant92a,Gant92b}.\ Authors
claim that during this decay the system remains homogeneous while going from
metallic to insulating phase.\ At the same time the metallic phase is
superconducting with $T_{c}\approx 5 K$ and remains such
up to metal---insulator transition.\ Close to it superconducting transition
becomes smeared,\ while incomplete transition persists even in the insulating
state.\ While these data are reminiscent of data on quench-condensed films of
$Sn$ and $Ga$\cite{Mark88},\
which were interpreted as reentrant superconductivity due
to sample inhomogeneities,\ it is stressed in Refs.\cite{Gant92a,Gant92b} that
in this new system situation is different and we are dealing with intrinsically
inhomogeneous superconductors state discussed in
Refs.\cite{BulSad86,BPS87,BPS89}.\ From our
point of view further studies of this system are necessary in order to show
unambiguously the absence of
structural imhomogeneities.\ Also rather peculiar characteristic of this system
is almost complete independence of the onset temperature of superconducting
transition on disorder.

The general conclusion is that in most cases of traditional
superconducting systems we can not find unambiguous demonstration of the
possibility of superconductivity in insulating state induced by disorder.\ At
the same we can see rather rich variety of data on superconductivity close to
metal---insulator transition which stimulate further studies.\ Some of the
anomalies of superconducting behavior discussed above can be successfully
explained by theories presented in this review,\ while the other require
further theoretical investigations.



\subsection{High---$T_{c}$ Superconductors}

Very soon after the discovery of high---temperature oxide
superconductors \cite{BM86,BM88} it was recognized that
localization effects has an important role to play in these systems.\ There are
many sources of disorder in these systems and the low level of conductivity
indicate
from the very beginning their closeness to Anderson transition.\ In the field
where there are hundreds of papers published on the subject it is impossible to
review or even to quote all of them.\ More or less complete impression of the
status of high---$T_{c}$ research can be obtained from Conference
Proceedings\cite{M2SHTSC}.\ Here we shall concentrate almost only on papers
which deal with disordering by fast neutron irradiation which we consider
probably the ``purest'' method to introduce disorder into
the system (allowing to neglect the complicated problems associated with
chemical substitutions).\
Also historically it was apparently the earliest method
used to study disorder effect in high---$T_{c}$ superconductors in a
controllable way \cite{Gosh87,GKS88}.\

There are several reasons for localization effects  to be
important in high---$T_{c}$ oxides:
\begin{itemize}
\item {\em Two---Dimensionality}.\ All the known high---$T_{c}$ systems (with
$T_{c}>30K$) are strongly anisotropic or quasi-two-dimensional conductors.\ We
have seen above that for such systems it is natural to expect the strong
enhancement of localization effects due to the
special role of spatial dimensionality $d=2$:\ in purely
two-dimensional case localization appears for infinitely
small disorder\cite{AALR,Sad81,LeeRam,Sad86}.\ The inplane conductivity scale
for metal---insulator transition in such systems as given by Eq. (\ref{mmc2d})
or Eq. (\ref{mmcq2d}) is larger than in isotropic case.\
Reasonable estimates show that the values of inplane ``minimal metallic
conductivity'' may exceed $10^{3}Ohm^{-1}cm^{-1}$.\ While due to continuous
nature
of Anderson transition there is no rigorous meaning of
minimal metallic conductivity,\ these estimates actually
define the scale of conductivity near the metal---insulator transition caused
by disorder.\ Then it is clear that most of the real samples of high---$T_{c}$
superconductors are quite close to Anderson transition
and even the very slight disordering is sufficient to
transform them into Anderson insulators\cite{GoSad89}.
\item {\em ``Marginal'' Fermi Liquid}.\ During our discussion of interaction
effects we have seen that there are serious reasons to believe that importance
of localization effects in high---$T_{c}$ oxides may be actually due to more
fundamental reasons connected with
anomalous electronic structure and interactions in these materials.\ The
concept of ``marginal'' Fermi liquid \cite{Varma89} leads to extreme
sensitivity of such system to disordering and the appearance of localized
states around the Fermi level at rather weak
disorder\cite{KotVar90,Var90}.
\end{itemize}
On the other hand high---$T_{c}$ systems are especially promising from the
point of view of the search for superconductivity in the Anderson insulator:
\begin{itemize}
\item High transition temperature $T_{c}$ itself may guarantee the survival of
superconductivity at relatively high disorder.
\item Due to small size of Cooper pairs in high---$T_{c}$ systems in
combination with high---$T_{c}$ (large gap !) we can easily satisfy the main
criterion for superconductivity in localized phase as given by Eq.
(\ref{BScrit}).
\item Being narrow band systems as most of the conducting
oxides high---$T_{c}$ systems are promising due to low values of the Fermi
energy $E_{F}$ which leads to less
effective $T_{c}$ degradation due to localization enhancement of Coulomb
pseudopotential $\mu^{\star}$.\
(Cf.\ Eq. (\ref{muedge})).
\end{itemize}

Anomalous transport properties of high---$T_{c}$ oxides are well
known\cite{Iye92}.\ Experimentally there are
two types of resistivity behavior of good single-crystals
of these systems.\ In highly conducting $ab$ plane of
$YBa_{2}Cu_{3}O_{7-\delta}$ and other oxides resistivity
of a high quality single-crystal always shows the notorious linear---$T$
behavior (by ``good'' we mean the
samples with resistivity $\rho_{ab}<10^{3}\ Ohm\ cm$).\
However,\ along orthogonal $c$ direction the situation is
rather curious:\ most samples produce semiconductor-like
behavior $\rho_{c}\sim 1/T$,\ though some relatively rare
samples (apparently more pure) show metallic-like $\rho_{c}\sim T$ (with strong
anisotropy $\rho_{c}/\rho_{ab}\approx 10^{2}$ remaining).\cite{Iye92,For92}
Metallic behavior in $c$ direction was apparently observed only in the best
samples of $YBa_{2}Cu_{3}O_{7-\delta}$ and almost in no other high---$T_{c}$
oxide.\ In
Fig.\ref{fig22} taken from Ref.\cite{Ito91} we show the temperature dependence
of $\rho_{c}$ in a number of high---$T_{c}$ systems.\ It is seen that
$\rho_{c}(T)$ changes between metallic and semiconducting behavior depending on
whether the resistivity is below or above the Ioffe---Regel limit defined for
quasi-two-dimensional case by Eq. (\ref{mmc2d}).\ Rather strange is the absence
of any
obvious correlation between the behavior of $\rho_{c}$
and $T_{c}$.

This unusual behavior leads us to the idea that most of
the samples of high---$T_{c}$ systems which are studied
in the experiment are actually already in localized phase
due to internal disorder which is always present.\ Surely,\ we realize that
such a drastic assumption contradicts the usual expectations 
and propose it just as
an alternative view open for further discussion.\  The attempted justification
of this idea
may be based upon the quasi-two-dimensional nature of these systems or  on
marginal Fermi liquid effects.\ In this case a
simple conjecture on the temperature behavior of resistivity of single-crystals
can be made which
qualitatively explains the observations \cite{Sad90,KotVar91}.\  In case of
localized states at
the Fermi level and for finite temperatures it is important to compare
localization length $R_{loc}$ with
diffusion length due to inelastic scattering $L_{\varphi}\approx
\sqrt{D\tau_{\varphi}}$,\  where $D$ is diffusion coefficient due to elastic
scattering on disorder,\ while $\tau_{\varphi}$ is phase coherence
time determined by inelastic processes.\ For $T>0$ this length $L_{\varphi}$
effectively replaces the sample size
$L$ in all expressions of scaling theory of localization when $L\gg
L_{\varphi}$,\ because on distances larger than $L_{\varphi}$ all information
on the nature of wave
functions (e.g.\ whether they are localized or extended)
is smeared out.\ Taking into account the usual low---temperature dependence
like $\tau_{\varphi}\sim T^{-p}$
(where $p$ is some integer,\ depending on the mechanism
of inelastic scattering) this can lead to a non---trivial
temperature dependence of conductivity,\ in particular to
a possibility of a negative temperature coefficient of
resistivity \cite{Im80}.\ Similar expressions determine the temperature
dependence of conductivity also for the localized phase until
$L_{\varphi}<R_{loc}$.\ In this
case electrons do not ``feel'' being localized and conductivity at high enough
$T$ will show metallic like
behavior.\ For localization to be important we must go
to low enough temperatures,\ so that $L_{\varphi}$ becomes greater than
$R_{loc}$.\  If disordered high---$T_{c}$ superconductors are in fact 
Anderson insulators with
very anisotropic localization length,\ $R^{ab}_{loc}\gg R^{c}_{loc}$ and both
localization lengths diminish as disorder grows,\ $L_{\varphi}$ is also
anisotropic and we can have three different types of temperature behavior of
resistivity\cite{Sad90}:
\begin{enumerate}
\item Low $T$ or strong disorder,\ when we have
\begin{equation}
L_{\varphi}^{ab}\approx\sqrt{D_{ab}\tau_{\varphi}}\gg R^{ab}_{loc} \qquad
L_{\varphi}^{c}\approx\sqrt{D_{c}\tau_{\varphi}}\gg R^{c}_{loc}
\end{equation}
This gives semiconductor-like behavior for both directions.
\item Medium $T$ or medium disorder,\ when
\begin{equation}
L_{\varphi}^{ab} < R^{ab}_{loc} \qquad
L_{\varphi}^{c} > R^{c}_{loc}
\end{equation}
and metallic behavior is observed in $ab$ plane,\ while
semiconducting temperature dependence of resistivity is
observed along $c$ axis.
\item High $T$ or low disorder,\ when
\begin{equation}
L_{\varphi}^{ab} < R^{ab}_{loc} \qquad
L_{\varphi}^{c} < R^{c}_{loc}
\end{equation}
and metallic behavior is observed in both directions.
\end{enumerate}
Here we do not speculate on the inelastic scattering mechanisms leading to the
concrete temperature behavior
in high---$T_{c}$ oxides,\ in particular on linear $T$
behavior in $ab$ plane or $1/T$ behavior in $c$ direction.\ Unfortunately too
little is known on these
mechanisms\cite{Iye92} to be able to make quantitative
estimates on the different types of behavior predicted
above.\ Of course detailed studies of such mechanisms are necessary
to prove the proposed idea and to explain the temperature dependence of
resistivity
in high-$T_{c}$ systems on its basis.\
However,\ most of the experimental data as we shall see
below at least do not contradict the idea of possibility
of Anderson localization in disordered high---$T_{c}$ cuprates.

Now let us consider the experiments on controllable disordering of
high---temperature superconductors.\ Already the first experiments on low
temperature ($T=80K$) fast neutron irradiation of ceramic samples of
high---$T_{c}$ systems \cite{Gosh88,Gosh89a,Gosh89b,Gosh89c,Gosh89d,Gosh89f}
has shown that the growth of structural disorder leads to a number of drastic
changes in their physical properties:
\begin{itemize}
\item continuous metal---insulator transition at very
slight disordering,
\item rapid degradation of $T_{c}$,
\item apparent coexistence of hopping conductivity and
superconductivity at intermediate disorder,
\item approximate independence of the slope of $H_{c2}$
at $T\sim T_{c}$ on the degree of disorder,
\item anomalous {\em exponential} growth of resistivity
with defect concentration.
\end{itemize}
These anomalies were later confirmed on single-crystals
and epitaxial films\cite{Gosh89e,Gosh90,Dyn89,Gosh91},\
and were interpreted\cite{GoSad89,Sad89} using the ideas
of possible coexistence of Anderson localization and
superconductivity.

In Fig.\ref{fig23} we show data\cite{GoSad89} on the dependence of the
superconducting transition temperature and resistivity (at $T=100K$,\ i.e.\
just before superconducting transition) on fast neutron fluence for
$YBa_{2}Cu_{3}O_{6.95}$.\ In all high---$T_{c}$ compounds
introduction of defects leads to strong broadening of
superconducting transition.\ The derivative $(dH_{c2}/dT)_{T_{c}}$ in ceramic
samples measured at the
midpoint of the superconducting transition  does not
change as $\rho_{100K}$ grows by an order magnitude.\
In Fig.\ref{fig24}\cite{GoSad89} we show the temperature dependence of
resistivity for samples of $YBa_{2}Cu_{3}O_{6.95}$ and
$La_{1.83}Sr_{0.17}CuO_{4}$
for different degrees of disorder.\ In all these materials the $\rho(T)$ curves
vary in the same way.\ In the fluence range $\Phi>10^{19}cm^{-2}$,\ where
superconductivity is absent,\ $\rho(T)$ follows a dependence which is
characteristic of conductivity via
localized states\cite{Mott74,MottDav}:
\begin{equation}
\rho(T)=\rho_{0}exp(Q/T^{1/4}) \qquad Q=2.1[N(E_{F})R^{3}_{loc}]^{-1/4}
\label{hopp}
\end{equation}
as shown in Fig.\ref{fig25}.\ (Mott's variable-range hopping conduction).

The most striking anomaly of resistivity behavior of all
high---$T_{c}$ systems under disordering is nonlinear,\
practically {\em exponential} growth of resistivity at
fixed temperature (e.g.\ $\rho(T=100K)$) with fluence,\
starting from the low fluences $\Phi<7\ 10^{18}cm^{-2}$,\
including superconducting
samples\cite{Gosh88,GoSad89,Gosh89a,Gosh89b,Gosh89c,Gosh89d}.\ These data are
shown in Fig.\ref{fig26}\cite{GoSad89} for the dependence of $\rho(T=80K)$ on
$\Phi$ obtained from measurements made
directly during the process of irradiation.\ For comparison the similar data
for $SnMo_{6}S_{8}$ are shown
which do not demonstrate such an anomalous behavior,\ its
resistivity is just proportional to $\Phi$ and saturates
at large fluences.\ We relate this exponential growth of
$\rho$ with the increase of $\Phi$ (i.e.\ of defect concentration) in all
high---$T_{c}$ systems to localization,\ which already appears for very small
degrees of disorder in samples with high values of $T_{c}$.\ As we have seen in
samples with much reduced or
vanishing $T_{c}$ localization is observed directly via
Mott's hopping in the temperature behavior of resistivity
given by Eq. (\ref{hopp}).

{}From these results it follows that the electronic system
of high---$T_{c}$ superconductors is very close to the
Anderson transition.\ The observed variation of $\rho$ as
a function both of fluence and of temperature can be described by the following
empirical formula\cite{Gosh88}:
\begin{equation}
\rho(T,\Phi)=(a+cT)exp(b\Phi/T^{1/4})
\label{hoppfl}
\end{equation}
Identifying the exponential factors in Eq. (\ref{hopp}) and Eq. (\ref{hoppfl})
it is possible to obtain a fluence
dependence of localization length (Cf.\ Ref.\cite{GoSad89}
and below).

Detailed neutron diffraction studies of structural changes in irradiated
samples were also performed \cite{Gosh88,GoSad89,Vor91}.\ These investigations
has
shown definitely that there are no oxygen loss in
$YBa_{2}Cu_{3}O_{6.95}$ during low temperature irradiation.\ Only some partial
rearrangement of oxygens
between positions $O(4)$ and $O(5)$ in the elementary cell occur as
radiation-induced defects are introduced.\
In addition, in all high---$T_{c}$ compounds the Debye---
Waller factors grow and the lattice parameters $a,b,c$
increase slightly\cite{GoSad89,Vor91}.\ The growth of
Debye---Waller factors reflect significant atomic shifts,\ both static and
dynamic,\ from their regular positions,\ which induce a random potential.\ This
disorder is pretty small from the structural point of view,\ the lattice is
only slightly distorted.\ However,\ we have seen that this small disorder is
sufficient to
induce metal---insulator transition and complete degradation of
superconductivity.\ The absence of oxygen
loss implies that there is no significant change in concentration of carriers
and we have really disorder---induced metal---insulator transition.\ This is
also confirmed by other methods\cite{Gosh89f,Podl91}.\ In Fig.\ref{fig27} we
show the data\cite{Gosh89f,Gosh90} on temperature dependence of the Hall
concentration of ceramic samples of irradiated and oxygen deficient
$YBa_{2}Cu_{3}O_{7-\delta}$.\ It is seen that disordering
weakens the anomalous temperature dependence of Hall effect,\ but Hall
concentration $n_{H}$ at low $T$ practically does not change in striking
difference with
data on oxygen deficient samples,\ where $n_{H}$ drops
several times.\ This also confirms the picture of disorder---induced
metal---insulator transition in radiation disordering experiments.\ Similar
Hall data
were obtained on epitaxial films\cite{Dyn89} and single-crystals\cite{Gosh91}.

Qualitatively same resistivity behavior was also obtained in the experiments on
radiation disordering of single-crystals\cite{Gosh89e,Gosh90} and epitaxial
films\cite{Dyn89}.\ Electrical resistivities of $YBa_{2}Cu_{3}O_{7-\delta}$
single crystals were measured at $T=80K$ directly during
irradiation by fast neutrons.\ The data are shown in Fig.\ref{fig28}.\ We can
see that $\rho_{ab}$ increases
exponentially with $\Phi$ (defect concentration) starting
from the smallest doses,\ while $\rho_{c}$ grows slower and only for
$\Phi>10^{19}cm^{-2}$ they grow with the
same rate.\ At large fluences both $\rho_{ab}$ and $\rho_{c}$ 
demonstrate\cite{Dav90} Mott's hopping $ln\rho_{ab,c}\sim
T^{-1/4}$.\ Similar data of Ref.\cite{Dyn89}
show $ln\rho\sim T^{-1/2}$ characteristic of Coulomb gap.\ We do not know the
reasons for this discrepancy between single-crystalline and epitaxial films
data (note
that another method of disordering by 1MeV $Ne^{+}$ ions
was used in Ref.\cite{Dyn89}).\ Anisotropy $\rho_{c}/\rho_{ab}$ at $T=80K$
drops rapidly (to the
values $\sim 30$ for $\Phi=10^{19}cm^{-2}$) and then practically does not
change and ``residual'' anisotropy
of the order of its room---temperature value in initial
samples remains.\ This means that temperature dependence of anisotropy weakens
in the disordered samples.\ Note,\ that unfortunately only the single-crystals
with ``semiconducting'' temperature dependence of resistivity along $c$ axis
were investigated up to now.

The upper critical fields of $YBa_{2}Cu_{3}O_{7-\delta}$
single-crystals (determined from standard resistivity measurements) for
different degrees of disorder are shown in Fig.\ref{fig29}\cite{Gosh90}.\
Temperature dependence of $H_{c2}$ in disordered samples is essentially
nonlinear,\ especially for samples with low
$T_{c}$.\ The estimated from high-field regions temperature derivative of
$H^{\bot}_{c2}$ (field along the $c$ axis) increases with disorder.\ However,\
similar derivative of $H^{\|}_{c2}$ (field along $ab$ plane) drops in the
beginning and then does not change.\ Anisotropy of $H_{c2}$ decreases with
disorder and in
samples with $T_{c}\sim 10K$ the ratio of $(H^{\|}_{c2})'/(H^{\bot}_{c2})'$ is
close to unity.\
According to Eq. (\ref{anislop}) this means the complete
isotropisation of the Cooper pairs.\ This is illustrated by
Fig.\ref{fig30}\cite{Dav91}.\ The remaining anisotropy
of resistivity may be connected with some kind of planar defects in the system.

In a recent paper \cite{Bozovic} Osofsky et al. presented the unique data on
the temperature dependence of the upper critical field of high-temperature
superconductor $Bi_{2}Sr_{2}CuO_{y}$ in wide temperature interval from
$T_{c}\approx 19K$ to $T\approx 0.005T_{c}$,\ which has shown rather anomalous
dependence with positive curvature at any temperature. The authors of
Ref.\cite{Bozovic} has noted that this type of behavior is difficult to
explain within any known theory. It is sharply different from the standard
behavior of BCS-model. It was demonstrated in Refs.\cite{KS93a,KS94}
that the observed dependence of $H_{c2}(T)$ can be satisfactorily explained
by localization effects in two-dimensional (quasi-two-dimensional) model in
the limit of sufficiently strong disorder. Measurements of
$H_{c2}$ in Ref.\cite{Bozovic} were performed on epitaxially grown films
of $Bi_{2}Sr_{2}CuO_{y}$, however it is quite possible that the films were
still disordered enough, which can be guessed from rather wide ($\sim 7K$)
superconducting transition.  Unfortunately the relevant data, in particular
on the value of conductivity of the films studied are absent. This gives us
some grounds to try to interpret the data obtained in Ref.\cite{Bozovic}
in the framework of rather strong disorder the effects of which are obviously
enhanced by the quasi-two-dimensional nature of high-temperature
superconductors.

The general discussion of the temperature dependence of the upper critical
field in two-dimensional and quasi-two-dimensional case with strong
localization effects was presented above in Section III.C.1.\ Note that we
mainly analyzed there the case of magnetic field perpendicular to the highly
conducting planes,\ which is precisely the case of Ref.\cite{Bozovic}.
We have seen\cite{KS93} that the anomalies of the upper critical field due
to the frequency dependence of diffusion coefficient appear only for
temperatures $T\ll\frac{e^{-1/\lambda}}{\tau}$. For higher
temperatures we obtained the usual behavior of "dirty" superconductors.
Also we have noted\cite{KS93} that superconductivity 
survives in a system with finite localization length if 
$T_{c}\gg\lambda\frac{e^{-1/\lambda}}{\tau}$,\ which is equivalent to
our criteria of the smallness of Cooper pair size compared
with localization length. This latter length is exponentially large in two-
dimensional systems with small disorder ($\lambda\ll 1$). The most
interesting (for our aims) limit of relatively strong disorder is defined by
$T_{c}\ll \frac{e^{-1/\lambda}}{\tau}$, so that in fact we are dealing with
pretty narrow region of $\lambda$'s when
$\lambda\frac{e^{-1/\lambda}}{\tau}\ll T_{c}\ll \frac{e^{-1/\lambda}}{\tau}$.
In this case we have seen that the upper critical field is practically
defined by Eq.(\ref{Hc33}) :
\begin{equation}
ln\left(\frac{\gamma}{2\pi}\frac{e^{-1/\lambda}}{\tau
T}\right)=(1+4\pi\frac{D_{0}}{\phi_{0}}\frac{\tau
H_{c2}}{e^{-1/\lambda}})ln\left(\frac{\gamma}{2\pi}\frac{e^{-1/\lambda}}{\tau
T_{c}}(1+4\pi\frac{D_{0}}{\phi_{0}}\frac{\tau H_{c2}}{e^{-1/\lambda}})\right)
\label{Hc3}
\end{equation}
($\gamma=1.781$)
from which we can directly obtain the $T(H_{c2})$---dependence.
The appropriate behavior of the upper critical field for two sets of parameters
is shown in Fig.\ref{fig30a}. The curve of $H_{c2}(T)$ demonstrates
positive curvature and $H_{c2}$ diverges for $T\rightarrow 0$. We have
seen that this weak (logarithmic) divergence is connected with our neglect
of the magnetic field influence upon diffusion. Taking this influence
into account we can suppress this divergence of $H_{c2}$ as $T\rightarrow 0$.
This is the main effect of broken time invariance and it is clear that it is
important only for extremely low temperatures\cite{KS93}. In the
following we neglect it.\ 
For the quasi-two-dimensional case  on the dielectric side of Anderson's
transition, but not too very close to it, the behavior of diffusion coefficient
is quite close to that of purely two-dimensional case, so that the upper
critical field can be analyzed within two-dimensional approach. Close to the
transition (e.g. over interplane transfer integral) both for metallic and
insulating sides and for parameters satisfying the inequality
$\lambda\frac{e^{-1/\lambda}}{\tau}\ll T_{c}\ll \frac{e^{-1/\lambda}}{\tau}$,
the temperature dependence of $H_{c2}$ is in fact again very close to
those in purely two-dimensional case considered above\cite{KS93}. Some
deviations appear only in a very narrow region of very low
temperatures\cite{KS93}.

In Fig.\ref{fig30a} we also show the experimental data for $H_{c2}$ from
Ref.\cite{Bozovic}.  Theoretical curve (1) is given for the parameters
which lead to rather good agreement with experiment in the low temperature
region.  The curve (2) corresponds to parameters giving good agreement in a
wide temperature region except the lowest temperatures. The cyclotron mass
$m$ was always assumed to be equal that of the free electron. In general we
observe satisfactory agreement between theory and experiment. Unfortunately,
the values of the ratio $\frac{e^{-1/\lambda}}{T_{c}\tau}$ for the second
curve, while corresponding to quite reasonable values of $\lambda$, lead to
nonrealistic (too small) values of $T_{c}\tau$, which are rather doubtful for
the system with relatively high $T_{c}$. For the first curve situation is
much better though the electron damping on the scale of $T_{c}$ is
still very large which corresponds to strong disorder. Note however, that the
detailed discussion of these parameters is actually impossible without the
knowledge of additional characteristics of the films studied in
Ref.\cite{Bozovic}.  In particular it is quite interesting to have an
independent estimate of $\lambda$. We also want to stress relatively
approximate nature of these parameters due to our two- dimensional
idealization. More serious comparison should be done using the expressions of
Ref.\cite{KS93} for the quasi-two-dimensional case, which again requires
the additional information on the system, in particular, the data on the
anisotropy of electronic properties.

In our opinion the relatively good agreement of experimental data of
Ref.\cite{Bozovic} with theoretical dependences obtained for the
two-dimensional (quasi-two- dimensional) case of disordered system with
Anderson localization illustrates the importance of localization effects for
the physics of high-temperature superconductors.\ However,\ we must note
that the similar anomalies of the temperature dependence of the upper
critical field were also observed in Ref.\cite{Mackenz} for the single
crystals of the overdoped $Tl_{2}Ba_{2}CuO_{6+\delta}$ which authors claim
to be extremely clean,\ so that apparently no explanation based upon strong
localization effects can be used.\ Similar data were recently obtained for
thin films of underdoped $YBa_{2}(Cu_{0.97}Zn_{0.03})_{3}O_{7-\delta}$
with pretty low transition temperatures\cite{Walker}.\ These films again seem
to be disordered enough to call localization effects as a possible
explanation of unusual positive curvature of $H_{c2}(T)$ dependence for all
temperatures.

Under irradiation localized moment contribution appears in the magnetic
susceptibility of high---$T_{c}$ oxides\cite{Gosh88,GoSad89}.\
In the temperature range from $T_{c}$ to $300K$  $\chi(T)$ is satisfactorily
described by Curie---Weiss type dependence:\ $\chi(T)=\chi_{0}+C/(T-\Theta)$.\
The value of $\chi_{0}$ and the Curie constant $C$ as a function of fluence
$\Phi$ are given in Fig.\ref{fig31}.\ The value of $C$ is proportional to
fluence.\ Note that the threefold larger slope of $C(\Phi)$ in
$YBa_{2}Cu_{3}O_{6.95}$ as compared with   $La_{1.83}Sr_{0.17}CuO_{4}$ is an
evidence that this Curie-law temperature
dependence is associated with localized moments forming on $Cu$ (there are
three times more coppers in the elementary cell of $Y$ compound than in $La$
compound).

The data presented above show that electronic properties of high---$T_{c}$ 
systems are quite different under
disordering from that of traditional superconductors \cite{GAC87,AAGE89} or
even some closely
related metallic oxides\cite{Gosh91,DG90}.\ We associate these anomalies with
the closeness
of the Anderson transition and believe that real samples of high---$T_{c}$
systems which always
possess some noticeable disorder may well be already in the state of the
Anderson insulator.\
However,\ we must stress that it is quite difficult to decide from the
experiments described above the precise position of the Anderson transition on
disorder scale.\ Some additional information on this problem may be obtained
from experiments on NMR relaxation in disordered state,\ using the approach
proposed rather long ago by Warren\cite{War71} and
later quantified theoretically in Refs.\cite{GK83,MVS85}.\ The study of NMR
relaxation rate
on $^{89}Y$ nuclei in radiationally disordered $YBa_{2}Cu_{3}O_{6.95}$ (which
opposite to
$Cu$ nuclei demonstrate Korringa behavior)\cite{Verkh92a,Verkh92b} has shown
the anomalies (a {\em maximum} in the so called Warren's enhancement factor) 
which according to Ref.\cite{MVS85} may indicate the Anderson 
transition  somewhere in the fluence interval
$\Phi = (1-2) 10^{19}cm^{-2}$.\ Unfortunately the number of samples in these
experiments was too limited to place the transition point more precisely,\
while superconductivity disappears exactly in this interval.\ In this sense we
still have no direct proof of coexistence
of superconductivity and localization in disordered high---$T_{c}$ oxides.\
However the method used in Refs.\cite{Verkh92a,Verkh92b} seems to be very
promising.\ Note that Knight
shift data of Refs.\cite{Verkh92a,Verkh92b} strongly indicate Coulomb gap
opening at the
Fermi level of strongly disordered oxides.\ Independently this conclusion was
reached in tunneling experiments of Ref.\cite{Srika} on a number of oxides
disordered by doping.

Using the experimental data on electrical resistivity of disordered samples of
$YBa_{2}Cu_{3}O_{6.95}$ and the relations given by Eq. (\ref{hopp}) and Eq.\
(\ref{hoppfl}) (assuming that exponentials there are identical) we can
calculate the change of localization
length $R_{loc}$ as a function of
fluence\cite{GoSad89,Gosh89a,Gosh89b,Gosh89c,Gosh89d}.\
This dependence is shown on Fig.\ref{fig32} along with fluence dependence of
$T_{c}$.\ It is
clearly seen that superconductivity is destroyed when localization length
$R_{loc}$ becomes
smaller than $\sim 30\AA$,\ i.e.\ it becomes of the order or smaller than a
typical size of the
Cooper pair in this system (Cf.\ Fig.\ref{fig30}) in complete accordance with
our basic criterion of Eq. (\ref{BScrit}).\ We can estimate the minimal value
of $R_{loc}$ for which superconductivity can still exist in a system of
localized electrons via Eq. (\ref{BScrit})\cite{GoSad89} taking the
free-electron value of $N(E_{F})\approx 5\ 10^{33} (erg cm^{3})^{-1}$ (for
carrier concentration of $\sim 6\ 10^{21}cm^{-3}$) and the gap value 
$\Delta\sim 5T_{c}$,\ corresponding to very
strong coupling\cite{GKS88}.\ We obtain the result shown in Fig.\ref{fig32}.\
In any case we can
see that criterion of Eq. (\ref{BScrit}) ceases to be fulfilled for $\Phi\sim
(5-7)10^{18}cm^{-2}$ in
remarkably good agreement with the experiment.

In the absence of accepted pairing mechanism for high temperature
superconductors it is very
difficult to speculate on the reasons for $T_{c}$ degradation in these
systems.\ If we assume that
the main mechanism of $T_{c}$ degradation is connected with the growth of
Coulomb effects
during disordering,\ as discussed above in this review,\ we can try to use
appropriate expressions
to describe the experimental data.\ Assuming superconductivity in the localized
phase we can
use Eq. (\ref{Tclocliz}),\ estimating $R_{loc}$ as above from empirical
relation Eq. (\ref{hoppfl})
and Eq. (\ref{hopp}) (or directly expressing the parameters entering Eq.
(\ref{Tclocliz}) via
experimental dependence of resistivity on fluence as described by Eq.
(\ref{hoppfl})\cite{GoSad89}).\ The results of such a fit (with the assumption
of $\mu\approx 1$) are also shown in Fig.\ref{fig32}.\ The agreement is also
rather satisfactory,\ the more rapid
degradation of $T_{c}$ for small degrees of disorder can be related to
additional contributions
to Coulomb repulsion within Cooper pairs neglected in the derivation of Eq.
(\ref{Tclocliz}).\ Surely we do not claim that this is a real explanation of
$T_{c}$ degradation in disordered high
temperature superconductors.\ However,\ note its relation to localized moment
formation under disordering which leads to the usual Abrikosov-Gorkov
mechanism of depairing due to spin-flip scattering on magnetic impurities.
According to Mott\cite{Mott71} (Cf.\ also Refs.\cite{Kam80,Sad86}) the
appearance of localized moments may be related to the presence of localized
states (single occupied states below the Fermi level as briefly discussed
above).\ We can then estimate the value of the effective magnetic
moment (in Bohr magnetons) in unit cell as\cite{GoSad89}:
\begin{equation}
\mu R^{-3}_{loc}\Omega_{0}=p^{2}_{theor}
\label{magmom}
\end{equation}
where $\Omega_{0}$ is the volume of a unit cell.\ For large degrees of disorder
($\Phi=2\ 10^{19}cm^{-2}$) and $R_{loc}\approx 8\AA$ with $\mu\approx 1$ we
obtain $p^{2}_{theor}=0.66$ for  $YBa_{2}Cu_{3}O_{6.95}$ in full agreement with
experiment.\ However,\ for smaller fluences $p_{theor}$ is considerably smaller
than the experimental value.\
Note,\ though ,\ that the estimate of Eq. (\ref{magmom}) is valid only for
small enough values of $R_{loc}$,\ i.e.\ when the Fermi level is well inside
the localized region.\ On the other hand,\  the
accuracy with which the Curie constant is determined in weakly disordered
samples is considerably less than in strongly disordered case.\ Of course,\ the
other mechanisms of local
moment formation,\ which were discussed above and can become operational even
before the metal---insulator transition can be important here.

Of course a plenty of works on localization effects in high-$T_{c}$ oxides
are being done using disorder induced by different types of chemical
substitutions in these systems.\ Of these we shall rather arbitrarily quote
Refs.\cite{Ell89,Inf90,Jay91,Mand91,Ciep92},\ which demonstrate the data
quite similar,\ though not necessarily identical,\ to those described above
on different types of systems and
obtained by different experimental methods.\ We note that the effects of
chemical chemical
disorder are always complicated by the inevitable changes of carrier
concentration due to doping effects.\ Still all these data indicate that
superconductivity in high-$T_{c}$ systems is
realized close to disorder induced metal---insulator transition,\ so that these
systems provide
us with plenty of possibilities to study experimentally the
general problems discussed in our review.\ More details can be found in the
extensive review paper\cite{Narlikar}.

Special attention should be payed to a recent study of angle resolved
photoemission in $Co$ doped
single-crystals of $Bi_{2}Sr_{2}CaCu_{2}O_{8+y}$\cite{Quitt}.  Doping
$Bi_{2}Sr_{2}CaCu_{2}O_{8+y}$ with $Co$ causes superconducting- insulator
transition, $Co$ doping decreases $T_{c}$ and causes increase in residual
resistivity. The changes in temperature behavior of resistivity from metallic
to insulating like correlate with the disappearance of the dispersing
band-like states in angle-resolved photoemission. Authors believe that
Anderson localization caused by the impurity potential of the doped $Co$
atoms provides a consistent explanation of all experimental features and
$T_{c}$ reduction is not caused by magnetic impurity pairbreaking effects but
by spatial localization of carriers with superconducting ground state being
formed out of spatially localized carriers. Similar data were also obtained
for some exceptional (apparently strongly disordered) samples of undoped
$Bi_{2}Sr_{2}CaCu_{2}O_{8+y}$\cite{Quittmann}. Of course it will be very
interesting to make similar type of experiments on neutron irradiated
samples where we are dealing with pure disorder.

Finally we must stress that in our opinion these data on rather strongly
disordered samples of high-temperature superconductors more or less definitely
exclude the possibility of $d$-wave pairing in these systems. As is well
known (and also can be deduced from our discussion in Appendix C) $d$-wave
pairing is much more sensitive to disordering and is completely suppressed
roughly speaking at the disorders measured by the energy scale
$1/\tau\sim T_{c0}$,\ which is at least an order of magnitude smaller than the
disorder necessary to induce the metal-insulator transition which can be
estimated as $1/\tau\sim E_{F}$. This apparently excludes the possibility to
observe any manifestations of localization effects in $d$-wave superconductors,
though these are clearly observed in high-$T_{c}$ systems. Of course,\ these
qualitative conclusions deserve further studies within the specific models of
microscopic mechanisms of high-temperature superconductivity.

We shall limit ourselves to this discussion of localization effects in high
temperature superconductors.\ Our conclusion is that these effects are
extremely important in these systems and some of the anomalies can be
successfully described by theoretical ideas formulated in this review.\ We must
stress that much additional work is needed both theoretical and experimental to
clarify the general picture of disorder effects in high-$T_{c}$
superconductors and we can expect that the future progress,\ especially with
the quality of samples,\ may provide some new and exciting results.


\newpage
\section{CONCLUSION}
We conclude our review trying to formulate the basic unsolved problems.\
{}From the theoretical point of view probably the main problem is to formulate
the theory of superconducting pairing in strongly disordered system along the
lines of the general theory of interacting Fermi systems.\ This problem is
obviously connected with the general theory of metal---insulator transition
in such approach,\ which as we mentioned during our brief discussion above is
rather far from its final form.\ Nevertheless,\ there were several attempts
to analyze superconducting transition within this framework
\cite{Fin84c,MaF86,KB91,KB92,BelKirk94}.\ In all cases,\ the authors limited
themselves to certain universality classes within the general renormalization
group approach of interaction theory of metal---insulator transition.\
Ref.\cite{Fin84c} dealt only with two---dimensional problem,\ while
Refs.\cite{MaF86,KB91,KB92,BelKirk94} also considered the bulk case.\ These
papers have demonstrated a large variety of possible behavior of
superconductivity under disordering,\ from  disorder---induced (triplet)
superconductivity\cite{KB91} to a complete destruction of it
close\cite{Fin84c,MaF86} or even long before the metal---insulator
transition\cite{KB92}.\ Our point of view is that at the moment it is rather
difficult to make any {\em general} conclusions from the results of these
approaches.\ In particular,\ we do not beliewe that the
present status of these theories is sufficient to prove
or disprove the general possibility of superconductivity
in Anderson insulators.\ However,\ it is obvious that further theoretical
progress in the problem of $T_{c}$
behavior under disordering will be largely possible only
within this general approach.\ In this sense our simplified discussion of
Coulomb effects and other mechanisms of $T_{c}$ degradation in this review is
only of qualitative nature.\ Still,\ more general approaches apparently do not
change our qualitative conclusions.\ These problems become even more
complicated if we address ourselves to the case of high temperature
superconductors,\ where we do not know precisely the nature of pairing
interaction in regular system.

Concerning the semiphenomenological approach to the theory of superconductivity
close to the Anderson transition we must stress the necessity of further
investigation of the region of strong statistical fluctuations with the aim of
more detailed study of their influence upon different physical properties,\
like e.g.\ the upper critical field,\ density of states,\ nuclear relaxation
etc.\ Obviously,\ all of them may be significantly changed in comparison with
predictions of what we called the statistical mean---field theory.\
Especially important are further studies of rather exotic
predictions of random fluxes in the ground state\cite{SpKiv91}.

Despite our explicit limitation to a discussion of superconductivity in {\em
bulk} disordered superconductors
we have to mention the extremely interesting problem of
universal conductivity at the superconductor---insulator
transition  at $T=0$ in two-dimensional systems which
attracted much attention recently\cite{Girv1,Girv2,Girv3,LiuGold}.\ It is
argued that the transition between the insulating and superconducting phases
of disordered two-dimensional system at zero temperature is of continuous
quantum nature, but the system behaves like a normal metal right at the
transition,\ i.e.\ the conductivity has a finite,\ nonzero value.\ This value
is {\em universal} and,\ apparently,\ equal to $(2e)^2/h$ (with $2e$ being
the Cooper pair charge).\ There is strong experimental
evidence\cite{Heb84,Heb85,Hav89,SJL90,Wang91,LiuGold} that a variety of
systems (metallic films,\ high-$T_{c}$ films,\ etc.) show the onset of
superconductivity to occur when their sheet resistance falls below a value
close to $h/4e^2\approx 6.45 k\Omega$.\ The theoretical analysis here is
based upon boson (Cooper pairs) approach to superconductivity and the main
conclusion is that in contrast to the case of localization of fermions in two
dimensions,\ bosons exhibit a superconductor to insulator
transition (as disorder grows) with the value of conductivity at the critical
point being independent of
microscopic details.\ A major theoretical problem arises
to describe a crossover to such behavior e.g.\ in quasi---two---dimensional
case of BCS superconductivity as interplane coupling goes to zero.

So we are not short of theoretical problems in this
important field of research.\ As to the experiments,\ certainly too much is
still to be done for unambigous
demonstration of exotic possibility of superconductivity of Anderson
insulators.

\begin{center}
{\bf Acknowledgments}
\end{center}

The author is grateful to all his collaborators during the research work on
superconductivity and localization,\ especially to Prof.\ L.\ N.\ Bulaevskii
whose insights were so important at early stages of this work.\ In recent years
I
have benefited very much from the joint work with Dr.\ E.\ Z.\
Kuchinskii.\
Useful discussions of experimental situation with Prof.\ B.\ N.\ Goshchitskii
and Dr.\ A.\ V.\ Mirmelstein are very much appreciated.

This research was supported in part by the Russian Academy of Sciences Program
on High---Temperature Superconductivity
under the Research Project $N^{o} 93-001$ as well as by the grant
of the Russian Foundation of Fundamental Research $N^{o} 93-02-2066$ and
International (Soros) Science Foundation grants $RGL000$ and $RGL300$.

\newpage
\appendix

\section{Spectral Densities and Criterion for Localization}

Convenient formalism to consider general properties of disordered system is
based
upon exact eigenstate representation for an electron in a random field created
by disorder.\ These eigenstates $\phi_{\nu}({\bf r})$ are formally defined by
the Schroedinger equation:
\begin{equation}
H\phi_{\nu}({\bf r})=\varepsilon_{\nu}\phi_{\nu}({\bf r}) \end{equation}
where $H$ is one---particle Hamiltonian of disordered system under
consideration,\
$\varepsilon_{\nu}$ are exact eigenvalues of electron energy in a random
potential.\
Obviously $\phi_{\nu}({\bf r})$ and $\varepsilon_{\nu}$ are dependent on
locations of scatterers ${\bf R}_{n}$ for a given realization of random field.

Let us define two---particle spectral densities \cite{BG79,Sad86}:
\begin{equation}
\ll\rho_{E}({\bf r})\rho_{E+\omega}({\bf r'})\gg^{F}=\frac{1}{N(E)}<\sum_{\nu
\nu'}\phi_{\nu}^{\star}({\bf r})\phi_{\nu'}({\bf r})\phi_{\nu'}^{\star}({\bf
r'})
\phi_{\nu}({\bf
r'})\delta(E-\varepsilon_{\nu})\delta(E+\omega-\varepsilon_{\nu'})>
\end{equation}
\begin{equation}
\ll\rho_{E}({\bf r})\rho_{E+\omega}({\bf r'})\gg^{H}=\frac{1}{N(E)}<\sum_{\nu
\nu'}|\phi_{\nu}({\bf r})|^{2}|\phi_{\nu'}({\bf
r'})|^{2}\delta(E-\varepsilon_{\nu})
\delta(E+\omega-\varepsilon_{\nu'})>
\end{equation}
where angular brackets denote averaging over disorder and
\begin{equation}
N(E)=<\sum_{\nu}|\phi_{\nu}({\bf r})|^{2}\delta(E-\varepsilon_{\nu})>
\end{equation}
is one---electron  (average) density of states.\ Obviously Eq. (A3) is just
a correlation function of {\em local} densities of states in a disordered
system.\ Spectral density given by Eq. (A2) determines electronic transport
\cite{BG79}.\ The following general properties are easily verified using the
completeness and orthonormality of $\phi({\bf r})$ functions:
\begin{equation}
\int d{\bf r} \ll\rho_{E}({\bf r})\rho_{E+\omega}({\bf
r'})\gg^{F}=\delta(\omega)
\qquad  \int d\omega \ll\rho_{E}({\bf r})\rho_{E+\omega}({\bf
r'})\gg^{F}=\delta({\bf r-r'})
\end{equation}
or for the Fourier---components:
\begin{equation}
\ll\rho_{E}\rho_{E+\omega}\gg_{{\bf q}=0}=\delta(\omega) \qquad
\int d\omega \ll\rho_{E}\rho_{E+\omega}\gg_{{\bf q}}=1
\end{equation}
and $\ll\rho_{E}\rho_{E+\omega}\gg_{{\bf q}}\geq 0$.\ From general definitions
given in Eqs. (A2) and (A3) it is clear that:
\begin{equation}
\ll\rho_{E}({\bf r})\rho_{E+\omega}({\bf r})\gg^{F}=\ll\rho_{E}({\bf
r})\rho_{E+\omega}({\bf r})\gg^{H}
\end{equation}
i.e.\ these spectral densities coincide for ${\bf r}={\bf r'}$.

Terms with $\varepsilon_{\nu}=\varepsilon_{\nu'}$ are in general present in
Eqs. (A2) and (A3).\ However,\ if these states are extended the appropriate
wave---functions $\phi_{\nu}({\bf r})$ are normalized on the total volume
$\Omega$ of the system and these contributions to Eqs. (A2) and (A3) are
proportional to to $\Omega^{-1}$ and vanish as $\Omega\rightarrow \infty $.\
Things change if states are localized.\ In this case states are normalized on
finite volume of the order of $\sim R_{loc}^{d}$.\ This leads to the appearance
of $\delta (\omega)$---contribution to spectral densities:
\begin{equation}
\ll\rho_{E}({\bf r})\rho_{E+\omega}({\bf r'})\gg^{F,H}=
A_{E}({\bf r-r'})\delta(\omega)+\rho_{E}^{F,H}({\bf r-r'}\omega)
\end{equation}
or in momentum representation:
\begin{equation}
\ll\rho_{E}\rho_{E+\omega}\gg^{F,H}_{{\bf q}}=A_{E}({\bf q})\delta(\omega)+
\rho_{E}^{F,H}({\bf q}\omega)
\end{equation}
where the second terms are regular in $\omega$.\ This singular behavior was
proposed as a general criterion for localization \cite{BG79}.\ It is easy to
show that
\begin{equation}
A_{E}({\bf r-r'})=\frac{1}{N(E)}<\sum_{\nu}\delta(E-\varepsilon_{\nu})
|\phi_{\nu}({\bf r)}|^{2}|\phi_{\nu}({\bf r'})|^{2}>
\end{equation}
\begin{displaymath}
A_{E}=A_{E}({\bf r-r'})|_{{\bf r=r'}}\sim R_{loc}^{-d}
\end{displaymath}
$A_{E}({\bf r-r'})$ represents the so called inverse participation ratio
\cite{Th74,Weg80}.\ Roughly speaking its value at ${\bf r=r'}$ is inversely
proportional to the number of atomic orbitals which effectively form quantum
state $\nu$.

These $\delta(\omega)$---singularities in spectral densities signal nonergodic
behavior of the system in localized state.\ This leads to a difference between
so called adiabatic and isothermal response functions
\cite{Kubo57,KazSad84,Sad86}.\
The intimate connection between localization and nonergodic behavior was
already
noted in the first paper by Anderson \cite{And58}.

{}From general properties given by Eqs. (A5) and (A6) for ${\bf q\rightarrow
0}$
in localization region we have \cite{BG79}:
\begin{equation}
\ll\rho_{E}\rho_{E+\omega}\gg^{F}_{{\bf q}}\approx
[1-R_{loc}^{2}q^{2}]\delta(\omega)+ \cdots
\end{equation}
where
\begin{equation}
R_{loc}^{2}=\frac{1}{2dN(E)}\int d^{d}r r^{2}
<\sum_{\nu}\delta(E-\varepsilon_{\nu})
|\phi_{\nu}({\bf r})|^{2}|\phi_{\nu}(0)|^{2}>
\end{equation}
defines the localization length.\ Delocalization leads to smearing of
$\delta(\omega)$---singularity for finite $q$.

Spectral densities of Eqs. (A2) and (A3) can be expressed via two---particle
Green's functions \cite{Sad86}.\ Using nonaveraged retarded and advanced
Green's
functions:
\begin{equation}
G^{R}({\bf rr'}E)=G^{A\star}({\bf rr'}E)=\sum_{\nu}\frac{\phi_{\nu}({\bf
r})\phi_{\nu}^{\star}({\bf r'})}
{E-\varepsilon_{\nu}+i\delta}
\end{equation}
we immediately get from Eqs. (A2) and (A3):
\begin{equation}
\ll\rho_{E}({\bf r})\rho_{E'}({\bf r'})\gg^{F}=\frac{1}{2\pi^{2}N(E)}Re\left\{<
G^{R}({\bf rr'}E')G^{A}({\bf r'r}E)>
-<G^{R,A}({\bf rr'}E')
G^{R,A}({\bf r'r}E)>\right\}
\end{equation}
\begin{equation}
\ll\rho_{E}({\bf r})\rho_{E'}({\bf r'})\gg^{H}=\frac{1}{2\pi^{2}N(E)}Re
\left\{<G^{R}({\bf rr}E')G^{A}({\bf r'r'}E)>
-<G^{R,A}({\bf rr}E')G^{R,A}({\bf r'r'}E)>\right\}
\end{equation}
In momentum representation Eq. (A14) is equivalent to:
\begin{equation}
\ll\rho_{E}\rho_{E+\omega}\gg^{F}_{{\bf q}}=\frac{1}{\pi N(E)}Im\left\{
\Phi^{RA}_{E}(\omega{\bf q})-\Phi^{RR}_{E}(\omega{\bf q})\right\}
\end{equation}
where
\begin{equation}
\Phi^{RA(R)}_{E}({\bf q}\omega)=-\frac{1}{2\pi i}\sum_{{\bf pp'}}<G^{R}({\bf
p_{+}p'_{+}}E+\omega)G^{A(R)}({\bf p'_{-}p_{-}}E)>
\end{equation}
and ${\bf p_{+-}=p^{+}_{-}}1/2{\bf q}$.\ It can be shown \cite{VW80,VW82}
that $\Phi^{RR(AA)}_{E}(\omega{\bf q})$ are nonsingular for small $\omega$ and
$q$.\ Accordingly $\delta(\omega)$---singularity signalling localization can
appear only from the first term in Eq. (A16).

\newpage
\section{Linearized Gap Equation in Disordered System}

Let us consider the derivation of linearized gap equation
Eq. (\ref{GapDS}) used to determine $T_{c}$ \cite{Dolg77,DolgSad}.\
Equation for Gorkovs's anomalous Green's function in an inhomogeneous
disordered
system (before any averaging procedure) at $T=T_{c}$ takes the following form:
\begin{equation}
(\varepsilon^{2}_{n}+\hat\varepsilon^{2}_{{\bf r}})F({\bf rr'}\varepsilon_{n})=
-T_{c}\sum_{m}V({\bf rr'}\varepsilon_{n}-\varepsilon_{m})F({\bf
rr'}\varepsilon_{m})
\end{equation}
where $\varepsilon_{n}=(2n+1)\pi T_{c}$ and $V({\bf
rr'}\varepsilon_{n}-\varepsilon_{m})$ is an effective
interelectron potential,\ $\hat\varepsilon_{{\bf r}}$ is one---electron energy
operator (energy zero is at the Fermi energy).\ Define
\begin{equation}
\Delta({\bf rr'})=-2\hat\varepsilon_{{\bf r}} cth\frac{\hat\varepsilon_{{\bf
r}}}{2T_{c}}T_{c}\sum_{n}
F({\bf rr'}\varepsilon_{n})
\end{equation}
and assume the following relation between $\Delta({\bf rr'})$ and $F({\bf
rr'}\varepsilon_{n})$:
\begin{equation}
F({\bf rr'}\varepsilon_{n})=\frac{1}{\varepsilon_{n}^{2}+\hat\varepsilon_{{\bf
r}}^{2}}T_{c}\sum_{m}V({\bf
rr'}\varepsilon_{n}-\varepsilon_{m})\frac{1}{\varepsilon_{m}^{2}+\hat\varepsilon_{{\bf r}}^{2}}\hat Q({\bf rr'}\varepsilon_{m})\Delta({\bf rr'})
\end{equation}
where $\hat Q$ is some unknown operator.\ Then after substitution of Eq. (B3)
into Eq. (B2) we get a BCS---like equation for $T_{c}$:
\begin{equation}
\Delta({\bf rr'})=-\hat U({\bf rr'})\frac{th\frac{\hat\varepsilon_{{\bf
r}}}{2T_{c}}}{2\hat\varepsilon_{{\bf r}}}\Delta({\bf rr'})
\end{equation}
where the operator of ``effective'' interaction is defined by:
\begin{eqnarray}
\hat U({\bf rr'})=2\hat\varepsilon_{{\bf
r}}cth\left(\frac{\hat\varepsilon_{{\bf
r}}}{2T_{c}}\right)T_{c}\sum_{n}\frac{1}{\varepsilon_{n}^{2}+\hat\varepsilon_{{\bf r}}^{2}}T_{c}\sum_{m}V({\bf rr'}\varepsilon_{n}-\varepsilon_{m})\times
\nonumber \\
\times\frac{1}{\varepsilon_{m}^{2}+\hat\varepsilon_{{\bf r}}^{2}}\hat Q({\bf
rr'}\varepsilon_{m})2\hat\varepsilon_{{\bf r}}cth\frac{\hat\varepsilon_{{\bf
r}}}{2T_{c}}
\end{eqnarray}
{}From Eqs. (B1)---(B3) we obtain the following equation for $\hat Q$ (we drop
${\bf rr'}$ for brevity):
\begin{eqnarray}
\hat
Q(\varepsilon_{n})=1-T_{c}\sum_{m}V(\varepsilon_{n}-\varepsilon_{m})\frac{1}{\varepsilon_{m}^{2}+\hat\varepsilon^{2}}\hat Q(\varepsilon_{m})+ \nonumber \\
+2\hat\varepsilon
cth\left(\frac{\hat\varepsilon}{2T_{c}}\right)T_{c}\sum_{n'}\frac{1}{\varepsilon^{2}_{n'}+\hat\varepsilon^{2}}T_{c}\sum_{m}V(\varepsilon_{n'}-\varepsilon_{m})\frac{1}{\varepsilon_{m}^{2}+\hat\varepsilon^{2}}\hat Q(\varepsilon_{m})
\end{eqnarray}
In case of weak coupling in the lowest order over  interaction in Eq. (B6) we
can leave only the first term
$\hat Q(\varepsilon_{n})=1$.\ Then Eq. (B5) reduces to
\begin{equation}
\hat U({\bf rr'})=2\hat\varepsilon_{{\bf
r}}cth\left(\frac{\hat\varepsilon_{{\bf r}}}{2T_{c}}\right)T_{c}
\sum_{n}\frac{1}{\varepsilon_{n}^{2}+\hat\varepsilon_{{\bf
r}}^{2}}T_{c}\sum_{m}V({\bf
rr'}\varepsilon_{n}-\varepsilon_{m})\frac{1}{\varepsilon_{m}^{2}+\hat\varepsilon_{{\bf r}}^{2}}2\hat\varepsilon_{{\bf r}}cth\frac{\hat\varepsilon_{{\bf r}}}{2T_{c}}
\end{equation}
and Eq. (B4) completely determines $T_{c}$.

Using the usual definition of superconducting gap:
\begin{equation}
\Delta({\bf rr'}\varepsilon_{n})=T_{c}\sum_{m}V({\bf
rr'}\varepsilon_{n}-\varepsilon_{m})F({\bf
rr'}\varepsilon_{m})=-(\varepsilon_{n}^{2}+\hat\varepsilon_{{\bf r}}^{2})F({\bf
rr'}\varepsilon_{n})
\end{equation}
it is easy to get:
\begin{equation}
\Delta({\bf rr'}\varepsilon_{n})=\hat Q({\bf rr'}\varepsilon_{n})\Delta({\bf
rr'})
\end{equation}
so that $\Delta({\bf rr'})$ represents the energy gap in the absence of
frequency dispersion,\ while $\hat Q$
describes the frequency dependence of the energy gap.

Cooper pairing takes place in the states which are time---reversed,\ thus in
the exact eigenstate representation of an electron in disordered system we
have:
\begin{equation}
\Delta({\bf rr'})=\sum_{\nu}\Delta_{\nu}\phi^{\star}_{\nu}({\bf
r'})\phi_{\nu}({\bf r})
\end{equation}
and Eq. (B4) gives
\begin{equation}
\Delta_{\nu}=-\sum_{\nu\nu'}\frac{1}{2\varepsilon_{\nu'}}
th\frac{\varepsilon_{\nu'}}{2T_{c}}U_{\nu\nu'}\Delta_{\nu'}
\end{equation}
where the kernel
\begin{equation}
U_{\nu\nu'}=\int d{\bf r}\int d{\bf r'}\phi^{\star}_{\nu}({\bf
r})\phi^{\star}_{\nu'}({\bf r'})\hat U({\bf rr'})\phi_{\nu}({\bf
r'})\phi_{\nu}({\bf r})
\end{equation}
has the form of ``Fock'' matrix element of an effective interaction.\ From Eq.
(B7) we have:
\begin{eqnarray}
U_{\nu\nu'}=T_{c}^{2}\frac{2\varepsilon_{\nu}\varepsilon_{\nu'}}{th\frac{\varepsilon_{\nu}}{2T_{c}}th\frac{\varepsilon_{\nu'}}{2T_{c}}}\sum_{n}\sum_{m}\frac{1}{\varepsilon_{n}^{2}+\varepsilon_{\nu}^{2}}\frac{1}{\varepsilon_{m}^{2}+\varepsilon_{\nu'}^{2}}\times \nonumber \\
\times\int d{\bf r}\int d{\bf r'} \phi^{\star}_{\nu}({\bf
r})\phi^{\star}_{\nu'}({\bf r'})V({\bf
rr'}\varepsilon_{n}-\varepsilon_{m})\phi_{\nu}({\bf r'})\phi_{\nu'}({\bf r})
\end{eqnarray}
It is convenient to rewrite Eq. (B11) introducing
summation over states belonging to some surface of constant energy with
subsequent integration over energies:
\begin{equation}
\Delta_{\nu}=-\int\limits_{-\infty}^{\infty}
dE'\frac{1}{2E'}th\frac{E'}{2T_{c}}\sum_{\nu'\in E'}{\cal
N}(E')U_{\nu\nu'(E)}\Delta_{\nu'(E')}
\end{equation}
where ${\cal N}(E)=\sum_{\nu}\delta(E-\varepsilon_{\nu})$.

Consider now averaging of the gap equation.\ Define
\begin{equation}
\Delta(E)=\frac{1}{N(E)}<\sum_{\nu}\Delta_{\nu}\delta(E-\varepsilon_{\nu})>
\end{equation}
i.e.\ the gap averaged over disorder and a surface of
constant energy $E=\varepsilon_{\nu}$.\ Here as usual we
denote $N(E)=<{\cal N}(E)>$.\ Suppose now that
$\Delta_{\nu}=\Delta(\varepsilon_{\nu})=\Delta(E=\varepsilon_{\nu})$,\ i.e.\
that $\Delta_{\nu}$ depends only on
energy $E=\varepsilon_{\nu}$,\ but not on the quantum numbers $\nu$.\ This is
similar to the usual assumption
of $\Delta({\bf p})$ depending only on $|{\bf p}|$ in a
homogeneous and isotropic system \cite{VIK}.

After the usual decoupling used e.g. in transforming Eq.
(\ref{Gap}) into Eq. (\ref{GAP}),\ i.e.\ assuming the self---averaging of the
gap,\ we obtain the following
linearized gap equation determining $T_{c}$:
\begin{equation}
\Delta(E)=-\int\limits_{-\infty}^{\infty}dE'K(E,E')\frac{1}{2E'}
th\frac{E'}{2T_{c}}\Delta(E')
\end{equation}
where
\begin{eqnarray}
K(E,E')=\frac{1}{N(E)}<\sum_{\nu\nu'}U_{\nu\nu'}\delta(E-\varepsilon_{\nu})\delta(E'-\varepsilon_{\nu'})>= \nonumber \\
=T_{c}^{2}\sum_{n}\sum_{m}\left[\frac{2E}{th(E/2T_{c})}\frac{1}{\varepsilon_{n}^{2}+E^{2}}\right]\left[\frac{2E'}{th(E'/2T_{c})}\frac{1}{\varepsilon_{m}^{2}+E'^{2}}\right]\times \nonumber \\
\times\int d{\bf r}\int d{\bf r'} V({\bf
r-r'}\varepsilon_{n}-\varepsilon_{m})\ll\rho_{E}({\bf r})\rho_{E'}({\bf
r'})\gg^{F}
\end{eqnarray}
where we have again introduced Gorkov---Berezinskii
spectral density defined in Eq. (A2).\ Effective interaction can be written as:
\begin{equation}
V({\bf r-r'}\varepsilon_{n}-\varepsilon_{m})=V_{p}({\bf
r-r'}\varepsilon_{n}-\varepsilon_{m})+V_{c}({\bf
r-r'}\varepsilon_{n}-\varepsilon_{m})
\end{equation}
i.e.\ as the sum of some kind of Boson---exchange attractive interaction
$V_{p}$ and Coulomb repulsion $V_{c}$,\ which leads to:
\begin{equation}
K(E,E')=K_{p}(E,E')+K_{c}(E,E')
\end{equation}
Assuming $V_{c}({\bf r-r'}\varepsilon_{n}-\varepsilon_{m})=v({\bf r-r'})$,\
i.e.\ static
approximation for Coulomb repulsion,\ we obtain:
\begin{equation}
K_{c}(E,E')=\int d{\bf r}\int d{\bf r'} v({\bf r-r'})
\ll\rho_{E}({\bf r})\rho_{E'}({\bf r'})\gg^{F}
\end{equation}
which coincides with Eq. (\ref{Kcsd}) used above in our
analysis of Coulomb repulsion within Cooper pairs in disordered systems.\ Above
we have used the approximation of Eq. (\ref{Kph}) to model $K_{p}$ due to
electron---phonon pairing mechanism (or similar model for some kind of
excitonic pairing).\ In this case Eq. (B16) reduces to
Eq. (\ref{GapDS}).

Note that $V_{c}({\bf r-r'}\varepsilon_{n}-\varepsilon_{m})$ may be taken also
as dynamically screened Coulomb interaction.\ Then we must use the appropriate
expressions for dielectric function $\epsilon({\bf q}\omega_{m})$ which may be
found using
the self---consistent theory of localization \cite{KazSad83,KazSad84}.\ Then
after some tedious calculations we can get the expressions for $K_{c}(E,E')$
which for small $|E-E'|$ practically coincide with those
used by us above for the case of static short---range interaction
\cite{DolgSad}.

\newpage

\section{Localization and $D$-wave Pairing}

There is a growing body of experimental evidence in high-$T_{c}$
superconductors that indicate that the pairing state is of $d_{x^2-y^{2\text{
}}}$ symmetry \cite{dexp,dexpr}. In superconductors with an anisotropic order
parameter, both magnetic and non-magnetic impurities are pair breaking.For
$d$--wave symmetry, the effect of
non-magnetic impurities is equivalent to magnetic impurities in $s$-wave
superconductors\cite{kalugin,radtke}.\ Effectively this means that
superconductivity
in such systems cannot persist until disorder becomes high enough to transform
the system into Anderson insulator.\
The situation is different for the so called extended $s$--wave
symmetry. This corresponds to an order parameter with uniform sign
 which could, in particular, vanish at certain  directions in momentum
space\cite{norman}.
Point impurities are not pair--breaking in this case, but they are
``pair--weakening'': for small impurity concentration
$T_{c}$ decreases linearly with disorder,
but the critical impurity concentration is formally infinite,\ i.e.\
Anderson's theorem works after essential isotropisation of the
gap\cite{markad}.

We shall present now some of thr relevant equations along the lines of our
discussion of the Anderson theorem in the main body of the review.\ Here we
partly follow Ref.\cite{Bals}.
We shall consider $d$-wave pairing on two-dimensional lattice induced by
the following interaction Hamiltonian:
\begin{equation}
\label{H1}
H_{int}=-g\sum_{{\bf r}}\hat{\Delta }_{{\bf r}}^{\dagger }\hat
{\Delta }_{{\bf r}}
\end{equation}
where ${\bf r}$ denotes lattice sites.
This Hamiltonian corresponds to an instantaneous anisotropic
attractive interaction with an implicit cutoff at a characteristic energy
$<\omega >$. In order to model $d_{x^2-y^{2\text{ }}}$ symmetry we choose
$\hat{\Delta }_{{\bf r}}^{\dagger }$ in the following form:
\begin{equation}
\label{del}\hat{\Delta }_{{\bf r}}^{\dagger }=\frac 1{\sqrt{2}}\sum_\delta
\epsilon _{{\bf\delta}} \left( c_{{\bf r}\uparrow }^{\dagger }c_{{\bf
r+\delta} \downarrow }^{\dagger }-c_{{\bf r}\downarrow }^{\dagger
}c_{{\bf r+\delta} \uparrow }^{\dagger }\right)
\end{equation}
with $\delta
=\pm {\bf e}_1,{\bf \pm e}_2$ being the lattice versors, and $\epsilon_{\pm
{\bf e}_1}=-\epsilon _{\pm {\bf e}_2}=1$.

Next we can perform the analysis similar to that used in deriving Eqs.
(\ref{lingap})---(\ref{TC}) and find that now we again have Eq.(\ref{TC})
determining the critical temperature $T_{c}$ with the kernel $K({\bf r
r'}\varepsilon_{n})$ in the exact eigenstates representation
taking the following form:

\begin{equation}
\label{ker2}K({\bf r r'}\varepsilon_{n})=gT<\sum_{\mu
\nu \delta \delta _{}^{\prime }}\epsilon _\delta \epsilon _{\delta ^{\prime
}}\frac{\phi _\mu ^{*}({\bf r})\phi _\nu ^{*}({\bf r}+\delta )\phi _\nu
({\bf r}^{\prime })\phi _\mu ({\bf r}^{\prime }+\delta^{\prime} )}{(\varepsilon _\nu
-i\varepsilon_n)(\varepsilon _\mu +i\varepsilon_n)}> =
\end{equation}
$$=\int_{-\infty}^{\infty} dE N(E)\int_{-\infty}^{\infty}
d\omega \frac{\ll d_{E}({\bf r})d_{E+\omega}({\bf
r'})\gg}{(i\varepsilon_{n}+E)(E+\omega-i\varepsilon_{n})}$$
where we have introduced the spectral density:
\begin{equation}
\ll d_{E}({\bf r}) d_{E+\omega}({\bf r'})\gg=
\frac{1}{N(E)}<\sum_{\mu\nu}\sum_{\delta\delta'}
\epsilon_{\delta}\epsilon_{\delta'}
\phi^{\star}_{\nu}({\bf r+\delta})\phi_{\mu}({\bf r})
\phi^{\star}_{\mu}({\bf r'})
\phi_{\nu}({\bf r^{\prime}+\delta^{\prime}})\delta(E-\varepsilon_{\nu})
\delta(E+\omega-\varepsilon_{\nu'})>
\label{dsd}
\end{equation}
Now we can rewrite Eq.(\ref{TC}) for $T_{c}$ as:
\begin{equation}
1=gT_{c}\int_{-\infty}^{\infty}dE N(E)\int_{-\infty}^{\infty}d\omega
\sum_{\varepsilon_{n}}\frac{g({\omega})}{(E+i\varepsilon_{n})(E+\omega-
i\varepsilon_{n})}
\label{TCd}
\end{equation}
where
\begin{equation}
g(\omega)=\int {\bf dr'}\ll d_{E}({\bf r})d_{E+\omega}({\bf r'})\gg=
\ll d_{E}d_{E+\omega}\gg _{{\bf q=0}}
\label{gd}
\end{equation}
No sum rules similar to that given by Eqs.(A5)-(A6) exist for the spectral
density of Eq.(\ref{dsd}). However, it can be easily expressed via the Green's
functions and we obtain the following relations similar to those obtained in
Appendix A:
\begin{equation}
\ll d_{E}d_{E+\omega}\gg _{{\bf q}}=\frac{1}{\pi N(E)}Im\left\{
\Phi^{RA}_{Ed}(\omega{\bf q})-\Phi^{RR}_{Ed}(\omega{\bf q})\right\}
\label{dddd}
\end{equation}
where
\begin{equation}
\Phi^{RA(R)}_{Ed}({\bf q}\omega)=-\frac{1}{2\pi i}\sum_{{\bf pp'}}
\gamma_{{\bf p}}^{d}
<G^{R}({\bf p_{+}p'_{+}}E+\omega)G^{A(R)}({\bf p'_{-}p_{-}}E)>\gamma_{{\bf
p'}}^{d}
\label{Phidanis}
\end{equation}
with the vertices $\gamma _{{\bf p}}^d=\cos p_x-\cos p_y$ for $d$--wave.
If from now on we ignore the
lattice effects then $\gamma _{{\bf p}}^d=\cos 2\theta _{{\bf p}}$,
which corresponds to a gap function $\Delta({\bf k})=\Delta(T)
\cos 2\theta_{{\bf p}}$,\ where $\theta_{{\bf p}}$ is the polar
angle in the plane\cite{norman}.\ Similar expressions will determine $T_{c}$
for the case of anisotropic $s$-wave pairing with the vertices
$\gamma_{{\bf p}}^{d}$ replaced by appropriate angle-dependent expressions
\cite{norman}.

Now we can write as usual:
\begin{equation}
g(\omega)=\frac{1}{\pi N(E)}Im\left\{\Phi^{RA}_{Ed}(\omega{\bf q}=0)\right\}=
\label{gomeg}
\end{equation}
$$=\frac{1}{\pi N(E)}Im\left\{-\frac{1}{2\pi i}\sum_{{\bf pp'}}
\cos 2\theta_{{\bf p}}
\Phi_{{\bf p p'}}^{RA}(E\omega{\bf q=}0)\cos 2\theta_{{\bf p'}}\right\} $$
Here $\Phi_{{\bf p p'}}^{RA}(E\omega{\bf q}=0)$ obeys the ${\bf q}=0$
limit of the Bethe-Salpeter equation Eq.(\ref{BS}) which is easily
transformed to the following kinetic equation\cite{VW80}:
\begin{equation}
\label{kin}(\omega -{\frac i\tau })\Phi^{RA}_{{\bf p}{\bf
p}^{\prime }}(E\omega)=-\Delta G_{{\bf p}}\left[ \delta ({\bf p}-{\bf
p}^{\prime })+\sum_{{\bf p}^{\prime \prime }}U^{E}_{{\bf p}{\bf p}^{\prime
\prime }}(\omega)\Phi^{RA}_{{\bf p}^{\prime \prime } {\bf p}^{\prime
}}(E\omega)\right] \
\end{equation}
with $\Delta G_{{\bf p}}\equiv G^{R}({\bf p}E+\omega)-G^{A}({\bf p}E)$.
If we replace in (\ref{kin}) the irreducible vertex by the bare vertex
$U_0=\rho V^2$, we obtain finally:
\begin{equation}
\label{gmetl}g(\omega )=\frac{1}{4\pi }\frac \tau {1+(\omega \tau )^2}\ .
\end{equation}
with the usual scattering rate $1/\tau = 2\pi\rho V^2 N(E)$.
Inserting (\ref{gmetl}) in (\ref{TCd}) and following the standard analysis
\cite{Genn} we obtain the well known expression for the critical temperature
variation\cite{radtke} $\ln (T_{c0}/T_{c})=\Psi (1/2+1/4\pi \tau T_c)-\Psi
(1/2)$ which is similar to the case of magnetic impurity scattering in
superconductors.\ However here the normal potential scattering rate is
operational leading to very fast degradation of $T_{c}$ ---
superconducting state is completely destroyed for $1/\tau>1.76T_{c0}$.
Actually this result does not depend on spatial dimensionality of the
system,\ i.e. the same dependence works in three dimensions.

Effectively this makes impossible to reach the Anderson transition before
superconductivity is destroyed:\ critical disorder for metal-insulator
transition is determined by $1/\tau\sim E_{F}\gg T_{c}$.\ The only hope
seems to be to analyze the quasi-two-dimensional case,\ where this critical
disorder can be reduced due to a small enough interplane transfer integral
$w$ as in Eqs.(\ref{wc})-(\ref{mobed2d}).\ Localization appears for
$w<w_{c}=\frac{\sqrt{2}}{\tau} exp(-\pi E_{F}\tau)$ and take as an estimate
some $1/\tau \approx T_{c0}$,\ so that superconductivity is still possible,\
we can arrive at the following criterion of coexistence of localization and
superconductivity:
\begin{equation}
w<T_{c0}exp(-\pi E_{F}/T_{c0})
\label{wcdwav}
\end{equation}
In typical situation even for high-temperature superconductors we have
$T_{c0}<0.1E_{F}$ and inequality in Eq.(\ref{wcdwav}) can be satisfied only
for extremely anisotropic systems with $w\ll T_{c0}$.\ Most known
superconductors apparently fail in this respect.\ This probably makes
$d$-wave pairing irrelevant for the main body of our review.\ It is then
quite difficult to reconcile the existing data on the closeness of e.g.\
radiationally disordered high-$T_{c}$ systems to the disorder induced
metal-insulator transition and all the evidence for $d$-wave pairing in
these systems.\ However,\ this reasoning does not apply to the case of
anisotropic $s$-wave pairing,\ where Anderson theorem
effectively works for large degrees of disorder\cite{norman}.\ In this
respect the experiments on disordering in high-$T_{c}$ systems can become
crucial in solving the problem of the nature of pairing (and thus of its
microscopic mechanisms) in these systems.

Still,\ even in the case of $d$-wave pairing localization effects may become
important and interesting,\ but for quite another problem --- that of
localization of BCS-quasiparticles within superconducting gap at relatively
small disorder\cite{Lee93,Balat94,Nerses94,Nerses94a}.\ It is
known that while in the pure $d$-wave superconductor density
of states close to the Fermi level is linear in energy $N(E)\sim E$ due to
the gap nodes at the Fermi surface,\ the impurity scattering makes it finite
at $E=0$\cite{kalugin}.\ In this sense the system becomes similar to the
{\it normal} metal and we can calculate\cite{Lee93} the low lying
quasiparticle contribution to conductivity $\sigma(\omega\rightarrow 0)$.\
This conductivity equals to:
\begin{equation}
\sigma\approx \frac{e^2}{2\pi\hbar}\frac{\xi_{0}}{a}
\label{Leecond}
\end{equation}
where $\xi_{0}=v_{F}/\pi\Delta_{0}$ is the superconducting coherence length
and $a$ is the lattice spacing (we assume $T=0$).\ The surprising thing is
that $\sigma$ {\it independent of the scattering rate} $1/\tau$,\ i.e.\ of
disorder.\ For two-dimensional case (applicable probably for high-$T_{c}$
systems) we know that all states are localized with localization controlled
by dimensionless conductance which now is equal to
$g=\sigma/(e^2/2\pi\hbar)=\xi_{0}/a$.\ The value of $g$ may be small enough
in high temperature superconductors due to the small values of $\xi_{0}$,\
which are typically only slightly larger than the lattice constant.\
This can make localization effects important with BCS-quasiparticles forming
a mobility gap in the vicinity of the Fermi level,\ leading to
anomalies in the low temperature behavior of microwave conductivity and the
penetration depth of a $d$-wave superconductor\cite{Lee93}.\

These results were first obtained\cite{Lee93} for the point-like impurity
scattering,\ later it was shown in Ref.\cite{Balat94} that the finite range
of the impurity potential can lead to the nonuniversal disorder-dependent
behavior of conductivity which becomes proportional to the normal state
scattering rate.\
Situation was further complicated by the claim
made in Refs.\cite{Nerses94,Nerses94a} that the more rigorous analysis leads
to the density of states of the impure $d$-wave superconductor behaving as
$N(E)\sim |E|^{\alpha}$ with $\alpha > 0$,\ but dependent on the type of
disorder.\ The renormalization group for the conductivity then apparently
leads to some kind of fixed point of intermediate nature,\ suggesting the
finite conductivity in two-dimensions.
All these aspects of disorder and localization for $d$-wave superconductors
deserve further intensive studies.



\newpage

\begin{figure}
\caption{Electron wave---function in a disordered system:
(a) --- extended state.
(b) --- localized state.}
\label{fig1}
\end{figure}

\begin{figure}
\caption{Electron density of states near the band edge in a disordered system.\
Dashed is the
region of localized states,\ $E_{c}$---is the mobility edge.}
\label{fig2}
\end{figure}

\begin{figure}
\caption{Qualitative form of $\beta_{d}(g)$ for different $d$.\
Dashed line shows the behavior
necessary to get discontinuous drop of conductivity at the mobility edge for
$d=2$.}
\label{fig3}
\end{figure}

\begin{figure}
\caption{Graphical representation of:
(a) --- two---electron Green's function $\Phi^{RA}_{{\bf pp'}}
(E{\bf q}\omega)$;
(b) --- equation for full vertex part $\Gamma^{E}_{{\bf pp'}}({\bf q}\omega)$;
(c) --- typical diagrams for irreducible vertex $U^{E}_{{\bf pp'}}({\bf q};
\omega)$;
(d) --- Bethe---Salpeter equation.
Dashed line denotes ``interaction'' $U_{0}({\bf p-p'})=\rho |V({\bf
p-p'})|^2$,\
where $\rho$ --- is density of scatterers,\ $V({\bf p-p'})$---is Fourier
transform of a
single scatterer potential.}
\label{fig4}
\end{figure}

\begin{figure}
\caption{ ``Maximally---crossed'' diagrams for irreducible vertex part of
Bethe---Salpeter equation (``Cooperon'').}
\label{fig5}
\end{figure}

\begin{figure}
\caption{Two equivalent forms of diagram for the correlator of local density of
states.\ Wavy lines denote diffusion propagator,\ i.e.\ the sum of ladder
diagrams.}
\label{fig6}
\end{figure}

\begin{figure}
\caption{Lowest order interaction corrections:
(a) Simplest Fock correction for self---energy in exact eigenstate
representation.
(b) Equivalent diagram in momentum representation.
(c) ``Triangular'' vertex defining diffusion renormalization.\
$U$---irreducible impurity scattering vertex,\ $\Gamma$---full impurity
scattering vertex.\
Wavy line denotes interelectron interaction.}
\label{fig7}
\end{figure}

\begin{figure}
\caption {Lowest order interaction corrections to conductivity.}
\label{fig7a}
\end{figure}

\begin{figure}
\caption{Dependence of dimensionless generalized diffusion coefficient on
dimensionless Matsubara frequency in metallic phase ($\alpha =0.5$),
obtained by numerical solution for different values of $\mu$: {\bf 1.} 0.24;
{\bf 2.}
0.6; {\bf 3.} 0.95; Dashed line --- the usual self-consistent theory of
localization, $\mu =0$.
At the insert: Dependence of static diffusion coefficient
($d=\frac{D(0)}{D_{0}}$) on disorder for $\mu =0.24$.}
\label{fig7b}
\end{figure}

\begin{figure}
\caption{Dependence of dimensionless generalized diffusion coefficient on
dimensionless Matsubara frequency in dielectric phase ($\alpha =-0.5$),
obtained by numerical solution for different values of
$\mu$: {\bf 1.} 0.12; {\bf 2.} 0.6; {\bf 3.} 1.2;
Dashed line --- the usual self-consistent theory of localization, $\mu =0$. }
\label{fig7c}
\end{figure}

\begin{figure}
\caption{Electron---phonon interaction and impurity scattering:
(a) Self---energy due to impurity scattering.
(b) Diagrams representing changes of (a) due to impurity vibrations.
(c) Diagrams for ``bare'' electron---phonon vertex
in case of vibrating impurities.}
\label{fig8}
\end{figure}

\begin{figure}
\caption{Electron---phonon vertex renormalization:
(a) Impurity ``ladder'' (diffusion) renormalization.
(b),\ (c),\ (d) Simplest corrections due to impurity vibrations.}
\label{fig9}
\end{figure}

\begin{figure}
\caption{Graphic representation of two---particle Green's functions
$\Psi_{E}({\bf q}\omega_{m})$ and $\Phi_{E}({\bf q}\omega_{m})$
(for $\omega_{m}=2\varepsilon_{n}$).\ There is no summation over
$\varepsilon_{n}$ in the loops.}
\label{fig10}
\end{figure}

\begin{figure}
\caption{Temperature dependence of the upper critical
field $H_{c2}$.
Numerical solution
for the dependence of
$h=\omega_{H}/T_{c}^{2/3}E^{1/3}$ on $T/T_{c}$ for
different values of $\theta=\omega_{c}/T_{c}$:\
1.\ $\theta=100$;\ 2.\ $\theta=10$;\ 3.\ $\theta=2\pi$;\
4.\ $\theta=3$;\ 5.\ $\theta=1$;\ 6.\ $\theta=0$ (Mobility edge).\
Metallic state,\ no magnetic field influence on diffusion.\
At the insert:\
Low temperature part of $h$ on $T/T_{c}$ close to the
Anderson transition.
Mobility edge ($\theta=0$) with magnetic field influence on diffusion.
Metallic phase ($\theta=0.1$),\ no magnetic field influence.
Mobility edge ($\theta=0$),\ no magnetic field influence.
Insulating phase ($\theta=0.1$),\ no magnetic field
influence.
Numerical cut---off was taken at $<~\omega~>~=~100~T_{c}$.}
\label{fig11}
\end{figure}

\begin{figure}
\caption{Temperature dependence of the upper critical field for two-dimensional
superconductor $(\frac{e^{-1/\lambda}}{\tau T_{c}}=0.4, \lambda=0.1,
h=\frac{\omega_{H}}{\pi\lambda T_{c}})$.\ $1$---no magnetic field influence
upon diffusion,\ $2$---with magnetic field influence upon diffusion,\
$3$---standard theory of "dirty" superconductors.}
\label{fig12a}
\end{figure}

\begin{figure}
\caption{Temperature dependence of the upper critical field for two-dimensional
superconductor $(\frac{e^{-1/\lambda}}{\tau T_{c}}=4, \lambda=0.126,
h=\frac{\omega_{H}}{\pi\lambda T_{c}})$.\ $1$---no magnetic field influence
upon diffusion,\ $2$---with magnetic field influence upon diffusion.}
\label{fig12b}
\end{figure}

\begin{figure}
\caption{Temperature dependence of the upper critical field for
quasi-two-dimensional
superconductor $(\frac{e^{-1/\lambda}}{\tau T_{c}}=4, \lambda=0.126,
h=\frac{\omega_{H}}{\pi\lambda T_{c}})$ for different values of the
interplane transfer integral around the critical value of $w_{c}$
corresponding to Anderson transition at a given disorder.\
$1$---purely two-dimensional behavior ($w=0$), $2$---dielectric side
close to Anderson transition ($L=|2ln(w/w_{c})|=0.7$),\ $3$---metallic side
close to Anderson transition ($L=2ln(w/w_{c})=0.7$),\ $4$---metallic state far
from
Anderson transition ($L=3$).\ Dashed line represents the behavior at the
Anderson transition ($L=0$).}
\label{fig12c}
\end{figure}

\begin{figure}
\caption{Diagrams for fluctuation conductivity.\ Wavy lines denote
fluctuation propagator,\ dashed lines---disorder scattering.}
\label{fig12}
\end{figure}

\begin{figure}
\caption{Qualitative form of instanton solution.}
\label{fig13}
\end{figure}

\begin{figure}
\caption{Qualitative structure of eigenvalues of $M_{L}$ (a) and $M_{T}$ (b)
operators.
$\varepsilon_{1}^{L}=0$ --- translation zero---mode
$\varepsilon_{0}^{T}\rightarrow 0$ for $\lambda\rightarrow 0$---transforms to
``rotation'' zero---mode.
The continuous part of the spectrum is shaded.}
\label{fig14}
\end{figure}

\begin{figure}
\caption{Fluence dependence of $T_{c}$ and $|dH_{c2}/dT|_{T_{c}}$ in
$SnMo_{5}S_{6}$.}
\label{fig15}
\end{figure}

\begin{figure}
\caption{Resistivity dependence of $T_{c}$ and $|dH_{c2}/dT|_{T_{c}}$ in
$Mo_{6}Se_{8}$.}
\label{fig16}
\end{figure}

\begin{figure}
\caption{Conductivity $\sigma$ and $T_{c}$ dependence on
the parameter $p_{F}l/\hbar$ in amorphous $InO_{x}$.\
$\sigma_{B}$ is estimated Drude conductivity.}
\label{fig17}
\end{figure}

\begin{figure}
\caption{$H_{c2}(T)$ in amorphous films of $In/InO_{x}$.\
Lines show standard theoretical dependences.}
\label{fig18}
\end{figure}

\begin{figure}
\caption{The dependence of activation energy of hopping conductivity
(triangles) and superconducting transition temperature $T_{c}$ (squares)
in amorphous films of $In/InO_{x}$
on disorder parameter $p_{F}l/\hbar$ as determined from room-temperature
conductivity and Hall measurements.Long-dashed line represents
$\Delta=1.76T_{c}$ following the BCS gap formula.The short-dashed line
best fits the insulating data points with $(p_{F}l/\hbar)_{c}\approx 0.35$
---the critical disorder of metal-insulator transition. A narrow region of
superconductivity within insulating phase can be inferred from these data.}
\label{fig18a}
\end{figure}

\begin{figure}
\caption{Disorder dependence of localization length (full curve) and
superconducting coherence length in amorphous $In/InO_{x}$ films.
Squares represent superconducting $\xi$ for metallic films while triangles
refer to insulating samples.}
\label{fig18b}
\end{figure}

\begin{figure}
\caption{Conductivity $\sigma$ and $T_{c}$ dependence on
gold concentration in amorphous $Si_{1-x}Au_{x}$ alloy.}
\label{fig19}
\end{figure}

\begin{figure}
\caption{$H_{c2}(T)$ in amorphous $Si_{1-x}Au_{x}$ alloy.}
\label{fig20}
\end{figure}

\begin{figure}
\caption{Temperature dependences of superconducting energy
gap $\Delta$ and of the resistance $R$ for amorphous
$Si_{0.79}Au_{0.21}$.}
\label{fig21}
\end{figure}

\begin{figure}
\caption{Temperature dependence of $\rho_{c}$ for different
high---$T_{c}$ cuprates.\ The dashed region
indicates the resistivity range corresponding to Ioffe---Regel limit.
}
\label{fig22}
\end{figure}

\begin{figure}
\caption{Dependence of the superconducting transition
temperature and resistivity (at $T=100K$) on neutron
fluence for ceramic $YBa_{2}Cu_{3}O_{6.95}$.\ Different notations correspond
to different methods of measurement and also evolution after annealing at
$300K$.}
\label{fig23}
\end{figure}

\begin{figure}
\caption{Temperature dependence of resistivity $\rho$ for
ceramic samples of $YBa_{2}Cu_{3}O_{6.95}$ (curves 1---3
and 5---8) and $La_{1.83}Sr_{0.17}CuO_{4}$ (curves 4,\ 9)
irradiated at $T=80K$ with different fluences:\ 1---$\Phi=0$;\
3,\ 6,\ 8 --- $\Phi=2.5 \mbox{ and } 7 10^{18}cm^{-2}$ plus annealing for
2 hours at $300K$;\ 2,\ 5,\ 7 --- irradiated with
$\Phi=2.5 \mbox{ and } 7\ 10^{18}cm^{-2}$ plus annealing for
2 weeks at $300K$;\ 4 --- $\Phi=0$;\ 9 --- $\Phi=5\ 10^{18}cm^{-2}$ plus
annealing for 2 hours at $300K$.}
\label{fig24}
\end{figure}

\begin{figure}
\caption{Dependence of $ln\rho$ on $T^{-1/4}$ for
$YBa_{2}Cu_{3}O_{6.95}$ irradiated with a fluence of
$\Phi=1.2\ 10^{19}cm^{-2}$ at $T=80K$ (curve 1),\ and
after 20-minute annealing at $T=150K(2);200K(3);250K(4);
300K(5)$ and two weeks annealing at $T=300K(7)$.\ Similar
dependences for  $La_{1.83}Sr_{0.17}CuO_{4}$ for $\Phi=2\ 10^{19}cm^{-2}$
annealed for 2 hours at $300K (6)$ and
for $La_{2}CuO_{4}$ for $\Phi=2\ 10^{19}cm^{-2}$ annealed for 2 hours at
$300K (8)$.}
\label{fig25}
\end{figure}

\begin{figure}
\caption{Dependence of $ln\rho$ on fluence $\Phi$ during
irradiation at $T=80K$:\ 1---$La_{2}CuO_{4}$;\ 2---$YBa_{2}Cu_{3}O_{6.95}$;\
3---single crystaline $\rho_{ab}$ in $YBa_{2}Cu_{3}O_{6.95}$;\
4---$La_{1.83}Sr_{0.17}CuO_{4}$;\ 5---$Bi-Sr-Ca-Cu-O$;\ 6---
$SnMo_{6}Se_{8}$.}
\label{fig26}
\end{figure}

\begin{figure}
\caption{Temperature dependence of Hall concentration for
the irradiated (left) and oxygen deficient (right) ceramic samples of
$YBa_{2}Cu_{3}O_{7-\delta}$.}
\label{fig27}
\end{figure}

\begin{figure}
\caption{Fluence dependence of $\rho_{ab}$ and $\rho_{c}$ at $T=80K$ during
fast neutron irradiation.}
\label{fig28}
\end{figure}

\begin{figure}
\caption{Temperature dependence of $H^{\|}_{c2}$ (upper curves) and
$H^{\bot}_{c2}$ (lower curves) for the
single-crystals of $YBa_{2}Cu_{3}O_{7-\delta}$ with
different degrees of disorder.}
\label{fig29}
\end{figure}

\begin{figure}
\caption{The dependence of coherence lengths determined
from $H_{c2}$ behavior under disordering on the critical
temperature $T_{c}$:\ $\xi_{\|}$---circles;\ $\xi_{\bot}$---black circles.}
\label{fig30}
\end{figure}

\begin{figure}
\caption{Temperature dependence of the upper critical field:\
theoretical curve (1) is given for the case of
$\frac{e^{-1/\lambda}}{T_{c}\tau}=2$,\ $\lambda=0.18$,\ while curve (2)
is for $\frac{e^{-1/\lambda}}{T_{c}\tau}=20$,\ $\lambda=0.032$.\
Squares represent the experimental data for $Bi_{2}Sr_{2}CuO_{y}$.}
\label{fig30a}
\end{figure}

\begin{figure}
\caption{Dependence of the Curie constant $C$ and the temperature-independent
part $\chi_{0}$
of magnetic susceptibility on neutron fluence $\Phi$ for
$La_{1.83}Sr_{0.17}CuO_{4}$ (black circles) and $YBa_{2}Cu_{3}O_{6.95}$
(circles).}
\label{fig31}
\end{figure}

\begin{figure}
\caption{Dependence of $T_{c}$ on fluence for $YBa_{2}Cu_{3}O_{6.95}$
(circles).\ The solid curve is the localization length calculated from hopping
conductivity.
Dashed
curve defines the minimum localization length at which superconductivity can
exist at given $T_{c}$.
Dashed-dotted curve
is theoretical fit using expressions described in the text.
}
\label{fig32}
\end{figure}

\newpage
\mediumtext
\begin{table}
\caption{Anderson transition and ferromagnet close to Curie point $T_{c}$. }
\label{Table 1}
\begin{tabular}{cc}
\multicolumn{1}{c}{Localization}&\multicolumn{1}{c}{Ferromagnet} \\
\tableline
$ E-E_{c} $  & $T-T_{c}$  \\
$ K_{F}  $   & $\chi_{\bot}$ \\
$ K_{H} $    & $\chi_{\|}$  \\
$-i\omega$   & $ H $  \\
$ N(E) $     & $ M $  \\
$ D_{E} $    & $\rho_{s}$  \\
$ \xi_{loc}$ & $ \xi $ \\
\end{tabular}
\end{table}
\begin{table}
\caption{Scaling dimensions in the theory of critical phenomena}\label{Table 2}
\begin{tabular}{cccc}
$\xi$  &  $q$  &  $M$  &  $H$ \\
\tableline
$-1$   &  $+1$ & $1/2(d-2+\eta)$  &  $1/2(d+2-\eta)$  \\
\end{tabular}
\end{table}

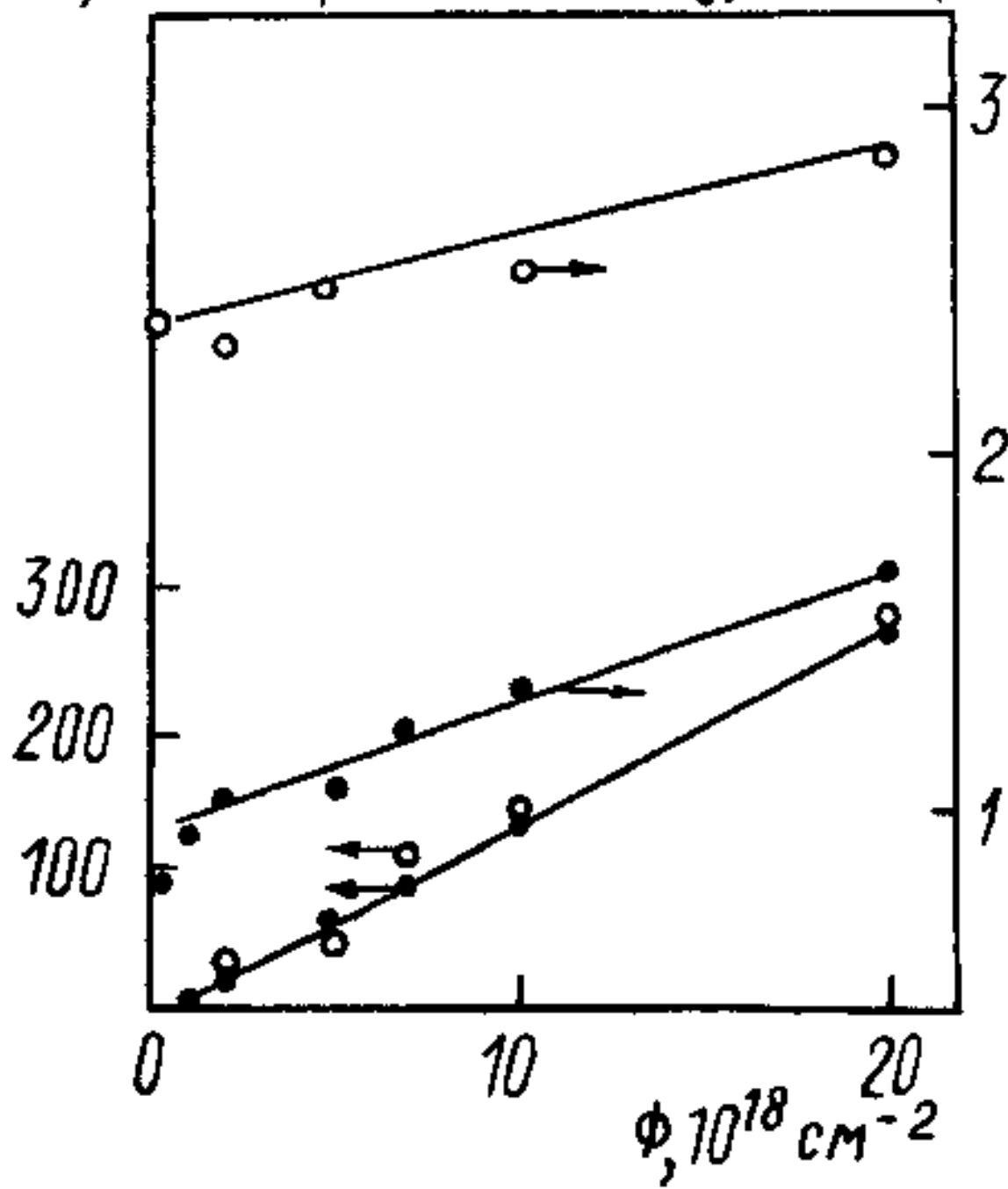
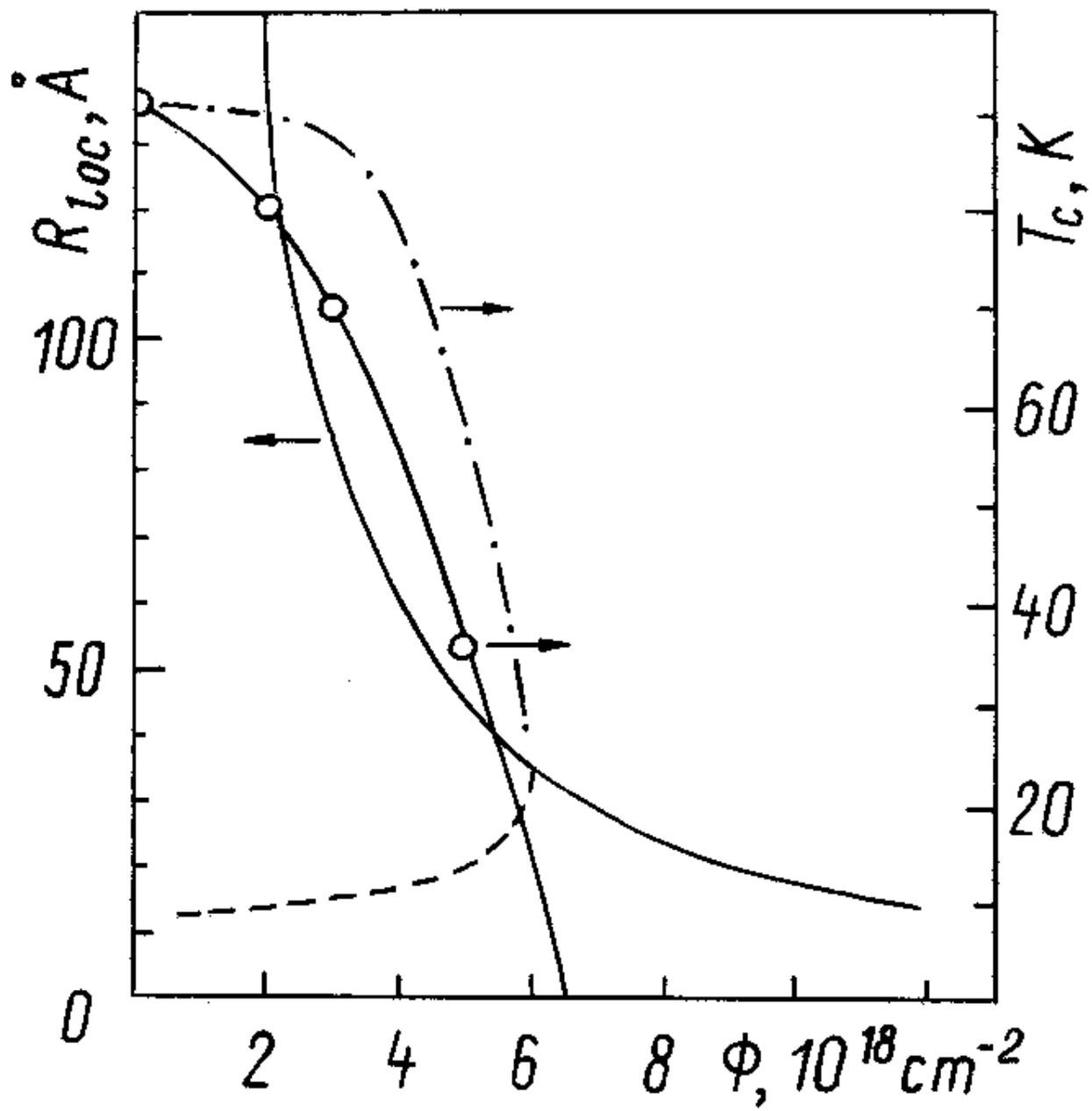
\end{document}